%% file: ssa-ppnp.tex
\documentclass[twoside,12pt]{article}
\usepackage{epsfig}

\usepackage{bm}
\usepackage{graphicx,color,here}
\usepackage{verbatim}

\newcommand{\be}{\begin{equation}}
\newcommand{\ee}{\end{equation}}
\newcommand{\bea}{\begin{eqnarray}}
\newcommand{\eea}{\end{eqnarray}}

\newcommand{\bfk}{\bm{k}}

\newcommand{\bfp}{\bm{p}}

\newcommand{\bfl}{\bm{\ell}}

\newcommand{\kt}{k_\perp}
\newcommand{\pt}{p_\perp}
\def\nostrocostruttino#1\over#2{\mathrel{\mathop{\kern 0pt \rlap
{\hbox{$#1$}}} \hbox{\kern-.125em $#2$}}}
\def\sumint{\nostrocostruttino \sum \over {\displaystyle\int}}
\begin{document}

\title{ \vspace{1cm} Azimuthal and Single Spin Asymmetries \\
in Hard  Scattering Processes}
\author{U.\ D'Alesio, F.\ Murgia \\
\\
INFN, Sezione di Cagliari and Dipartimento di Fisica,
Universit\`a di Cagliari,\\
C.P. 170, I-09042 Monserrato (CA), Italy\\
}

\maketitle
\begin{abstract}
In this article we review the present understanding of azimuthal and single spin asymmetries
for inclusive and semi-inclusive particle production in unpolarized and polarized hadronic
collisions at high energy and moderately large transverse momentum. After summarizing the
experimental information available, we discuss and compare the main theoretical approaches
formulated in the framework of perturbative QCD. We then present in some detail a
generalization of the parton model with inclusion of spin and intrinsic transverse momentum
effects. In this context, we extensively discuss the phenomenology of azimuthal and single
spin asymmetries for several processes in different kinematical configurations. A comparison
with the predictions of other approaches, when available, is also given. We finally emphasize
some relevant open points and challenges for future theoretical and experimental
investigation.
\end{abstract}
\eject
\tableofcontents
\input{introduction-rev}
\input{exp-rev}
\input{theo-appr-rev}
\input{formalism-rev}
\input{pheno-rev}
\input{conclusions-rev}

\bibliography{spiresppnp}    

\bibliographystyle{myphys}

\end{document}

%% file: introduction-rev.tex
\section{Introduction}
\label{intro}

Polarization phenomena in hadronic processes deeply challenge high-energy physics theories
and perturbative Quantum Chromodynamics (pQCD). It has been known since a long time that
experimental results for several polarized processes can hardly be explained by means of
available theoretical approaches. Some relevant examples are: $i)$ The
large spin-spin correlation
%polarization transfer
measured in $pp$ elastic scattering and its unusual behaviour as a function of the relevant
kinematical variables; $ii)$ The so-called ``spin crisis", and the fact that quarks only
account for a small fraction of the total proton spin; $iii)$ The large transverse
polarization of hyperons produced in unpolarized hadronic collisions;  $iv)$ The large
transverse single spin asymmetries (SSA) measured in inclusive pion and kaon production in
polarized hadronic collisions and in semi-inclusive deeply inelastic scattering (SIDIS). Due
to space limitation, in this work we will not cover at all the topics $i)$ and $ii)$. The
rich, low-energy aspects of topics $iii)$ and $iv)$, which may be confronted
only with nonperturbative approaches, %and models,
will also remain uncovered. A comprehensive overview on these subjects and on related topics
can be found in several review papers, addressing both their theoretical and experimental
aspects~\cite{Bourrely:1980mr,Pondrom:1985aw,Anselmino:1994gn,Troshin:1995si,
Stiegler:1991qi,Heller:1996pg,Nurushev:1997sx,Felix:1999tf,Liang:2000gz,
Lampe:1998eu,Bunce:2000uv,Barone:2001sp,Leader:2001gr,Filippone:2001ux,
Barone:2003fy,Bass:2004xa,Vogelsang:2007zz}. Here we will focus on the high-energy regime and
the treatment in the context of pQCD approaches of transverse SSA's and, marginally, of the
transverse polarization of spin 1/2 particles in unpolarized processes.

The SSA for the inclusive process $A^\uparrow B\to C+X$ is defined as
 \be \label{intro:an}
A_N = \frac{d\sigma^\uparrow-d\sigma^\downarrow}{d\sigma^\uparrow+d\sigma^\downarrow} =
\frac{d\Delta\sigma}{2d\sigma^{\rm unp}}\,,
 \ee
where $d\sigma^{\uparrow(\downarrow)}$ stands for the invariant differential cross section,
$E_Cd^3\sigma^{\uparrow(\downarrow)}/d^3\bm{p}_C$, for the production of hadron $C$, with
4-momentum $p_{C}^{\mu}=(E_{C},\bm{p}_{T},p_{L})$, in the scattering of a hadron $A$ off an
unpolarized hadron $B$, with $A$ upwards (downwards) transversely polarized w.r.t.~the
production plane. More precisely, in the $AB$ center of mass frame, with $A$ moving along the
$+Z$ direction and $C$ produced in the $XZ$ plane, the upwards (downwards) transverse spin
directions are respectively along $\pm\,Y$. The SSA, or analyzing power, $A_N$, is also
denoted as ``left-right'' asymmetry since, by rotational invariance, the number of left going
hadrons $C$  when $A$ is downwards polarized, $d\sigma^\downarrow$, is the same as the number
of right going hadrons $C$ with $A$ upwards polarized. Notice that left (right) is defined
w.r.t.~the plane spanned by the momentum and spin directions of the polarized hadron $A$
looking downstream its direction of motion.

The transverse polarization measured in the $AB\to C^\uparrow + X$ process, where $C$ is a
spin 1/2 hadron, is defined, analogously to $A_N$ in
Eq.~(\ref{intro:an}),
as $P_T= (d\sigma^\uparrow - d\sigma^\downarrow)/d\sigma^{\rm unp}$,
where now the
arrows in $d\sigma^{\uparrow(\downarrow)}$ stand for the transverse spin directions for the
final hadron $C$.

Most of these processes show peculiar and unexpected dependences on the relevant kinematical
variables: the rapidity, $y=(1/2)\ln[(E_C+p_L)/(E_C-p_L)]$, or the pseudorapidity
$\eta=-(1/2)\ln\tan(\theta/2)$, the Feynman variable, $x_F=p^{}_L/p_L^{\,\rm{max}}$, and
$p_T$, where $\theta$ and $p_L$ ($p_T$) are respectively the production angle and the
longitudinal (transverse) momentum of the observed particle w.r.t.~the direction of the
colliding beams. The magnitude and behaviour of these polarized observables are very
difficult to account for in pQCD approaches, as well as in nonperturbative models, and have
motivated a huge amount of theoretical work. Although significant progress has been done, as
it will be illustrated in the rest of this review, much remains to be clarified. Moreover, a
new family of second-generation experiments are starting to produce new sets of data with
larger statistics, at higher energies and with an enlarged coverage of the kinematical
variables. For these reasons, we will try to collect the main ideas on SSA's and
give an account of their present status, %of their understanding,
both from the experimental and, in more depth, the theoretical side.

The study of SSA's is strongly related to our understanding of the structure of hadrons and
their spin and orbital angular momentum content in terms of partons. This leads naturally to
a special attention to the role played by parton distribution and fragmentation functions in
pQCD approaches. In this review we will then present, in some detail, a new class of parton
distributions, beyond the usual collinear ones, resulting from the inclusion of spin and
intrinsic transverse momentum effects. We will also show how they can help in describing the
observed SSA results. In particular, two spin and transverse momentum dependent (TMD)
mechanisms, introduced in the early 90s, will play a special role: the Sivers effect,
describing the asymmetry in the azimuthal distribution of an unpolarized parton inside a
transversely polarized hadron~\cite{Sivers:1989cc,Sivers:1990fh}; the Collins effect,
describing the azimuthal asymmetry in the fragmentation of a transversely polarized quark
into an unpolarized hadron~\cite{Collins:1992kk,Collins:1993kq}.

Among the topics not covered in this review, that are of some relevance in the context of
SSA's, we mention the generalized parton distributions (GPD) and the dihadron fragmentation
functions (DiFF). GPD's represent another generalization of ordinary parton distributions and
indeed could provide a unified parameterization of many different aspects of hadron physics.
We refer the interested reader to the following papers
\cite{Mueller:1998fv,Ji:1996ek,Ji:1996nm,Radyushkin:1997ki,Ji:1998pc,
Diehl:2003ny,Belitsky:2005qn,Diehl:2006js}, which emphasize the different aspects of the
physics encoded in these quantities. We also mention their possible connection to the TMD
distributions, discussed in Refs.~\cite{Burkardt:2002hr,Burkardt:2002ks,
Burkardt:2003je,Burkardt:2005hp,Diehl:2005jf,Meissner:2007rx}. Concerning DiFF's (describing
the probability that a parton hadronizes into two hadrons plus anything else) we recall their
role in spin studies, in particular for the extraction of the transversity distribution
function in double inclusive processes in $pp$ collisions and
SIDIS~\cite{vanderNat:2005yf,Schill:2007ku},  and the azimuthal correlations in $e^+e^-$
annihilations. Some seminal papers on the DiFF's, with their definitions and properties,
are~\cite{Efremov:1992pe,Collins:1993kq,Collins:1994ax,Artru:1995zu,Jaffe:1997hf,
Bianconi:1999cd,Bacchetta:2003vn,Ceccopieri:2007ip}.

The outline of this review is the following: in section~\ref{exp} we present an overview of
experimental results on SSA's, including also the azimuthal asymmetries observed in
unpolarized collisions, that are intimately connected to our main subject. In
section~\ref{theo} we discuss QCD theoretical approaches to SSA's, which represent nowadays
the most reliable attempts towards the understanding of the observed results. Their most
recent developments are also presented. We will discuss in particular two pQCD approaches:
one based on the extension of the collinear factorization theorems, at leading twist, with
the inclusion of intrinsic transverse momentum effects; another one based on a twist-three
collinear formalism. As a representative example, a short account of a semiclassical model
will be also given. In section~\ref{form} we will give a detailed description of the first
pQCD approach within the helicity formalism. Its phenomenological implications and a
comparison with the results of other theoretical approaches are collected in
section~\ref{pheno}. Finally, perspectives and conclusions are gathered in
section~\ref{conclusion}.

This work is to a great extent based on our longstanding and fruitful collaboration with
M.~Anselmino, M.~Boglione, and E.~Leader; without their contribution it would have been
impossible. We are also grateful to A.~Kotzinian, S.~Melis, and A.~Prokudin for valuable
collaboration in the last years. Many people have contributed with enlightening discussions
and clarifying remarks to this work, before and during its completion; in particular it is a
pleasure to thank A.~Bacchetta, D.~Boer, C.~Bomhof, L.~Gamberg, Y.~Koike, A.~Metz,
P.~Mulders, F. Pijlman, P.~Schweitzer, W.~Vogelsang, F.~Yuan, for their theoretical insights;
H.~Avakian, L.~Bland, F.~Bradamante,
%Gerry Bounce,
M.~Grosse Perdekamp, D.~Hasch, S.~Heppelmann, J.~Lee, A.~Martin,
L.~Nogach, R.~Seidl, F.~Videbaek, for
their kind support on the experimental side and for providing us with preliminary results of
their experimental collaborations.

%% file: exp-rev.tex
\section{Experimental results on azimuthal and transverse single spin asymmetries}
\label{exp}

A comprehensive account of all experimental information available today on polarization
phenomena in high-energy hadronic physics is well beyond the scope of this review. Forced by
space limitation, in the following pages we will only summarize the most significant results,
limiting ourselves to those concerning the main subjects of this paper: transverse
single-spin and azimuthal asymmetries. The selection presented somehow reflects our personal
perception. We apologize for those papers and topics not included and discussed here. The
interested reader may fortunately find complete and comprehensive accounts on this subject in
Refs.~\cite{Bourrely:1980mr,Pondrom:1985aw,Liang:2000gz,Leader:2001gr}.

Whenever possible we will refer to published work. For ongoing experiments we will also quote
available preliminary results. We will not discuss details concerning experimental setups,
detector performances, etc., referring to the quoted papers for all technical aspects. In
order to avoid duplicates, we will not show plots of the results in this section, deferring
instead to section~\ref{pheno}, where a selection of relevant results, and their comparison
with theoretical predictions and fits, will be presented.
\subsection{Transverse single spin asymmetries for $p^{\uparrow} p\to h + X$ processes}
\label{exp:ssa}

Starting from the late 70s and during the 80s a number of experimental collaborations
presented early results on SSA's for the $p p\to h + X$ process.
For its impact on the later developments we mention,
for instance, the work on the transverse $\Lambda$ polarization
by Bunce {\it et al.}~\cite{Bunce:1976yb}.
SSA's significantly different from zero
were found for the $p^{\uparrow} p\to \pi + X$ process
in the forward production region (large positive
$x_F$), at relatively small transverse momentum, $p_T$. These results were in accord with
previous extensive measurements performed for similar exclusive channels (e.g., $\pi p\to \pi
p$) and were interpreted as purely nonperturbative effects. Additional measurements on the
$p_T$ dependence of these results showed interesting behaviours, difficult to understand and
explain. All these experiments were fixed-target ones, at relatively low c.m.~energies and
low $p_T$ (below 1.0-1.5 GeV/$c$). Therefore, no serious attempt was made to understand these
effects in terms of pQCD involving hard elementary scattering. Moreover, a
well-known paper by Kane, Pumplin and Repko~\cite{Kane:1978nd}, reinforced the common wisdom
that spin effects should become negligible at high-energy scales.

\subsubsection{E704-Fermilab results }
\label{exp:ssa:e704}

During the 90s, the E581/E704 Collaborations at Fermilab (henceforth indicated as the E704
Collaboration) set the stage at a different level, because of the larger c.m.~energy
available ($\sqrt{s}\simeq 20$ GeV), the improved separation between the $x_F$ and $p_T$
kinematical dependence of the SSA's and the small but significant extension of the $p_T$
range covered. The E704 Collaboration collected SSA data for $p p$ and $\bar{p} p$ collisions
using (for the first time) polarized secondary $p$, $\bar{p}$ beams and for several produced
particles (mainly pions, but also photons, $\Lambda$ and $\eta$ particles). The experiment
covered essentially two different kinematical regions: $0.2 \leq x_F \leq 0.6$  and $p_T$ in
the range $0.2\div2.0$ GeV$/c$ (the forward production, or beam fragmentation, region);
$|x_F|\leq0.15$ and $p_T$ values up to 4.0 GeV/$c$ (the central rapidity region).
 The main results gathered by the E704 Collaboration can be summarized as follows:\\
1) Large SSA's (up to 30-40\% in magnitude) have been measured for inclusive pion production
in the large positive $x_F$ region, for polarized proton and antiproton beams colliding with
an unpolarized hydrogen target~\cite{Adams:1991rw,Adams:1991cs,Adams:1991ru,Bravar:1996ki}.
For both neutral and charged pions the SSA was found to be almost zero at low $x_F$, up to
$x_F\sim0.3$, where it starts raising with increasing $x_F$, reaching values of $\sim15\%$
and $30-40\%$ in size for $\pi^0$'s and $\pi^{\pm}$'s respectively [see
Fig.~\ref{pheno:anmax:e704}, p.~\pageref{pheno:anmax:e704}]. The measured asymmetries are
roughly similar in magnitude but opposite in sign for positive and negative pions (and
positive in the $p^\uparrow p\to\pi^+ + X$ case); for charged pions, they change sign (being
somehow reduced in size) when going from polarized proton to polarized antiproton beams. The
asymmetry for neutral pions is always positive and almost half of that of the nearest charged
pion case, in good agreement with isospin conservation in the fragmentation process.
Information gathered on the $p_T$ dependence of the SSA's was less detailed. In
Ref.~\cite{Bravar:1996ki} the $p_T$ dependence of the SSA for charged pions in the
$\bar{p}^\uparrow p\to\pi^{\pm} + X$ process was measured at moderate $p_T$ ($0.2\leq
p_T\leq1.5$ GeV$/c$) over a wide $x_F$ range, $0.2\leq x_F\leq0.9$. A threshold for the onset
of the asymmetry was observed around $p_T\sim 0.5$ GeV$/c$, below which the SSA was
essentially zero and above which it increases in magnitude with $p_T$. In general, in all
data sets considered, the magnitude of the SSA at large positive $x_F$ was found to be more
pronounced for the largest $p_T$ values reached.
No data on the corresponding unpolarized cross sections were presented.\\
2) In the central rapidity region, $|x_F|\leq 0.15$ and $1.0<p_T<4.5$ GeV$/c$, the inclusive
and semi-inclusive SSA's for neutral pions, using transversely polarized 200 GeV$/c$ proton
and antiproton beams off a hydrogen target were measured~\cite{Adams:1994yu}. These
asymmetries were found to be consistent with zero within the experimental uncertainties over
the full $p_T$ range covered, both for polarized proton and antiproton beams [see, e.g.,
Fig.~\ref{pheno:an:midrap}~( left panel), p.~\pageref{pheno:an:midrap}]. In this case,
measurements of the corresponding invariant unpolarized cross sections were also provided.\\
3) The SSA for inclusive direct-photon production, $p^\uparrow p\to\gamma + X$, was measured
in the central rapidity region at $2.5<p_T<3.1$ GeV$/c$~\cite{Adams:1995gg}. Data on the
corresponding unpolarized cross section at $2.5<p_T<3.8$ GeV$/c$ were also provided. Only two
data points were measured for the SSA, that was found to be consistent with zero within
(large) experimental uncertainties. \\
4) Some interesting results are also available for SSA's in inclusive $\Lambda$ and
$\eta$-meson production~\cite{Bravar:1995fw,Adams:1997dp}, showing non-vanishing, sizable
asymmetries in kinematical regions similar to those considered above.

The E704 results triggered renewed interest on SSA's, given the larger c.m.~energy available
as compared to that of previous experiments and the large, unambiguous effects measured.
However, a clear separation between the $x_F$ and $p_T$ dependence of the results was not
reached and the covered $p_T$ range was still limited, although the largest SSA's were
measured at the highest $p_T$. For all the 90s these results were the only ones available at
sufficiently high c.m.~energies. The advent of new experimental setups started a new stage in
the investigation of transverse spin effects in hadronic processes.

Transverse SSA's were also measured, although again at lower total c.m.~energies, by
experimental collaborations at IHEP (Protvino). In Ref.~\cite{Apokin:1990ik} they have shown
the transverse SSA for $\pi^0$'s inclusively produced in the collisions of a 40 GeV$/c$
$\pi^-$ beam off transversely polarized protons and deuterons, in the region $|x_F|\leq 0.2$
and $1.6\leq p_T\leq 3.2$ GeV$/c$.  The SSA's were found to be approximately equal for the two
reactions, compatible with zero or slightly positive for $1.0\leq p_T \leq 2.0$ GeV$/c$,
large in magnitude and negative for $2.4\leq p_T\leq 3.2$ GeV$/c$. In
Ref.~\cite{Abramov:1996vp} SSA's were measured for inclusive hadron production in collisions
of a transversely polarized 40 GeV$/c$ proton beam with an unpolarized liquid hydrogen
target. The asymmetries were measured for $\pi^{\pm}$, $K^{\pm}$, protons and antiprotons,
produced in the central region, $0.02\leq x_F\leq 0.10$ and $0.7\leq p_T\leq 3.4$ GeV$/c$.
Within experimental errors, the measured asymmetries for $\pi^{\pm}$, $K^{\pm}$ and $\bar{p}$
show a linear dependence on $x_T=2p_T/\sqrt{s}$ and change sign near $x_T=0.37$. For protons,
a negative SSA, independent of $x_T$, was found.
\subsubsection{BNL-RHIC results}
\label{exp:rhic}

Starting from the end of the 90s, very interesting results on transverse SSA's and
unpolarized cross sections for inclusive single particle production in polarized hadronic
collisions came up and are currently collected by the experimental collaborations (STAR,
PHENIX, BRAHMS) using the Relativistic Heavy Ion Collider (RHIC) at the Brookhaven National
Laboratories (BNL).

For the first time proton-proton collisions, with one (both) beam(s) transversely
(longitudinally) polarized, can be studied at c.m.~energies as large as 200 GeV, with the
possibility to reach $\sqrt{s}=500$ GeV in the near future. This represents a significant
advancement with respect to all previous (fixed target) experiments. The setups of the three
RHIC experiments allow to cover a large kinematical range, including particularly the
previously almost unexplored negative $x_F$ region.

Let us also remind a precursor
experiment at the BNL Alternate Gradient Syncrotron (AGS). The
E925~Collaboration~\cite{Krueger:1998hz,Allgower:2002qi} measured transverse SSA's for charged
pions using an incident 22-GeV$/c$ transversely polarized proton beam on hydrogen and carbon
targets, in the range $0.55<x_F<0.75$ and $0.6 <p_{T}<1.2$ GeV/$c$. Although still at low
energies and $p_T$, this experiment gave a first, independent
confirmation of the large SSA's
measured by the E704~Collaboration.

We now give a summary of the main high-energy RHIC results on SSA's:\\
1) The STAR collaboration measured the transverse SSA for forward neutral pion production in
$p^\uparrow p$ collisions at $\sqrt{s}=200$ GeV, average pseudorapidity
$\langle\eta\rangle=3.8$ and $30 <E_{\pi}\simeq x_F\sqrt{s}/2<55$ GeV, corresponding to 1.0
$<\langle p_T\rangle<2.4$ GeV/$c$~\cite{Adams:2003fx}. The unpolarized cross section was also
measured at $\langle\eta\rangle=3.3$, 3.8, 4.0~\cite{Adams:2003fx,Adams:2006uz} and found to
be consistent with next-to-leading order (NLO) pQCD
calculations~\cite{Vogelsang:2007zz,deFlorian:2007aj}. In agreement with the E704 results,
the SSA is small for $x_F$ values below about 0.3 and becomes large and positive at higher
$x_F$. The spin-dependent azimuthal asymmetry for almost back-to-back dijets in $p^\uparrow
p$ collisions at $\sqrt{s}=200$ GeV was measured over a wide pseudorapidity range for the two
jets and found to be consistent with zero within experimental
uncertainties~\cite{Abelev:2007ii}.\\
Preliminary results for $\pi^0$ SSA's, based on a larger data sample, for both positive and
negative $x_F$ in the range $0.2<|x_F|<0.6$, collected at two average pseudorapidities,
$\langle\eta\rangle=3.3$ (corresponding to $1.6<\langle p_T\rangle<3.7$ GeV$/c$) and
$\langle\eta\rangle=3.7$ (corresponding to $1.3<\langle p_T\rangle<2.8$ GeV$/c$) have been
presented~\cite{Nogach:2006gm}. They essentially confirm the previous data published in the
positive $x_F$ region and show a SSA consistent with zero for negative $x_F$ [see
Fig.~\ref{pheno:an:xf:sivers}~(right panel), p.~\pageref{pheno:an:xf:sivers}]. A preliminary
study of the $p_T$ dependence of the neutral pion SSA for $x_F>0.4$ and in $x_F$-bins for
$0.25<x_F<0.56$, using combined data at $\langle\eta\rangle=3.3, 3.7, 4.0$, in the range
$1<p_T<4$ GeV$/c$ was also performed~\cite{Nogach:2006gm}. Data show a non trivial dependence
of the SSA on $p_T$, which seems not to behave as a smooth decreasing function of $p_T$ [see
Fig.~\ref{pheno:an:star:pt:sivers}, p.~\pageref{pheno:an:star:pt:sivers}]. Other interesting
STAR preliminary results concern the differential cross section for inclusive charged pion
production at mid-rapidity and for inclusive neutral pion production at forward and
mid-rapidity in $pp$ collisions at $\sqrt{s}=200$
GeV~\cite{Simon:2006xt}.\\
2) The PHENIX Collaboration has measured the unpolarized cross section and the transverse SSA
for neutral pions and charged hadrons produced at mid-rapidity ($|\eta|<0.35$) in $p^\uparrow
p$ collisions at $\sqrt{s}=200$ GeV~\cite{Adler:2003pb,Adler:2005in,Adare:2007:zzdd}. For
neutral pions (charged hadrons) the unpolarized cross section in the range $0.5<p_T<20$
GeV$/c$ ($0.5<p_T<7.0$ GeV$/c$) was found to be in good agreement, above $\sim2$ GeV$/c$,
with pQCD calculations at NLO accuracy~\cite{Vogelsang:2007zz,deFlorian:2007aj}. The SSA for
both neutral pions and charged hadrons was found to be consistent with zero within errors in
the range $1.0<p_T<5.0$ GeV$/c$~\cite{Adler:2005in}. These results are in agreement with
those found by the E704 Collaboration at lower energy and in similar kinematical
conditions~\cite{Adams:1994yu} [see also Fig.~\ref{pheno:an:midrap}~(right panel),
p.~\pageref{pheno:an:midrap}], extending them to larger $p_T$ values, previously unexplored.
The invariant unpolarized cross section for direct-photon production at mid-rapidity was also
measured up to $p_T=16$ GeV$/c$ in $\sqrt{s}= 200$ GeV $pp$
collisions~\cite{Adler:2005qk,Adler:2006yt}. Data are well described by NLO theoretical
predictions for $p_T>5$ GeV$/c$; in addition, the ratio of isolated photons to all
nonhadronic decay photons is well described by pQCD for $p_T>7$ GeV$/c$. Other recent PHENIX
preliminary results concern~\cite{Aidala:2007bb}: measurements of the SSA for forward
neutrons and mid-rapidity $J/\psi$'s at $\sqrt{s}=200$ GeV; for the neutron case, large flat
asymmetries for positive $x_F$ and SSA's consistent with
zero in the negative $x_F$ range were observed.\\
3) The BRAHMS Collaboration has recently presented preliminary
results for invariant unpolarized cross sections for charged
hadrons, $\pi^\pm$, $K^\pm$, proton, antiproton production in $p p$
collisions at $\sqrt{s}=200$ GeV measured at forward rapidities
($y=2.95$ and 3.3) and $p_T$ up to 4 GeV$/c$~\cite{Arsene:2007jd}.
Other interesting preliminary results have also been
presented~\cite{Lee:2007zzh} concerning: the invariant cross section
for $p p\to\pi^- + X$ at $\sqrt{s}=62.4$ GeV and rapidity $y\sim2.7,
3.3$; the SSA for $\pi^\pm$, $K^\pm$ at $\sqrt{s}=62.4$ GeV, for
both negative and positive $x_F$ in the range $0.25<|x_F|<0.6$, and
$p_T$ up to $\sim1.5$ GeV$/c$; the SSA for $\pi^\pm$, $K^\pm$, $p$
and $\bar{p}$ at $\sqrt{s}=200$ GeV, $0.15<x_F<0.3$ and
$0.2<p_T<3.5$ GeV$/c$. Among the main features of these results we
mention: the almost flat $x_F$-behaviour of $A_N$ for $\pi^\pm$ at
200 GeV and $\theta=2.3^0$ [see Fig.~\ref{pheno:an:brahms} (left
panel), p.~\pageref{pheno:an:brahms}]; the comparable and sizable
$K^+$ and $K^-$ SSA's observed at 200 and 62 GeV [see
Fig.~\ref{pheno:ankaon:brahms}, p.~\pageref{pheno:ankaon:brahms}]; a
large $A_N$ for antiproton production and an almost vanishing SSA
for the proton case.

\subsection{Azimuthal asymmetries in polarized semi-inclusive deeply inelastic scattering}
\label{exp:sidis}
Starting from the end of the 90s the HERMES Collaboration at DESY, and later the COMPASS
Collaboration at CERN and the CLAS Collaboration at Jlab, have performed a series of
measurements of beam and target azimuthal asymmetries in semi-inclusive particle production
in the deeply inelastic collisions of longitudinally polarized leptons off both
longitudinally and transversely polarized proton and deuteron targets. As it will be
discussed in detail in the following sections, these azimuthal asymmetries with respect to
the virtual photon direction, in the $\gamma^*$-target c.m.~frame, can give a lot of
information. The azimuthal angles under consideration refer e.g.~to the transverse component
of the observed hadron momentum ($\phi_h$) and of the initial hadron polarization vector
($\phi_S$) w.r.t.~the virtual photon direction, measured from the leptonic plane [see
Fig.~\ref{pheno:planessidis}, p.~\pageref{pheno:planessidis}]. By integrating the polarized
cross section or the asymmetries, opportunely multiplied by some circular function of the
azimuthal angles, one can disentangle several interesting contributions to single spin
asymmetries and other polarized observables. These  moments of the azimuthal asymmetries are
usually indicated by $A^{W(\phi)}_{S_BS_T}$, where $S_B,\,S_T=U,L,T$ refer to unpolarized
($U$), longitudinally ($L$) and (for the target) transversely ($T$) polarized beam, target
respectively, while $W(\phi)$ refers to some appropriate $\phi$-dependent circular function,
$W=1$, $\sin\phi$, $\cos\phi$, $\sin2\phi$, etc. As a phenomenologically relevant example,
$A_{UT}^{\sin(\phi_h \pm \phi_S)}$ is given as~\cite{Bacchetta:2004jz} \be
\label{exp:sidis:aut} A_{UT}^{\sin(\phi_h \pm \phi_S)} \equiv 2 \langle \sin(\phi_h \pm
\phi_S) \rangle = \frac{\int d\phi_h d\phi_S \sin(\phi_h \pm \phi_S) [d\sigma(\phi_S) -
d\sigma(\phi_S+\pi)]}{\int d\phi_h d\phi_S
  [d\sigma(\phi_S) + d\sigma(\phi_S+\pi)]}\,.
\ee
Notice that in the literature sometimes the factor two in the above
equation is not included.

Throughout this section and in the sequel, we will make use of the
well-known DIS kinematical variables: $P$, $\ell$, $\ell'$,
$q=\ell-\ell'$ are respectively the four-momenta of the target
nucleon, the initial and final leptons and the virtual photon
exchanged; $Q^2=-q^2$; $x_{B}=Q^2/(2P\cdot q)$ is the Bjorken
variable; $\nu= q\cdot P/M$ is the lepton energy loss in the target
nucleon rest frame; $y=(P\cdot q)/(P\cdot \ell)$ is the so-called
inelasticity; $W$ is the c.m.~energy of the virtual photon-target
nucleon system. Additionally, we also adopt the usual SIDIS
variables: $P_{h}$ is the four-momentum of the final observed
hadron, and $\bm{P}_{h\perp}$ its transverse component w.r.t.~the
virtual photon direction of motion, with magnitude
$P_{h\perp}\equiv|\bm{P}_{h\perp}|$; $z_h=(P\cdot P_{h})/(P\cdot q)$
is the usual hadronic variable.

\subsubsection{HERMES results}
\label{exp:sidis:hermes}
In its first stage, HERMES was equipped with a longitudinally polarized hydrogen target. This
gave access to measurements of the quark helicity distributions (in DIS) and  to the
leading-twist azimuthal asymmetry $A^{\sin2\phi_h}_{UL}$, associated with the transverse
momentum dependent (TMD) quark distribution in a longitudinally polarized hadron convoluted
with the Collins function. In the virtual photon-target center of mass frame, even a
longitudinally polarized (in the Lab. frame) proton target acquires a (kinematically) small
($\sim 1/Q$) transverse spin component. Therefore, early attempts to measure transverse SSA's
and some related effects (namely the Sivers and Collins effects) through the azimuthal
asymmetry $A^{\sin\phi_h}_{UL}$ were performed. Even if these transverse-spin contributions
mix up with concurrent effects due to subleading-twist quark distribution and fragmentation
functions, first indications for interesting physics were found. Concerning SSA's, however,
the most interesting HERMES results came out with the advent of a transversely polarized
hydrogen target. This allowed to extract the first unambiguous indications of non-vanishing
leading-twist transverse SSA's in SIDIS, and to disentangle the possible contributions due to
the Sivers and Collins effects.

The kinematical regime covered by the HERMES experiment is approximately the following (some
differences are possible for specific measurements; we refer to the original papers for more
detail):\\
$1 < Q^2 < 15$ (GeV/$c)^2$; 0.023 $< x_{B}<$ 0.4; $y<0.85$; $W^2>
10$ GeV$^2$; $2<E_h<15$ GeV, $0.2<z_h<0.7$; $P_{h\perp}>50$ MeV/$c$.
Notice that in some cases more selective cuts on the last two
variables are used.

We now briefly summarize the main results of the HERMES experiment, quoting only those
results that are relevant in the context of this review paper (a full account of all HERMES
results may be found in the papers quoted in the bibliography):\\
1) First evidence of azimuthal SSA's for charged~\cite{Airapetian:1999tv} and
neutral~\cite{Airapetian:2001eg} pion production in deeply inelastic scattering of polarized
positrons off a longitudinally polarized hydrogen target; $A^{\sin\phi_h}_{UL}$ was found to
be significant and comparable for $\pi^+$ and $\pi^0$ production, and consistent with zero
for $\pi^-$. Indications of an increasing of this analyzing power with increasing of $x_B$,
and with $P_{h\perp}$ up to $\sim0.8$ GeV$/c$, were also found.
$A^{\sin2\phi_h}_{UL}$ was found to be consistent with zero for both $\pi^+$ and $\pi^-$.\\
2) Measurement of azimuthal SSA's in SIDIS production of pions and kaons on a longitudinally
polarized deuterium target~\cite{Airapetian:2002mf}; the dependences of these asymmetries on
$x_B$, $P_{h\perp}$, and $z_h$ were investigated; positive asymmetries for $\pi^+$ and
$\pi^0$ and an indication of a positive asymmetry for $\pi^-$ mesons were found; the
asymmetry for $K^+$ was measured to be comparable with that for $\pi^+$ mesons;\\
3) Measurement of the azimuthal SSA's
$A_{UT}^{\sin(\phi_h\pm\phi_S)}$ in SIDIS production of charged
pions and kaons on a transversely polarized hydrogen
target~\cite{Airapetian:2004tw,Diefenthaler:2005gx}. For the first
time two different phenomena, possibly associated with the Sivers
and Collins effects, indistinguishable in previous data were
disentangled. The $\langle\sin(\phi_h-\phi_S)\rangle$ moment was
measured to be positive and nonzero for $\pi^+$ and consistent with
zero for $\pi^-$ mesons [see Fig.~\ref{pheno:siv:hermes-compass}
(left panel), p.~\pageref{pheno:siv:hermes-compass}]. The
$\langle\sin(\phi_h+\phi_S)\rangle$ moment  was found to be positive
for $\pi^+$ and negative and comparable in magnitude for $\pi^-$
mesons [see Fig.~\ref{pheno:col:hermes-compass} (left panel),
p.~\pageref{pheno:col:hermes-compass}]. For charged kaons the
measured $A_{UT}^{\sin(\phi_h-\phi_S)}$ asymmetry came out to be
positive, showing large values for $K^+$;
$A_{UT}^{\sin(\phi_h+\phi_S)}$ for $K^-$, even if with large error
bars, presented large positive values [see Fig.~\ref{pheno:col:kaons}, p.~\pageref{pheno:col:kaons}].\\
4) Extraction of subleading-twist effects in SSA's in SIDIS
production of charged pions on a longitudinally polarized hydrogen
target~\cite{Airapetian:2005jc}. By combining measurements of
azimuthal asymmetries with both longitudinally and transversely
polarized hydrogen targets (see previous point) this contribution
was found to be significantly positive for $\pi^+$ mesons, therefore
dominating the asymmetries on a longitudinally polarized target (see
points 1) and 2) above). The subleading-twist contribution for
$\pi^-$ mesons was found to be small.\\
5) Measurement of the subleading-twist beam spin asymmetry $A^{\sin\phi_h}_{LU}$ in the
azimuthal distribution of pions produced in SIDIS~\cite{Airapetian:2006rx}.
$A^{\sin\phi_h}_{LU}$ for $\pi^+$'s was found to be consistent with zero for low $z_h$ and
rising with $z_h$ up to about 0.02; for $\pi^-$'s, $A^{\sin\phi_h}_{LU}$
was found to be consistent with zero within the statistical accuracy; for neutral pions, the
asymmetry was found to be roughly constant and of the order of 0.03, decreasing to zero only
in the lowest/highest $z_h$ regions.\\
6) Interesting preliminary results on azimuthal SSA's in SIDIS production of charged kaons on
a transversely polarized hydrogen target have been also presented (see, e.g.,
Ref.~\cite{Diefenthaler:2006vn}).
\subsubsection{COMPASS results}
\label{exp:sidis:compass}

 The COMPASS Collaboration at CERN has performed measurements of transverse
azimuthal asymmetries for charged hadrons produced in deeply
inelastic scattering of the CERN-SPS 160 GeV$/c$ muon beam off a
transversely polarized deuteron target. The high-energy lepton beam
allows to reach large values of $Q^2$ and a very large coverage of
$W$. This can be very helpful in disentangling genuine TMD effects
from perturbatively generated contributions. The large $x_B$ region
covered, well below the valence region, allows to better estimate
the first moments of the TMD distributions and test related sum
rules. Moreover the use of a transversely polarized (isospin scalar)
deuteron target well complements with HERMES results, in particular
concerning flavour separation of the TMD distributions and
fragmentation functions. COMPASS also plans to perform measurements
with a transversely polarized proton target in the near future.

The asymmetries $A_{UT}^{\sin(\phi_h\pm\phi_S)}$ for charged hadrons, and their dependence on
$x_B$ and, for leading hadrons, $z_h$, $P_{h\perp}$, were measured and found to be compatible
with zero within statistical errors~\cite{Alexakhin:2005iw} [see
Fig.~\ref{pheno:siv:hermes-compass} (right panel), p.~\pageref{pheno:siv:hermes-compass} and
Fig.~\ref{pheno:col:hermes-compass} (right panel), p.~\pageref{pheno:col:hermes-compass}].
The kinematical region covered was: $Q^2>1$~(GeV/$c$)$^2$, $W>5$ GeV, $0.003<x_B<0.3$,
$0.1<y<0.9$. Concerning the observed hadron, the cuts $P_{h\perp}>0.1$ GeV/$c$, $z_h>0.2$
($z_h>0.25$) have been imposed for all (leading) hadrons.

Recently, new high-precision measurements for the same observables and in the same
kinematical region but with larger statistical significance have been
published~\cite{Ageev:2006da}. These results extend and essentially confirm those of
Ref.~\cite{Alexakhin:2005iw}, with statistical errors a factor of two
smaller.

Interesting preliminary results have also been presented on $A_{UT}^{\sin(\phi_h\pm\phi_S)}$
for charged pions and kaons, in the same kinematical configuration and using data collected
in 2003-2004~\cite{Martin:2007au,Bradamante:2007ex}. Again, all transverse spin effects
investigated are found to be compatible with zero.

Very recently COMPASS has also presented preliminary results~\cite{Kotzinian:2007uv}, in the
same kinematical region, on the first measurement of the other six possible transverse
azimuthal asymmetries (see, e.g., Ref.~\cite{Bacchetta:2006tn}),
$A_{UT}^{\sin(3\phi_h-\phi_S)}$, $A_{LT}^{\cos(\phi_h-\phi_S)}$ (leading twist), and
$A_{UT}^{\sin\phi_S}$, $A_{UT}^{\sin(2\phi_h-\phi_S)}$, $A_{LT}^{\cos\phi_S}$,
$A_{LT}^{\cos(2\phi_h-\phi_S)}$ (subleading twist). All these asymmetries are compatible with
zero within statistical errors.
\subsubsection{JLAB-CLAS results}
\label{exp:sidis:clas}
The JLAB-CLAS Collaboration has performed measurements of both longitudinal beam and target
spin asymmetries for pion production in SIDIS off protons using an electron beam with energy
4.3 GeV and 5.7 GeV. More precisely, they have measured:\\
1) The $\sin\phi_h$ moment of the subleading beam spin asymmetry,
$A_{LU}^{\sin\phi_h}$, for the process $e p\to e^\prime \pi^+ + X$
above the baryon resonance region, in the range $0.15<x_B<0.4$,
$0.5<z_h<0.8$, using a polarized electron beam of 4.3
GeV~\cite{Avakian:2003pk} and of 5.7 GeV~\cite{Avakian:2005bh} (in
this last case only preliminary results are available). The measured
beam SSA is positive for a positive electron helicity, varying
roughly in the range $0.02\div0.08$; it increases with
$z_h$ and shows no significant dependence on the $x_B$-range;\\
2) The $\sin\phi_h$, $\sin2\phi_h$ moments of the longitudinal target spin asymmetry,
$A_{UL}^{\sin\phi_h}$ and $A_{UL}^{\sin2\phi_h}$, for both charged and neutral pions, in the
range $0.15<x_B<0.4$, $0.5<z_h<0.8$, using an electron beam of 5.7 GeV (preliminary results,
see Refs.~\cite{Avakian:2005bh,Avakian:2005ps}). Data for $\pi^+$'s show clear $\sin\phi_h$
and $\sin2\phi_h$ modulations, leading to $\langle\sin\phi_h\rangle=0.058\pm0.011$ and
$\langle\sin2\phi_h\rangle=-0.041\pm0.011$ (errors are statistical).

\subsection{Azimuthal asymmetries in unpolarized processes}
\label{exp:unp}
Several azimuthal asymmetries in unpolarized cross sections for SIDIS, Drell-Yan and $e^+e^-$
processes have been measured. Among other explanations of their origin, intrinsic parton
motion is one of the most accredited. Therefore, these asymmetries are intimately related to
transverse single spin asymmetries and in particular to TMD parton distribution and
fragmentation functions. For this reason, we find useful here to summarize some of the most
relevant results.
\subsubsection{SIDIS processes}
\label{exp:unp:sidis}
The most general azimuthal dependence of the cross section for SIDIS off an unpolarized
target can be written as (see, e.g., Ref.~\cite{Ahmed:1999ix} and references therein):
 \be
 \frac{d\sigma^{\ell N\to\ell' h X}}{d\phi_h}\propto A+B\cos\phi_h+C\cos2\phi_h
 +D\sin\phi_h+E\sin2\phi_h\,,
 \label{exp:unp:phidep}
 \ee
\noindent where the angle $\phi_h$, as mentioned above, is the azimuthal angle of the
observed hadron momentum around the virtual photon direction, measured starting from the
leptonic plane [see Fig.~\ref{pheno:planessidis}, p.~\pageref{pheno:planessidis}].

In collinear pQCD, $A$ is of order ${\cal O}(\alpha_s^0)$; the first
corrections to $A$ and the  $B$-$E$ contributions come from higher
order terms, in particular $B$, $C$ are of order ${\cal
O}(\alpha_s)$, $D$, $E$ are of order ${\cal O}(\alpha_s^2)$.
Furthermore, the T-odd contributions $D$ and $E$ are present only
for parity violating weak interactions or for electromagnetic
interactions with longitudinally polarized lepton beams.
Nonperturbative transverse momentum effects and TMD functions may
give leading-twist contributions to $B$ and $D$ terms which may be
non-negligible even at very large $Q^2$.

The EMC Collaboration at CERN measured the azimuthal distribution
[see Fig.~\ref{pheno:cahn}, p.~\pageref{pheno:cahn}] and its
$\langle\cos\phi_h\rangle$, $\langle\cos2\phi_h\rangle$,
$\langle\sin\phi_h\rangle$ moments for charged hadron production in
SIDIS with a muon beam of energy 280 GeV off a liquid hydrogen
target~\cite{Arneodo:1986cf}. Both the $x_F$ and (for the moments)
$P_{h\perp}$ dependences were measured. The main kinematical cuts
considered were: $Q^2 \geq 4$ (GeV/$c)^2$, $40\leq W^2\leq 450$
GeV$^2$, $\nu \geq 20$ GeV, $y\leq 0.8$, $P_{h\perp}\geq 0.2$
GeV/$c$. The main results found were: A strong dependence of the
$\langle\cos\phi_h\rangle$ moment on $x_F$, which has a small and
positive value for negative $x_F$, becomes negative at $x_F \sim 0$
and falls below $-0.1$ for $x_F> 0.5$. A small dependence of the
$\langle\cos\phi_h\rangle$, $\langle\cos2\phi_h\rangle$ moments on
$P_{h\perp}$. The values of $\langle\cos2\phi_h\rangle$,
$\langle\sin\phi_h\rangle$ as functions of $x_F$ were found to be
close to zero. Previous results limited to the forward hemisphere
were also available~\cite{Aubert:1983cz}.

The Fermilab E665 Collaboration measured the azimuthal asymmetry of
forward charged hadrons produced in SIDIS with a 490 GeV muon beam
off proton and deuteron targets~\cite{Adams:1993hs}. The main
kinematical cuts imposed were: $Q^2>3.0$ (GeV/$c)^2$, $100<W^2<900$
GeV$^2$, $60<\nu<500$ GeV, $x_B>0.003$, $0.1<y<0.85$,
$P_h>8$~GeV/$c$. They measured, among other observables, the
$\langle\cos\phi_h\rangle$ moment as a function of the minimum
hadron transverse momentum up to 2.5 GeV$/c$, imposing $W^2>300$
GeV$^2$ and $x_F>0.2$. Moment values are negative, starting from
about $-0.04$ at the lowest cutoff in $P_{h\perp}$, decreasing up to
about $-0.12$ for a $P_{h\perp}$ cutoff of $\sim 1.5$ GeV$/c$ and
then increasing again (but with larger errors) to $\sim-0.03$ for
the largest considered cutoff in $P_{h\perp}$ of $2.5$ GeV$/c$.

EMC and E665 results are consistent with theoretical models including parton transverse
momenta arising both from intrinsic nonperturbative motions and NLO pQCD radiative
corrections. Apparently, the nonperturbative contributions dominate in the kinematical
regimes covered by these experiments.

The ZEUS Collaboration at HERA-DESY measured the azimuthal angle distribution for charged
hadrons produced in neutral current deeply inelastic positron-proton
scattering~\cite{Breitweg:2000qh}. The dependence of the moments of the distribution,
$\langle\cos\phi_h\rangle$, $\langle\cos2\phi_h\rangle$ on (the minimum value of) the
transverse momentum of the charged hadrons observed, $P^{\rm{min}}_{h\perp}$, was also
studied. At low $P^{\rm{min}}_{h\perp}$ a clear $\langle\cos\phi_h\rangle$ contribution was
observed, while as the value of $P^{\rm{min}}_{h\perp}$ increased, a
$\langle\cos2\phi_h\rangle$ term became evident; $\langle\sin\phi_h\rangle$ was found to be
consistent with zero independently of the $P^{\rm{min}}_{h\perp}$  value chosen; ii) For
hadrons produced at large transverse momenta the $\langle\cos\phi_h\rangle$ moment was
measured to be negative, in agreement with pQCD predictions; the $\langle\cos2\phi_h\rangle$
moment was measured for the first time and found to be non-zero and positive, with increasing
magnitude as a function of $P^{\rm{min}}_{h\perp}$. These results were interpreted as giving
clear evidence for a pQCD contribution to the azimuthal asymmetry, in contrast with the
results of EMC and E665 Collaborations in different kinematical regimes.

The ZEUS~Collaboration has also performed a study of the azimuthal asymmetry for inclusive
jet production in neutral current deeply inelastic $e^+p$ scattering. Again, NLO pQCD
calculations seem to give a good description of the observed results~\cite{Chekanov:2002sz}.

More recently, the ZEUS~Collaboration presented new results on the dependence of azimuthal
moments on the pseudorapidity and the minimum transverse energy of the final-state
hadrons~\cite{Chekanov:2006gt}. Both neutral and charged hadrons were considered over an
extended phase space as compared to Ref.~\cite{Breitweg:2000qh}. The value of
$\langle\cos\phi_h\rangle$ was found to be negative for $\eta<-2$, becoming positive for
larger $\eta$. NLO pQCD predictions reproduce reasonably well this behaviour but fail in
describing the magnitude of the asymmetries. However, the predicted values of
$\langle\cos2\phi_h\rangle$ agree with the data. A deviation of $\langle\sin\phi_h\rangle$
from zero at the level of three standard deviations was observed, while
$\langle\sin2\phi_h\rangle$ was consistent with zero.

\subsubsection{Drell-Yan processes}
\label{exp:unp:dy}
Concerning the Drell-Yan (DY) process, it can be shown that after averaging over the initial
hadron polarizations, summing over the final lepton spins, and applying invariance principles
(permutation symmetry, gauge invariance, parity conservation and unitarity), the most general
form of the lepton angular distribution, in the lepton-pair c.m.~frame reads
 \be
 \frac{1}{\sigma}\,\frac{d^2\sigma}{d\cos\theta d\phi}\propto
 1+\lambda\cos^2\theta+\mu\sin2\theta\cos\phi+(\nu/2)\sin^2\theta\cos2\phi\>,
 \label{exp:dydistr}
 \ee
where $\theta$ and $\phi$ are respectively the polar and azimuthal angle identifying the
lepton-pair direction of motion [see also Fig.~\ref{pheno:dy:cs},  p.~\pageref{pheno:dy:cs}];
the coefficients $\lambda$, $\mu$, $\nu$ are in general functions of the other relevant
kinematical variables: the total c.m.~energy $\sqrt{s}$, the lepton-pair invariant mass $M$,
its transverse momentum $q_T$, and its fraction of longitudinal momentum $x_F$. The values of
these coefficients depend on the choice of the reference frame, the so-called Collins-Soper
(CS) frame being often adopted (other popular choices are the Gottfried-Jackson (GJ) frame
and the $u$-channel frame). When the angular distributions in both $\cos\theta$ and $\phi$
are measured in a given frame, one can easily compute the coefficients in any other one.
Notice however that this is not the case when only the $\cos\theta$ distribution is measured.

In the naive parton model for the Drell-Yan process, the assumption of massless quarks
implies that the virtual photon is transversely polarized, so that $\lambda=1$, $\mu=\nu=0$,
and one has $d\sigma/d\cos\theta\sim 1+\cos^2\theta$. A general relation by Lam and
Tung~\cite{Lam:1978pu,Lam:1978zr,Lam:1980uc}, analogous to the Callan-Gross relation in DIS,
is expected to hold in any reference frame: $\lambda=1-2\nu$. This relation is not modified
by first-order QCD corrections~\cite{Lam:1980uc}, but can be influenced by parton intrinsic
motions and other nonperturbative effects.

Concerning experimental results, we limit ourselves to mention some measurements that, as will
be discussed in section~\ref{pheno}, are relevant in the context of this review.

The NA3~\cite{Badier:1981ti} and NA10~\cite{Falciano:1986wk,Guanziroli:1987rp} Collaborations
at CERN have measured the angular distributions of high-mass muon pairs produced by 140
GeV$/c$ and 194 GeV$/c$ $\pi^-$ beams impinging on a tungsten target, and by a 286 GeV$/c$
$\pi^-$ beam on deuterium and tungsten targets. The dependence of the parameters $\lambda$,
$\mu$, $\nu$ on the c.m.~energy $\sqrt{s}$, the invariant mass $M$, transverse momentum $q_T$
and rapidity $y$ of the lepton pair, and the fractional momentum of the parton in the pion,
$x_1$, was measured. The kinematical region covered was indicatively (see
Ref.~\cite{Guanziroli:1987rp} for more details): $4.0\leq M\leq 8.5$ GeV$/c^2$ and $M\geq 11$
GeV$/c^2$, $0.32\leq q_T\leq 2.60$ GeV$/c$, $-0.14\leq y\leq 0.71$, $0.28\leq x_1\leq 0.80$.
No evidence for a c.m.~energy dependence or a nuclear dependence of the angular distribution
parameters was found. The value of $\lambda$ was found close to 1, as predicted by the naive
parton model, and substantially independent of any kinematical variable, apart from the high
$x_1$ region, where indications of a decrease of $\lambda$ with $x_1$, as suggested by
higher-twist effects~\cite{Berger:1979du,Berger:1979xz}, were found. The value of $\mu$ was
found close to zero in the CS frame, as expected if both annihilating partons contribute
equally to the transverse momentum of the lepton pair. The most important result of this
analysis was the strong dependence of $\nu$ on $q_T$, in clear disagreement with pQCD
expectations. In particular, starting from $\nu\sim 0$ at the lowest $q_T$ values, $\nu$ was
found to grow up to $0.15\div0.30$ (depending on $\sqrt{s}$) at larger $q_T$ values [see
Fig.~\ref{pheno:unp-an:dy}~(left panel), p.~\pageref{pheno:unp-an:dy}]. Since $\lambda\sim
1$, this result also implied a clear violation of the Lam-Tung relation.

The E615~Collaboration at Fermilab has measured the  cross section, the transverse momentum
distribution and the full angular distribution in $\cos\theta$ and $\phi$ as a function of
$M$, $x_F$, and $q_T$, for muon pairs produced by a 252 GeV $\pi^-$ beam interacting with a
tungsten target~\cite{Conway:1989fs}. The dependence of the parameters $\lambda$, $\mu$,
$\nu$, on the fractional momentum of the parton in the pion, $x_{1}$, was also measured. The
kinematical region covered was: $4.05< M< 8.55$ GeV$/c^2$, $0<q_T<5.0$ GeV$/c$,
$0.2<x_{1}<1.0$. Again, a clear violation of the Lam-Tung relation, and a strong dependence
of $\nu$ on $q_T$ were found. In this case also, starting from $\nu\sim 0$ at the lowest
$q_T$ values, $\nu$ was found to grow up to $\sim0.54(0.73)$ in the GJ(CS) frame at the
largest $q_T$ values.

Recently, the FNAL E866/NuSea Collaboration has presented results on the angular
distributions of DY dimuons produced using an 800 GeV/$c$ proton beam on a deuterium
target~\cite{Zhu:2006gx}, in the kinematical range $4.5<M<15.0$ GeV/$c^2$, $q_T<4$ GeV/$c$,
and $0<x_F<0.8$. No significant $\cos2\phi$ dependence was found, in contrast with
pion-induced DY data discussed above. These results put constraints on theo\-retical models
that predict a large $\cos2\phi$ dependence originating from QCD vacuum effects and suggest
that the sea-quark Boer-Mulders functions~\cite{Boer:1997nt} are much smaller than those for
valence quarks.
\subsubsection{$e^+e^-\to\pi\pi+X$ processes}
\label{exp:unp:belle}
Recently the Belle Collaboration at the KEKB asymmetric-energy $e^+e^-$ storage rings has
published~\cite{Abe:2005zx} very interesting results on the inclusive production of hadron
pairs (namely charged pion pairs) in $e^+e^-$ annihilation. Using two different
reconstruction methods, they have found evidence of statistically significant azimuthal
asymmetries. The first method is based on the reconstruction of the thrust axis in
the $e^+e^-$ c.m.~frame: combining two hadrons from different hemispheres in jet-like events,
with azimuthal angles $\phi_1$ and $\phi_2$ defined w.r.t.~the plane spanned by the lepton
momenta and the thrust axis [see Fig.~\ref{pheno:kin:epem} (left),
  p.~\pageref{pheno:kin:epem}],
a $\cos(\phi_1+\phi_2)$ modulation of the dihadron yield has
been observed [see Fig.~\ref{pheno:belle}~(left panel), p.~\pageref{pheno:belle}]. An
alternative method does not require the knowledge of the thrust axis. In this case
the dihadron yields are measured as a function of one angle, $\phi_0$, that is the angle
between the plane spanned by the momentum vector of the first hadron and the lepton momenta
and the plane defined by the two hadron momenta [see Fig.~\ref{pheno:kin:epem} (right),
  p.~\pageref{pheno:kin:epem}]. In this case a $\cos(2\phi_0)$ modulation
appears [see Fig.~\ref{pheno:belle}~(right panel), p.~\pageref{pheno:belle}]. As it will be
discussed in section~\ref{pheno:epem}, these results are related and give access to the
product of two Collins functions.

The kinematical cuts imposed in both methods are the following: $-0.6<\cos(\theta_{\rm lab})<
0.9$, where $\theta_{\rm lab}$ is the polar angle in the laboratory frame; $z_{1,2}>0.2$,
where $z_i=2E_{h_i}/Q$, and $Q$ is the c.m. energy; $Q_T< 3.5$ GeV/$c$, where $Q_T$ is the
transverse momentum of the virtual photon in the rest frame of the hadron pair. Moreover, to
cancel acceptance effects double ratios of normalized rates for unlike-sign ($U$) over
like-sign ($L$) pion pairs ($UL$ double ratios) have been considered. At large $z_{1,2}$
values the measured asymmetries are significantly different from zero [see
Fig.~\ref{pheno:belle}, p.~\pageref{pheno:belle}]. In Ref.~\cite{Ogawa:2006bm} preliminary
results on the double ratios of unlike-sign, $U$, over all charged ($C$) pion pairs have been
presented. These $UC$ double ratios can give useful additional information on the Collins
function. Belle results have been confirmed by recent preliminary data with enlarged
statistics and much smaller error bars~\cite{Ogawa:2007zz}.

%% file: theo-appr-rev.tex
\section{Theoretical approaches to single spin asymmetries}
\label{theo}

We present here some of the approaches developed to explain the interesting features of the
SSA data discussed so far. We will try to follow a historical perspective, starting with the
former attempts to describe the SSA's observed in hadronic collisions and then entering into
the latest developments.

The first approach we address is based on a pQCD factorization scheme with
inclusion of spin and intrinsic
transverse momentum dependent (TMD) effects: we will generally refer to it
as the TMD approach.
In particular we start with its application to inclusive hadron production in
hadron-hadron collisions and its formulation
%a model based on
as a phenomenological generalization of the parton model.
% with
%inclusion of spin and intrinsic transverse momentum effects,
%with its original application to inclusive hadron production in
%hadron-hadron collisions.
This form of the TMD approach will be also referred to as
%We will refer to it as
the {\em Generalized Parton Model} (GPM) and
will find a more detailed description within the helicity formalism in
section~\ref{form}.
We then discuss more recent, and significant, QCD developments of the TMD
approach with inclusion of initial and final state colour interactions
which modify, to some extent, the partonic interpretation of the GPM
approach.
%We will discuss similarities and connections as well the main
%differences with the GPM approach.
Some details on the impact parameter picture for SSA's, which gives a
useful physical interpretation of the TMD approach, are also presented.

The second approach we consider extends the QCD collinear
factorization theorems to higher-twist contributions:
the so-called {\em twist-three approach}.
We recall its basic ingredients, giving some details of its
application to SSA's in $pp\to \pi +X$ and its later extension to other
processes.

The issues of validity, applicability and possible overlapping of these two approaches, for
different kinematical regions and processes, as well as the status of factorization in pQCD,
are also discussed.  In particular, we will see how the TMD approach applies naturally to
leading twist asymmetries, like SSA's in Drell-Yan processes and SIDIS, or azimuthal
asymmetries in $e^+e^-\to h_1 h_2 + X$, at low transverse momentum, where also a second large
scale (the virtuality of the exchanged boson) is present. With some caution, it can be also
adopted in double inclusive production of hadrons (jets, or photon-jet) in hadronic
collisions.

On the other hand, the collinear higher-twist approach
applies in principle to subleading asymmetries, like SSA's in $pp\to h +X$
and $pp\to \gamma +X$, where a single hard scale (the transverse
momentum of the observed particle) appears.
For these cases the GPM can be viewed as an effective and, as we will
 see,  phenomenologically successful, TMD description.

For completeness we will give some details of an alternative
description of SSA's based  on a semiclassical model: the so-called {\em orbiting
valence quark model}.

Among other attempts to describe SSA's, even if on a more qualitative ground, we mention:
$i)$ a model discussed in Ref.~\cite{Kochelev:1999nd}, for $pp\to \pi +X$, and in
Ref.~\cite{Ostrovsky:2004pd}, for SIDIS, where the quark chirality flip
and the required T-odd phase (see below) %needed to have a non zero SSA
are associated with instanton fluctuations in the QCD
vacuum; $ii)$ a soft rescattering mechanism~\cite{Hoyer:2005ev} with a
Pauli coupling that leads in SIDIS to a
$\sin(\phi_h+\phi_S)$ dependence (like that usually ascribed to the
Collins effect in the TMD approach);
$iii)$ a coherence mechanism~\cite{Hoyer:2006hu}, where the
entire hadron contributes to the scattering process, aimed to
describe the sizable values of $A_N$ observed in $pp$ collisions at
large $x_F$.

%We warn once again the reader that this presentation will be conditioned by the authors
%perspective and by no means it can be considered
%exhaustive.  Therefore, for technicalities and a comprehensive
%discussion we refer the reader to the original bibliography.

For technicalities and a comprehensive discussion of the topics
addressed here we refer the reader to the original bibliography.

\subsection{Parton model approaches including intrinsic transverse motion}
\label{theo:kt}

It has been known for a long time, especially in the fixed target regime,  that unpolarized
cross sections for inclusive particle production in high-energy hadron-hadron collisions at
moderate $p_T$ (few GeV/$c$) cannot be properly described in the collinear factorized pQCD
approach, neither at LO nor at NLO accuracy~\cite{Bourrely:2003bw}. Only recently, improved
NLO calculations including threshold resummation effects have been developed, see
e.g.~Ref.~\cite{deFlorian:2005yj}. These results show how for inclusive cross sections
integrated over all hadron rapidity range the agreement with data is much less problematic.
On the other hand the rapidity-dependent case is still under study.

Even more dramatic is the failure of the description of the Drell-Yan low $q_T$ spectrum,
where $q_T$ is the lepton-pair transverse momentum. Indeed, in this case no transverse
momentum can be generated in the collinear LO approximation. All these facts have raised,
starting from former work of Feynman and collaborators~\cite{Field:1976ve,Feynman:1978dt},
the interest in considering the role of intrinsic transverse momentum.

In the context of transverse SSA's
the intrinsic transverse motion of partons has played an even more essential role.
Indeed, as shown in Ref.~\cite{Kane:1978nd}, in  collinear pQCD
$A_N$ appears only as the immaginary part of interference terms
between spin-flip and no-spin-flip partonic scattering
amplitudes. Since at LO these are real
and helicity is conserved for massless partons, it was natural to
expect $A_N\simeq \alpha_s m/\sqrt s$.

Contrary to these expectations several experimental observations, as seen in
section~\ref{exp}, show sizable SSA's in the high-energy regime.

The cross section for a generic inclusive process $A\,B \to C+X$,
in a phenomenological approach based on the
generalization of the factorization theorem with the inclusion of
intrinsic motions, $\bm{k}_\perp$, of partons inside a nucleon and of
final hadrons relatively to the fragmenting parton,  reads, schematically:
\be
d\sigma \propto \sum_{a,b,c}  \hat f_{a/A}(x_a,\bm{k}_{\perp a}) \otimes
 \hat f_{b/B}(x_b, \bm{k}_{\perp b}) \otimes
d\hat\sigma^{ab \to c d}(x_a, x_b, \bm{k}_{\perp a}, \bm{k}_{\perp b})
\otimes  \hat D_{C/c}(z, \bm{k}_{\perp C}) \,,
\label{theo:gen}
\ee
where $\otimes$ stands for appropriate convolutions both in the light-cone momentum
fractions, $x_{a,b},\, z$, and in the $\bm{k}_\perp$'s. The $\hat f$'s and the $\hat D$'s are
the TMD parton distribution (PDF) and fragmentation functions (FF), respectively;
$\bm{k}_{\perp a,b}$ is the transverse momentum of  parton $a,b$ w.r.t.~the hadron $A,B$
momentum and $\bm{k}_{\perp C}$ the transverse momentum of hadron $C$ w.r.t.~the fragmenting
parton direction of motion.

The above QCD factorization scheme -- with unintegrated $\bm{k}_\perp$
dependent distribution and fragmentation functions -- has never been
formally proven for the process under consideration, but only for the Drell-Yan
process~\cite{Collins:1977iv}
and for two-particle inclusive
production in $e^+e^-$ annihilation~\cite{Collins:1981uk}
(somehow a time-reversed Drell-Yan process).
Factorization at low transverse momentum
for SIDIS has been recently addressed
\cite{Ji:2004wu,Ji:2004xq} and
proved at the same level as for the two above mentioned processes.
Recent developments on the status of the more problematic
$k_\perp$-factorization in $AB\to CD +X$  will be discussed in
section~\ref{theo:develop}.

The study of spin asymmetries requires the extension of Eq.~(\ref{theo:gen}) to the polarized
case. The inclusion of $\bfk_\perp$ effects and TMD functions can in principle lead to
sizable SSA's; however, if the TMD functions are symmetric in the azimuthal direction,  $A_N$
would be still negligible. It was then originally suggested by
Sivers~\cite{Sivers:1989cc,Sivers:1990fh} that there could exist a correlation between the
azimuthal distribution of an unpolarized parton and the spin of its parent hadron. That is,
we could have the following nonvanishing asymmetry:
 \be \Delta \hat f_{a/A^\uparrow}(x,
 \bm{k}_{\perp}) \equiv \hat f_{a/A^\uparrow}(x, \bm{k}_{\perp}) - \hat f_{a/A^\downarrow} (x,
 \bm{k}_{\perp}) = \hat f_{a/A^\uparrow}(x, \bm{k}_{\perp})- \hat f_{a/A^\uparrow}(x, -
 \bm{k}_{\perp}) \,, \label{theo:sivasy}
  \ee
where the last relation comes from rotational invariance and $\uparrow (\downarrow)$ stands
for the upwards (downwards) transverse spin direction of the hadron $A$. It is easy to see
that $\Delta \hat f_{a/A^\uparrow}$ must be, by parity conservation, proportional to
$(\bm{p}_A\times \bm{k}_\perp )\cdot \bm{S}_A$, where $\bm{p}_A, \bm{S}_A$ are the momentum
and the spin of the hadron $A$, and $\bm{k}_\perp$ the transverse momentum of the parton.
This function %, the so called Sivers function,
is $\bm{k}_\perp$-odd and chiral-even, as will
be clear in section~\ref{form}.
It is also naively time reversal odd (T-odd).
Notice that the above asymmetry can be rephrased in terms of the
number density of unpolarized partons inside a transversely polarized hadron
($\hat{\bm{p}}_A, \hat{\bm{k}}_\perp$ being unit vectors and
$k_\perp=|\bm{k}_\perp|$):
\be
\hat f_{a/A^\uparrow}(x, \bm{k}_{\perp}) =  f_{a/A}(x, k_{\perp}) +
\frac{1}{2} \, \Delta^N f_{a/A^\uparrow}(x, k_{\perp}) \>
(\hat{\bm{p}}_A\times \hat{\bm{k}}_\perp )\cdot
\bm{ S}_A \,, \label{theo:polden}
\ee
where $ \Delta^N f_{a/A^\uparrow}$ (related to
$f_{1T}^\perp$~\cite{Boer:1997nt}, see Ref.~\cite{Bacchetta:2004jz}) is referred to as the {\em Sivers
function}.

In this first formulation the numerator of $A_N$, see Eq.~(\ref{intro:an}), assumed to be
generated by the Sivers effect, was computed according to the following
expression:
\be
d\Delta\sigma^{\rm Sivers} \propto \sum_{a,b,c} \Delta\hat
f_{a/A^\uparrow}(x_a,\bm{k}_{\perp a}) \otimes
 f_{b/B}(x_b) \otimes
d\hat\sigma^{ab \to c d}(x_a, x_b, \bm{k}_{\perp a}) \otimes
D_{C/c}(z) \,;
\label{theo:ansiv}
\ee
that is, by keeping TMD effects only where the collinear approximation
would otherwise
give zero (remember that the Sivers function vanishes for
$k_\perp=0$ and  is $\bm{k}_\perp$-odd).

In 1993 Collins proposed a proof of the
vanishing of the Sivers function~\cite{Collins:1992kk}.
By imposing, in the light-cone gauge (this point would have become
crucial), parity and time reversal invariance, he
showed that strong interactions forbid such an asymmetry.
However, on the basis of possible initial state interactions between the
two colliding hadrons, in Ref.~\cite{Anselmino:1994tv} it was argued that
 this asymmetry might be allowed.
By a reasonable parameterization of this function a good description of E704 data was then
obtained~\cite{Anselmino:1994tv,Anselmino:1998yz}.

In Refs.~\cite{Collins:1992kk,Collins:1993kq} another possible asymmetric azimuthal angular
dependence was proposed: a left-right asymmetry in the fragmentation of a transversely
polarized quark into a spinless (or, more generally, unpolarized) hadron. This implies a new
TMD and spin dependent twist-two function ($\bm{p}_q$ and $\bm{s}_q$ being the momentum and
the spin of the quark and $\bm{p}_C \simeq z \bm{p}_q + \bm{k}_\perp$ the hadron momentum),
\be
 \hat D_{C/q^\uparrow}(z, \bm{k}_{\perp}) =  D_{C/q}(z, k_{\perp}) +
\frac{1}{2} \, \Delta^N D_{C/q^\uparrow}(x, k_{\perp}) \>
 (\hat{\bm{p}}_q\times \hat{\bm{k}}_\perp )\cdot
\bm{ s}_q \>, \label{theo:polfrag}
\ee
where
%$ D_{C/q}(z, k_{\perp})$ is the unpolarized TMD FF and
$\Delta^N D_{C/q^\uparrow}$ (or $H_{1}^\perp$ in Refs.~\cite{Mulders:1995dh,Boer:1997nt}, see
also Ref.~\cite{Bacchetta:2004jz}) is the {\em Collins  function}. Again we have \be
\Delta\hat D_{C/q^\uparrow}(z, \bm{k}_{\perp}) \equiv \hat D_{C/q^\uparrow}(z,
\bm{k}_{\perp})- \hat D_{C/q^\downarrow}(z, \bm{k}_{\perp}) = \hat D_{C/q^\uparrow}(z,
\bm{k}_{\perp})- \hat D_{C/q^\uparrow}(z, - \bm{k}_{\perp}) \,, \ee
%the so called Collins function,
where $\uparrow (\downarrow)$ stands for the upwards (downwards) transverse spin direction of
the fragmenting quark. In this case, due to the rescattering of the hadron with the remnants
of the jet (or, in more formal words, since the final hadron, contrary to the incoming one,
cannot be represented as a plane wave state), time-reversal invariance does not imply any
constraint, allowing the existence of such an asymmetry.

This new effect in the fragmentation mechanism was then proposed as a transverse quark
polarimeter with the aim of extracting information on the transverse polarization of a quark,
i.e.,  as a tool to measure the transversity distribution. This function, denoted by
$h_{1q}(x)$ (or $\Delta_T q$,  $\delta q$), and defined as
 \be h_{1q}(x) =
f_{q^\uparrow/p^\uparrow}(x) - f_{q^\downarrow/p^\uparrow}(x) \,, \label{theo:transversity}
 \ee
is the difference between the probabilities to find a quark polarized along the transverse
proton polarization and against it. This chiral-odd function, in contrast with the
unpolarized and the helicity distribution functions (both chiral-even), is much more
difficult to measure.
Indeed, it cannot be accessed in %inclusive deeply inelastic scattering
DIS,  since it implies a
helicity flip at the parton level whereas QED and QCD for massless quarks
conserve helicity. It has therefore to be accompanied by
another chiral-odd quantity.
As originally proposed by Ralston and Soper in Ref.~\cite{Ralston:1979ys},
the most promising source of information on $h_{1q}$,
is the double transverse spin
asymmetry, $A_{TT}$, in Drell-Yan processes, where it appears quadratically,
but unfortunately no data are still available.

Collins idea was to consider a new process where another
chiral-odd function could appear: this is the SIDIS process
with a transversely polarized proton target, $e p^\uparrow\to e' h +X$, where
the azimuthal dependence of the final hadron distribution, w.r.t.~the lepton
scattering plane,  would
probe at the same time the transversity distribution function, the
transverse spin dependence of the hard scattering and, eventually, the
spin dependence in the fragmentation. This process, as we will discuss in
section~\ref{pheno}, has become, indeed, the first source of
phenomenological information on the transversity distribution.

%Another proposed tool to access the transversity distribution via its coupling
%to the Collins function is the study of the azimuthal
%correlation in pion pairs produced in proton-proton collisions with a
%single transversely polarized hadron.

A calculation of the Collins function, based
on a sigma model for pions with an effective Lagrangian incorporating
chiral symmetry breaking, showed that
$\Delta \hat D_{C/q^\uparrow}(z, \bm{k}_{\perp})$ could be
different from zero~\cite{Collins:1992kk}.

%\begin{comment}
%led to the following
%analyzing power
%\be
%\frac{ \hat D_{C/q^\uparrow}(z, \bm{k}_{\perp}) - \hat D_{C/q^\downarrow}(z,
%\bm{k}_{\perp}) }{ \hat D_{C/q^\uparrow}(z, \bm{k}_{\perp}) +
%\hat  D_{C/q^\downarrow}(z, \bm{k}_{\perp}) } \equiv
%\frac{\Delta \hat D_{C/q^\uparrow}(z, \bm{k}_{\perp})}{2 D_{C/q}(z,
%k_{\perp})} \simeq \frac{k_\perp}{k_\perp^2+M^2}\,{\rm Im} (AB^*),
%\ee
%which then could be significantly different from zero.
%Notice that $A,B$ are complex scalar coefficients of the dressed quark
%propagator ($A=B=1$ for an on-shell quark).
%\end{comment}

This idea was then extended to the study of $A_N$ in $p^\uparrow
p\to\pi +X$  collisions.
In the  factorized approach discussed above,
the numerator of the SSA generated
by the Collins effect would read
\be
d\Delta\sigma^{\rm Collins} \propto \sum_{a,b,c} h_{1a}(x_a) \otimes
 f_{b/p}(x_b) \otimes
d\Delta\hat\sigma^{ab \to c d}(x_a,x_b,\bm{k}_\perp) \otimes  \Delta \hat
D_{\pi/c^\uparrow}(z, \bm{k}_\perp)\,,
\label{theo:ancol}
\ee
where $d\Delta\hat\sigma^{ab \to c d}$ is the partonic double
transverse spin
asymmetry related to the partonic spin transfer, $D_{NN} =
d\Delta\hat\sigma/d\hat\sigma$.
Notice once again that only the leading $\bfk_\perp$-dependence is
kept.

A first phenomenological attempt to explain the SSA in terms
of such effect was presented in Ref.~\cite{Artru:1995bh}.
The azimuthal asymmetry in
the fragmentation process was generated via the Lund string mechanism
\cite{Andersson:1983ia}.
According to this model, in the breaking of the string spanned
between the  scattered quark  and the target, the quark
and the antiquark of every pair created
acquire a transverse momentum (w.r.t.~the direction of the string)
and a polarization which are strongly correlated. Namely,
the antiquark polarization goes like
$
\bm{P}_{\bar q} \simeq - \hat{\bm{z}} \times \bm{k}_\perp
$
where $\hat{\bm{z}}$ is the unit vector along the string direction.

In order to form a pion, the polarized scattered quark (coming from the hard scattering) has
to combine with the antiquark (belonging to the string pair) into a spin singlet. This
implies a correlation between the $q$ and $\bar q$ polarizations,
 resulting in an asymmetric azimuthal distribution of the pion
w.r.t.~the quark polarization. In this model the Collins effect is introduced only for
leading, or ``first-rank'', pions (those containing the original quark spanning the string):
the azimuthal distribution of higher-rank (subleading) pions is assumed to be symmetric. It
is worth to mention that even starting from Eq.~(\ref{theo:ancol}), that is a pQCD inspired
formula, the authors of Ref.~\cite{Artru:1995bh} adopted a partonic spin-transfer $D_{NN}$
equal to one (as it is for very small scattering angles) and a simple power-like
$p_T$-dependent partonic cross-section based on the observed inclusive pion spectra. By
assuming that the transverse polarization of the $u$ and $d$ quarks in the proton are $+1$
and $-1$ at $x=1$, respectively, they obtained qualitative agreement with the E704 data,
although underestimating their size.

Among the peculiar features of this model we recall that for spin-1 meson (e.g. the $\rho$)
production the spin-$\bfk_\perp$ correlation gives a very different Collins effect; as a
consequence, $A_N(\rho)\simeq -(1/3) A_N(\pi)$ \cite{Czyzewski:1996ih}. No SSA is expected
for $K^-$ production (which has no valence quark in common with the polarized proton) since
the subleading fragmentation mechanism is assumed to be symmetric.

A more detailed phenomenological study of the Collins effect in $pp$ collisions, based on the
factorized scheme of Eq.~(\ref{theo:ancol}), was performed in Ref.~\cite{Anselmino:1999pw},
where the authors adopted the helicity formalism with inclusion of TMD effects. We postpone
the details of this approach to section~\ref{form}. Here we only recall that by a reasonable
parameterization of the Collins function and the transversity distribution a good description
of the E704 data was obtained. As we will discuss in section~\ref{pheno} a complete treatment
of the kinematics, taking into account the full $\bm{k}_\perp$ dependences in the factorized
formula of Eq.~(\ref{theo:ancol}), plays a big role and leads to a strong revision of this
conclusion. More generally we will see how the correct $\bfk_\perp$ kinematics is crucial in
the TMD approach to inclusive particle production.

Almost in the same years, a first %complete and comprehensive
systematic classification of the twist-two and
twist-three transverse momentum
and spin dependent parton distribution and fragmentation functions was presented in
Refs.~\cite{Mulders:1995dh,Boer:1997nt}.
%The authors extended this expansion up to twist-three level;
%introducing new TMD functions besides the three entering the
%collinear fatorization:
%$g_T(x)$, $e(x)$ and $h_L(x)$.
For the twist-three case these functions have
no partonic interpretation since they
are related to quark-quark-gluon matrix elements.
Further improvements beyond leading
twist in the above classification were reached in
Ref.~\cite{Metz:2004je}.
This work has been recently
reassessed and completed in Ref.~\cite{Bacchetta:2006tn} where some
new developments (see also Refs.~\cite{Bacchetta:2004zf,Goeke:2005hb})
have been taken into account; in particular,
a systematic presentation of the relevant observables for all combinations of
lepton and hadron polarizations at small transverse momentum and twist-three accuracy has
been given. For space reasons, and given its relevance, in the
following we will restrict to the leading-twist TMD sector. We refer the reader to the above
mentioned papers for a more detailed discussion.

The starting point in this classification is the correlator of quark fields ($\psi$) for a
polarized nucleon (with spin $S$) entering the DIS processes in the diagrammatic expansion of
the hard scattering amplitude,
\be \label{theo:phi-ij} \Phi_{ij}(P,S,k) = \int \frac{d^4y}{(2\pi)^4} e^{i k\cdot y} \langle
P,S|\bar\psi_j(0) \psi_i(y) |P,S\rangle\,, \ee and the analogous correlator for the
fragmentation sector \be \label{theo:delta} \Delta_{ij}(P_h,S_h,k) = \sum_X \int
\frac{d^4y}{(2\pi)^4} e^{i k\cdot y} \langle 0| \psi_i(y)|P_h,S_h,X\rangle\langle P_h,S_h,X|
\bar\psi_j(0) |0\rangle\,. \ee

By integrating over $k^-= (k^0-k^3)/\sqrt{2}$, where $k^\mu=xP^\mu +k_\perp^\mu$ and
$k_\perp^\mu=(0, \bm{k}_\perp, 0)$, and keeping the leading twist terms, one gets, for the
distribution sector,
 \bea \Phi(x,\bm{k}_{\perp}, S) &=& \frac{1}{2} \left[
 f_1\rlap{/}{n}_+ +
f_{1T}^\perp \frac{\epsilon _{\mu\nu\rho\sigma}
\gamma^\mu n_+^\nu k_{\perp }^\rho
S_T^\sigma}{M} +
\left( S_L g_{1L} +
\frac{\bm{k}_{\perp }\cdot\bm{S}_T}{M} g_{1T} \right)
\gamma ^5 \rlap{/}{n}_+
\right. \nonumber \\ &+&
\left. h_{1T} \,i\sigma_{\mu\nu} \gamma^5 n_+^\mu S_T^\nu +
\left( S_L h_{1L}^\perp + \frac{\bm{k}_{\perp }
\cdot\bm{S}_T}{M}h_{1T}^\perp\right)
\frac{i\sigma_{\mu\nu} \gamma^5 n_+^\mu k_{\perp }^\nu}{M} %\right.
%\nonumber \\
%&+&\left.
+ h_1^\perp \frac{\sigma _{\mu\nu} k_{\perp }^\mu n_+^\nu}{M}
\right]\,, \label{theo:phi}
\eea
where $n_{\pm}$ are auxiliary light-like vectors and $M$ is the
nucleon mass.
All functions above depend on $x$ and $|\bm{k}_\perp|$.
Few words on the above notation are helpful: $f$, $g$ and $h$ stand
for unpolarized, longitudinally polarized, and transversely
polarized quarks, respectively:
the subscript ``1'' stands for leading twist; the apex $\perp$
indicates the explicit presence of transverse momenta with a
noncontracted index~\cite{Mulders:1995dh}; the
subscripts $L$, $T$ stand for the longitudinal and transverse
polarization of the hadron. Therefore, for instance, $f_1$ is the TMD
unpolarized PDF, $f_{1T}^\perp$ is the Sivers function and $h_{1T}$,
$h_{1T}^\perp$ are related to
the TMD transversity distribution.
For relations among different notations see also
Refs.~\cite{Bacchetta:2004jz,Barone:2001sp,Anselmino:2005sh}.

By appropriate Dirac projections, $\Phi^{[\Gamma]} = {\rm Tr}(\Gamma \Phi)$, one can single
out the various sectors of distribution functions. In particular, $\Gamma=(n_-)_\alpha
\gamma^\alpha/2$ projects out the $f_1$ sector, and $\Gamma=\frac{1}{2} i
\sigma_{\mu\nu}(n_-)^\mu \frac{(S_T)^\nu}{2} \gamma^5$ gives the $h_1$ sector. Among the
others we only mention, in particular, the function $h_1^\perp$, also known as the
Boer-Mulders function, which has been proposed in Ref.~\cite{Boer:1999mm} as a possible tool
to explain the azimuthal asymmetry observed in unpolarized Drell-Yan processes.

The partonic meaning of the eight functions appearing in Eq.~(\ref{theo:phi}) and their
relations with those defined in a complementary formalism, the helicity approach,
 will be discussed in more detail in section~\ref{form}.

The same structure can be obtained for the fragmentation sector,
starting from the corresponding correlator, Eq.~(\ref{theo:delta}). In
this case one gets, among the others,
the Collins function, $\Delta^N D_{C/q^\uparrow}$,
and a chiral-even TMD fragmentation function
of an unpolarized quark into a transversely polarized
hadron (denoted consistently as $\Delta^N D_{C^\uparrow/q}$).
This function, also known as the {\em polarizing} FF,
could play a role
 in describing the
observed transverse polarization of $\Lambda$ hyperons
produced in unpolarized $pp$
collisions~\cite{Anselmino:2000vs}.

For the sake of clarity we give here the relation between two standard notations for the most
relevant TMD functions  appearing in the sequel (the Sivers and Boer-Mulders distributions;
the Collins and polarizing fragmentation functions): \bea \Delta^N
f_{q/p^\uparrow}(x,|\bfk_\perp|) = - \frac{2 |\bfk_\perp|}{M}\, f_{1T}^{\perp
q}(x,|\bfk_\perp|) && \Delta^N f_{q^\uparrow/p}(x,|\bfk_\perp|) = - \frac{|\bfk_\perp|}{M}\,
h_{1}^{\perp q}(x,|\bfk_\perp|)\nonumber\\
\Delta^N D_{h/q^\uparrow}(z,|\bfk_{\perp h}|) = \frac{2 |\bfk_{\perp h}|}{zM_h}\,
H_{1}^{\perp q}(z,|\bfk_{\perp h}|) &&
\Delta^N D_{h^\uparrow/q}(z,|\bfk_{\perp h}|) =  \frac{|\bfk_{\perp h}|}{zM_h}\,
D_{1 T}^{\perp q}(z,|\bfk_{\perp h}|)\,.
\label{theo:notation}
\eea
It is also customary to consider $k_\perp$-moments of the TMD
distributions. The most relevant ones in the PDF sector are defined as
(similar relations hold also for the FF case, see e.g.~Eq.~(\ref{pheno:deltaDz}))
%($ \Delta^N f_{q/p^\uparrow}= -\frac{2k_\perp}{M_N} f_{1T}^{\perp}$)
%
\be
\label{theo:siv:transv:mom}
        f_{1T}^{\perp(1)}(x) \equiv \int\!{\rm d}^2{\bfk}_\perp\;
        \frac{\bfk_\perp^2}{2 M^2}\;f_{1T}^{\perp}(x,|\bfk_\perp|)
        \quad\quad
f_{1T}^{\perp (1/2)}(x)\equiv \int d^2 \bfk_\perp\frac{|\bfk_\perp|}{2M}
f_{1T}^{\perp }(x,|\bfk_\perp|) \ .
\ee

\subsubsection{QCD developments}
\label{theo:develop}

An important and clear improvement of the TMD approach,  in particular concerning the
distribution sector, came with the discovery of a mechanism to generate a transverse spin
asymmetry at leading twist~\cite{Brodsky:2002cx}. Here, by considering the deeply inelastic
lepton-proton scattering process, the authors showed that final-state interactions due to
gluon exchange between the outgoing quark and the target spectator system lead to single spin
asymmetries at leading twist in pQCD; {\em i.e.}, the rescattering corrections are not
power-law suppressed. The existence of such SSA's requires a phase difference between two
amplitudes coupling the proton target with $J^z_p = \pm {1\over 2}$ (where $J_p$ is the total
proton angular momentum) to the same final state. These same amplitudes are necessary to
produce a nonzero proton anomalous magnetic moment. The calculation of
Ref.~\cite{Brodsky:2002cx} was performed in a field-theoretic model in which QCD, with
massive quarks, is supplemented by a coloured scalar diquark field and an elementary proton
field. Soon afterwards this mechanism was reanalyzed by Collins  \cite{Collins:2002kn} who
proved that it is compatible with factorization and due to a spin asymmetry in the
$\bfk_\perp$ distribution of quarks in a transversely polarized hadron: the ``Sivers
asymmetry''. The earlier statement by Collins, that the Sivers asymmetry has to vanish
because of time-reversal invariance, gets invalidated by the path-ordered exponential of the
gluon field in the operator definition of TMD parton densities. What the time-reversal
argument shows is instead that the Sivers distribution is reversed in sign in Drell-Yan
processes w.r.t.~the SIDIS case. This result, which would imply a violation of naive
universality of parton densities, offers a clear test of our understanding of the TMD
approach in pQCD.

The origin of this sign change can be understood by looking at the gauge
link entering the correlator.
Therefore it is crucial to start with
the properly defined gauge-invariant
parton density for a hadron with momentum
$P$ and transverse spin $S$ ($\uparrow,\downarrow$),
that has the following operator definition
\begin{eqnarray}
  \label{theo:pdfdef}
%\hat  f_{q/A^\uparrow}(x, \bm{k}_\perp)
%  &=& \int \frac{dy^- \, d^2 \bm{y}_T }{(2\pi)^3}
%    e^{-ixp^+y^- + i\bm{k}_\perp\cdot\bm{y}_T}
%\,
%   \langle P,S| \bar\psi(0,y^-,\bm{y}_T) W_{y\infty}^{\dag}
%     \frac{\gamma^+}{2}
%     W_{0\infty} \psi(0) |P,S\rangle.
\hat  f_{q/A^\uparrow}(x, \bm{k}_\perp)
  &=& \int \frac{d\xi^- \, d^2 {\bm{\xi}}_\perp }{(2\pi)^3}
    e^{-ixp^+\xi^- + i\bm{k}_\perp\cdot\bm{\xi}_\perp}
\,
   \langle P,S| \bar\psi(0,\xi^-,{\xi}_\perp) W_{\xi\infty}^{\dag}
     \frac{\gamma^+}{2} W_{0\infty} \psi(0) |P,S\rangle\,.
\end{eqnarray}

Here, light-front coordinates are used:
%$\xi^\mu =(\xi^+,\xi^-,\xi_\perp) =
$\xi^{\pm}=(\xi^0 \pm \xi^z)/\sqrt2$ and
$\xi_\perp^\mu = (0,\bm{\xi}_\perp, 0)$.
The symbol $W_{\xi\infty}$ indicates a Wilson-line operator, also called
gauge-link, going out from the point $\xi$ to future infinity.
These links are path-ordered exponentials, defined
as~\cite{Ji:2002aa, Belitsky:2002sm}
 \be
W_{\xi\infty} = {\cal P} \exp \Big(-ig_s \int_{\xi^-}^\infty \!
dz^-\,\hat{A}^+ (z^-, \xi_\perp)  \Big)
\cdot {\cal P} \exp \Big(-ig_s \int_{\xi_\perp}^\infty \! dz_\perp
\cdot\hat{A}_{\perp} (z^- =\infty, z_\perp) \Big) \, ,
\label{theo:link}
\ee
where $\hat{A^\mu}=\sum_a t^a\,A_a^\mu$, with
$t^a=\lambda^a/2$ being the Gell-Mann colour matrices.

In particular, in the light-cone gauge ($\hat A^+=0$) the first piece in
Eq.~(\ref{theo:link}) reduces to one and we are left with the pure transverse gauge link,
depending on the nonvanishing vector potential at infinity. As extensively discussed in
Ref.~\cite{Belitsky:2002sm}, this link in the transverse direction is crucial to maintain the
gauge invariance of TMD parton distributions under residual gauge transformations; at the
same time, it is responsible for the final(initial) state interactions in
SIDIS(DY)~\cite{Brodsky:2002cx,Collins:2002kn}. In nonsingular gauges, like covariant gauges,
where the vector potential at infinity vanishes, in Eq.~(\ref{theo:link}) only the gauge link
along the light-like direction, $z^\mu\simeq(0,\bm{0},z^-)$,  survives. In this case, as
discussed in Ref.~\cite{Collins:1981uk} and reassessed in Ref.~\cite{Collins:2003fm} severe
light-cone divergences arise from contributions of virtual gluons with vanishing light-cone
plus momentum. A standard way to cut them off is the use of a Wilson line slightly out of the
light-like direction ($z^+\sim 0$) ~\cite{Collins:1989gx}. Notice that this complication does
not affect model calculations of TMD distributions at lowest order.

The original proof on the vanishing of the Sivers
function~\cite{Collins:1992kk}, obtained
applying space- and time-reversal to the quark fields in the
operator definition of the parton densities, was then incomplete,
because it ignored the presence of Wilson lines.
Under time-reversal the future-pointing Wilson lines are replaced by
past-pointing Wilson lines
%see Fig.~\ref{theo:figlink},
so that the correct version
of the proof gives~\cite{Collins:2002kn}
\begin{equation}
  \label{theo:proof}
\hat f_{q/A^\uparrow}(x, \bm{k}_\perp) |_{{\rm future-pointing}~W}
=
%  P(x, \trans{k}, \trans{s}, \zeta) |_{{\rm future-pointing}~W}
%  = P(x, \trans{k}, -\trans{s}, \zeta) |_{{\rm past-pointing}~W}
\hat f_{q/A^\downarrow}(x, \bm{k}_\perp)|_{{\rm past-pointing}~W} \,.
\end{equation}

Note the change in direction of the transverse spin-vector ($\uparrow$
goes into $\downarrow$).  Since
the past-pointing Wilson lines are appropriate for factorization in
the Drell-Yan process \cite{Collins:2002kn, Brodsky:2002rv},
the correct result is not that the Sivers asymmetry, Eq.~(\ref{theo:sivasy}),
vanishes, but that it has opposite signs in DIS and in Drell-Yan:
\begin{equation}
  \label{theo:disdy}
\Delta \hat f_{q/A^\uparrow}(x, \bm{k}_{\perp}) |_{\rm DIS} =
- \Delta \hat f_{q/A^\uparrow}(x, \bm{k}_{\perp}) |_{\rm DY} \,.
% f^\perp_{1T}(x,k_T,\zeta)|_{\rm DIS} = -f^\perp_{1T}(x,k_T,\zeta)|_{\rm DY} .
\end{equation}

The TMD spin-independent parton distribution keeps its universality.

A complementary study for the fragmentation sector has been developed in
Refs.~\cite{Metz:2002iz, Collins:2004nx}, where the  authors compare SIDIS processes and
$e^+e^-$ annihilation into two hadrons. The relevant result is that spin and $\bfk_\perp$
dependent fragmentation functions are universal, also in sign. This finding has played a
crucial role in the first extraction of the transversity distribution from a global analysis
of the azimuthal asymmetries observed in $e^+e^-\to\pi\pi +X$ and SIDIS (see
sections~\ref{pheno:epem}, \ref{pheno:sidis:ssa}).

The role of the gauge links in TMD distributions has then been further and systematically
investigated in a series of papers \cite{Ji:2002aa, Belitsky:2002sm, Ji:2004wu, Ji:2004xq,
Idilbi:2004vb, Idilbi:2005er},  which focus mainly on the gauge invariant definition of TMD
distributions and the proper factorization in QCD at leading twist for SIDIS and Drell-Yan
processes at low transverse momentum. We only mention here that according to the $k_\perp$
factorization for SIDIS, in the convolution of $k_\perp$ dependent PDF's and FF's with the
hard scattering parts another term appears: a soft factor coming from soft gluon radiations
and defined by a matrix element of Wilson lines in QCD vacuum \cite{Ji:2004wu,
Collins:2004nx}. One of its effects is that the transverse momentum of the observed hadron in
SIDIS could be generated not only
by %by a three
%sources:
the intrinsic motion of the quarks in the nucleon and in the
quark fragmentation process but also via soft gluon radiation.
Notice that, at the present stage,
all phenomenological studies have neglected this extra unknown factor.

In Ref.~\cite{Idilbi:2004vb} the energy evolution of the spin and TMD distributions is also
discussed. The authors, following Ref.~\cite{Collins:1981uk}, show how the Collins-Soper
equation in the impact parameter space can be applied also to these distributions. This
equation, intimately related to the regularization of the light-cone singularities of TMD
distributions in covariant gauges, allows to resum the large logarithms arising in
perturbative calculations for SIDIS (or DY) process at low  transverse momentum.

Another interesting issue related to the TMD azimuthal asymmetries is the
potential suppression due to Sudakov factors coming from soft gluon
radiation. This problem has been analyzed
in Ref.~\cite{Boer:2001he} for a class of processes, like the
Drell-Yan process or two-hadron production in $e^+e^-$ annihilation,
where two energy scales (for DY:  the transverse momentum of the
lepton pair, $q_T$, and its invariant mass, $Q$)
are present. In order to extend the TMD
factorization picture beyond the region of very small $q_T$ (where
tree-level formulas are sufficient) to
the region where the transverse momentum becomes moderate (still
$q_T\ll Q$) one has to include the Sudakov factors by proper
resummation. This implies a broadening of the $q_T$ distribution and a
suppression effect, that becomes more important with rising
energy~\cite{Boer:2001he}.

The scale dependence, at LO, of the first moments of the TMD distribution
and fragmentation functions has been studied in
Ref.~\cite{Henneman:2001ev}, by using Lorentz invariance and the QCD
equations of motion in the large $N_c$ limit.
A phenomenological study on the
$Q^2$-evolution (via resummation of soft and collinear parton
emissions at LO) of unpolarized TMD distributions in SIDIS has been presented
in Refs.~\cite{Ceccopieri:2005zz,Ceccopieri:2007ek}.

Before completing this short overview of the TMD approach we want to mention the recent study
of the gauge links in more complicated processes, like $pp \to h_1 h_2 + X$, where the
elementary scattering involves many subprocesses and hadrons are present both in the initial
and final state.

Indeed, in the simplest hadronic scattering processes, such as SIDIS, DY and
$e^+e^-$-annihilation, only a limited number of different gauge-link structures appear. These
are the future and past-pointing Wilson lines mentioned above. When going beyond these
processes one may encounter more complex gauge-link structures. In
Refs.~\cite{Bomhof:2004aw,Bacchetta:2005rm} the paths that can appear in general scattering
processes have been analyzed (some technical aspects are presented in
Ref.~\cite{Bomhof:2006dp}). It was then shown  that the presence of gauge links can lead to
modified hard parts, referred to as \emph{gluonic pole cross sections}, for SSA's in
$p^\uparrow p{\rightarrow}\pi\pi +X$~\cite{Bacchetta:2005rm, Bomhof:2006ra}. By taking
suitable $k_\perp$-moments of the SSA's (with proper weights) the net effect coming from
these highly non-trivial structures is to modify the usual sum of Feynman diagrams
representing the standard partonic cross sections into gauge invariant weighted sums, with
$SU(3)$ colour factors as weights (analogous to the $\pm 1$ factors discussed above for SIDIS
and DY).

Very recently the double inclusive hadron (or jet) production in hadron-hadron
collisions with a small $q_T$ imbalance, has received a special
attention, in particular concerning the validity of the $k_\perp$
factorization.
In Ref.~\cite{Collins:2007nk} by adopting a simplified
Abelian model, Collins and Qiu argue that
 factorization is violated, even in the extended version
of the TMD approach including gluonic pole cross sections.
On the other hand, another approach \cite{Qiu:2007ar, Qiu:2007ey},
based on a careful analysis of one-gluon radiation
contributions, led apparently to the opposite result.
Some light on this apparent discrepancy has been shed in
Refs.~\cite{Collins:2007jp, Vogelsang:2007jk} and further clarified in
Ref.~\cite{Bomhof:2007xt},
showing a mutual consistency among the different approaches.

Summarizing, the main findings are: $i)$ by expanding the gauge link in the correlators
entering the colour gauge invariant approach~\cite{Bacchetta:2005rm, Bomhof:2006ra} at first
order in the coupling constant one recovers the results of Refs.~\cite{Qiu:2007ar,
Qiu:2007ey}; $ii)$ by extending, and properly adapting, the calculations of
Refs.~\cite{Qiu:2007ar, Qiu:2007ey} to two-gluon exchange, a non factorizable term in the TMD
unpolarized cross section appears, in agreement with the results found in
Ref.~\cite{Collins:2007nk, Collins:2007jp}. This extra piece can be taken into account by
redefining the gauge link entering the TMD parton distribution~\cite{Vogelsang:2007jk}. A
similar complication (i.e.~a non factorizable hard part, or equivalently, a redefinition of
the gauge link entering the TMD distribution) for the SSA (Sivers effect) is expected at the
next order of perturbation theory. In Ref.~\cite{Bomhof:2007xt}, by assuming factorization at
the diagrammatic level, the authors obtain a manifest gauge invariant expression for the TMD
cross sections in terms of gluonic pole cross sections and TMD distributions with
process-dependent Wilson lines. In this sense the TMD distributions are in general non
universal. In this approach, beside confirming the results of Refs.~\cite{Qiu:2007ar,
Qiu:2007ey}, gluonic pole cross sections in TMD unpolarized cross sections as well as
ordinary partonic cross sections in unweighted SSA's arise together with
universality-breaking parts, in agreement with Refs.~\cite{Collins:2007jp,Vogelsang:2007jk}.
However, these pieces are shown to vanish for the integrated and weighted observables
considered in Refs.~\cite{Bacchetta:2005rm,Bomhof:2006ra,Bacchetta:2007sz,Bomhof:2007su}.
The identification of these process dependent pieces %, i.e.~the fact that they can be calculated,
could be an important step towards a unified picture of TMD
factorization for hadronic processes. Further work is still needed.

\subsubsection{The impact parameter picture}
\label{theo:impact}

This physical picture, mostly applied to the Sivers effect,
has been discussed in a series of papers by Burkardt and
collaborators~\cite{Burkardt:2002ks,Burkardt:2003je,Burkardt:2005hp,Burkardt:2003yg}.
It is based on two main ingredients: the distortion in the transverse
plane of the distribution of partons inside a polarized target and the
attractive(repulsive) final(initial) state interactions in SIDIS(DY).
The first aspect is strongly related to the information one can
extract from the generalized parton distributions (GPD).

GPD's are generalizations of ordinary parton distributions to non-forward matrix elements of
a lightlike correlation function. They can be accessed for instance in deeply virtual Compton
scattering, $\gamma^*p\to\gamma p'$, where a finite momentum is transferred to the proton.
GPD's depend on two longitudinal momentum fractions ($x$ and $x+\xi$, where $\xi$ is the
so-called skewness parameter) and on the invariant momentum transfer to the proton,
$t=\Delta^2$. In the simpler case where $\xi=0$ the Fourier transform of the GPD w.r.t.~$t$
gives the parton distribution in position space,
$q(x,\bm{b})$, where %$x$ is the longitudinal momentum fraction and
$\bm{b}$ is the impact parameter giving the transverse
distance of the struck quark from the center of mass of the proton.

When the proton is transversely polarized (for instance along the $+Y$ axis) two GPD's
describe the distribution of unpolarized quarks, namely $H(x,0,t)$ (related to the
unpolarized PDF) and $E(x,0,t)$, whose Fourier transform gives the distortion of the
distribution.

By relating the integral of $E_q(x,0,0)$ to the $q$-flavour contribution ($\kappa_q$) to the
proton anomalous magnetic moment one can get an estimate of such distortions. On this basis
one can expect an opposite sign of the distortion for $u$ and $d$ quarks, with $u$ quarks
mostly displaced along $-X$. Notice that this left-right asymmetry in $b$-space is T-even, in
contrast with the left-right asymmetry in momentum space.

Let us come to the second ingredient in this picture, by considering SSA's for the production
of $\pi^+, \pi^0$  in SIDIS. Due to charge factors and the fact that $u\to \pi^+, \pi^0$
fragmentation is favoured, most of $\pi^+, \pi^0$ mesons come from an initial $u$ quark in
the polarized proton. We now recall that any left-right asymmetry in the transverse momentum
space at the quark level requires final state interactions (FSI). Intuitively one then
expects that the FSI's are on average attractive, since it costs energy to build up the
string of gauge fields that connects the struck quark with the spectators. More precisely, if
the photon, moving along the $-Z$ axis, collides with a nucleon polarized along the $+Y$
axis, from the previous results we expect that a $u$ quark tends to be displaced in the $-X$
direction in impact parameter space. If the FSI's are attractive,  this quark experiences a
force along the $+X$ direction. This translates into the preferred direction of $\pi^+,
\pi^0$ mesons along $+X$, in agreement with model calculations of the Sivers function as well
its phenomenological extractions. In the Drell-Yan process the initial state interactions,
expected to be repulsive, imply an opposite sign for the Sivers function.

%\subsubsection{Models for TMD distributions}
\subsubsection{Models and constraints for TMD distributions}
\label{theo:models}

The Sivers function, being a naively T-odd entity,
requires the interference
between two complex amplitudes with different phases.  Spectator models at
tree level cannot provide these nontrivial phases. However they can
arise as soon as a gluon is exchanged between the struck quark and the target
spectator~\cite{Brodsky:2002cx}.
In other words the presence of the gauge link, which
insures the colour gauge invariance of parton distributions, at the
same time could provide the nontrivial, relevant, phases.

\begin{figure}[h!]
\begin{center}
\begin{minipage}[t]{14 cm}
%\hspace{1cm}
\includegraphics[width=6cm, angle=-90]{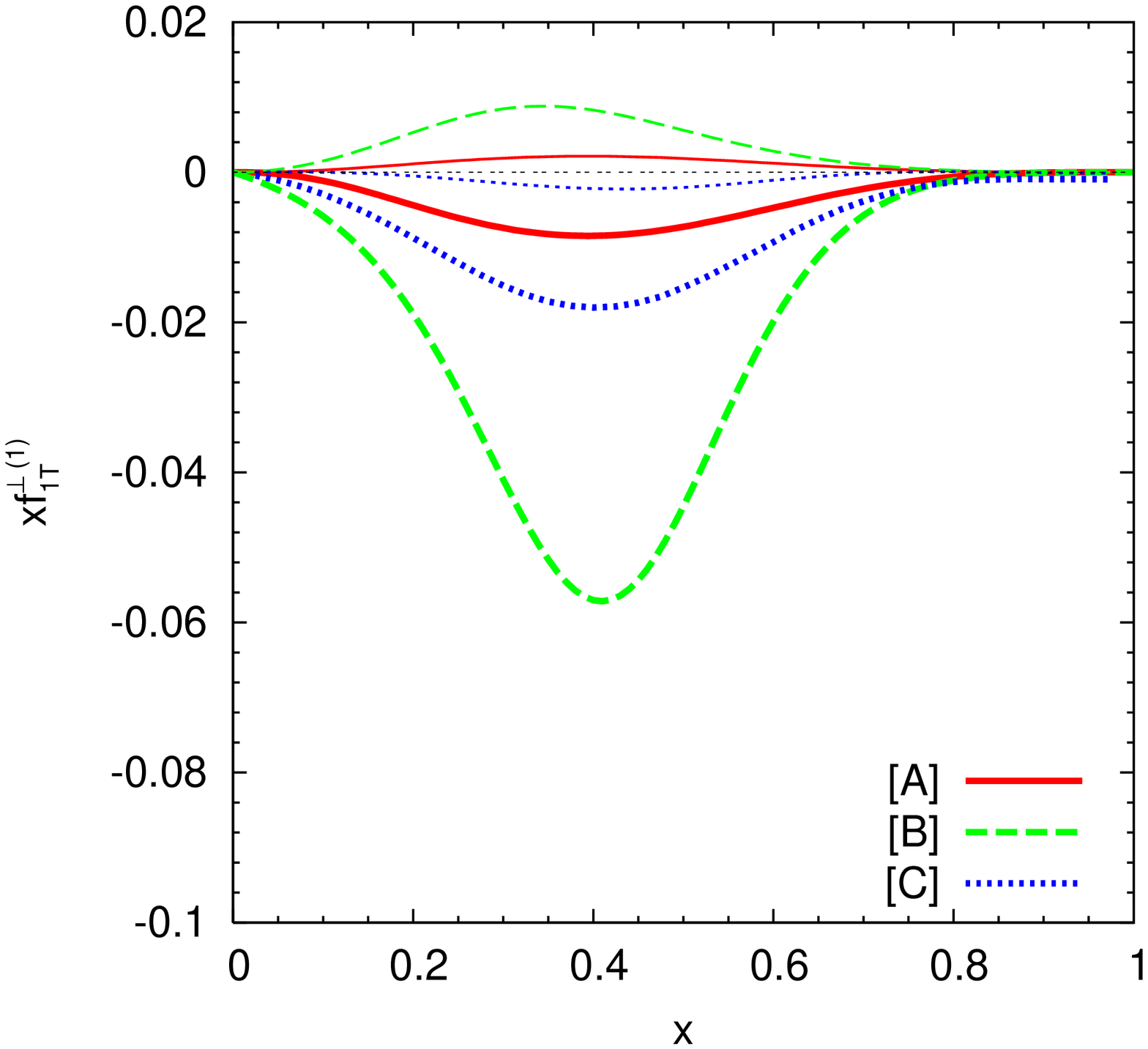}
\hspace{1cm}
\includegraphics[width=6cm, angle=-90]{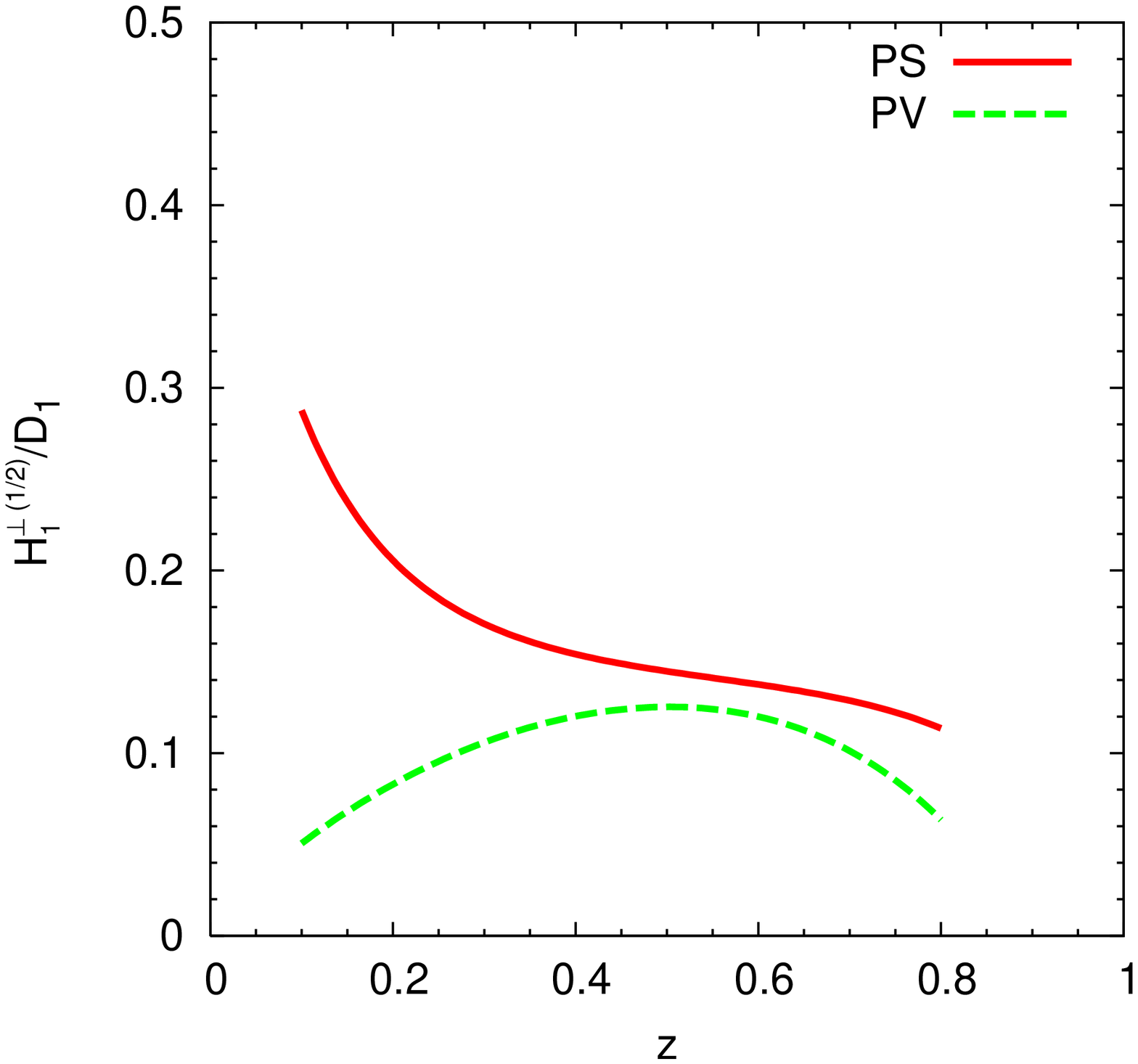}
%
%(b)\hspace{0.cm}\includegraphics[width=7cm]{figures/.ps}
\end{minipage}
\begin{minipage}[t]{16.5 cm}
\caption{Left panel: Comparison of models calculations of $xf_{1T}^{\perp(1)}$ for up (thick
lines) and down (thin lines) quark
  Sivers function according to Refs.~\cite{Yuan:2003wk} [A], \cite{Bacchetta:2003rz}
[B], and \cite{Cherednikov:2006zn} [C]. Right panel: Model calculations of
$H_{1}^{\perp(1/2)}$ for the favoured Collins function normalized to the unpolarized
FF~\cite{Amrath:2005gv}. Results for pseudoscalar (solid line) and pseudovector (dashed line)
pion-quark coupling with gluon loops are shown. \label{theo:siv:model}}
\end{minipage}
\end{center}
\end{figure}

Starting from the work of Brodsky, Hwang and Schmidt (BHS)
\cite{Brodsky:2002cx},
different models for the Sivers function as well as for
 the Boer-Mulders function
have been developed. In Ref.~\cite{Boer:2002ju} the authors extended the BHS calculation,
based on a scalar spectator diquark model of the nucleon with a point-like form factor, to DY
processes. They also computed the Boer-Mulders function for $u$ quarks, finding $h_1^{\perp
u}= f_{1T}^{\perp u}$. The same relation was found in
Refs.~\cite{Goldstein:2002vv,Gamberg:2003ey}, still with a scalar spectator but adopting a
Gaussian form factor. In both cases a negative $f_{1T}^\perp$ for $u$ quarks came out. The
MIT bag model was adopted by Yuan in Ref.~\cite{Yuan:2003wk}, where he still found a negative
Sivers function for $u$ quarks and $f_{1T}^{\perp d} = -(1/4)\, f_{1T}^{\perp u}$ (model [A]
in Fig.~\ref{theo:siv:model}, left panel). Moreover he obtained $h_{1}^{\perp u} = (3/2)\,
f_{1T}^{\perp u}$ and $h_{1}^{\perp d} = -3\, f_{1T}^{\perp d}=(3/4)\,f_{1T}^{\perp u}$. In
Ref.~\cite{Bacchetta:2003rz} the Sivers function has been calculated in a spectator model
with scalar and axial-vector diquarks (model [B] in Fig.~\ref{theo:siv:model}, left panel),
with yet a different choice for the form factor w.r.t.~Ref.~\cite{Gamberg:2003ey}. The $d$
quark Sivers function turns out to have the opposite sign compared to the $u$ quark one and
to be much smaller in size; again $h_1^\perp\simeq f_{1T}^\perp$.

Analogous results have been obtained in Ref.~\cite{Lu:2004au}
both with  scalar and vector diquarks in a light-cone $SU(6)$
quark-diquark model.
The instanton liquid model for  QCD vacuum with
 MIT bag model wave functions
for quarks was adopted in Ref.~\cite{Cherednikov:2006zn} (model [C] in
Fig.~\ref{theo:siv:model}, left panel).
In this model, the two independent perturbative and instanton terms sum up to give
the total contribution to the Sivers function.
For  $d$ quarks the two terms almost cancel
each other, leading to a small, negative Sivers function.
On the contrary, in the case of the
$u$-quark Sivers function, the instanton contribution and
the perturbative one add together in fair agreement with the other models.

{}From these models the emerging common result is the sign of the Sivers function for $u$
quarks. It is worth to stress that while this is in agreement with the phenomenological
extractions, a clear discrepancy is found concerning the overall size. In this respect, the
study in Ref.~\cite{Brodsky:2006hj}, based on the QCD counting rules, could help in putting
useful constraints on the Sivers function, which should be one power of $(1-x)$ suppressed
relative to the unpolarized PDF.
Concerning the Boer-Mulders function new results have
appeared~\cite{Gamberg:2007wm,Burkardt:2007xm}. With the exception of a previous
calculation~\cite{Bacchetta:2003rz}, all models, including the new findings in the diquark
spectator model~\cite{Gamberg:2007wm} predict a negative Boer-Mulders function for both $u$
and $d$ quarks. The same conclusion has been reached in Ref.~\cite{Burkardt:2007xm} through
the calculation of the chirally odd GPD $\bar E_T$ (which has opposite sign
w.r.t.~$h_1^\perp$~\cite{Burkardt:2005hp}) in different models both for nucleon and pion
targets.

Let us now consider the Collins function. Four model calculations for the fragmentation of a
transversely quark into a pion have been presented so far in the  literature
\cite{Bacchetta:2001di, Bacchetta:2002tk, Gamberg:2003eg, Bacchetta:2003xn} (besides the
former attempts by Collins~\cite{Collins:1992kk} and Artru~\cite{Artru:1995bh}, discussed
above). All of them produce the necessary imaginary parts by adopting a simple model for the
fragmentation process at tree level and inserting one-loop corrections. Two possibilities for
the tree-level amplitudes and two possible kinds of one-loop corrections have been
investigated, for a total of four different models: pseudoscalar pion-quark coupling with
pion loops~\cite{Bacchetta:2001di} and with gluon loops~\cite{Gamberg:2003eg,Amrath:2005gv};
pseudovector pion-quark coupling with pion loops~\cite{Bacchetta:2002tk} and with gluon
loops~\cite{Bacchetta:2003xn}. For gluon loop cases see Fig.~\ref{theo:siv:model} (right
panel). We refer the reader to Ref.~\cite{Amrath:2005gv} for further details.

In the above-mentioned approaches disfavoured fragmentation functions (e.g.~$d\to\pi^+$)
vanish, something apparently not supported by data, see section~\ref{pheno}. In principle,
these functions could be calculated in these models by considering diagrams with the emission
of two pions, one of which goes unobserved.

The different results obtained for the Collins function do not allow to draw any conclusion
even about its sign. The fact that more than one diagram with a different structure
contributes to the Collins function makes it also difficult, if not impossible, to interpret
the resulting effect, as we do for the Sivers function, in terms of simple
attractive/repulsive interactions.

Some useful constraints on TMD distribution and fragmentation functions in terms of bounds
and sum rules are available. In Ref.~\cite{Bacchetta:1999kz} by considering the combined
Dirac $\otimes$ target spin space and requiring the positivity of the spin-matrix
eigenvalues, they found, beside the well-known Soffer bound~\cite{Soffer:1994ww} ($2|h_1|\le
(f_1+g_{1L})$), new bounds on the eight TMD distributions entering the correlator in
Eq.~(\ref{theo:phi}).

The first sum rule for a T-odd TMD function was discussed in Ref.~\cite{Schafer:1999kn}: the
authors considered the Collins function and, by imposing conservation of the intrinsic
transverse momentum in parton fragmentation, obtained
 \be
 \label{theo:Col:rule}
\sum_h\int dz\, z\, H_1^{\perp (1)q}(z) = \sum_h\int dz \int d^{\,2}\bm{k}_\perp \,
\frac{|\bm{k}_\perp|}{M_h}\;\Delta^{N}D_{h/q^\uparrow}(z,|\bm{k}_\perp|) = 0\, ,
 \ee
where $H_1^{\perp (1)q}(z)$ is defined analogously as in Eq.~(\ref{theo:siv:transv:mom}) and
the sum is over all hadrons.

More recently a somewhat corresponding sum rule for the Sivers function has been derived by
Burkardt~\cite{Burkardt:2004ur}. This rule states that the net transverse momentum due to the
Sivers mechanism must vanish; more formally
\begin{equation}
 \langle\bm{k}_{\perp}\rangle =
 \sum_{a=q,\bar q,g} \;\langle\bm{k}_{\perp}\rangle_a =
 \int\!dx \int d^{\,2}\bm{k}_\perp \,\bm{k}_\perp
 \sum_{a=q,\bar q,g} \Delta \hat{f}_{a/p^{\uparrow}}(x,\bm{k}_\perp) = 0 \, .
 \label{theo:Siv:rule}
\end{equation}
Even if intuitively expected, the nontrivial fact is its validity in presence of final state
interactions, which might spoil the simple partonic interpretation. The Burkardt sum rule
(BSR) has been also checked explicitely in the diquark model calculations of the Sivers
function~\cite{Goeke:2006ef}. Its role in SSA phenomenology will be discussed in
Section~\ref{pheno}.

\subsection{Twist-three effects in collinear pQCD}
\label{Qiu}

The origin of this approach could be traced back to the early paper by Kane et
al.~\cite{Kane:1978nd}, where it was shown how in a collinear
QCD parton model any single spin asymmetry
should be proportional to the quark mass. Although
 this result cannot explain the large values of $A_N$
observed, it was read as a signal of a higher twist effect. Another step towards this
interpretation was developed by Efremov and
Teryaev~\cite{Efremov:1981sh,Efremov:1983eb,Efremov:1984ip}
%~\cite{Efremov:1994dg}
who pointed out that a nonvanishing SSA can be obtained in pQCD beyond the leading power
expansion. However, the first consistent calculation in collinear pQCD of a sizable SSA in
the forward region (large $x_F$) was given by Qiu and Sterman in their paper on hadronic
direct photon production~\cite{Qiu:1991pp,Qiu:1991wg} and their subsequent application to
pion production~\cite{Qiu:1998ia}. The asymmetries are evaluated in the framework of pQCD
generalized factorization theorems with the introduction of new twist-three quark-gluon
correlator functions convoluted with ordinary twist-two parton distribution functions and a
short-distance hard scattering part. The new functions, being expectation values between the
hadronic states of three field operators,
 do not have a simple partonic interpretation. More precisely,
for the numerator of $A_N$ in the process $A^\uparrow B \to C + X$ ($C$ unpolarized) they
obtain, neglecting higher power corrections (we keep here the original
notation of Ref.~\cite{Qiu:1998ia}),
\bea
\label{theo:htw1}
d\Delta\sigma_{A^\uparrow B\to C+X} & = & \sum_{abc}\phi_{a/A^\uparrow}^{(3)}(x_1,x_2)\otimes
\phi_{b/B}(x')\otimes H_{a+b\to c}\otimes D_{c\to C}(z)\nonumber\\
&+&
\sum_{abc}\delta q_{a/A}^{(2)}(x)\otimes
\phi_{b/B}^{(3)}(x_1',x_2')\otimes H^{''}_{a+b\to c}\otimes D_{c\to
 C}(z) \\
&+&
\sum_{abc}\delta q_{a/A}^{(2)}(x)\otimes
\phi_{b/B}(x')\otimes H^{'}_{a+b\to c}\otimes D_{c\to
  C}^{(3)}(z_1,z_2)\nonumber\,,
\eea
where  $a,b,c$ represent parton flavours ($q,\bar q, g$), and
$\phi_{b/B}(x')$ and $D_{c\to C}(z)$ are ordinary twist-two parton
distribution and fragmentation functions, respectively; $\delta
q_{a/A}^{(2)}(x)$ is the twist-two transversity distribution function.
The new pieces, namely twist-three functions (see apexes), depend on
two light-cone variables and can give three different contributions, from
the initial polarized hadron, the initial unpolarized hadron, the
final unpolarized hadron.
As in ordinary pQCD factorization, only the hard scattering functions
$H$  are calculable and their %explicit
forms depend on the definition
of the twist-three distributions.

In order to present this formalism we consider in some detail the first term (first
line) in Eq.~(\ref{theo:htw1}).

Keeping in mind that the asymmetries are larger in the forward region,  one can restrict to
the valence flavour approximation for $a$ in $\sum_{abc}$, and then consider only a valence
twist-three distribution. Another important simplification comes from the fact that the
$H_{a+b\to c}$ part involves two contributions: one proportional to $\delta(x_1-x_2)$ and
another one proportional to $\delta(x_i)$. The first case sets the gluon momentum in the
correlator to zero; in the second case, one of the quarks carries a vanishing momentum. These
are referred to as {\it soft gluon} and {\it soft fermion} poles respectively. Moreover,
while soft gluon poles enter with derivatives of the twist-three distribution, soft fermion
ones do not. If, as for ordinary twist-two functions, the derivatives of twist-three
functions are enhanced near the edges of phase space, then soft gluon poles could give the
leading contribution to the large asymmetries for $x_F\to 1$.

To get the master formula in this approach one has to start with the factorized expression in
a diagrammatic representation of the process \be d\Delta\sigma  \approx \sum_a\int\!\frac{d^4
k_1}{(2\pi)^4} \frac{d^4k_2}{(2\pi)^4} T_a(k_1,k_2)S_a(k_1,k_2) \,, \ee where $k_{1,2}$ are
valence quark momenta, $T_a$ is proportional to the non perturbative matrix element of the
quark-quark-gluon operator between polarized initial hadron states (a twist-three matrix
element) and $S_a$ refers to the rest of the process. Expanding $S_a$ in the collinear
approximation enables to reduce the four-dimensional integrals to convolutions in the
momentum fractions of partons with $k_i\simeq x_iP$. Notice that in this approach a nonzero
spin dependence is found only from pole terms in the hard scattering. Without them the
symmetries of the strong interactions would forbid any SSA. Keeping the first nonvanishing
term in $d\Delta\sigma$, that is the leading pole structure of $S_a$, one gets \be
d\Delta\sigma  \approx \sum_a\int\! dx_1 dx_2\, i\epsilon^{\rho s_T n\bar n} \frac{\partial
S_a}{\partial k_2^\rho}(x_1,x_2) T_{a,F}(x_1,x_2)\,, \ee where $n,\bar n$ are light-like unit
vectors and
\be
T_{a,F}(x_1,x_2) =
\int\! \frac{dy_1^- dy_2^-}{4\pi} \,
     \mbox{e}^{ix_1P^+y_1^- + i(x_2-x_1)P^+ y_2^-}
\langle P,\bm{s}_T |\bar{\psi}_a(0)\gamma^+
 \left[\epsilon^{s_T\sigma n \bar{n}}\, F_{\sigma}^{\ +}(y_2^-)\right]
 \psi_a(y_1^-) |P,\bm{s}_T\rangle \ ,
\label{s3e18}
\ee
with $F_{\sigma}^{\ +}$ the gluon field strength.
Since $T_{a,F}$ is real only the poles of $S_a$ can contribute.

The last step is to
factorize the remaining function $S_a$
into a perturbatively calculable partonic term,
$H_{a+b\rightarrow c}$, a corresponding target parton
distribution, $\phi_{b/B}$, and a fragmentation function,
$D_{c\rightarrow C}$:
\begin{eqnarray}
S_a(k_1,k_2)
&\approx & \sum_{ bc}\int\! \frac{dx'}{x'} \,
\phi_{b/B}(x')\,\int\! dz\,
            H_{a+b\rightarrow c}(k_1,k_2,x',p_c)\,
            D_{c\rightarrow C}(z)\ ,
\end{eqnarray}
where $\sum_{bc}$ runs over all parton flavours.

Finally, the  full factorized expression for $d\Delta\sigma$  reads
 \be d\Delta\sigma
= \frac{1}{2s} \sum_{abc}
\!\int\!\! dz\, D_{c\rightarrow C}(z)
\!\int\! \frac{dx'}{x'}\, \phi_{b/B}(x')
\!\int\!\! dx_1 dx_2\, T_{a,F}(x_1,x_2)
\!\left[
 i\epsilon^{\rho s_T n \bar{n}}\,
 \frac{\partial}{\partial k_{2}^{\rho}}
      H_{a+b\rightarrow c}(x_1,x_2,x',z) \right]_{k_2^\rho=0} \ ,
\label{s3e21}
\ee
where the integration over either $x_1$ or $x_2$
can be done by using the poles in $H_{a+b\rightarrow c}$.

Notice that this results in a factorization formula with only
a single momentum fraction for each  hadron,
similar to that for the spin-averaged cross section
with $\phi_{a/A}(x)$ replaced by
$T_{a,F}(x,x)$.

A way to understand why $T_{a,F}$ is expected to be
the dominant contribution is the following.
{}From the sources of the $k_i$ ($i=1,2$) dependence
in $H_{a+b\rightarrow c}(k_1,k_2,x',p_c)$ (on-shellness of the
unobserved final parton $d$, on- and off-shellness of the
propagators and the explicit dependence in the
numerator)  we get two kinds of terms:
one proportional to $(\partial/\partial x) T_{a,F}(x,x)$
and one proportional to $T_{a,F}(x,x)$.

One has to remember that this approach was intended to address the description of the
asymmetries in the forward region, where $x_F$ is large. Here the dominant contribution comes
from large net momentum fraction $x$ for the polarized beam hadron, coupled with relatively
small momentum fraction $x'$ from the partons of the unpolarized target hadron.  Since all
distributions vanish for large $x$ as $(1-x)^{\beta}$, with $\beta>0$, $(\partial/\partial
x)T_{a,F}(x,x) \gg T_{a,F}(x,x)$ when $x\rightarrow 1$.  Therefore, in the forward region,
terms proportional to derivatives of the distributions $T_{a,F}$ dominate. Moreover it is
only the matrix element $T_{a,F}$ that inherits derivative terms, as a result of the
collinear expansion involving soft gluon poles. Soft fermion poles have no such derivatives
at LO and also do not correspond to the valence quark approximation. In this spirit only the
derivative term was kept and the corresponding hard part was calculated.

A recent development of this approach has been discussed in Ref.~\cite{Kouvaris:2006zy}.
Still neglecting the soft fermion poles, the complete soft-gluon pole structure, including
the non derivative term, has been computed. The main result of this work is that one can
incorporate the two terms  in a simple way since they have a common hard scattering part.
This comes from a remarkable property of the hard scattering functions recently proven by
Koike and Tanaka \cite{koike:2007rq}. On the other hand the phenomenological role of this
extra piece in the forward region is found to be negligible, as assumed in  former papers.

Without giving further  details of the calculation (see
\cite{Qiu:1998ia,Kouvaris:2006zy}),
we only mention that by
properly exploiting the pole structure of $H_{a+b\rightarrow
c}(k_1,k_2)$  (initial and final-state poles),
one gets for the numerator of the SSA, restoring all factors,
\begin{eqnarray}
\label{theo:finalcr} E_C\frac{d^3\Delta\sigma}{d^3\bm{p}_C} &=& \frac{\alpha_s^2}{s}\,
\sum_{a,b,c}
    \int_{z_{\rm min}}^1 \frac{dz}{z^2}\, D_{c\to C}(z)
   \, \int_{x'_{\rm min}}^1 \frac{dx'}{x'}\, \frac{1}{x's + t/z} \,
\phi_{b/B}(x') \\
&\times& \sqrt{4\pi\alpha_s}\,
    \left(\frac{\epsilon^{p_C s_T n \bar{n}}}{z\hat{u}}\right)
\,   \frac{1}{x}
\left[T_{a,F}(x,x)-x\left(\frac{d}{dx}T_{a,F}(x,x)\right)\right]
    H_{a+b\rightarrow c}(\hat{s},\hat{t},\hat{u})\ ,
\nonumber
\end{eqnarray}
where $s,t,u$ ($\hat s,\hat t,\hat u$) are the standard hadronic (partonic) Mandelstam
variables, %$x\equiv -(x'u/z)/(x's+t/z) $
 and $x$, $x'_{\rm min}$, $z_{\rm min}$ are fixed by the usual elastic
 scattering condition.

The $H_{a+b\rightarrow c}$ are the final hard-scattering functions
and read
\be
H_{a+b\rightarrow c}\,=\,
H^I_{a+b\rightarrow c}(\hat{s},\hat{t},\hat{u}) +
        H^F_{a+b\rightarrow c}(\hat{s},\hat{t},\hat{u}) \,\left(1+
        \frac{\hat{u}}{\hat{t}}\right) \, ,
\label{theo:Hfin} \ee where $H^I_{a+b\rightarrow c}$ ($H^F_{a+b\rightarrow c}$) denotes the
contributions due to  initial (final) -state interactions. An important feature is that they
are given in terms of the same Feynman diagrams needed to calculate the usual unpolarized
partonic cross sections for the $ab\to cd$ process but with different colour factors due to
the extra initial (final) state interactions, see Appendix of Ref.~\cite{Kouvaris:2006zy}.
Two aspects of the final result in Eq.~(\ref{theo:finalcr}) deserve a word: the denominator
$\hat u$ gives explicitly the power suppression of the SSA and the extra factor
$\sqrt\alpha_s$ w.r.t. the unpolarized cross section comes from the additional interaction
with a gluon field.

It is also worth to mention that both in the original and the more recent
works the contributions involving
the three-gluon twist-three correlation function, which would enter the
$gg$ partonic process and could probably be relevant in the mid-rapidity
region, have been ignored.

In order to perform a phenomenological study one has to rely on some parameterizations of the
$T_{a,F}$ twist-three functions. Since the operator defining it contains the same quark field
operators as the ordinary twist-two parton distribution functions, the most natural ansatz
is: \be T_{a,F}(x,x) = N_a(x) \phi_a(x) \,. \ee In
Refs.~\cite{Qiu:1991pp,Qiu:1991wg,Qiu:1998ia} a simple functional form with $N_a(x) =
\kappa_a \lambda$ was adopted, where $\lambda$ has the dimension of a mass and
$\kappa_{u,d}=\pm 1$ (for a proton). In the recent developments of this formalism
\cite{Kouvaris:2006zy}, with more data having become available, a more general form has been
adopted, i.e. $N_a(x) = N_a x^{\alpha_a}(1-x)^{\beta_a}$. In both cases the main features of
the data for $A_N$ in pion production in $pp$ as well as in $\bar pp$ collisions can be
described rather well. Notice, however,  that all these papers use leading order, leading
twist expressions for the unpolarized cross section in the denominator of $A_N$.

This approach, applied to SSA's in prompt photon production,
by adopting the same twist-three functions as extracted from
the analysis of pion data (positive for $up$ and negative for $down$
quarks)  gives a sizable negative asymmetry.
This result, as we will discuss in section~\ref{pheno},
could be of extreme relevance in
discriminating between the GPM approach, based on TMD distribution and
fragmentation functions, and the twist-three formalism.

A complementary study has been also devoted to the role of the other two sources of SSA's in
$pp\to \pi + X$ in the twist-three approximation, namely the chiral-odd contributions in
Eq.~(\ref{theo:htw1}) (second and third lines). The twist-three distribution,
$\phi_{b/B}^{(3)}(x_1',x_2')$ and its relative hard scattering parts have been discussed in
Refs.~\cite{Kanazawa:2000hz,Kanazawa:2000kp}.
This contribution, expected to be relevant in the
negative $x_F$ region, comes out to be negligible over all ranges in
$x_F$ due to the smallness of
the hard scattering partonic asymmetry.
In Ref.~\cite{Koike:2002gm} a detailed analysis
of the chiral-odd twist-three fragmentation function  was performed,
showing how, with some difficulties, a description of E704 data can be obtained.

Remaining in the context of inclusive particle production in hadron-hadron collisions, this
approach has been also applied to the study of the transverse $\Lambda$ polarization in
unpolarized $pp$ collisions. In this case two twist-three effects could contribute: the
twist-three chiral-odd distribution $\phi_{a/p}^{(3)}(x_1,x_2)$,
 coupled with a twist-two  unpolarized PDF and a twist-two chiral-odd
fragmentation function (analogous to the twist-two transversity PDF); a twist-three
chiral-even FF, that, following the same notation adopted in
Eq.~(\ref{theo:htw1}), would read $D_{c\to\Lambda^\uparrow}^{(3)}(z_1,z_2)$, coupled with two
twist-two unpolarized PDF's. The first contribution has been studied in
Ref.~\cite{Kanazawa:2000cx}, where by adopting a simple ansatz on  $\phi_{a/p}^{(3)}(x_1,x_2)$,
namely a proportionality relation with the chiral-odd transversity distribution, the main
features of the data are reproduced, although some discrepancies still remain.

The twist-three formalism has been also applied to SIDIS azimuthal asymmetries
\cite{Eguchi:2006qz,Eguchi:2006mc} at large $p_T\simeq Q$ (with $Q$ the virtuality of the
exchanged photon). In these papers the authors show that among the two possible mechanisms,
the first coming from the twist-three chiral-even distribution and the second from the
twist-three chiral-odd fragmentation (coupled with the transversity PDF), the last one is
negligible. In particular they calculated the soft-gluon pole contributions for both cases.
They also improved their calculation for the first and most relevant case, identifying all
pole contributions (soft-fermion and hard-gluon poles included) and providing a systematic
collinear expansion for this approach in SIDIS.

Before concluding this section we mention some attempts to connect the TMD approach to the
twist-three formalism. A first study has been presented in Ref.~\cite{Boer:2003cm}, showing
that the inclusion of transverse gauge links, which lead to $T$-odd effects, is connected to
the Qiu-Sterman mechanism. More precisely, it was shown that the first $k_\perp$ moment of
the Sivers function is equal to the twist-three chiral-even distribution, $T_{a,F}$. This
connection has been also addressed in Ref.~\cite{Ma:2003ut}.

Recently a unified picture in terms of the twist-three approach and the TMD
formalism~\cite{Ji:2006ub,Ji:2006vf,Ji:2006br,Koike:2007dg} both in SIDIS and Drell-Yan
processes has been presented. The main result is that although the two mechanisms have their
own domain of validity, they describe the same physics in the kinematic region where they
overlap. More precisely, focusing on SIDIS for the sake of clarity,  when the transverse
momentum of the observed hadron, $p_T$, and the photon virtuality, $Q$, are much larger than
$ \Lambda_{\rm QCD}$, the spin-dependent cross section can be calculated in terms of a
twist-three quark-gluon correlation. On the other hand, when $p_T\simeq \Lambda_{\rm QCD}\ll
Q$, single spin asymmetries can be generated from a spin-dependent TMD quark distribution.
There is, however, a common kinematic region, $\Lambda_{\rm QCD} \ll p_T \ll Q$, where both
mechanisms should work. In this region $p_T$ is large, so that the asymmetry is a twist-three
effect, but at the same time $p_T \ll Q$, so that the TMD factorization formalism also
applies. The same arguments work for the Drell-Yan process by simply replacing the $p_T$ of
the observed hadron with the transverse momentum of the lepton pair. The resulting connection
unifies the physical pictures for the underlying dynamics of transverse SSA's and imposes
important constraints on phenomenological studies.

We finally stress that both in the TMD  and in the twist-three approach, on-shell massless
partons are considered in the soft as well as in the hard parts. The role of off-shellness,
still largely unknown, has been recently emphasized and investigated in
Refs.~\cite{Watt:2003vf,Linnyk:2006mv,Collins:2007ph} (see also references therein).

%\subsection{Nonperturbative approaches}
\subsection{A semiclassical model}
\label{theo:np}

Among the nonperturbative models proposed to describe the
SSA's observed in inclusive hadron production in
hadron-hadron collisions we mention a picture based on a semiclassical
geometrical approach: the {\em orbiting valence quark model}
\cite{Liang:2000gz, Liang:1992hw, Boros:1993ps, Liang:1993rz, Boros:1995yt,
Boros:1996ar, Dong:2003ch}.
The main ideas behind this approach can be summarized as follows:
$i)$ The constituents
of a polarized hadron are assumed to perform an orbital motion about
the polarization axis.
Their relativistic ground state wave functions
 allow to calculate the polarization of valence quarks in a
polarized proton. The results show that there is a flavour dependent asymmetry in
the valence-quark polarization. The sea quarks come out unpolarized.
$ii)$
In the fragmentation region the main contribution to hadron
production comes from quark-antiquark annihilation via a
direct-formation mechanism.
$iii)$ Since hadrons are extended objects, a surface effect is
expected in such production processes.

Let us consider a proton-proton collision along the $Z$ axis in the rest frame of the
polarized proton. In the transversely upward ($+X$) polarized hadron, $u$ quarks, being on
average polarized along the hadron polarization, have an orbital motion such that they go
preferably left ($-Y$) at the front surface and right ($+Y$) at the back surface of the
hadron. When the polarized target interacts with the unpolarized projectile this preferred
direction is preserved at the front surface (when the projectile and the target start
overlapping) while it is partially spoiled at the back surface (when they start separating
again) because of the colour interactions acting during the transition of the projectile
through the target. The net result is that in the fusion mechanism of a $u$ quark from the
polarized proton and a suitable antiquark from the unpolarized proton, the formed meson goes
preferably left, i.e. along $-Y$. More generally this means that, when a polarized proton
collides, only colour singlet $q \bar q$ pairs formed at the front surface can get an extra
momentum due to the orbiting motion of valence quarks. This determines the observed asymmetry
in the distribution of the produced mesons.

In this model one then expects that for $p^\uparrow p \to \pi + X$ in the
fragmentation region of the polarized proton, where mainly valence quarks
contribute, $\pi^+$ and $\pi^0$ mesons go left, whereas $\pi^-$'s go right. For
the case of a polarized antiproton  $\pi^+$'s and $\pi^-$'s reverse
their direction, whereas $\pi^0$'s keep going left. The left-right
asymmetry is more significant at large $x_F$. According to this
mechanism, also Drell-Yan lepton pairs  should exhibit a left-right asymmetry.

One of the main features of this model is that in order to have a
nonzero SSA the produced hadron must have at least a common valence
quark with the polarized hadron.
In particular, for kaon production in $p^\uparrow p$
collisions one expects that for $K^+(u\bar s)$ $A_N$ is similar (in
size and sign) to $A_N(\pi^+)$, whereas for $K^-(\bar u s)$ it should vanish.

This model has been also applied to the description of the transverse polarization of
hyperons [$P_T(Y)$] produced in unpolarized hadron-hadron
collisions~\cite{Boros:1995fv,Liang:1997rt}.

For a detailed and comprehensive description of this approach and its
phenomenological applications see Ref.~\cite{Liang:2000gz} and
references therein.

%% file: formalism-rev.tex
\section{Spin asymmetries in the generalized parton model}
\label{form} In this section we will present and discuss in detail
the generalized, spin and $\bm{k}_\perp$ dependent, parton model
approach to polarization effects in high-energy hadronic collisions.
As discussed in section~\ref{theo}, there are several approaches
available in the literature. In what follows we will adopt the one
developed in a series of papers, starting from the middle of 90s.
This approach is an extension of early attempts to deal with parton
intrinsic motion and unintegrated, transverse momentum dependent
(TMD) parton distribution (PDF) and fragmentation functions (FF). It
consistently includes spin and polarization effects, requiring the
introduction of a new class of leading twist TMD distributions. Like
any theoretical approach, the formalism we will use has both
advantages and drawbacks. The main advantages are the following: a)
It is a QCD-improved, generalized parton model approach, with direct
inclusion, in the spirit of the parton model itself, of spin and
transverse momentum ($\bm{k}_\perp$) effects; b) It retains the
simple partonic interpretation of all soft, leading-twist TMD
distributions; c) Using the helicity formalism, it very clearly
shows the connection among the polarization states of the particles
involved and the role of spin effects in the soft and hard
processes; d) It fully includes kinematical $\bm{k}_\perp$ effects,
without approximations. It has of course also some drawbacks: a) For
some of the processes considered, factorization has not been proven
explicitly yet. It is rather taken as a reasonable assumption for
this generalized scheme; b) It only accounts for leading-twist soft
distributions (in this sense its application to subleading-twist
SSA's can be considered as an effective approach); c) It does not
properly account for initial and final state interactions, which
have been shown to be crucial for the nonvanishing of naively
time-odd PDF's and FF's. Despite these problems, its
phenomenological success is presently remarkable. In our opinion it
is also the most convenient approach for a simple illustration of
the pQCD-based formalisms and of their main phenomenological
applications. In section~\ref{pheno} we will discuss and compare the
results and predictions of this approach with available experimental
results and with those of other theoretical approaches. In
particular, we will emphasize measurements which could help in
disentangling among different formalisms and are at reach of present
experiments.

\subsection{Polarized cross sections and helicity formalism}
 \label{form:hel}

Let us consider, to be definite, the doubly polarized hadronic process $A(S_A)+B(S_B)\to C+X$
at high energy and moderately large transverse momentum, $p_T$. Here $p_T$ is the transverse
momentum of the observed hadron with respect to the direction of the colliding beams. $A$,
$B$, $C$ are generic hadrons, respectively with spin state $S_A$, $S_B$, $S_C$ ($S_h=$ 0,
1/2, 1). Using the helicity formalism the invariant differential cross section for the
production of an unpolarized particle in the polarized processes under consideration can be
expressed through the following {\it master formula}:
\bea \label{form:genunp} && \frac{E_C \, d\sigma^{(A,S_A) + (B,S_B) \to C + X}} {d^{3}
\bm{p}_C} = \sum_{a,b,c,d, \{\lambda\}} \! \int \frac{dx_a \, dx_b \, dz}{16 \pi^2 x_a x_b
z^2  s} \;
 d^2 \bm{k}_{\perp a} \, d^2 \bm{k}_{\perp b}\, d^3 \bm{k}_{\perp C}\, \delta(\bm{k}_{\perp C}
\cdot \hat{\bm{p}}_c) \, J(\bm{k}_{\perp C}) \\
&&
\times\, \rho_{\lambda^{\,}_a,\lambda^{\prime}_a}^{a/A,S_A} \,
 \hat f_{a/A,S_A}(x_a,\bm{k}_{\perp a}) \> \rho_{\lambda^{\,}_b, \lambda^{\prime}_b}^{b/B,S_B} \,
 \hat f_{b/B,S_B}(x_b,\bm{k}_{\perp b})\, \hat M_{\lambda^{\,}_c, \lambda^{\,}_d; \lambda^{\,}_a,
\lambda^{\,}_b} \, \hat M^*_{\lambda^{\prime}_c, \lambda^{\,}_d;
\lambda^{\prime}_a,\lambda^{\prime}_b} \> \delta(\hat s + \hat t + \hat u) \> \hat
D^{\lambda^{\,}_C,\lambda^{\,}_C}_{\lambda^{\,}_c,\lambda^{\prime}_c}(z,\bm{k}_{\perp C}) \>.
\nonumber
 \eea
Notice that with generic $\lambda^{\,}_C, \lambda^{\prime}_C$ helicities (i.e.~replacing
$\sum_{\lambda_C} D^{\lambda^{\,}_C,\lambda_C}_{\lambda^{\,}_c,\lambda^{\prime}_c}$ with
$D^{\lambda^{\,}_C,\lambda^{\prime}_C}_{\lambda^{\,}_c,\lambda^{\prime}_c}$) on the r.h.s.~of
the above equation, we get the (unnormalized) helicity density matrix for the final particle
$C$ that allows to evaluate its polarization state.  This is particularly relevant for the
case of transverse hyperon polarization in unpolarized hadronic collisions. However, since
here we want to focus on the basic ideas of the approach, for simplicity we will consider the
production of unpolarized particles in the collisions of spin 1/2 hadrons. A detailed
treatment of spin $1/2$ particle production using the helicity formalism in a generalized
parton model is under completion~\cite{Dalesio:2007u}. We will shortly discuss the formalism
for the spin 1/2 case in section~\ref{form:lambda}, and the phenomenology for transverse
$\Lambda$ polarization in section~\ref{pheno:lam}.

\indent Eq.~(\ref{form:genunp}) gives the cross section for the polarized hadronic process
under consideration as a factorized convolution of all possible hard elementary QCD
processes, $ab\to cd$, with soft, spin and $\bm{k}_\perp$ dependent PDF's and FF's. Let us
briefly comment on the notation and physical content of this equation. Further details can be
found in the original
papers~\cite{Anselmino:2005sh,D'Alesio:2004up,Anselmino:2004ky} and in references therein.\\
(i) $A$ and $B$ are initial, spin 1/2 hadrons in pure spin states denoted by $S_A$ and $S_B$
and corresponding polarization (pseudo)vectors $\bm{P}^A$ and $\bm{P}^B$, respectively;\\
(ii) Unless differently stated, we always consider the c.m.~frame of the initial
colliding hadrons, the $AB$ c.m.~frame or hadronic c.m.~frame; in this
frame, hadron $A$, $B$,
move respectively along the $\pm{Z}_{\rm{cm}}$-axis direction and the final hadron $C$ is
produced in the $({X}{Z})_{\rm{cm}}$ plane, with $(p_C)_{{X}_{\rm{cm}}}>0$;\\
(iii) The notation $\{\lambda\}$ indicates a sum over all helicity indices; $x_a$, $x_b$ and
$z$ are the usual light-cone momentum fractions of partons $a,b$ inside the initial hadrons
$A,B$ and of hadron $C$ produced in the fragmentation of parton $c$; $\bm{k}_{\perp a}$
($\bm{k}_{\perp b}$) are the (two-dimensional) transverse momenta of parton $a$($b$) with
respect to hadron $A$($B$); $\bm{k}_{\perp C}$ is the generic three-momentum of hadron $C$ in
the hadronic c.m.~frame; notice that the delta function $\delta(\bm{k}_{\perp
C}\cdot\hat{\bm{p}}_c)$ constrains $\bm{k}_{\perp C}$ to be
effectively two-dimensional and always orthogonal to parton $c$ three-momentum, $\bm{p}_c$;\\
(iv) The factor $(16\pi^2x_a x_b s)^{-1}$, where $s=(p_A+p_B)^2$ is the total energy of the
colliding beams, collects phase space factors related to the elementary cross sections and
the corresponding flux factor for the noncollinear (when viewed in the hadronic c.m. frame)
parton scattering process;\\
(v) The phase-space factor $J(\bm{k}_{\perp C})/z^2$ is the invariant TMD Jacobian factor for
massless particles connecting the parton momentum $\bm{p}_c$ with the hadron momentum
$\bm{p}_C$, (see e.g. Appendix~A of Ref.~\cite{D'Alesio:2004up});\\
(vi) $\rho_{\lambda^{\,}_a,\lambda^{\prime}_a}^{a/A,S_A}$ is the
helicity density matrix of parton $a$ inside hadron $A$ with spin
state $S_A$, and analogously for
$\rho_{\lambda^{\,}_b,\lambda^{\prime}_b}^{b/B,S_B}$; notice that
the helicity density matrix always describes the polarization state
of the particle in its helicity rest frame; for a massless particle
no rest frame exists and the helicity frame is defined as the
standard frame, reached from the hadronic c.m. frame, in which the
particle four-momentum is $p^{\mu}=(p,0,0,p)$~\cite{Leader:2001gr};\\
(vii) $\hat f_{a/A,S_A}(x_a,\bm{k}_{\perp a})$ is the spin and TMD distribution function of
the unpolarized parton $a$ inside the polarized hadron $A$, and similarly for parton $b$
inside hadron $B$; these leading-twist TMD distributions generalize the usual partonic
distributions in the collinear configuration; notice that in this paper, for the sake of
clarity, we adopt the following convention: for \emph{all} $\bm{k}_\perp$ dependent
distribution and fragmentation functions and soft amplitudes we will use a ``hat'' to
indicate full, vectorial dependence on $\bm{k}_\perp$, while we will not use it for
dependence on $k_\perp=|\bm{k}_\perp|$, with factorized azimuthal dependence; notice also
that this notation is partially at variance with those adopted in the papers where the
original formulation of the GPM was developed, which by the way slightly differ
from paper to paper;\\
(viii) The $\hat M_{\lambda^{\,}_c, \lambda^{\,}_d; \lambda^{\,}_a, \lambda^{\,}_b}$'s are
the helicity amplitudes for the elementary process $ab\to cd$, opportunely normalized
according to the kinematical phase-space factors explicitly shown in Eq.~(\ref{form:genunp});\\
(ix)
$\hat{D}^{\lambda^{\,}_C,\lambda^{\prime}_C}_{\lambda^{\,}_c,\lambda^{\prime}_c}(z,\bm{k}_{\perp
C})$ is the product of nonperturbative \emph{fragmentation amplitudes} for the polarized
fragmentation process $c\to C + X$.

We will now consider in more detail the quantities which describe how the polarization of the
initial hadrons is transferred, through the soft and hard elementary processes, to the final
partons and to the produced hadron (if its polarization state is observed).
\subsection{Spin and TMD parton distribution functions at leading twist}
\label{form:distr}
\indent The quantity
$\rho_{\lambda^{\,}_a,\lambda^{\prime}_a}^{a/A,S_A}\,\hat
f_{a/A,S_A}(x_a,\bm{k}_{\perp a})$ in Eq.~(\ref{form:genunp})
encodes complete information on the polarization state of parton $a$
with spin $s_a$, as determined by the polarization state of the
parent hadron $A$ (which is fixed by the experimental conditions
considered) and by the soft, nonperturbative processes related to
the hadron structure (analogously for parton $b$ inside hadron $B$).
Indeed, introducing soft, nonperturbative \emph{helicity amplitudes}
for the inclusive process $A\to a+ X$, $\hat {\cal
F}_{\lambda^{\,}_a,\lambda_{X_A};\lambda^{\,}_A}(x_a,\bm{k}_{\perp_a})$,
we can write:
 \be
 \rho_{\lambda^{\,}_a, \lambda^{\prime}_a}^{a/A,S_A} \> \hat f_{a/A,S_A}(x_a,\bm{k}_{\perp a})
= \sum _{\lambda^{\,}_A, \lambda^{\prime}_A} \rho_{\lambda^{\,}_A,
\lambda^{\prime}_A}^{A,S_A} \sumint_{X_A, \lambda_{X_A}} \!\!\!\!\! \hat{\cal
F}_{\lambda^{\,}_a, \lambda^{\,}_{X_A}; \lambda^{\,}_A} \, \hat{\cal
F}^*_{\lambda^{\prime}_a,\lambda^{\,}_{X_A}; \lambda^{\prime}_A} \equiv \sum
_{\lambda^{\,}_A, \lambda^{\prime}_A} \rho_{\lambda^{\,}_A, \lambda^{\prime}_A}^{A,S_A} \>
\hat F_{\lambda^{\,}_A, \lambda^{\prime}_A}^{\lambda^{\,}_a,\lambda^{\prime}_a}
\>,\label{form:defF}
 \ee
where $\sumint_{X_A, \lambda_{X_A}}$ stands for a spin sum and phase space integration over
the undetected system of hadron $A$ remnants, $\rho_{\lambda^{\,}_A,
\lambda^{\prime}_A}^{A,S_A}$ is in turn the helicity density matrix of hadron $A$,
 \be
 \rho_{\lambda^{\,}_A, \lambda^{\prime}_A}^{A,S_A} = \frac{1}{2}\,{\left(
\begin{array}{cc}
1+P^A_Z & P^A_X - i P^A_Y \\
 P^A_X + i P^A_Y & 1-P^A_Z
\end{array}
\right)}_{\!\!\!A,S_A}= \frac{1}{2}\,{\left(
\begin{array}{cc}
1+P^A_L & P^A_T \, e^{-i\phi_{S_A}}  \\
P^A_T \, e^{i\phi_{S_A}} & 1-P^A_L
\end{array}
\right)}_{\!\!\!A,S_A} \,, \label{form:rho-A}
 \ee
$\bm{P}^A=(P^A_T\,\cos\phi_{S_A},P^A_T\,\sin\phi_{S_A},P^A_L)$ is its polarization vector and
$\phi_{S_A}$ the corresponding azimuthal angle, defined in the helicity reference frame of
hadron $A$. We have also defined
 \be
  \hat F_{\lambda^{\,}_A,
  \lambda^{\prime}_A}^{\lambda^{\,}_a,\lambda^{\prime}_a}
 (x_a,\bm{k}_{\perp a}) \equiv \> \sumint_{X_A,\lambda_{X_A}} \!\!\!\!\!\!
 \hat{\cal F}_{\lambda^{\,}_a,\lambda^{\,}_{X_A};\lambda^{\,}_A}(x_a,\bm{k}_{\perp a}) \,
\hat{\cal F}^*_{\lambda^{\prime}_a,\lambda^{\,}_{X_A}; \lambda^{\prime}_A}(x_a,\bm{k}_{\perp
a}) \>. \label{form:defFF}
 \ee
The product of nonperturbative helicity amplitudes $\hat F_{\lambda^{\,}_A,
\lambda^{\prime}_A}^{\lambda^{\,}_a,\lambda^{\prime}_a}$ has a simple physical interpretation
and is directly related to the leading twist part of the well known hand-bag diagrams for the
hadronic correlator in deeply inelastic scattering processes. The distribution amplitudes
$\hat{\cal F}_{\lambda^{\,}_a,\lambda_{X_A};\lambda^{\,}_A}$ depend on the parton light-cone
momentum fraction $x_a$ and on its intrinsic transverse momentum $\bm{k}_{\perp a}$, with
modulus $k_{\perp a}$ and azimuthal angle $\phi_a$ (all defined in the parent hadron helicity
frame), in the following way:
 \be
 \hat{\cal F}_{\lambda^{\,}_a,\lambda^{\,}_{X_A}; \lambda^{\,}_A}(x_a, \bm{k}_{\perp a}) = {\cal
 F}_{\lambda^{\,}_a,\lambda^{\,}_{X_A}; \lambda^{\,}_A}(x_a, k_{\perp a}) \> {\rm
 exp}[i\lambda^{\,}_A \phi_a] \>.
 \label{form:dampphi}
 \ee
Rotational and parity invariance imply that only eight functions as
defined in Eq.~(\ref{form:defFF}) are independent. Notice
that since some parity relations are different according to the parton type
involved (quark or gluon), the two cases must be treated separately.
For quarks, keeping in mind that upper (lower) helicity indices refer
to the parton (hadron) respectively, one can easily associate the eight functions
 \be
 F^{++}_{++},\quad F^{--}_{++},\quad F^{+-}_{+-},\quad F^{-+}_{+-},\quad
 F^{++}_{+-},\quad F^{--}_{+-},\quad F^{+-}_{++},\quad F^{+-}_{--}\>,
 \label{form:find}
 \ee
to the eight independent, leading twist, polarized TMD quark distribution functions:\\
1) The combinations $F^{++}_{++}\pm F^{--}_{++}$ are respectively related to the
unpolarized and longitudinally polarized PDF's, $f_{a/A}$ and
$\Delta_L f_{a/A}$ (denoted as $f_1$ and $g_{1L}$ in Ref.~\cite{Boer:1997nt});\\
2) The combinations $F^{+-}_{+-}\pm F^{-+}_{+-}$ are in turn related to the two possible TMD
contributions to the transversity PDF, $\Delta_T \hat f_{a/A}$
 (corresponding to $h_{1T}$ and $h_{1T}^\perp$ in Ref.~\cite{Boer:1997nt});\\
3) The combinations $F^{++}_{+-}\pm F^{--}_{+-}$ are respectively related to the probability
of finding an unpolarized (longitudinally polarized) quark inside a transversely polarized
hadron, that is to the well-known Sivers function~\cite{Sivers:1989cc,Sivers:1990fh}
$\Delta^N f_{a/A^\uparrow}$%(x_a,k_{\perp a})
(or %analogous to
$f_{1T}^{\perp}$~\cite{Bacchetta:2004jz}),
 and to the function $\Delta \hat f_{a,s_z/A,S_T}$,
(related to $g_{1T}$ of Ref.~\cite{Boer:1997nt});\\
4) Finally, the combinations $F^{+-}_{++}\pm F^{+-}_{--}$ are related to the probability of
finding a transversely polarized quark respectively inside an unpolarized hadron (that is, to
the Boer-Mulders function~\cite{Boer:1997nt}, $\Delta^N f_{a^\uparrow/A}$ or $h_1^\perp$) or
inside a longitudinally polarized hadron (that is, to the PDF $\Delta \hat f_{a,s_T/A,S_L}$
or $h_{1L}^\perp$~\cite{Boer:1997nt}).

One can also see that:
the functions $F^{\pm\pm}_{++}$, are obviously purely real; the $F^{\pm\mp}_{+-}$ are purely
real (purely imaginary) for quarks (gluons); the last four functions in Eq.~(\ref{form:find})
are complex but not independent, and their sums and differences can be also expressed as the
real and imaginary parts of $F^{++}_{+-}$ and $F^{+-}_{++}$.

Some additional comments are in order here to clarify the physical content of these
functions and summarize their main features: among the eight functions, the only ones that
survive in the collinear configuration (that is, under $\bm{k}_\perp$ integration) are the
three usual leading twist PDF's, $f_{a/A}$, $\Delta_L f_{a/A}$, $\Delta_T f_{a/A}$. Those
which are off-diagonal in the parton helicity indices, like $\Delta_T f_{a/A}$, are chiral
odd; therefore, they cannot be directly measured in DIS processes with on-shell, massless
partons. In any physical observable, they must combine with some other helicity-odd term
(e.g., in SIDIS, with a chiral-odd FF, like the Collins function, or with higher twist terms
due to parton masses and off-shellness); those functions proportional to the imaginary part
of the $F$'s are in turn naively T-odd, and require initial/final state interactions in order
to be nonvanishing~\cite{Brodsky:2002cx,Collins:2002kn} (see also section~\ref{theo}).

The most general expression for the helicity density matrix of quark $a$ inside a %spin 1/2
hadron, $A$, with polarization state $S_A$ is
\be
\rho_{\lambda^{\,}_a, \lambda^{\prime}_a}^{a/A,S_A} = {\left(
\begin{array}{cc}
\rho_{++}^{a} & \rho_{+-}^{a} \\
\rho_{-+}^{a} & \rho_{--}^{a}
\end{array}
\right)}_{\!\!\!\!A,S_A} \!\!\!\!\!\!\! = \frac{1}{2}\,{\left(
\begin{array}{cc}
1+P^a_z & P^a_x - i P^a_y \\
 P^a_x + i P^a_y & 1-P^a_z
\end{array}
\right)}_{\!\!\!\!A,S_A} \!\!\!\!\!\!\! = \frac{1}{2}\,{\left(
\begin{array}{cc}
1+P^a_L & P^a_T \, e^{-i\phi_{s_a}} \\
P^a_T \, e^{i\phi_{s_a}} & 1-P^a_L
\end{array}
\right)}_{\!\!\!\!A,S_A}\!\!\! , \label{rho-a}
\ee
where
$\bm{P}^{a}=(P^a_x,P^a_y,P^a_z)=(P^a_T\cos\phi_{s_a},P^a_T\sin\phi_{s_a},P^a_L)$
is the quark polarization vector and $\phi_{s_a}$ its azimuthal
angle, in the quark helicity frame (to which the $x$,$y$,$z$ axes
refer). In specific calculations, when performing the sum over the
helicity indices $\lambda^{\,}_a$, $\lambda^{\prime}_a$, and
$\lambda^{\,}_b$, $\lambda^{\prime}_b$ in Eq.~(\ref{form:genunp}),
one obtains terms of the form ($j=x,y,z$)
 \be
 P^a_j\,\hat f_{a/A,S_A}=\hat f^a_{s_j/S_A}-\hat f^a_{-s_j/S_A}\equiv \Delta\hat f^a_{s_j/S_A}\,,
 \label{form:delfj}
 \ee
where, as often in the sequel, we do not show the functional dependence of the $\hat f$'s on
$x$ and $\bm{k}_\perp$. These terms refer to polarized quarks inside hadron $A$.
Additionally, we can have also terms corresponding to unpolarized quarks inside polarized
hadrons (the Sivers asymmetry)
 \be
 \Delta \hat f_{a/S_A}=\hat f_{a/S_A}-\hat f_{a/-S_A}\,.
 \label{form:delfSA}
 \ee
Parity and rotational invariance allows this function to be nonzero for transversely
polarized hadrons~only.

By employing Eq.~(\ref{form:defF}), the relations among the functions in
Eq.~(\ref{form:find}) and those defined in Eqs.~(\ref{form:delfj}), (\ref{form:delfSA}) can
be more precisely cast in the following form:
 \bea
 f_{a/A} &=& f_{a/A,S_L} = \left( F^{++}_{++} + F^{--}_{++} \right)\nonumber \\
 \hat f_{a/A,S_T} &=& f_{a/A} + \frac{1}{2}\,\Delta\hat f_{a/S_T} =
 \left( F^{++}_{++} + F^{--}_{++} \right) + 2 \, {\rm Im} F^{++}_{+-}
 \sin(\phi_{S_A} -\phi_a)\nonumber\\
 P_x^a \, f_{a/A,S_L} &=& \Delta f_{s_x/S_L} = 2 \, {\rm Re} F^{+-}_{++}\nonumber\\
 P_x^a \, \hat f_{a/A,S_T} &=& \Delta \hat f_{s_x/S_T} = \left( F^{+-}_{+-} + F^{-+}_{+-} \right) \,
 \cos(\phi_{S_A}-\phi_a)
 \label{form:fff}\\
 P_y^a \, f_{a/A,S_L} &=& P_y^a \, f_{a/A} = \Delta f_{s_y/S_L} = \Delta f_{s_y/A}
 = -2 \, {\rm Im}  F^{+-}_{++} \nonumber\\
 P_y^a \, \hat f_{a/A,S_T} &=& \Delta \hat f_{s_y/S_T} =
\Delta f_{s_y/A} + \Delta^{\!-}\hat{f}_{s_y/S_T} =
 -2 \, {\rm Im}  F^{+-}_{++} + \left( F^{+-}_{+-} - F^{-+}_{+-}
 \right) \, \sin(\phi_{S_A}-\phi_a)
 \nonumber \\
 P_z^a \, f _{a/A,S_L} &=& \Delta f_{s_z/S_L} =
 \left( F^{++}_{++} - F^{--}_{++} \right) \nonumber\\
 P_z^a \, \hat f _{a/A,S_T} &=& \Delta \hat f_{s_z/S_T} =
 2 \, {\rm Re} F^{++}_{+-} \cos(\phi_{S_A} -\phi_a) \,, \nonumber
 \eea
where $\phi_{S_A}$ and $\phi_a$ are the azimuthal angles of the hadron polarization vector
and of the quark transverse momentum $\bm{k}_{\perp a}$ respectively, in the hadronic
c.m.~frame. Notice that $P^a_x f_{a/A}=\Delta f_{s_x/A}=0$, and that the function $\Delta
\hat f_{s_y/S_T}$ can be decomposed into two terms, one independent of the hadron transverse
polarization, $\Delta f_{s_y/A}$, and a term that changes sign when the hadron polarization
is reversed, $\Delta^{\!-}\hat{f}_{s_y/S_T}=(1/2)[\Delta \hat f_{s_y/S_T}-\Delta \hat
f_{s_y/-S_T}]$. Analogous relations hold for quark $b$ inside hadron $B$.

As already said, apart from the well-known PDF's surviving in the collinear configuration,
two of the remaining functions are of particular phenomenological relevance for the study of
azimuthal and single spin asymmetries: the Sivers function, which may be responsible for
several transverse single spin asymmetries; the Boer-Mulders function $\Delta
f_{s_y/A}$ that, besides contributing with additional terms to the
transverse SSA's,
may be responsible
for azimuthal asymmetries in unpolarized processes, e.g.~in Drell-Yan. Appendix C of
Ref.~\cite{Anselmino:2005sh} gives a detailed comparison of the above notation with the one of the
Amsterdam group~\cite{Boer:1997nt},
often adopted in the literature (see also %Ref.~\cite{Bacchetta:2004jz}
Eq.~(\ref{theo:notation})).

For an on-shell massless particle with spin 1 the helicity density
matrix is formally very similar to that of a spin 1/2 particle,
considered above for the quark case. Therefore, most of the
discussion and treatment presented there can be directly translated
to the gluon case. There are two main differences: a) gluons cannot
carry any transverse spin; there is however a formal analogy between
\emph{transversely} polarized quarks and \emph{linearly} polarized
gluons; b) the complete specification of the gluon polarization
state requires the introduction of a polarization tensor $T_{ij}$.
Below, we briefly summarize the results for the gluon sector,
emphasizing where they differ in some respect from the quark case.

The helicity density matrix for an on-shell gluon inside hadron $A$
in a spin state $S_A$ can be cast  as
 \be
\rho_{\lambda_g^{\,}, \lambda^{\prime}_g}^{g/A,S_A}= \frac{1}{2}\,{\left(
\begin{array}{cc}
1+P_{z}^{g} &
{\cal T}_1^g-i{\cal T}_2^g \\
 {\cal T}_1^g+i{\cal T}_2^g & 1-P_{z}^{g}
\end{array}
\right)}_{\!\!\!\!A,S_A} \!\!\!\!\!\!\! = \frac{1}{2}\,{\left(
\begin{array}{cc}
1+ P^g_{circ}&
- P^g_{lin} \, e^{-2i\phi}\\
- P^g_{lin} \, e^{2i\phi} & 1-P^g_{circ}
\end{array}
\right)}_{\!\!\!\!A,S_A} \label{form:rho-gl} \!\!\!.
 \ee

$P_{circ}^g$ corresponds to $P_{z}^g$, the gluon longitudinal polarization. The off-diagonal
elements of Eq.~(\ref{form:rho-gl}) are related to the linear polarization of the gluons in
the transverse $(xy)$ plane at an angle $\phi$ and $\phi+\pi/2$ w.r.t.~the $x$ axis. The $x$,
$y$ and $z$ axes refer to the standard gluon helicity frame, as reached from the hadronic
c.m.~frame, in which its momentum is $p^\mu = (p,0,0,p)$. $P_{lin}^{g}$ can be expressed in
terms of the parameters ${\cal T}_1^g$ and ${\cal T}_2^g$, which are closely related to the
Stokes parameters used in classical optics; they play a role formally analogous to that of
the $x$ and $y$-components of the quark polarization vector in the quark sector. The use of
the parameters ${\cal T}_1^g$ and ${\cal T}_2^g$ makes the gluon distribution functions
formally similar to those for the quarks and considerably simplifies the formulae for the
spin asymmetries.

The nonperturbative helicity amplitudes $\hat {\cal
F}_{\lambda^{\,}_a,\lambda_{X_A};\lambda^{\,}_A}(x_a,\bm{k}_{\perp
a})$ and their products $\hat
F^{\lambda^{\,}_a,\lambda^{\prime}_a}_{\lambda^{\,}_A,\lambda^{\prime}_A}$
now refer to the process $A\to g+X$ and their parity properties are
then slightly different. The rest of the calculation goes along the
same way, i.e. adopting the eight functions in
Eq.~(\ref{form:find}), with the difference that the functions
$F^{+-}_{+-}$ and $F^{-+}_{+-}$ are now purely imaginary quantities.
%
\begin{comment}
The analogue of Eq.~(\ref{form:fff}) is then
%
 \bea
 f_{g/A} &=& f_{g/A,S_L} = \left( F^{++}_{++} + F^{--}_{++} \right)\nonumber \\
 \hat f_{g/A,S_T} &=& f_{g/A} + \frac{1}{2}\,\Delta\hat f_{g/S_T} =
 \left( F^{++}_{++} + F^{--}_{++} \right) + 2 \, {\rm Im} F^{++}_{+-}
 \sin(\phi_{S_A} -\phi_a)\nonumber\\
 {\cal T}_1^g \, f_{g/A,S_L} &=& \Delta f^g_{{\cal T}_1/S_L} =
 \Delta f^g_{{\cal T}_1/A} = 2 \, {\rm Re} F^{+-}_{++}\nonumber\\
 {\cal T}_1^g \, \hat f_{g/A,S_T} &=& \Delta \hat f^g_{{\cal T}_1/S_T} =
 2 \, {\rm Re}  F^{+-}_{++} + {\rm Im}\left( F^{+-}_{+-} + F^{-+}_{+-}
 \right) \, \sin(\phi_{S_A}-\phi_a)
 \nonumber \\
  {\cal T}_2^g \, f_{g/A,S_L} &=& \Delta f^g_{{\cal T}_2/S_L}
 = -2 \, {\rm Im}  F^{+-}_{++} \label{form:ffg}\\
{\cal T}_2^g \, \hat f_{g/A,S_T} &=& \Delta \hat f^g_{{\cal T}_2/S_T} =
  -\,{\rm Im}\left( F^{+-}_{+-} - F^{-+}_{+-} \right) \,\cos(\phi_{S_A}-\phi_a)
 \nonumber\\
 P_z^g \, f _{g/A,S_L} &=& \Delta f^g_{s_z/S_L} =
 \left( F^{++}_{++} - F^{--}_{++} \right) \nonumber\\
 P_z^g \, \hat f _{g/A,S_T} &=& \Delta \hat f^g_{s_z/S_T} =
 2 \, {\rm Re} F^{++}_{+-} \cos(\phi_{S_A} -\phi_a) \,. \nonumber
 \eea
%
\end{comment}

Without giving the detailed expressions of the gluon TMD distributions
(see Ref.~\cite{Anselmino:2005sh}) we only mention
that the Sivers function exists also for gluons and in this approach
maintains the same physical interpretation as for the quark case. There is also an analogue
of the quark Boer-Mulders function, referred to in the sequel as the Boer-Mulders-like gluon
function, ${\cal T}_1^g f_{g/A}=\Delta f^g_{{\cal T}_1/A}$. However, the physical
interpretation of this function is more involved: it represents the difference between the
probabilities of finding, inside an unpolarized hadron, a gluon with linear polarization in
the plane containing the gluon and hadron three-momenta and in the direction orthogonal to
this plane. Similarly, there exists an analogue of the quark transversity distribution,
involving again linearly polarized gluons. A more detailed comparison with the corresponding
notation of the Amsterdam group~\cite{Mulders:2000sh} can be found in Appendix C of
Ref.~\cite{Anselmino:2005sh} (see also Ref.~\cite{Meissner:2007rx}).

\subsection{Spin and TMD fragmentation functions at leading twist}
\label{form:frag}

Leading twist fragmentation functions can be described within the
same formalism presented for PDF's in the previous section. Let us
define the nonperturbative helicity \emph{fragmentation amplitudes},
$\hat{{\cal D}}_{\lambda^{\,}_C,\lambda^{\,}_{X}; \lambda^{\,}_c}$,
for the process $c\to C+X$ in which parton $c$, coming from the
elementary hard scattering process $ab\to cd$, hadronizes into a jet
containing the detected final hadron $C$, with light-cone momentum
fraction $z$ and transverse momentum $\bm{k}_{\perp C}$. We then
define the product of fragmentation amplitudes
 \be
 \hat D^{\lambda^{\,}_C,\lambda^{\prime}_C}_{\lambda^{\,}_c,\lambda^{\prime}_c}
 (z, \bm{k}_{\perp C})
 = \> \sumint_{X, \lambda_{X}}
 {\hat{\cal D}}_{\lambda^{\,}_C,\lambda^{\,}_{X};\lambda^{\,}_c}(z, \bm{k}_{\perp C}) \,
 {\hat{\cal D}}^*_{\lambda^{\prime}_C,\lambda^{\,}_{X}; \lambda^{\prime}_c}
 (z, \bm{k}_{\perp C})
 \, . \label{form:ddamp}
 \ee
\indent Denoting by $\phi_C^H$ the azimuthal angle of hadron $C$ momentum in the parton $c$
helicity frame, one has
 \be
 \hat{\cal D}_{\lambda^{\,}_C,\lambda^{\,}_{X}; \lambda^{\,}_c}(z, \bm{k}_{\perp C}) =
 {\cal D}_{\lambda^{\,}_C,\lambda^{\,}_{X}; \lambda^{\,}_c}(z, k_{\perp C}) \>
 e^{i\lambda^{\,}_c \phi_C^H} \>. \label{form:fragphi}
 \ee

In the sequel, apart from a short discussion of transverse $\Lambda$ polarization in
sections~\ref{form:lambda}, \ref{pheno:lam}, we will always refer to unpolarized final
hadrons. In this case, the generalized FF % fragmentation function
${\hat D}_{\lambda^{\,}_{c},\lambda^{\prime}_c}^{\lambda^{\,}_C,\lambda^{\prime}_C} (z,
\bm{k}_{\perp C})$ simplifies to
 \be
 {\hat D}_{\lambda^{\,}_{c},\lambda^{\prime}_c}^{\,C/c}(z, \bm{k}_{\perp C}) \equiv
  \sum_{\>\lambda^{\,}_{C}}
\hat D^{\lambda^{\,}_C,\lambda^{\,}_C}_{\lambda^{\,}_c,\lambda^{\prime}_c} (z, \bm{k}_{\perp
C}) = {D}_{\lambda^{\,}_{c},\lambda^{\prime}_c}^{\,C/c} (z, k_{\perp C})
 \> e^{\,i(\lambda^{\,}_{c} - \lambda^{\prime}_c)\,\phi_C^H} \>,
 \label{form:ddunp}
 \ee
and its azimuthal independent term fulfills the simplified parity relations
 \be
 {D}_{-\lambda^{\,}_{c},-\lambda^{\prime}_c}^{\,C/c}(z, k_{\perp C}) = (-1)^{\,2s_c} \>
  (-1)^{\lambda^{\,}_c + \lambda^{\prime}_c} \>
  {D}_{\lambda^{\,}_{c},\lambda^{\prime}_c}^{\,C/c}(z, k_{\perp C}) \,,
 \label{form:ddupar}
 \ee
which again imply some differences between the quark ($s_c=1/2$) and the gluon ($s_c=1$)
case.

\subsubsection{Quark and gluon fragmentation into an unpolarized hadron}
\label{form:frag:qunp}
{}From parity and rotational constraints
we can easily see
that for quark fragmentation into an unpolarized hadron there are only two independent
fragmentation functions at leading twist:
 \be
 \hat{D}_{\pm\pm}^{C/q}(z,\bm{k} _{\perp C}) =
 D_{\pm\pm}^{\,C/q}(z,k _{\perp C}) \equiv
{D}_{C/q}(z, k _{\perp C}) %\>,
 \label{form:dupp}
 \ee
%
%for equal helicity indices, and
%
 \be
{\hat D}_{\pm\,\mp}^{C/q}(z,\bm{k} _{\perp C}) = D_{\pm\,\mp}^{\,C/q}(z,k _{\perp C})
\,e^{\,\pm\,i\phi_C^H} = -\,D_{\mp\,\pm}^{\,C/q}(z,k _{\perp C}) \,e^{\,\pm\,i\phi_C^H}\>,
 \label{form:dupm1}
 \ee
with
 \be
 [\,D_{+-}^{\,C/q}(z,k _{\perp C})\,]^* = - D_{+-}^{\,C/q}(z,k _{\perp C})\>.
 \label{form:dupm2}
 \ee
%
%for unequal helicity indices.
$D_{C/q}(z, k _{\perp C})$ is the unpolarized TMD fragmentation
function. When integrated over the intrinsic transverse momentum it gives the usual
unpolarized FF in the collinear configuration, $D_{C/q}(z)$,
 \be
 D_{\,C/q}(z) = \frac{1}{2}\, \sum_{\,\lambda^{\,}_q} \int d^2\bm{k}_{\perp C} \,
 D^{\,C/q}_{\lambda^{\,}_q,\lambda^{\,}_q}(z, k_{\perp C}) = \frac{1}{2}\,
 \sum_{\lambda^{\,}_q,\,\lambda^{\,}_C} \int d^2\bm{k}_{\perp C} \,
 D^{\lambda^{\,}_C,\lambda^{\,}_C}_{\lambda^{\,}_q,\lambda^{\,}_q}(z, k_{\perp C}) \,.
 \label{form:dunp}
 \ee

$D^{\,C/q}_{+-}(z,k_{\perp C})$ is a purely imaginary quantity, related to the Collins FF,
$\Delta^{N}D_{C/q^\uparrow}(z, k_{\perp C})$,
 \be
 -2\,i\,D^{\,C/q}_{+-}(z, k_{\perp C}) = 2\,{\rm Im}D^{\,C/q}_{+-}(z, k_{\perp C}) =
 \Delta^{N}D_{C/q^\uparrow}(z, k_{\perp C})\>,
 \label{form:dcol}
 \ee
which gives the difference between the number densities of unpolarized hadrons $C$ produced
in the fragmentation of a quark polarized transversely to the hadronization plane (containing
the quark, $\bm{p}_{\,q}$, and hadron, $\bm{p}_{\,C}$, three-momenta) in the
$\bm{p}_{\,q}\times\bm{p}_{\,C}$ direction and in the opposite one.

We can summarize these results in a form similar to Eq.~(\ref{form:fff}):
 \bea
 D_{\,C/q} = D_{\,C/q,s_L}=D^{\,C/q}_{++} &\hspace*{1cm}&
 \Delta \hat{D}_{C/q,s_T} = -2\,{\rm Im}D^{\,C/q}_{+-}\,\sin(\phi_{s_q}-\phi^H_C)\>,
 \label{form:ddd}
 \eea
where $\phi_{s_q}$ and, as already explained, $\phi^H_C$ are the azimuthal angles of the
fragmenting quark polarization vector and of $\bm{k}_{\perp C}$ in the quark helicity frame,
as reached from the hadronic c.m.~frame.

 The gluon generalized fragmentation functions
$D^{\,C/g}_{\pm\pm}(z,k_{\perp C})$ with equal helicity indices obey the same relations as
for the quark case, see Eq.~(\ref{form:dupp}) with $q\to g$, whereas, due to the different
parton spins and parity properties, the gluon analogues of Eqs.~(\ref{form:dupm1}),
(\ref{form:dupm2}) for unequal helicity indices read
 \bea
 {\hat D}_{\pm\mp}^{\,C/g}(z,\bm{k} _{\perp C}) &=& D_{\pm\mp}^{\,C/g}(z,k _{\perp C})
 \,e^{\,\pm\,2i\phi_C^H} =
D_{\mp\pm}^{\,C/g}(z,k _{\perp C}) \,e^{\,\pm\,2i\phi_C^H} \nonumber \\
{[\,D_{+-}^{\,C/g}(z,k _{\perp C})\,]}^{*} &=& D_{+-}^{\,C/g}(z,k _{\perp C})\,.
 \label{form:dgpm}
 \eea
\indent Therefore, $D_{+-}^{\,C/g}(z,k _{\perp C})$ is now a purely real quantity. Similarly
to what happens for the gluon parton distributions, $D^{C/g}_{+-}(z,k _{\perp C})$ is related
to the fragmentation process of a \emph{linearly polarized} gluon into an unpolarized hadron.
In analogy to Eq.~(\ref{form:dcol}) we define for gluons a Collins-like FF:
 \be
 2\,D_{+-}^{\,C/g}(z,k _{\perp C}) = 2\,{\rm Re}D_{+-}^{\,C/g}(z,k _{\perp C}) \equiv
\Delta^N \hat{D}_{C/{\cal T}_1^g} (z,k _{\perp C}) \label{form:dcolg}\>.
 \ee
\indent
In the gluon helicity frame, in which the
fragmentation process takes place in the $(xz)$ plane,
this quantity gives the difference between the number densities of unpolarized hadrons $C$
resulting from the fragmentation of gluons linearly polarized along
the $x$-direction and along the $y$-direction.

\subsubsection{Quark fragmentation into a spin 1/2 hadron}
 \label{form:lambda}
\indent In this section we shortly describe the helicity formalism for the polarized TMD
fragmentation functions for the production of spin 1/2 particles. We will limit to the case
of quark partons. Clearly, in this case there is a strong analogy with the polarized TMD
distribution functions for parton quarks inside a spin 1/2 hadron, discussed in
section~\ref{form:distr}. The extension to the case of gluon fragmentation can be easily
performed along the same lines, using the results presented in the previous sections.

The polarization state of a spin 1/2 hadron $C$, produced in the fragmentation of an on-shell
massless quark is described by its helicity density matrix
 \be
 \rho_{\lambda^{\,}_{C},\lambda^{\prime}_C}^{\,C}{\hat D}_{\,C/q,s_q}(z,\bm{k} _{\perp
C}) = \sum_{\lambda^{\,}_{q},\lambda^{\prime}_q} {\hat
D}_{\lambda^{\,}_{q},\lambda^{\prime}_q}^{\lambda^{\,}_C,\lambda^{\prime}_C}(z, \bm{k}_{\perp
C})\, \rho_{\lambda^{\,}_{q},\lambda^{\prime}_q}^{q,s_q}\>,
 \label{form:rholam}
 \ee
and depends on the polarization state of the fragmenting quark,
through its own helicity density matrix
$\rho_{\lambda^{\,}_{q},\lambda^{\prime}_q}^{q,s_q}$, and on the
nonperturbative dynamics of the polarized hadronization process,
through the generalized fragmentation functions introduced in the
previous sections. The polarization state, and the helicity density
matrix, of the fragmenting quark depend in turn on its production
process.

Analogously to the PDF sector, by using rotational and parity
invariance we can easily see that at leading twist there are eight
independent polarized TMD fragmentation functions, which can be
connected to the generalized functions
 \be
 D_{++}^{++}\,,\quad D^{--}_{++}\,,\quad D^{+-}_{+-}\,,\quad D^{-+}_{+-}\,,\quad
 D^{++}_{+-}\,,\quad D^{--}_{+-}\,,\quad D^{+-}_{++}\,,\quad D^{+-}_{--}\>,
 \label{form:ddlam}
 \ee
through relations very similar to those of Eqs.~(\ref{form:fff}) for the PDF case.

Even if we will discuss spin 1/2 particle production only marginally, let us stress the following points:\\
1) There are only three TMD fragmentation functions that do not vanish under $\bm{k}_{\perp
C}$ integration, that is the three usual FF's of the collinear approach: the unpolarized,
longitudinally polarized and transversely polarized FF's (like in the PDF case, the last one
has two TMD contributions; only the dominant one survives under $\bm{k}_{\perp C}$ integration);\\
2) There is a term corresponding to the Collins effect, describing the fragmentation of a
transversely polarized quark into an unpolarized spin 1/2 hadron, as already discussed in
section~\ref{form:frag:qunp};\\
3) There is a very interesting new distribution, $\Delta{D}_{S_{_{Y}}/q}=
\Delta^ND_{C^\uparrow/q}$, related to the fragmentation of an unpolarized quark into a
transversely polarized (in its helicity frame)
spin 1/2 hadron. This TMD function, first introduced in Ref.~\cite{Mulders:1995dh}, needs not to vanish, and is in
a sense the analogue of the Sivers PDF in the fragmentation sector. As it will be discussed
in section~\ref{pheno}, it can play a relevant role for the description of transverse
$\Lambda$ polarization in unpolarized hadronic collisions within a pQCD approach. Its
phenomenological relevance was pointed out for the first time in
Ref.~\cite{Anselmino:2000vs}, where it was also named, given its role, as \emph{polarizing}
fragmentation function.
\subsection{Helicity amplitudes for the elementary process $ab\to cd$}
 \label{form:ampl}
Another essential ingredient of the master formula,
Eq.(\ref{form:genunp}), are the helicity amplitudes for the hard
elementary process $ab\to cd$. These enter in the combination $\hat
M_{\lambda^{\,}_c,\lambda^{\,}_d; \lambda^{\,}_a, \lambda^{\,}_b} \,
\hat M^*_{\lambda^{\prime}_c, \lambda^{\,}_d;
\lambda^{\prime}_a,\lambda^{\prime}_b}$, which is diagonal only in
the helicity indices of the unobserved parton $d$ (fragmenting into
a jet). For particles $a$, $b$, $c$, both diagonal and non diagonal
helicity terms play a role, in different combinations corresponding
to all the allowed contributions to a given polarized process. These
contributions will be discussed in more detail in
section~\ref{form:kern}, where the structure of
Eq.~(\ref{form:genunp}) will be explicitly given for some of the
terms appearing in the case of doubly polarized processes. Helicity
amplitudes are normalized so that the unpolarized cross section,
\emph{for a collinear collision}, is given by
 \be
 \frac{d\hat\sigma^{ab \to cd}}{d\hat t} = \frac{1}{16\pi\hat s^2}\frac{1}{4}
 \sum_{\lambda^{\,}_a, \lambda^{\,}_b, \lambda^{\,}_c, \lambda^{\,}_d}
  |\hat M_{\lambda^{\,}_c, \lambda^{\,}_d; \lambda^{\,}_a, \lambda^{\,}_b}|^2\,.
 \label{form:mmnorm}
 \ee
\indent It is important to keep in mind that all hadronic polarization observables are
measured in the hadronic c.m.~frame. On the other hand, in the helicity formalism the
polarization state of each particle involved in the process (including partons) is given via
its helicity density matrix, defined in the appropriate helicity frame, as reached by the
chosen overall hadronic frame. Therefore, one must carefully connect all helicity frames with
the hadronic c.m.~frame. A similar connection must be found between the helicity amplitudes
$\hat{M}^0$, given in the \emph{canonical partonic c.m.~frame} (that is, a frame in which
partons $a$ and $b$ collide head-on moving along the $\pm z$-axis direction respectively,
creating particles $c$, $d$, laying in the $xz$ plane), and their expressions in the hadronic
c.m.~frame, $\hat{M}$, which enter the master formula for cross sections. While this
connection is irrelevant for any process involving only unpolarized elementary scatterings,
it plays an essential role for the \emph{polarized} case, which in principle introduces
azimuthal factors that are not Lorentz invariants. As we will see in more detail in
section~\ref{pheno}, these factors, in addition to those coming from the polarized PDF's and
FF's discussed previously, are crucial for a proper understanding of each term contributing
to spin and azimuthal asymmetries.

Let us now start from the hadronic c.m.~frame for the process $AB\to C+X$. Neglecting for
simplicity all masses, the hadronic four-momenta are given by
 \be
 p_A^\mu = \frac{\sqrt{s}}{2}\,(1,0,0,1)\,,\quad
 p_B^\mu = \frac{\sqrt{s}}{2}\,(1,0,0,-1)\,,\quad
 p_C^\mu = (E_C,p_T,0,p_L)\>,
 \label{form:hadr4mom}
 \ee
with $E_C=\sqrt{p_T^2+p_L^2}$, and $s=(p_A+p_B)^2$. As for partons, we introduce the usual
light-cone momentum fractions $x_a=p_a^+/p_A^+$, $x_b=p_b^+/p_B^+$, $z=p_C^+/p_c^+$ and
transverse momenta $\bm{k}_{\perp a}$, $\bm{k}_{\perp b}$ and $\bm{k}_{\perp C}$.

Due to the simplified kinematics, in the $\bm{k}_\perp$-integrated, collinear configuration
the hadronic reaction plane (the $(XZ)_{\rm{cm}}$ plane of our overall c.m.~frame) and the
partonic scattering plane coincide. The hadronic and partonic c.m.~frames are connected by a
simple Lorentz boost along the direction of the colliding beams. Clearly, this situation does
not apply to a generalized parton model including all kinematical TMD effects. The elementary
scattering, $a(\bm{p}_a)+b(\bm{p}_b)\to c(\bm{p}_c)+d(\bm{p}_d)$, is not planar anymore in
the generic $\bm{k}_\perp$ configuration and takes place out of the $(XZ)_{\rm{cm}}$ plane.

There are two different ways to find the elementary helicity amplitudes in the hadronic
frame:\\
1) One can directly compute the amplitudes $\hat M$ in this frame, with the given kinematical
configuration, using well known effective calculational techniques. As a result, the
amplitudes will be given in terms of azimuthal phase factors directly measured in the
hadronic frame, having therefore a simple interpretation. However, they will have
complicated, not very illuminating, expressions.
Moreover, their general properties, e.g. under parity transformations, will be not so evident.\\
2) One can alternatively use the amplitudes $\hat{M}^0$ in the canonical partonic frame, that
are well known and have very simple expressions at leading order, together with their parity
properties, which are also well known. As a second step, one can relate the amplitudes
$\hat{M}^0$ to the amplitudes $\hat{M}$ by considering the Lorentz transformation connecting
the partonic and hadronic frames and its action on the helicity states of the particles
involved. This method, which will be used in the sequel, leads to simple expressions for the
amplitudes $\hat{M}$, related by simple azimuthal phase factors to the amplitudes
$\hat{M}^0$. However, the azimuthal phases  will be given as very involved
combinations of all angular variables coming from the kinematics and the Lorentz
transformation.

Following this procedure,  the elementary scattering amplitudes computed in the hadronic
c.m.~system are related to those
computed in the partonic c.m.~system (in the $XZ$ plane) %, $\phi_c=0$)
by:
 \be
 \hat M_{\lambda^{\,}_c, \lambda^{\,}_d; \lambda^{\,}_a, \lambda^{\,}_b} \! =
 \hat M^0 _{\lambda^{\,}_c,\lambda^{\,}_d; \lambda^{\,}_a, \lambda^{\,}_b}
 \, e^{-i (\lambda^{\,}_a \xi _a + \lambda^{\,}_b \xi _b -
          \lambda^{\,}_c \xi _c - \lambda^{\,}_d \xi _d)}
 \, e^{-i [(\lambda^{\,}_a - \lambda^{\,}_b) \tilde \xi _a -
         (\lambda^{\,}_c - \lambda^{\,}_d) \tilde \xi _c]}
 \, e^{i(\lambda^{\,}_a - \lambda^{\,}_b)\phi^{\prime\prime}_c}\>,
 \label{form:M-M0}
 \ee
where $\xi_i$, $\tilde \xi_i$ and $\phi^{\prime\prime}_c$ are phases
acquired as a result of the boost and rotations~\cite{Leader:2001gr,Anselmino:2005sh}.

The parity properties of the canonical c.m.~amplitudes $\hat M^0$ are
the usual ones~\cite{Leader:2001gr}:
 \be
 \hat M^0_{-\lambda^{\,}_c, -\lambda^{\,}_d; -\lambda^{\,}_a, -\lambda^{\,}_b} =
 \eta_a \eta_b \eta_c \eta_d (-1)^{s_a + s_b - s_c - s_d} \>
 (-1)^{(\lambda^{\,}_a - \lambda^{\,}_b) - (\lambda^{\,}_c - \lambda^{\,}_d)}
\hat M^0_{\lambda^{\,}_c, \lambda^{\,}_d; \lambda^{\,}_a, \lambda^{\,}_b} \>,
 \label{form:parM0}
\ee
where $\eta_i$ is the intrinsic parity factor for particle $i$. From Lorentz and rotational
invariance properties~\cite{Leader:2001gr} one can also obtain useful expressions relating,
up to an overall phase factor, irrelevant here, the canonical amplitudes for processes which
only differ by the exchange of the two initial partons, $a \leftrightarrow b$, or of the two
final partons, $c \leftrightarrow d$:
 \bea
\!\!\hat M_{\lambda^{\,}_c, \lambda^{\,}_d; \lambda^{\,}_b, \lambda^{\,}_a}^{0, \, ba \to cd}
 (\theta) = \hat M_{\lambda^{\,}_c, \lambda^{\,}_d; \lambda^{\,}_a, \lambda^{\,}_b}^{0, \, ab \to cd}
 (\pi - \theta) \, e^{-i\pi(\lambda^{\,}_c - \lambda^{\,}_d)} \label{form:ex-ab} &&\!\!
\hat M_{\lambda^{\,}_d, \lambda^{\,}_c; \lambda^{\,}_a, \lambda^{\,}_b}^{0, \, ab \to dc}
 (\theta) = \hat M_{\lambda^{\,}_c, \lambda^{\,}_d; \lambda^{\,}_a, \lambda^{\,}_b}^{0, \, ab \to cd}
 (\pi -\theta) \, e^{-i \pi(\lambda^{\,}_a - \lambda^{\,}_b)}, \mbox{}%\label{form:ex-cd}
 \eea
where the scattering angle $\theta$ is defined in the canonical partonic c.m.~frame.
\subsection{Kernels for doubly polarized processes}
\label{form:kern}
The computation of the invariant differential cross section corresponding to any polarized
hadronic process $A(S_A)+B(S_B)\to C+X$, see Eq.~(\ref{form:genunp}), requires the evaluation
and integration, for each elementary process $ab\to cd$, of the general kernel
 \bea
 &&\Sigma(S_A,S_B)^{ab\to cd} =\nonumber\\
 &&\sum _{\{\lambda\}} \rho_{\lambda^{\,}_a,
 \lambda^{\prime}_a}^{a/A,S_A} \, \hat f_{a/A,S_A}(x_a,\bm{k}_{\perp a}) \>
 \rho_{\lambda^{\,}_b, \lambda^{\prime}_b}^{b/B,S_B} \,
 \hat f_{b/B,S_B}(x_b,\bm{k}_{\perp b})
 \,\hat M_{\lambda^{\,}_c, \lambda^{\,}_d; \lambda^{\,}_a, \lambda^{\,}_b}
 \, \hat M^*_{\lambda^{\prime}_c,\lambda^{\,}_d; \lambda^{\prime}_a, \lambda^{\prime}_b} \>
 \hat D^{\lambda^{\,}_C,\lambda^{\,}_C}_{\lambda^{\,}_c,\lambda^{\prime}_c}(z,\bm{k}_{\perp C})\,.
 \label{form:genkern}
 \eea
\indent All the ingredients of the kernels have been separately discussed in detail in the
previous sections. We have stressed the fact that, including $\bm{k}_\perp$ effects, the
non-collinearity of soft processes, together with the non-planarity of the elementary
scattering processes, generate non trivial azimuthal phase factors that must be properly
taken into account in the calculations. As it will be shown in section~\ref{pheno}, these
phase factors have direct consequences on the behaviour of spin and azimuthal asymmetries in
term of the kinematical variables and can be experimentally tested.

Before giving explicit kernels, Eq.~(\ref{form:genkern}), for some representative partonic
channels contributing  to the doubly polarized process $A(S_A)+B(S_B)\to C+X$, we note that
for on-shell, massless partons there are only three nonvanishing, independent helicity
amplitudes for the process $ab\to cd$:
 \bea
 \hat{M}_{++;++} &=& \hat{M}^0_{++;++}\,e^{\,i\varphi_1} =
 \hat{M}^0_1\,e^{\,i\varphi_1}\nonumber\\
 \hat{M}_{-+;-+} &=& \hat{M}^0_{-+;-+}\,e^{\,i\varphi_2} =
 \hat{M}^0_2\,e^{\,i\varphi_2} \label{form:m123} \\
 \hat{M}_{-+;+-} &=& \hat{M}^0_{-+;+-}\,e^{\,i\varphi_3} =
 \hat{M}^0_3\,e^{\,i\varphi_3}\>,\nonumber
 \eea
where the phases $\varphi_i$, $i=1,2,3$, can be obtained from Eq.~(\ref{form:M-M0}) by
inserting the proper values for the helicities $\lambda_j$, $j=a,b,c,d$ and are such that
they change sign by helicity inversion. The $(+)$ and $(-)$ subscripts correspond
respectively to $(\pm1/2)$ helicities for quarks and to $(\pm1)$ helicities for gluons. For
shortness, in the kernels we will not show explicitly the PDF and FF dependences on the
parton light-cone fractions and transverse momenta. Kernels for processes related to those
presented by the exchange $a\leftrightarrow b$, and/or $c\leftrightarrow d$ can be easily
obtained by using Eq.~(\ref{form:ex-ab}) and exchanging properly some labels. The elementary
amplitudes $\hat{M}^0_i$, $i=1,2,3$, and their products, can be found, e.g., in
Ref.~\cite{Anselmino:2005sh}.

\vspace*{10pt}

\noindent  (1) $\qquad q_a q_b \to q_c q_d$ processes
 \bea
 && 2\,\Sigma(S_A,S_B)^{q_a q_b \to q_c q_d} =\nonumber\\
 &&\biggl\{\,\hat f_{a/S_A}\, \hat f_{b/S_B}
 \left(|\hat M^0_1|^2 + |\hat M^0_2|^2 + |\hat M^0_3|^2\right)
 +\,\Delta\hat f^a_{s_z/S_A}\,\Delta\hat f^b_{s_z/S_B}
\left(|\hat M^0_1|^2 - |\hat M^0_2|^2 - |\hat M^0_3|^2\right) \nonumber \\
 && \quad\ +\> 2\, \biggl[ \left( \Delta\hat f^a_{s_x/S_A}\,\Delta\hat f^b_{s_x/S_B}
 + \Delta\hat f^a_{s_y/S_A}\,\Delta\hat f^b_{s_y/S_B}\right) \, \cos(\varphi_{3}-\varphi_{2})
 \nonumber\\
 &&\quad\ -\> \left(\Delta\hat f^a_{s_x/S_A}\,\Delta\hat f^b_{s_y/S_B} -
 \Delta\hat f^a_{s_y/S_A}\,\Delta\hat f^b_{s_x/S_B} \right) \,
 \sin(\varphi_{3}-\varphi_{2})\,
 \biggr]\hat M^0_2 \, \hat M^0_3 \biggr\}\,D_{C/c} \label{form:krqq} \\
 &-& \biggl\{\,\biggl[\,\Delta\hat f^a_{s_x/S_A}\,\hat f_{b/S_B}
 \, \sin(\varphi_1 - \varphi_2 + \phi_C^H) - \Delta\hat f^a_{s_y/S_A}\,\hat f_{b/S_B}
 \, \cos(\varphi_1 - \varphi_2 + \phi_C^H)\biggr]\,\hat M^0_1 \, \hat M^0_2\nonumber\\
 &&\quad\ + \>\biggl[\hat f_{a/S_A}\,\Delta\hat f^b_{s_x/S_B}\sin(\varphi_1 - \varphi_3 + \phi_C^H)
 \,-\,\hat f_{a/S_A}\,\Delta\hat f^b_{s_y/S_B}\cos(\varphi_1 - \varphi_3 + \phi_C^H)
 \biggr]\,\hat M^0_1 \, \hat M^0_3\,\biggr\}\,\Delta^N D_{C/c^\uparrow}\>.\nonumber
 \eea
\indent This case includes all channels involving quarks and
antiquarks like e.g.~$\bar{q}_a \bar{q}_b\to \bar{q}_c\bar{q}_d$,
$q_a \bar{q}_b\to q_c\bar{q}_d$.

\vspace*{10pt}

 \noindent  (2) $\qquad q g \to q g$ processes
\bea
 &&2\,\Sigma(S_A,S_B)^{q g \to q g} = \nonumber\\
 && \biggl[\,\hat f_{q/S_A}\, \hat f_{g/S_B}
 \left(|\hat M^0_1|^2 + |\hat M^0_2|^2\right)
 +\,\Delta\hat f^q_{s_z/S_A}\,\Delta\hat f^g_{s_z/S_B}
\left(|\hat M^0_1|^2 - |\hat M^0_2|^2\right)\,\biggr]\,D_{C/q}  \label{form:krqg} \\
 &-&\biggl[\,\Delta\hat f^q_{s_x/S_A}\,\hat f_{g/S_B}
 \, \sin(\varphi_1 - \varphi_2 + \phi_C^H) - \Delta\hat f^a_{s_y/S_A}\,\hat f_{b/S_B}
 \, \cos(\varphi_1 - \varphi_2 + \phi_C^H)\biggr]\,\hat M^0_1 \, \hat M^0_2\,
 \,\Delta^N D_{C/c^\uparrow}\>. \nonumber
 \eea
\indent Notice that in this case the amplitude $\hat{M}^0_3$ is zero
because of helicity conservation.

\vspace*{10pt}

 \noindent (3) $\qquad g g \to g g$ processes
\bea
 && 2\,\Sigma(S_A,S_B)^{g g \to g g} = \nonumber\\
 &&\biggl\{\,\hat f_{g/S_A}\, \hat f_{g/S_B}
 \left(|\hat M^0_1|^2 + |\hat M^0_2|^2 + |\hat M^0_3|^2\right)
 +\,\Delta\hat f^g_{s_z/S_A}\,\Delta\hat f^g_{s_z/S_B}
\left(|\hat M^0_1|^2 - |\hat M^0_2|^2 - |\hat M^0_3|^2\right) \nonumber \\
 && \>\>\ +\> 2\, \biggl[ \left( \Delta\hat f^g_{{\cal T}_1/S_A}\,\Delta\hat f^g_{{\cal T}_1/S_B}
 + \Delta\hat f^g_{{\cal T}_2/S_A}\,\Delta\hat f^g_{{\cal T}_2/S_B}\right)
 \, \cos(\varphi_{3}-\varphi_{2}) \nonumber\\
 &&\>\>\ -\> \left(\Delta\hat f^g_{{\cal T}_1/S_A}\,\Delta\hat f^g_{{\cal T}_2/S_B} -
 \Delta\hat f^g_{{\cal T}_2/S_A}\,\Delta\hat f^g_{{\cal T}_1/S_B} \right) \,
 \sin(\varphi_{3}-\varphi_{2})\,
 \biggr]\hat M^0_2 \, \hat M^0_3 \biggr\}\,D_{C/g} \label{form:krgg} \\
 &+&\!\!\!\biggl\{\,\biggl[\,\Delta\hat f^g_{{\cal T}_1/S_A}\,\hat f_{g/S_B}
 \, \cos(\varphi_1 - \varphi_2 + 2\phi_C^H) + \Delta\hat f^g_{{\cal T}_2/S_A}\,\hat f_{g/S_B}
 \, \sin(\varphi_1 - \varphi_2 + 2\phi_C^H)\biggr]\,\hat M^0_1 \, \hat M^0_2\nonumber\\
 &&\>\>\ + \>\biggl[\hat f_{g/S_A}\,\Delta\hat f^g_{{\cal T}_1/S_B}
 \cos(\varphi_1 - \varphi_3 + 2\phi_C^H)
 \,+\,\hat f_{g/S_A}\,\Delta\hat f^g_{{\cal T}_2/S_B}\sin(\varphi_1 - \varphi_3 + 2\phi_C^H)
 \biggr]\,\hat M^0_1 \, \hat M^0_3\,\biggr\}\,\Delta^N D_{C/{\cal T}^g_1}\>.\nonumber
 \eea
\indent A few comments are in order here:\\
1) The explicit expressions of the polarized TMD quark distributions, for specific $S_A$ and
$S_B$ combinations, can be read from Eq.~(\ref{form:fff}) (for gluons see
Ref.~\cite{Anselmino:2005sh}). Notice that in general these functions still depend on the
azimuthal phases (given in the hadronic frame) of the hadron polarization vector and of the
parton transverse momentum. These angles will therefore appear in the final azimuthal
phase factors of a specific polarized cross section;\\
2) Since for the fragmentation process into an unpolarized hadron there are only two
independent TMD functions, the unpolarized FF and the Collins function, and only the last one
depends on the azimuthal phase $\phi^H_C$, this dependence has been explicitly factored
out in the azimuthal phase factors;\\
3) When considering processes obtained from those listed above by the exchange
$a\leftrightarrow b$, and/or $c\leftrightarrow d$, it is important to recall that this
concerns also the phases $\varphi_i$, $i=1,2,3$ already factored out from the amplitudes
$\hat{M}^0_i$, see Eq.~(\ref{form:M-M0}).\\
4) The unintegrated kernels still depend on the parton transverse momenta $\bm{k}_{\perp a}$,
$\bm{k}_{\perp b}$, $\bm{k}_{\perp C}$. As a result, we have seen that the partonic process
is not planar and does not lay on the hadronic production plane. Therefore, the kernels still
contain terms that, under $\bm{k}_\perp$ integration, must vanish for some specific asymmetry
(as it has been checked both analytically and numerically). As an example, there are terms
corresponding to longitudinal SSA's, that are forbidden, for the overall hadronic process, by
parity and rotational invariance. This is somehow analogous to the case of single spin
azimuthal asymmetries measured (in the virtual photon-target proton reference frame) in SIDIS
off longitudinally polarized (in the laboratory reference frame) protons. Since the two
ref.~frames are not coplanar, the longitudinal proton polarization vector has a transverse
component, of order $M/Q$, in the $\gamma^*-p$ c.m.~frame.

To conclude, given its phenomenological interest we show, as an example, the kernel for the
$q_a q_b\to q_c q_d$ channel for the transverse polarization of spin 1/2 particles in
unpolarized $AB$ collisions.
 \bea
 \label{numSSATqcm}
 &&P_{Y_{\rm{cm}}}\Sigma(0,0)^{q_a q_b \to q_c q_d}=\\
 &&\frac{1}{2}\Big\{f_{a/A} \, f_{b/B} \, \left[ \, |{\hat M}_1^0|^2 + |{\hat M}_2^0|^2 +
 |{\hat M}_3^0|^2 \right] + 2\, \Delta \hat f^a_{s_y/A} \, \Delta \hat f^b_{s_y/B}\,
 \cos(\varphi_3 -\varphi_2)\, {\hat M}_2^0 \, {\hat M}_3^0\Big\}
 \Delta D_{S_{Y}/c}^{C/c}\cos\tilde{\phi}\nonumber\\
 &+& \hat f_{a/A}\,\Delta \hat f^b_{s_y/B}\,\hat M^0_1 \, \hat M^0_3\,
 \Big\{\Delta^{\!-}{\hat D}_{S_{Y}/s_T}^{C/c}\cos(\varphi_1 - \varphi_3 + \phi_C^H)
 \cos\tilde{\phi} - \Delta D_{S_{X}/s_T}^{C/c}
 \sin(\varphi_1 - \varphi_3 + \phi_C^H)\sin\tilde{\phi}\Big\}\nonumber\\
 &+&\Delta \hat f^a_{s_y/A}\,\hat f_{b/B}\hat M^0_1 \, \hat M^0_2\,
 \Big\{\Delta^{\!-} D_{S_{Y}/s_T}^{C/c}\cos(\varphi_1 - \varphi_2 + \phi_C^H)
 \cos\tilde{\phi} - \Delta{D}_{S_{X}/s_T}^{C/c}
 \sin(\varphi_1 - \varphi_2 + \phi_C^H)\sin\tilde{\phi}\Big\}\,,\nonumber
 \eea
where the polarized fragmentation functions for the process $c \to C +X$ are the analogues of
the quark distributions discussed in section~\ref{form:distr}. $P_{Y_{\rm{cm}}}$ is the
transverse polarization of the spin 1/2 final particle w.r.t.~the production plane, which is
the $(XZ)_{\rm{cm}}$ plane of the hadronic c.m.~frame. The angle $\tilde{\phi}$ projects the
transverse polarization of the particle, as given in its helicity frame, on the transverse
direction in the hadronic frame just defined above. Notice that
$\cos\tilde{\phi}=Y_{\rm{cm}}\cdot(\hat{\bm{p}}_{c}\times\hat{\bm{k}}_{\perp
  C})$.

%% file: pheno-rev.tex
\section{Phenomenology of single spin asymmetries}
\label{pheno}

In this section we will consider in more detail, and from a phenomenological point of view,
the theoretical computations aiming to a description of unpolarized and single-polarized
cross section data for various high-energy processes of interest. We will consider in
particular the generalized parton model approach in the helicity formalism (GPM), presented
in section~\ref{form}, making when possible a comparison with analogous results obtained
within the other theoretical approaches discussed in section~\ref{theo}.

Following an historical perspective, we will first consider single and, to a lesser extent,
double inclusive particle production in $pp$ collisions. We will give an overview on how spin
and $\bm{k}_\perp$ effects play a role in explaining some former and well-known observations,
as well as the most recent data collected at RHIC on SSA's. Consideration of several final
particles, like pions, kaons, heavy mesons, hyperons, and photons, in different kinematical
configurations, will guide us in showing the most interesting aspects of SSA's and their
description in the GPM approach. Let us recall however that for the above processes
factorization in the TMD approach is still under investigation and, for the single inclusive
case, SSA's appear at subleading-twist level.

We will then move to a class of processes for which the $\bfk_\perp$ factorization has been
proved and SSA's show up already at leading twist. First we will discuss the Drell-Yan
process, where both the unpolarized and single-polarized cases present interesting features
to be analyzed in the context of the TMD approach. Finally, $e^+e^-$ and SIDIS processes will
be also addressed, with emphasis on some recent data collected by the Belle, HERMES and
COMPASS Collaborations. We will then give a discussion of the latest extractions of the
Sivers function, the transversity distribution and the Collins function.

We will show how rich this phenomenology can be, focusing also on the information one can
gather from a combined careful study of all these cases.

In the theoretical estimates at LO accuracy the MRST01 set~\cite{Martin:2002dr} is adopted
for the collinear unpolarized PDF, $f_{a/p}(x)$. Where it appears together with the helicity
PDF, $\Delta_L f_{a/p}$, the GRV98~\cite{Gluck:1998xa} and the GRSV2000~\cite{Gluck:2000dy}
sets or, in some cases, the MRST01 and the LSS01~\cite{Leader:2001kh} sets are consistently
used. For the FF ($D_{h/c}(z)$), the KKP~\cite{Kniehl:2000fe} and the Kretzer
(K)~\cite{Kretzer:2000yf} sets are adopted.

\subsection{Unpolarized cross sections and SSA's in $pp\to C +X$}

The inclusive production of large $p_T$ particles in high-energy hadronic collisions has been
for a long time a crucial testing ground for perturbative QCD. We will focus here on the role
of intrinsic transverse momentum effects, utilizing the helicity formalism for the
computation of (un)polarized cross sections.

\subsubsection{Pion, kaon and heavy meson production}
\label{pheno:pp}

Pion production is by far the most representative case: a relatively large sample of data is
already available, and more are going to come in the near future; moreover, pion SSA's may be
considered as the original motivation behind the GPM approach.

A crucial and distinguishing ingredient of this approach is the proper inclusion of full
$\bm{k}_\perp$ kinematics and its interplay with the azimuthal phase factors appearing in the
expressions of the polarized cross sections. We fix for the $pp\to \pi + X$ inclusive process
the following configuration: we adopt the $pp$ c.m.~frame (the hadronic frame), with the $Z$
axis along the direction of the colliding protons, and the pion produced in the $XZ$ plane
with positive $(p_\pi)_X$. For the proton moving along $+Z$, the transverse polarization is
defined along the $Y$ axis with $\uparrow(\downarrow)\equiv \pm Y$ [$\phi_{S_A}= \pm \pi/2$
in Eq.~(\ref{form:fff})].

Let us start with the unpolarized case. The corresponding differential cross section for
inclusive pion production is given in terms of the kernels in Eq.~(\ref{form:genkern}),
$\Sigma(0,0)= 1/2 [\,\Sigma(\uparrow,0)$ $+ \Sigma(\downarrow,0)\,]$, one for each active
partonic channel, $ab\to cd$. As discussed in Ref.~\cite{Anselmino:2005sh}, the main new
result of the GPM approach is the appearance of additional terms besides the usual
contribution from the unpolarized $\bfk_\perp$ dependent PDF's and FF's, coupled with the
unpolarized partonic cross sections. This term, once integrated over all $\bm{k}_\perp$'s,
would be the only one in the factorization scheme in collinear configuration. It is also the
only contribution considered in previous approaches using the parton model with TMD effects
but neglecting the new spin and $\bfk_\perp$ dependent, leading-twist, PDF's and
FF's~\cite{Contogouris:1978kh,Wong:1998pq,Wang:1998ww,Zhang:2001ce}. In fact, focusing here
on the $qq\to qq$ channel, the additional contributions come from: a) The convolution of two
(chiral-odd) Boer-Mulders functions with the double transverse spin asymmetry for the $qq$
interaction; 2) The convolution of one Boer-Mulders function and the Collins function (again,
two chiral-odd quantities) coupled with the partonic spin-transfer factor. Analogously, all
the other partonic channels contain some extra pieces as compared to the naive TMD extension
of the collinear factorized approach.

The study of so many additional contributions, given in terms of new unknown distributions,
might look as hopeless from the phenomenological point of view. It is however the very
structure behind this scheme, and its exact kinematical formulation, to offer a way out.

%%%%%%ppnp style
\begin{figure}[th!b]
\begin{center}
\begin{minipage}[t]{12 cm}
\epsfig{file=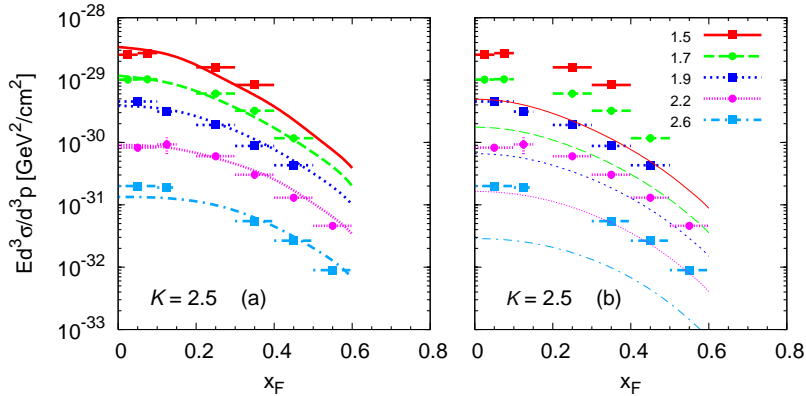,scale=0.4,angle=-90}
\end{minipage}
\begin{minipage}[t]{16.5 cm}
\caption{ Invariant differential cross section for $pp\to\pi^0+X$ at $\sqrt s = 19.4$ GeV and
for different $p_T$ values (in GeV$/c$) vs.~$x_F$~\cite{Donaldson:1977yz}. The corresponding
results in the GPM approach with $\bm{k}_\perp$ effects (a) and in the collinear partonic
configuration (b) are also shown. All curves are rescaled by a fixed $K$-factor, $K=2.5$.
Unpolarized PDF and FF sets: MRST01-KKP. \label{pheno:fnal} }
\end{minipage}
\end{center}
\end{figure}

Once the integrations over the nonobservable intrinsic motions have been performed, the many
azimuthal phases appearing in the elementary interactions, due to the noncollinear
configuration, lead to large cancelations of most of the contributions involving the new
unknown TMD functions. However, the inclusion of $\bm{k}_\perp$ effects in the standard
unpolarized contribution in collinear configuration could be relevant for a good description
of the unpolarized cross section data measured in inclusive particle production in different
kinematical configurations. These cross sections are otherwise often underestimated by a
factor of up to one order of magnitude~\cite{D'Alesio:2004up}. As discussed in
Ref.~\cite{deFlorian:2005yj} threshold resummation effects might reduce to some extent this
discrepancy.

As an example, in Fig.~\ref{pheno:fnal} we show the invariant differential cross section for
inclusive $\pi^0$ production in $pp$ collisions, at $\sqrt s = 19.4$ GeV and for different
$p_T$ values, as a function of $x_F$. In the left panel we plot the estimated cross sections,
obtained with inclusion of $\bm{k}_\perp$ effects, while the corresponding results in the
usual collinear partonic configuration are plotted in the right panel (all curves are
rescaled by a fixed $K$-factor, $K=2.5$). With the aim of giving a consistent treatment of
both SSA's and unpolarized cross sections at the same level of accuracy (namely leading order
pQCD calculations with leading twist soft functions) a large collection of data on Drell-Yan,
direct photon production and pion production in $pp$, $pA$ collisions has been considered.
NLO corrections to the unpolarized cross sections, usually parameterized as $K$-factors,
having values around 2-3, are assumed to cancel, at least partly, in the SSA. Some of the
relevant parameters of this approach, namely the Gaussian widths of $\bm{k}_\perp$-dependent
PDF's and FF's, could then be reasonably fixed. For details see Ref.~\cite{D'Alesio:2004up}.

For higher-energy kinematical configurations, like e.g.~that of the RHIC-BNL experiments, the
dominance of the standard unpolarized term, $f_{a/p} \otimes f_{b/p} \otimes D_{\pi/c}$ (but
with inclusion of $\bm{k}_\perp$ effects), w.r.t.~all other additional contributions, comes
out again in a neat way. In this case, however, the enhancing factor coming from the
$\bfk_\perp$ effects is much less important.

In this context, in Ref.~\cite{Bourrely:2003bw} it has been claimed that experimental data on
pion SSA's observed at fixed target experiments (moderate c.m.~energies) cannot be explained
by pQCD, since the collinear approach (LO and NLO) fails to reproduce the corresponding
unpolarized cross sections. On the contrary, at high energy a good description can be
obtained at NLO accuracy. As shown in Fig.~\ref{pheno:fnal} the inclusion of $k_\perp$
effects reduces the gap between theoretical estimates in pQCD and data on unpolarized cross
sections even for fixed target experiments. Therefore, a consistent treatment in the GPM
approach of both unpolarized cross sections and SSA's for pion production can be given.

Let us now consider the numerator of $A_N$, focusing on two partonic channels:

\noindent (1) $\qquad q_a q_b \to q_c q_d$ processes
 \bea
 &&\hspace*{-1.6cm}[\Sigma(\uparrow,0) - \Sigma(\downarrow,0)]^{q_a q_b
 \to q_c q_d} =
\frac{1}{2} \, \Delta \hat f_{a/p^\uparrow} (x_a,\bm{k}_{\perp a}) \, f_{b/p}(x_b,
 k_{\perp b}) \, \left[\,|{\hat M}_1^0|^2 + |{\hat M}_2^0|^2 + |{\hat M}_3^0|^2 \right] \,
 D _{\pi/c} (z, k_{\perp \pi}) \nonumber \\
 && \hspace*{1cm}
+\; \left[ \Delta^- \hat f^a_{s_y/\uparrow} (x_a, \bm{k}_{\perp a})\,
 \cos(\varphi_1 -\varphi_2 + \phi_\pi^H)
 - \Delta \hat f^a_{s_x/\uparrow}(x_a,\bm{k}_{\perp a})\,
 \sin(\varphi_1 -\varphi_2 + \phi_\pi^H) \right] \,\nonumber \\
 && \hspace*{1.5cm}
\times \, f_{b/p}(x_b, k_{\perp b})\, {\hat M}_1^0 \, {\hat M}_2^0 \, \Delta^N
 {D}_{\pi/c^\uparrow} (z, k_{\perp \pi})\,
 \nonumber \\
 &&  \hspace*{1cm}
+\;2\,\left[ \Delta^- \hat f^a_{s_y/\uparrow} (x_a, \bm{k}_{\perp a}) \,
 \cos(\varphi_3 -\varphi_2) -\Delta \hat f^a_{s_x/\uparrow} (x_a,\bm{k}_{\perp a}) \,
 \sin(\varphi_3 -\varphi_2) \right] \,
 \label{num-asym-qq} \\
 && \hspace*{1.5cm}
\times \, \Delta f^b_{s_y/p}(x_b,k_{\perp b})\, {\hat M}_2^0 \, {\hat M}_3^0 \,
 D _{\pi/c} (z, k_{\perp \pi})
 \nonumber \\
 && \hspace*{1.cm}
+\; \frac{1}{2} \, \Delta \hat f_{a/p^\uparrow} (x_a, \bm{k}_{\perp a})\, \Delta
 f^b_{s_y/p} (x_b, k_{\perp b}) \, \cos(\varphi_1 -\varphi_3 + \phi_\pi^H) \, {\hat
 M}_1^0 \, {\hat M}_3^0 \, \Delta ^N {D} _{\pi/c^\uparrow} (z,
 k_{\perp \pi}) \,; \nonumber
 \eea
 \noindent  (2) $\qquad qg \to qg$ processes
 \bea
 &&\hspace*{-1.6cm}[\Sigma(\uparrow,0) - \Sigma(\downarrow,0)]^{q g \to q g} =
\frac{1}{2} \, \Delta \hat f_{q/p^\uparrow} (x_q, \bm{k}_{\perp q}) \, f_{g/p}(x_g,
 k_{\perp g}) \, \left[ \, |{\hat M}_1^0|^2 + |{\hat M}_2^0|^2 \right] \,
 D _{\pi/q} (z, k_{\perp \pi}) \nonumber \\
 && \hspace*{1.cm}
 +\; \left[ \Delta^- \hat f^q_{s_y/\uparrow} (x_q, \bm{k}_{\perp q}) \,
 \cos(\varphi_1 -\varphi_2 +\phi_\pi^H)
 - \Delta \hat f^q_{s_x/\uparrow} (x_q,\bm{k}_{\perp q}) \,
 \sin(\varphi_1 -\varphi_2 + \phi_\pi^H) \right] \,
 \nonumber \\
 &&\hspace*{1.5cm}
\times \, f_{g/p}(x_g, k_{\perp g})\, {\hat M}_1^0 \, {\hat M}_2^0 \, \Delta^N
 {D}_{\pi/q^\uparrow}(z, k_{\perp \pi})\,.
 \label{num-asym-gq}
 \eea

Let us recall that the above equations have been obtained by properly exploiting the
phases coming from the scattering helicity amplitudes and  the fragmentation process,
while the phases entering the spin and TMD initial distributions, if present, are still
implicitly included in their vectorial $\bm{k}_\perp$ dependence.

Again we briefly comment on the $qq\to qq$ case. In the first term of Eq.~(\ref{num-asym-qq})
we find the contribution due to the Sivers effect; in the second term we have the Collins
effect coupled with the two pieces of the TMD transversity distribution. Notice that these
contributions, already familiar from previous work adopting a simplified $\bm{k}_\perp$
kinematics, enter now with a complete phase structure without any simplification. Two extra
terms appear: the third one contains the Boer-Mulders effect coupled again with the TMD
transversity distribution; finally, the fourth term comes from a combination of the Sivers,
Collins and Boer-Mulders effects. Notice that in the $qg\to qg$ case only the Sivers effect
and, separately, the transversity + Collins effect, could be active. For the $gg\to gg$
process one gets an analogous structure as in the $qq\to qq$ case, with a corresponding
number of effects (involving linear gluon polarizations instead of transverse quark
polarizations).

The complex structure in terms of azimuthal phases plays an even more crucial role in the
numerator of the SSA. To show that one can saturate all unknown polarized TMD distribution
functions, using the corresponding positivity bounds. In some cases this is certainly an
overestimate: e.g., for the transversity distribution it violates the Soffer
bound~\cite{Soffer:1994ww}.

\begin{figure}[th!]
\begin{center}
\begin{minipage}[t]{14 cm}
\epsfig{file=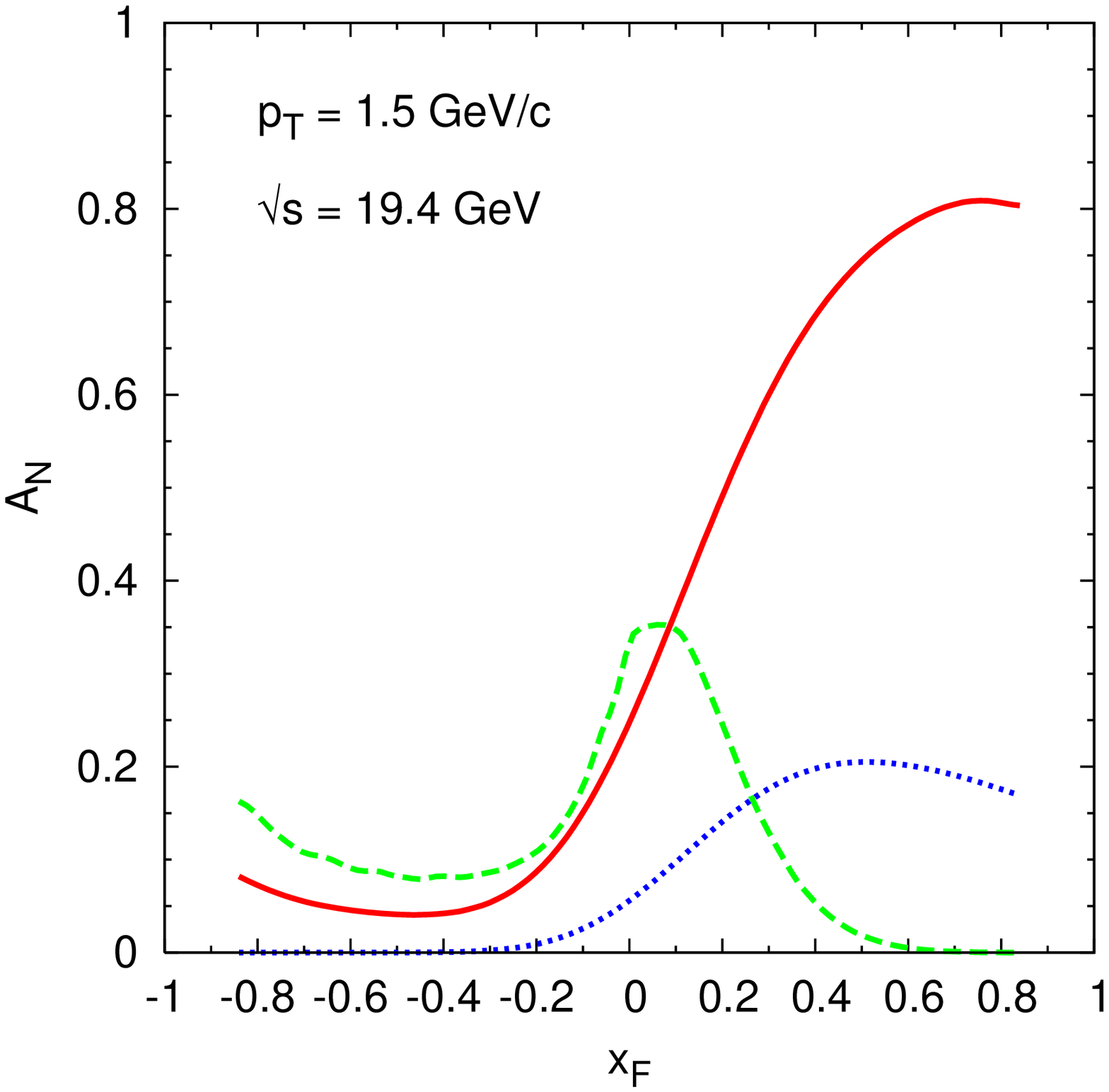,scale=0.35,angle=-90}
\hspace*{-1cm}
\epsfig{figure=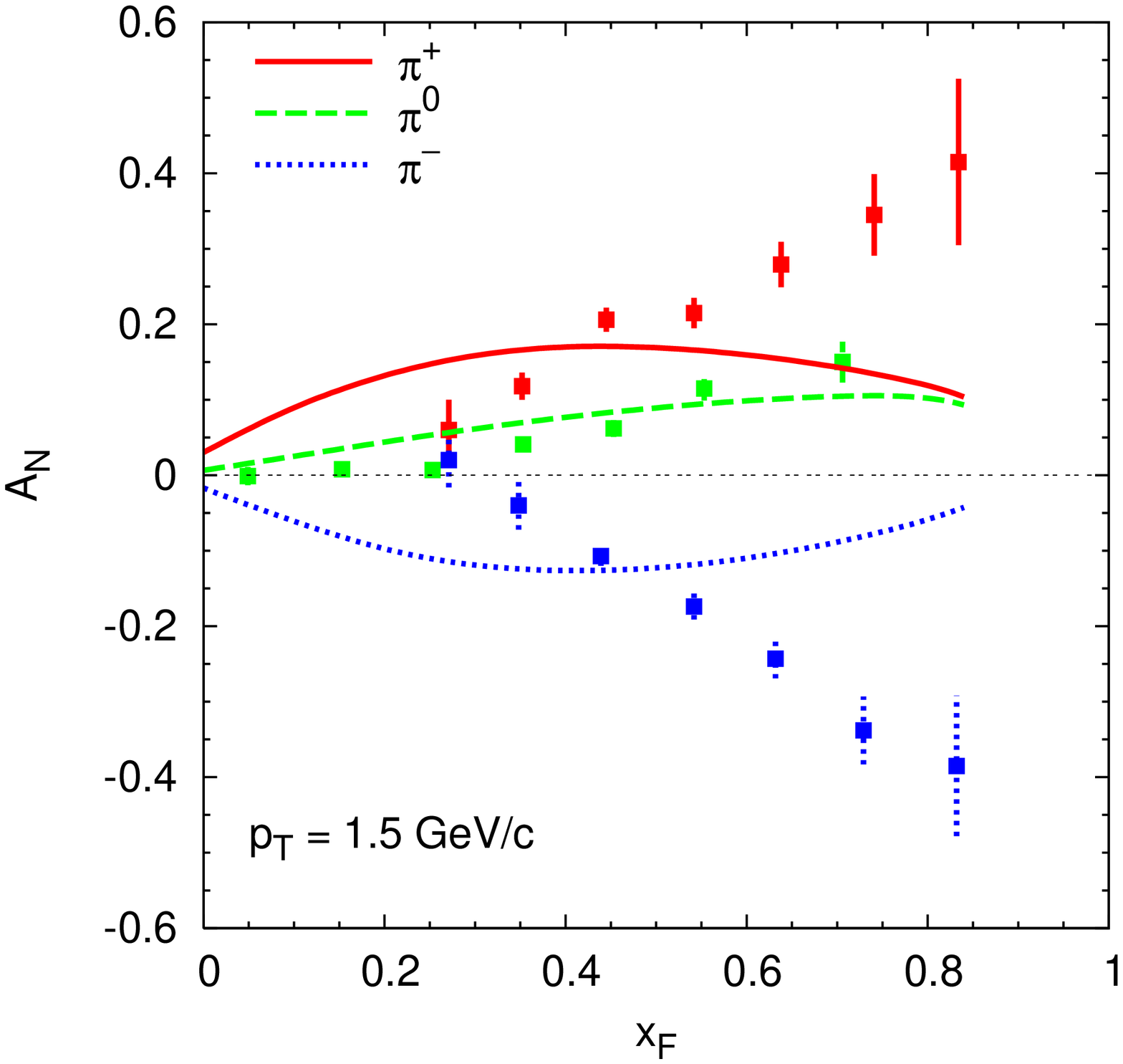,angle=-90,scale=0.35}
\end{minipage}
\begin{minipage}[t]{16.5 cm}
\caption{
\label{pheno:anmax:e704} Left panel:
Maximized contributions to $A_N$, plotted as a
function of $x_F$, for $p^\uparrow p \to \pi^+ + X$ processes and
E704 kinematics.
Solid line: quark Sivers mechanism alone; dashed line: gluon Sivers mechanism alone;
dotted line: transversity $\otimes$ Collins effect. All other contributions
are negligible. Unpolarized PDF and FF sets: MRST01-K.
Right panel:
Values of $A_N$ vs.~$x_F$ (E704 kinematics),
due to the Collins mechanism alone, obtained by saturating all proper bounds on the
unknown soft functions. Data are from
Refs.~\cite{Adams:1991rw,Adams:1991cs,Adams:1991ru}.
(Un)polarized PDF and FF sets: (MRST01)LSS01-K.
}
\end{minipage}
\end{center}
\end{figure}

In Fig.~\ref{pheno:anmax:e704} (left panel) we plot all different maximized contributions to
$A_N$ for the $p^\uparrow p \to \pi^+ + X$ process in the E704 experimental configuration,
for which very large values of $A_N$ have been
measured~\cite{Adams:1991rw,Adams:1991cs,Adams:1991ru}. Notice that, whenever different terms
could combine with different signs ({\it e.g.}, Sivers or Collins functions for different
quark flavours), their contributions have been summed always with {\it the same sign}, in
order to avoid any kind of cancelations not resulting from azimuthal factors and phase-space
integrations. One sees that potentially the Sivers mechanism is largely dominant, some role
could be played by the Collins mechanism, while all other contributions are
negligible~\cite{Anselmino:2005sh}.

%%%our style

However, a clear-cut statement on the relevance of the Collins
effect in $A_N(p^\uparrow p\to\pi+X)$ would require the combined
study of the SSA for both charged pions.

In Ref.~\cite{Anselmino:2004ky} a detailed study on this issue was
performed, by imposing the proper bounds for the TMD transversity
distributions~\cite{Bacchetta:1999kz} with the LSS01 set for the
helicity PDF's. This analysis showed how the phases involved, when
the $\bm{k}_\perp$ kinematics is treated carefully, are crucial,
leading to a strong suppression of the asymmetry generated by the
Collins mechanism, see Fig.~\ref{pheno:anmax:e704} (right panel).

The role of the azimuthal phases can be checked more explicitly by performing the simple
exercise of setting all of them to zero  (somewhat an analogue of the former simplified
studies with planar kinematics): in  such a case, as expected, one obtains considerably
larger results. However, the competing contributions from $qq\to qq$ and $qg\to qg$ processes
(due to the opposite sign in the partonic spin transfer factor), still prevent to get $A_N$
values as large as those observed at large $x_F$. This result is in contrast with former
believes, that the remarkably large SSA measured {\it e.g.} by the E704 experiment could be
generated by the Collins mechanism alone~\cite{Anselmino:1999pw}. In fact, in these former
studies, only the leading $\bm{k}_\perp$ effects were taken into account in a simplified
planar configuration for the hard scattering.

It is worth to mention here that the Collins effect is not completely out of reach in
hadronic collisions: indeed it has been recently shown that by measuring the azimuthal
distribution of a pion inside a jet inclusively produced in $p^\uparrow p$ collisions, one
can select this mechanism (see Refs.~\cite{Yuan:2007nd} and \cite{Dalesio:2007xx}).

{}From Fig.~\ref{pheno:anmax:e704} (left panel) one also sees that the contribution of the
gluon Sivers mechanism for E704 kinematics at negative $x_F$ could be important. Indeed, very
similar results can be obtained for the $p^\uparrow \bar p \to \pi + X$ process in the
kinematical configuration of the proposed PAX experiment at GSI~\cite{Barone:2005pu}, where
at large negative $x_F$ one could gain some information on the poorly known gluon Sivers
function.

The STAR experiment at RHIC-BNL has also measured non zero values of $A_N$  in $p^\uparrow p
\to \pi^0 + X$ processes at forward rapidity~\cite{Adams:2003fx,Gagliardi:2006uz}. Again, in
the STAR kinematical regime the Sivers mechanism gives the largest contribution to $A_N$.
There might remain some contribution from the Collins mechanism, while all other
contributions are negligible. In this case, however, in contrast with the lower energy case,
at negative $x_F$ all contributions are vanishingly small.

Let us now try to explain some of the above results, namely, the
different behaviour of the Sivers mechanism in different kinematical
configurations. The azimuthal phase factor $\cos\phi_a$ entering the
Sivers function, see section~\ref{form}, plays a crucial role and
deserves some additional comments. The only other term depending on
$\phi_a$ in the contribution of the Sivers effect to the numerator
of $A_N$ is the partonic cross section, in particular via the
corresponding Mandelstam variable $\hat{t}$. At large positive $x_F$
and moderately large $p_T$ the (average) values of $\hat{t}$ are
relatively small. Therefore, the (dominant) $\hat{t}$-channel
contributions, proportional to $1/\hat t^{\,2}$, depend sizably on
$\phi_a$, so that $A_N$ is not necessarily suppressed. This explains
the potentially large effect at large $x_F$. Instead, for negative
values of $x_F$, all partonic Mandelstam variables are much less
dependent on $\phi_a$, so that one is roughly left with the
$d^2\bm{k}_{\perp a}\,\cos\phi_a$ integration alone, which cancels
the potentially large Sivers contribution. As a consequence, the
possibility of gaining information on the gluon Sivers distribution
from the recent STAR and BRAHMS data at negative values of $x_F$ is
frustrated. Notice that, due to the much lower values of $\sqrt{s}$
involved, this suppression caused by the $\cos\phi_a$ dependence is
much less effective for the kinematical regimes of E704 and of the
proposed PAX~\cite{Barone:2005pu} experiment.

On the basis of the above results it is clear that in the GPM approach
the Sivers effect is the only mechanism
which could by itself describe the large $A_N$ values observed by the
E704, STAR and BRAHMS collaborations in the forward region.

In Ref.~\cite{D'Alesio:2004up}, in order to perform numerical estimates and carry out a
comparison with available experimental data a simple factorized form for the Sivers function
was introduced:
 \be
 \Delta^Nf_{q/p^\uparrow}(x, k_\perp) = 2 \, {\mathcal N}^S_q(x)
 \,f_{q/p}(x)\, {\mathcal H}(k_\perp) \,\frac{\beta^2}{\pi}\, e^{-\beta^2 k_\perp^2}\,,
 \label{pheno:sivpar}
 \ee
where $\beta^2=1/\langle k^2_\perp\rangle$ is a flavour-independent
width parameter defining the Gaussian $k_\perp$ shape of the
unpolarized PDF (as extracted from the unpolarized cross section
analysis: $\langle k^2_\perp\rangle=0.64$ (GeV$/c)^2$
\cite{D'Alesio:2004up}). Notice that the natural positivity bounds
are automatically fulfilled  by imposing ${\mathcal N}^S_q$ and
${\mathcal H}$ to be separately smaller than unity in magnitude in
the full $x$ and $k_\perp$ range respectively (see section
\ref{pheno:sidis:ssa}).

To start with, only valence $u$ and $d$ quark Sivers functions (QSF) have been considered,
neglecting (for what concerns the Sivers effect only) possible contributions from antiquarks
and gluons. The fact that sizable values of $A_N$ set up at large positive $x_F$ somehow
supports this choice.

In order to describe the E704 data invoking the Sivers effect alone, one needs a positive
(negative) $u$ ($d$) quark Sivers function, $\Delta^{\!N}f_{q/p^\uparrow}(x,k_\perp)$ [recall
the opposite sign between $\Delta^{\!N}f_{q/p^\uparrow}$ and $f_{1T}^{\perp q}$, see
Eq.~(\ref{theo:notation})]. Therefore, in the subsequent fragmentation process the unfavoured
flavour contributions (e.g. $d\to\pi^+$) enter with opposite sign (all other factors in the
kernel being positive) w.r.t.~the favoured ones (e.g. $u\to\pi^+$) in the charged pion SSA's.
This implies that the extraction of the quark Sivers functions is sensitive to the relative
weight of the leading and non-leading terms in a specific set of FF's.

%%%%%%ppnp style
\begin{figure}[h!tb]
\begin{center}
\begin{minipage}[t]{16 cm}
\vspace*{-2cm} \epsfig{figure=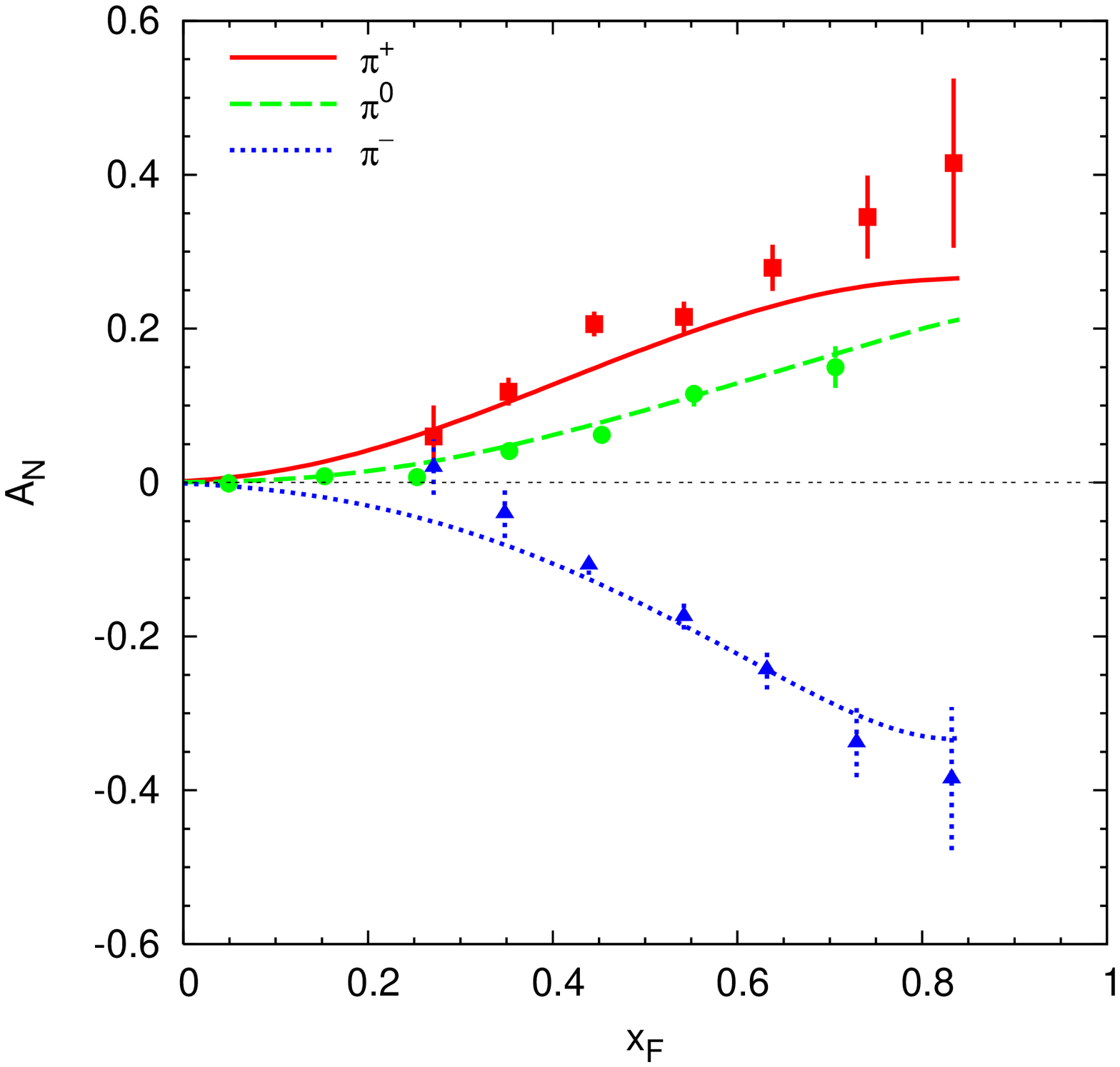,angle=-90,scale=0.35} \hspace*{0.3cm}
\epsfig{figure=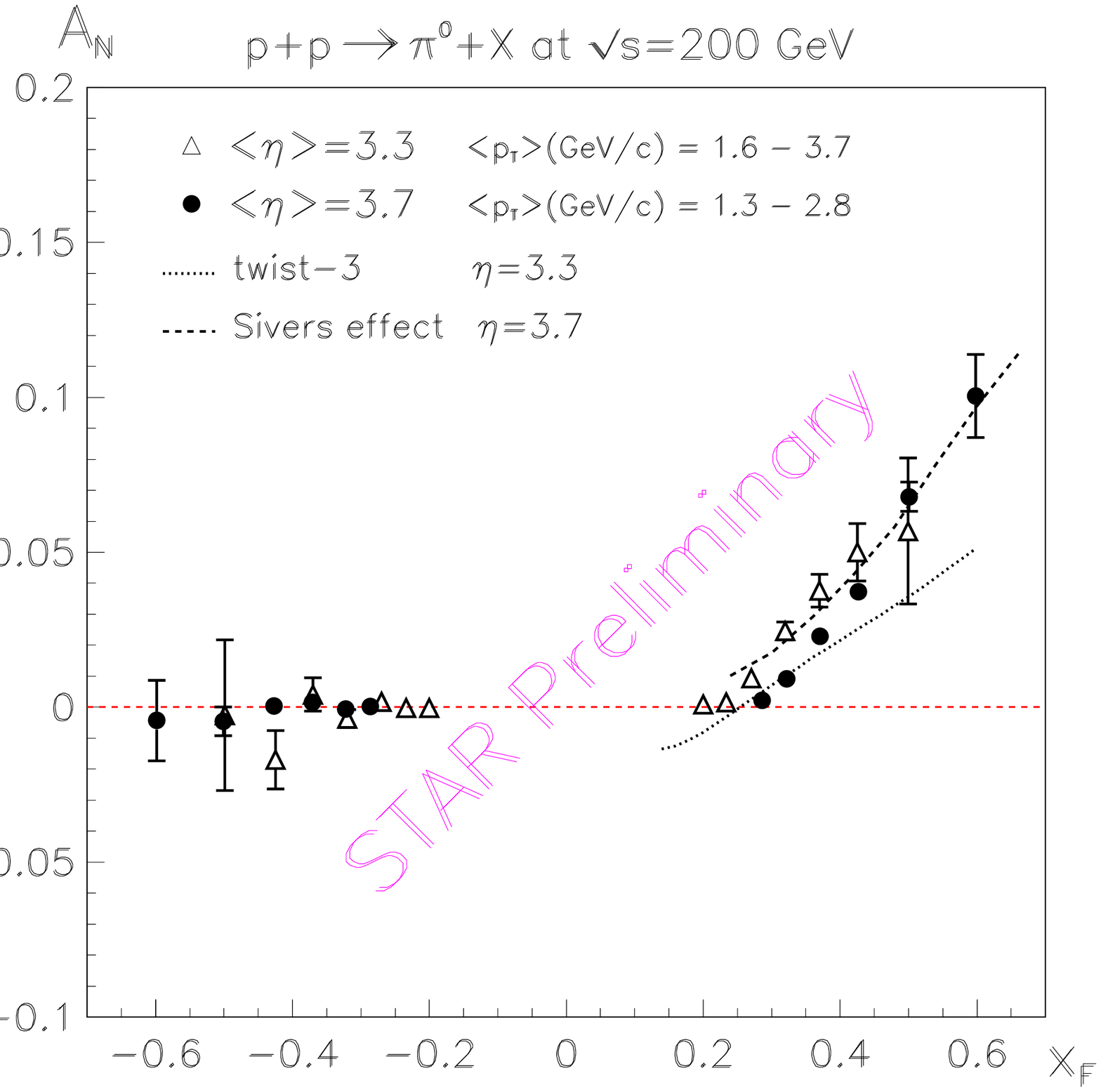,bb= 10 520 640 760,angle=0,scale=0.35}
\end{minipage}
\begin{minipage}[t]{16.5 cm}
\caption{Left:  $A_N(p^\uparrow p \to \pi + X)$
 vs.~$x_F$, at $\sqrt s = 19.4$ GeV
and fixed $p_T=1.5$ GeV/$c$ (Sivers effect); unpolarized PDF and FF sets: MRST01-KKP; data
are from \cite{Adams:1991rw,Adams:1991cs,Adams:1991ru}. Right: $A_N(p^\uparrow p \to \pi^0 +
X)$ at $\sqrt s = 200$ GeV; data are from~\cite{Nogach:2006gm}. A comparison with
 the Sivers effect in the GPM approach (dashed line)~\cite{D'Alesio:2004up} and
twist-three calculations (dotted line)~\cite{Kouvaris:2006zy} is also shown.
 \label{pheno:an:xf:sivers}}
\end{minipage}
\end{center}
\end{figure}

Under these assumptions, a good description of E704 data was obtained, see
Fig.~\ref{pheno:an:xf:sivers} (left panel). Although somehow expected, the most striking
feature of the Sivers functions so obtained was the marked valence-like behaviour (that is,
$\Delta^{\!N}f_{q/p^\uparrow}(x,k_\perp)$ is strongly suppressed at small $x$). This
analysis, which consistently treats the full $\bm{k}_\perp$ kinematics in the calculation of
both the numerator and the denominator of $A_N$, also confirms all the main features of the
Sivers contribution and of the Sivers function parameterizations found in previous papers
based on a simplified planar kinematics~\cite{Anselmino:1994tv,Anselmino:1998yz}.

\begin{figure}[h!tb]
\begin{center}
\begin{minipage}[t]{9 cm}
\epsfig{figure=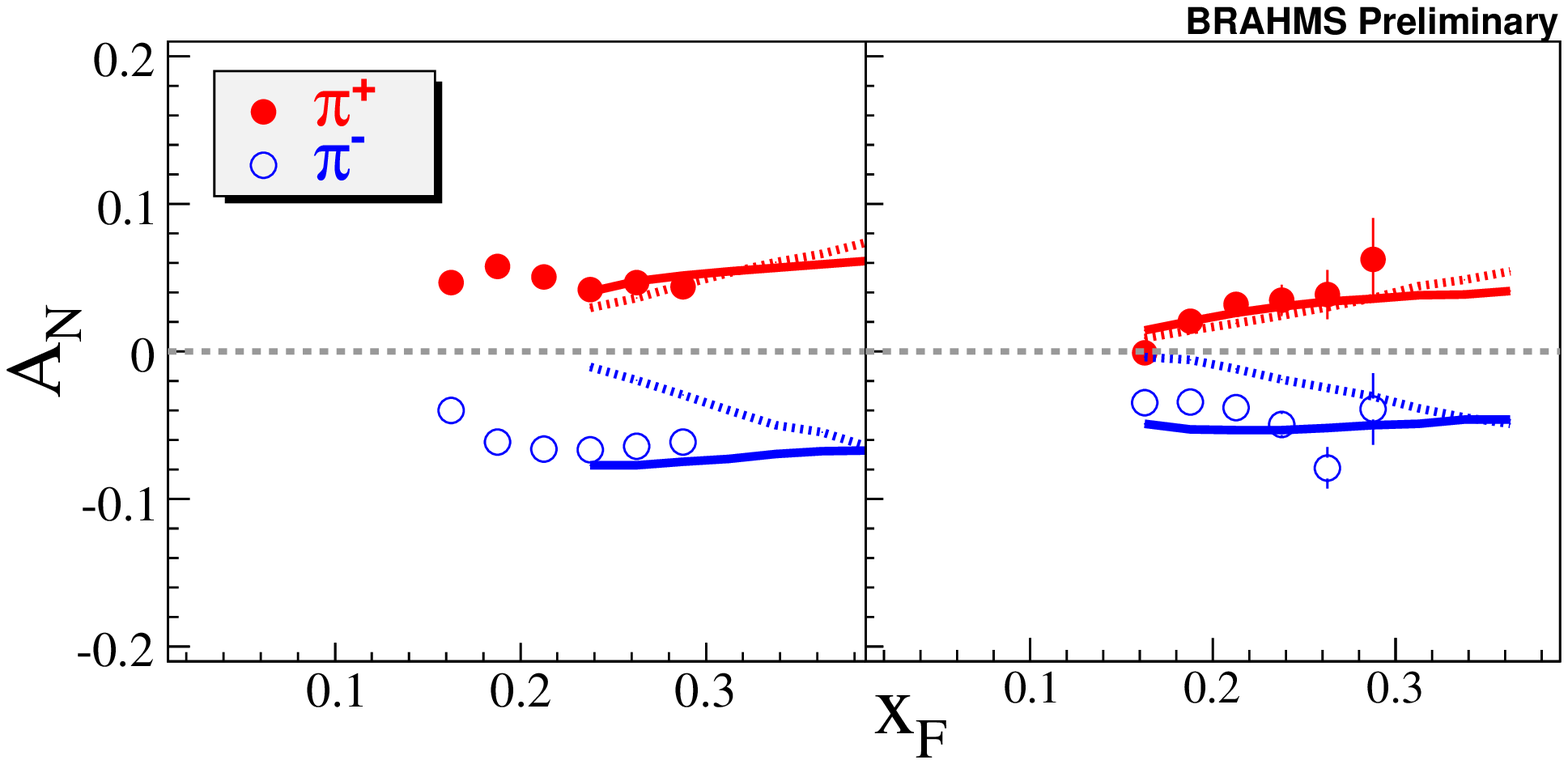,scale=0.45, bb=  300 50 700 300}
  \hspace*{-20pt}
\epsfig{figure=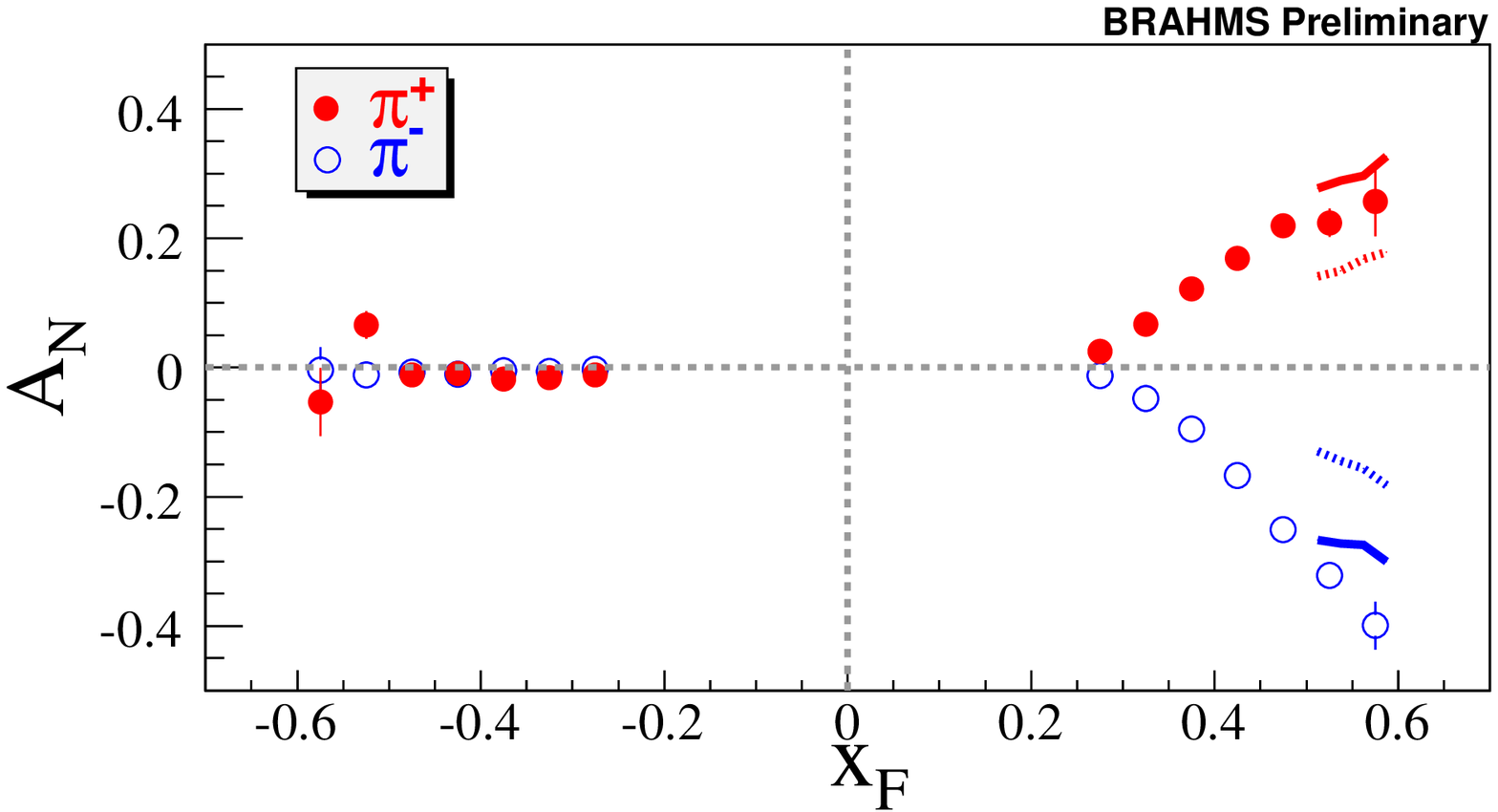,scale=0.45, bb=  100 50 10 300}
\end{minipage}
\begin{minipage}[t]{16.5 cm}
\caption{
Left panel: $A_N(p^\uparrow p \to \pi^\pm +X)$
 at $\sqrt s = 200$ GeV for two scattering angles,
2.3$^0$ (left) and 4$^0$ (right). Right panel: $A_N$ at $\sqrt s = 62$ GeV. Dotted line: GPM,
Sivers effect; solid line: twist-three approach. Curves are calculated at $p_T> 1$ GeV$/c$.
Data are from \cite{Lee:2007zzh}.
 \label{pheno:an:brahms}}
\end{minipage}
\end{center}
\end{figure}

The so extracted Sivers function has been adopted to give predictions at much larger energies
as those reached at RHIC \cite{D'Alesio:2004up}. The results have been found in good
agreement with the subsequent published data \cite{Adams:2003fx} both in the positive and in
the negative $x_F$ region. Latest STAR results, with enhanced statistics, have confirmed the
peculiar $x_F$ behaviour of $A_N$. In Fig.~\ref{pheno:an:xf:sivers} (right panel) we show the
theoretical estimates obtained by adopting the Sivers functions as extracted from E704 data
(dashed line) and compare them with preliminary data from STAR
collaboration~\cite{Nogach:2006gm}. As predicted theoretically, sizable pion SSA's, with
features similar to those found in the E704 kinematical regime, are confirmed experimentally
even at such large c.m.~energies. In Fig.~\ref{pheno:an:xf:sivers} (right panel) we also show
theoretical results in the twist-three approach (dotted line). In order to compare the two
phenomenological studies some words are mandatory: in Ref.~\cite{Kouvaris:2006zy}, adopting
the twist-three approach a {\em global} fit on all available E704, STAR and BRAHMS data has
been performed, leading to a good description of SSA results (see also
Fig.~\ref{pheno:an:brahms}); LO cross sections in collinear configuration have been adopted
to compute the denominator of $A_N$; in order to describe both low and high energy SSA data,
calculations for E704 kinematics have been rescaled by a factor of 2.

Further data for charged pion SSA's have been collected by the
BRAHMS collaboration~\cite{Lee:2007zzh}, covering a different
kinematical region (most of the data is at lower $p_T$) and at two
energy values. In this case, adopting the parameterization of the
quark Sivers functions extracted by fitting E704 data, theoretical
results underestimate the high-energy data (dotted lines in
Fig.~\ref{pheno:an:brahms}). This indicates that an extension of the
GPM fit procedure including all data is required to better fix the
Sivers function. Indeed the global fit based on the twist-three
approach gives a better description (Fig.~\ref{pheno:an:brahms},
solid lines).

%%%%%%ppnp style

\begin{figure}[h!tb]
\begin{center}
\begin{minipage}[t]{12 cm}
\epsfig{figure=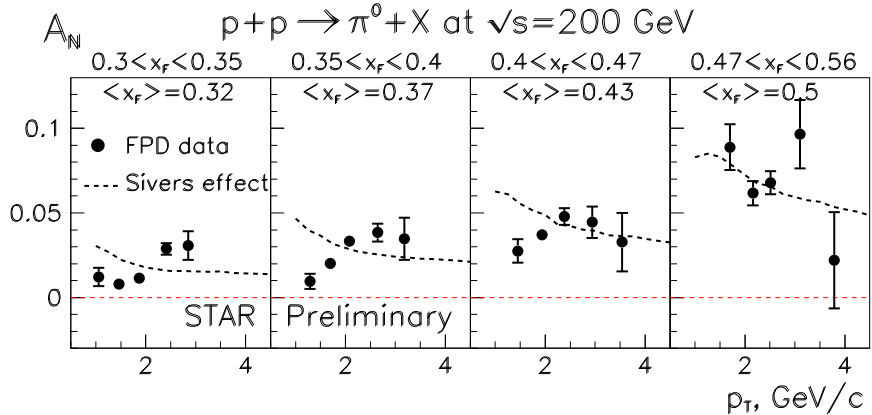,scale=0.6}
\end{minipage}
\begin{minipage}[t]{16.5 cm}
\caption{
 $A_N(p^\uparrow p \to \pi^0 +X)$  at $\sqrt s = 200$ GeV
as a function of $p_T$ and for different $x_F$ bins.
Data are from \cite{Nogach:2006gm}. Predictions for the Sivers effect in the GPM approach
 are also shown.
\label{pheno:an:star:pt:sivers}}
\end{minipage}
\end{center}
\end{figure}

In Fig.~\ref{pheno:an:star:pt:sivers} we show preliminary STAR data on $A_N(p^\uparrow p \to
\pi^0 +X)$, as a function of $p_T$~\cite{Nogach:2006gm} in different $x_F$ bins and compare
them with theoretical predictions based on the Sivers effect alone. With the possible
exception of the lowest $p_T$ points, at small $x_F$, $A_N$ is reasonably reproduced, taking
into account that the GPM approach (based on pQCD), should be better suited for $p_T$ bigger
than 1-2 GeV/$c$. Let us also remind that the theoretical curves are (genuine) predictions
obtained from a simple parameterization of the quark Sivers functions, aimed to describe the
main features of the large $x_F$ E704 data (at much lower energy). A similar $p_T$-behaviour
is also expected in the twist-three approach.

%%%%%%ppnp style

We mention here that a preliminary study adopting the parameterizations of the Sivers
function extracted from the azimuthal asymmetries observed in SIDIS (see
section~\ref{pheno:sidis:ssa}) leads to a much better agreement with BRAHMS data, still
preserving a fair description of STAR data~\cite{Dalesio:2007xx}.

Let us focus now on the mid-rapidity region. As already mentioned above, in this region only
the Sivers effect could generate a sizable SSA in the GPM approach. In particular, at the
large c.m.~energies reached at RHIC, the potentially dominant contribution should come from
the gluon Sivers function (GSF). This region is indeed covered by the PHENIX experiment,
whose recent data are consistent with a vanishing SSA~\cite{Adler:2005in}. This potential
gluon dominance, together with the (almost) vanishing of all possible contributions to $A_N$
other than the Sivers effect, allow to interpret the data -- showing tiny values of $A_N$ --
in terms of useful constraints (upper bounds) on the magnitude of the GSF. As $p_T$ grows,
however, $x_{a}^{\rm{min}}$, the minimum value of the parton light-cone momentum fraction in
the polarized proton, increases and the dominance of the gluonic channels becomes less
prominent.

\begin{figure}[h!b]
\begin{center}
\begin{minipage}[t]{14 cm}
\epsfig{figure=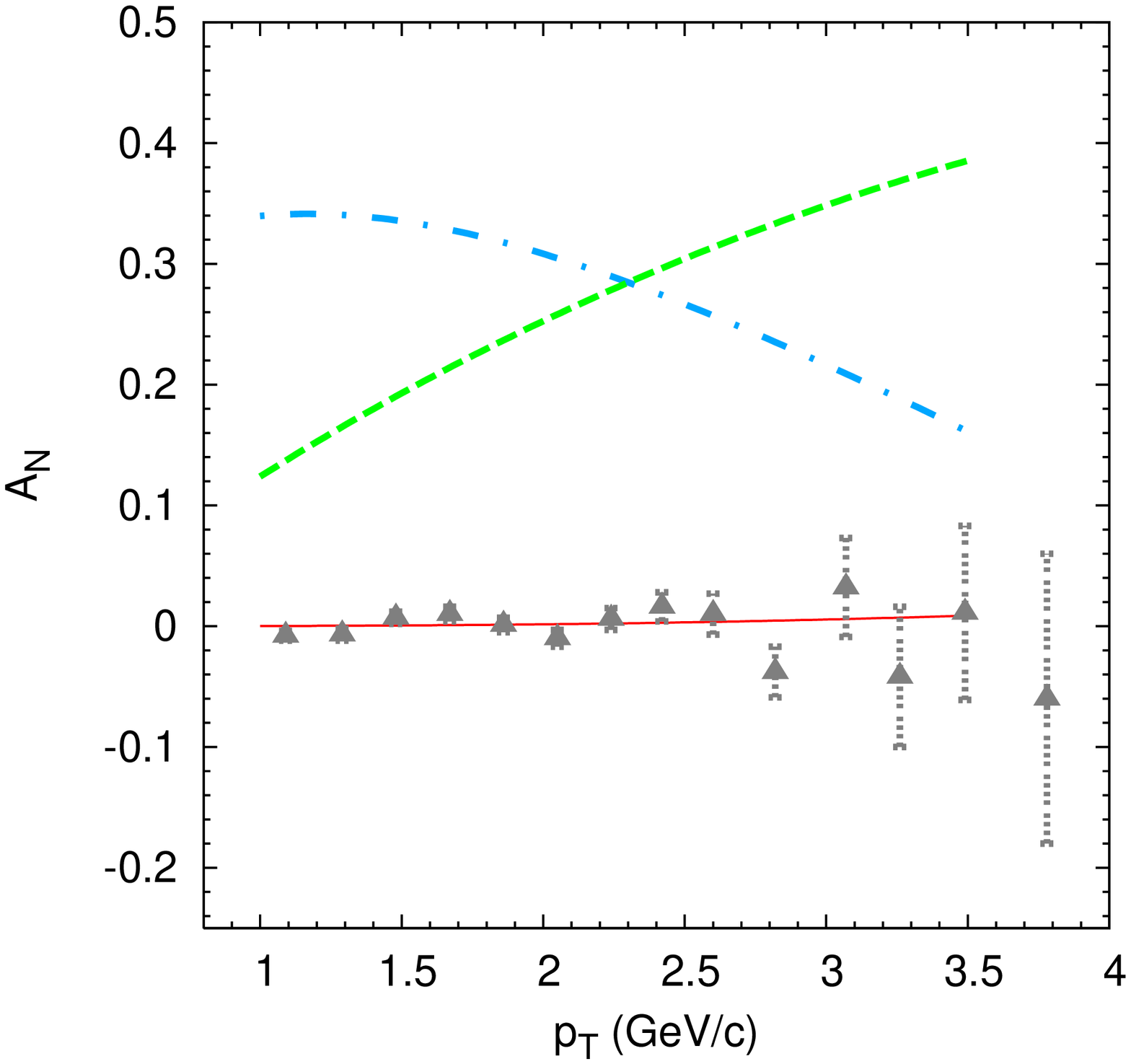,angle=-90,scale=0.3}
\hspace*{-1cm}
\epsfig{figure=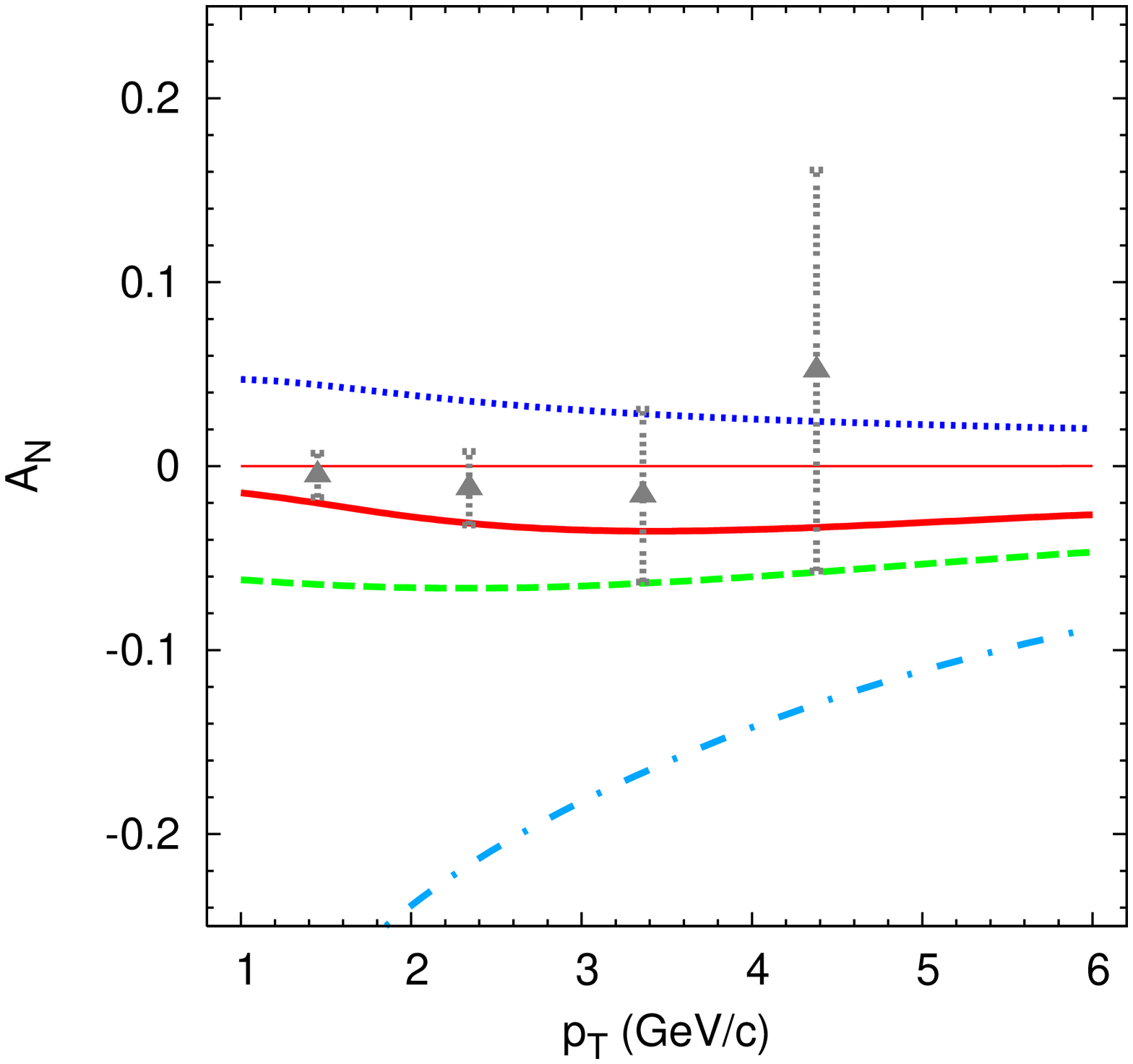,angle=-90,scale=0.3}
\end{minipage}
\begin{minipage}[t]{16.5 cm}
\caption{
$A_N(p^\uparrow p\to \pi^0 +X)$ at mid-rapidity vs.~$p_T$.
Left panel: E704 data~\cite{Adams:1994yu} at $\sqrt s \simeq 20$ GeV.
Maximized contributions, in magnitude, from the quark (dashed line)
and gluon (dot-dashed line) Sivers functions.
Right panel: PHENIX data~\cite{Adler:2005in} at $\sqrt s = 200$ GeV.
Contributions of: maximized GSF (dot-dashed line);
sea (maximized) + valence quark Sivers functions (dotted line);
constrained GSF (dashed line).  Overall total Sivers effect (thick,
solid line).
In both plots the thin, solid lines show the predictions using the
 $u$ and $d$ Sivers functions extracted in
Ref.~\cite{D'Alesio:2004up}. Unpolarized PDF and FF sets: MRST01-KKP.
\label{pheno:an:midrap}}
\end{minipage}
\end{center}
\end{figure}

For the same reason, another set of data, for comparable rapidity and $p_T$ ranges
\cite{Adams:1994yu} but at much lower energy, $\sqrt{s}\simeq 20$ GeV, cannot give
significant constraints on the GSF. In this case, in fact, even at the smallest $p_T$ values
$x_{a}^{\rm{min}}$ remains large and a possible mixing with quark initiated contributions
cannot be excluded, see Fig.~\ref{pheno:an:midrap} (left panel). Different scenarios have
been considered in Ref.~\cite{Anselmino:2006yq}, where a significant constraint on the GSF
has been obtained, see Fig.~\ref{pheno:an:midrap} (right panel). For instance, even allowing
for the largest possible cancelations coming from maximized, sea-quark Sivers contributions
(still totally unknown), in the small $x$ region (below 0.05), where the gluon plays a
relevant role, the GSF is bounded to be less then 30\% in size of its positivity bound (twice
the unpolarized gluon distribution).

%%%%%%ppnp style

As an indirect tool to constrain the gluon Sivers function one could also consider the
Burkardt sum rule of Eq.~(\ref{theo:Siv:rule})~\cite{Burkardt:2004ur}. As discussed in
section~\ref{theo}, this rule states that the net transverse momentum due to the Sivers
mechanism must vanish. Strict use of the BSR would require integration over the full $x$
range, including the poorly known low $x$ region, of each single parton Sivers function,
which might even result in divergences. For this reason at the present stage it can be only
adopted as a useful cross-check. Similar conclusions on the smallness of a possible GSF have
been reached in a different context in Ref.~\cite{Brodsky:2006ha}. Let us also mention that
in Ref.~\cite{Efremov:2004tp} the BSR, together with the large $N_c$ limit of QCD, was used
to predict a suppression of the gluon Sivers function w.r.t.~the non-singlet quark Sivers
distribution at not too small $x$.

Another interesting process to study is inclusive kaon production in $p^\uparrow p\to K + X$,
in particular for charged kaons, where hopefully one could learn more on T-odd mechanisms in
the sea-quark sector.

\begin{figure}[h!tb]
\begin{center}
\begin{minipage}[h!t]{8 cm}
\epsfig{figure=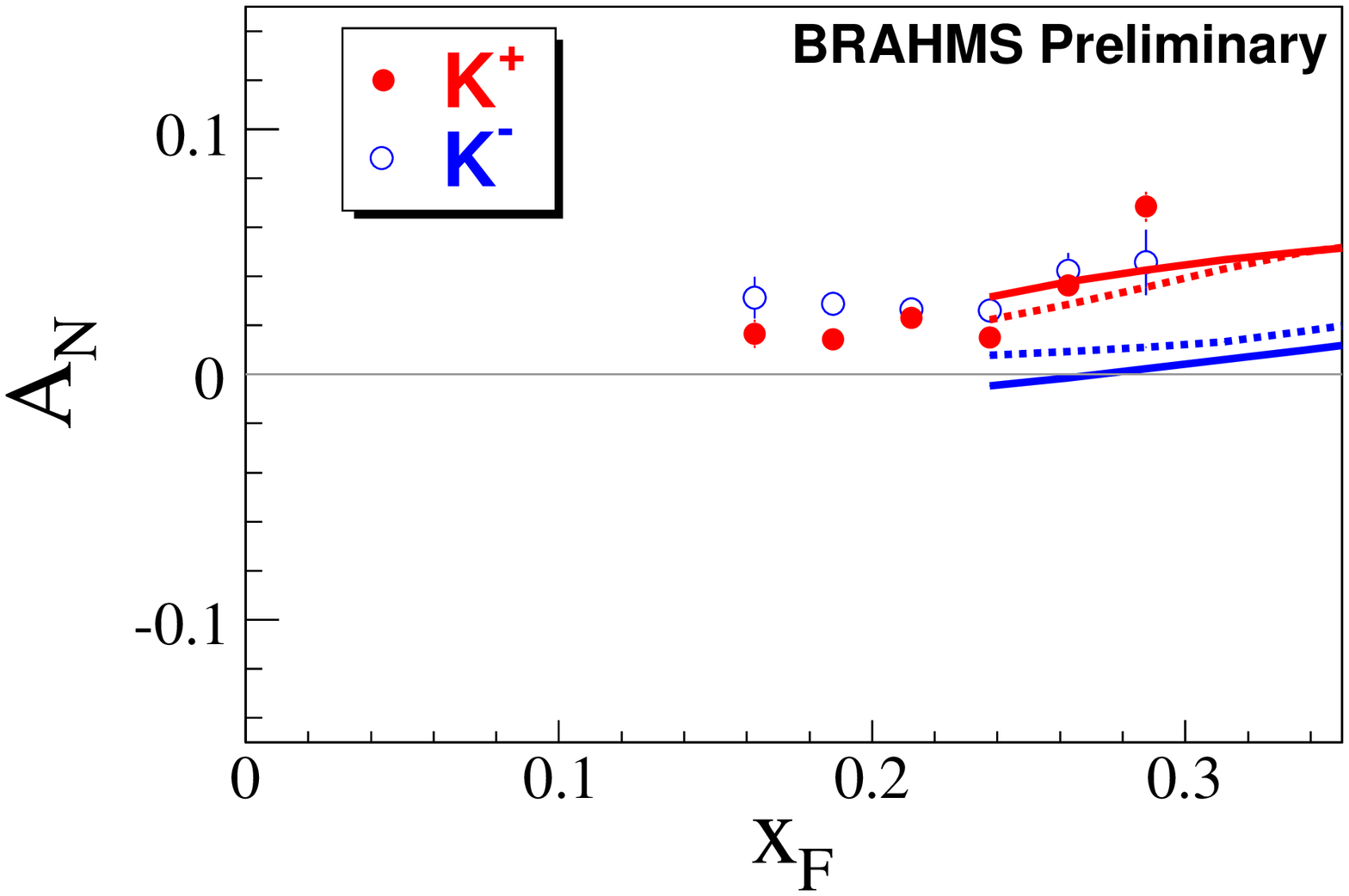,scale=0.35, bb=  300 20 700 370}
\epsfig{figure=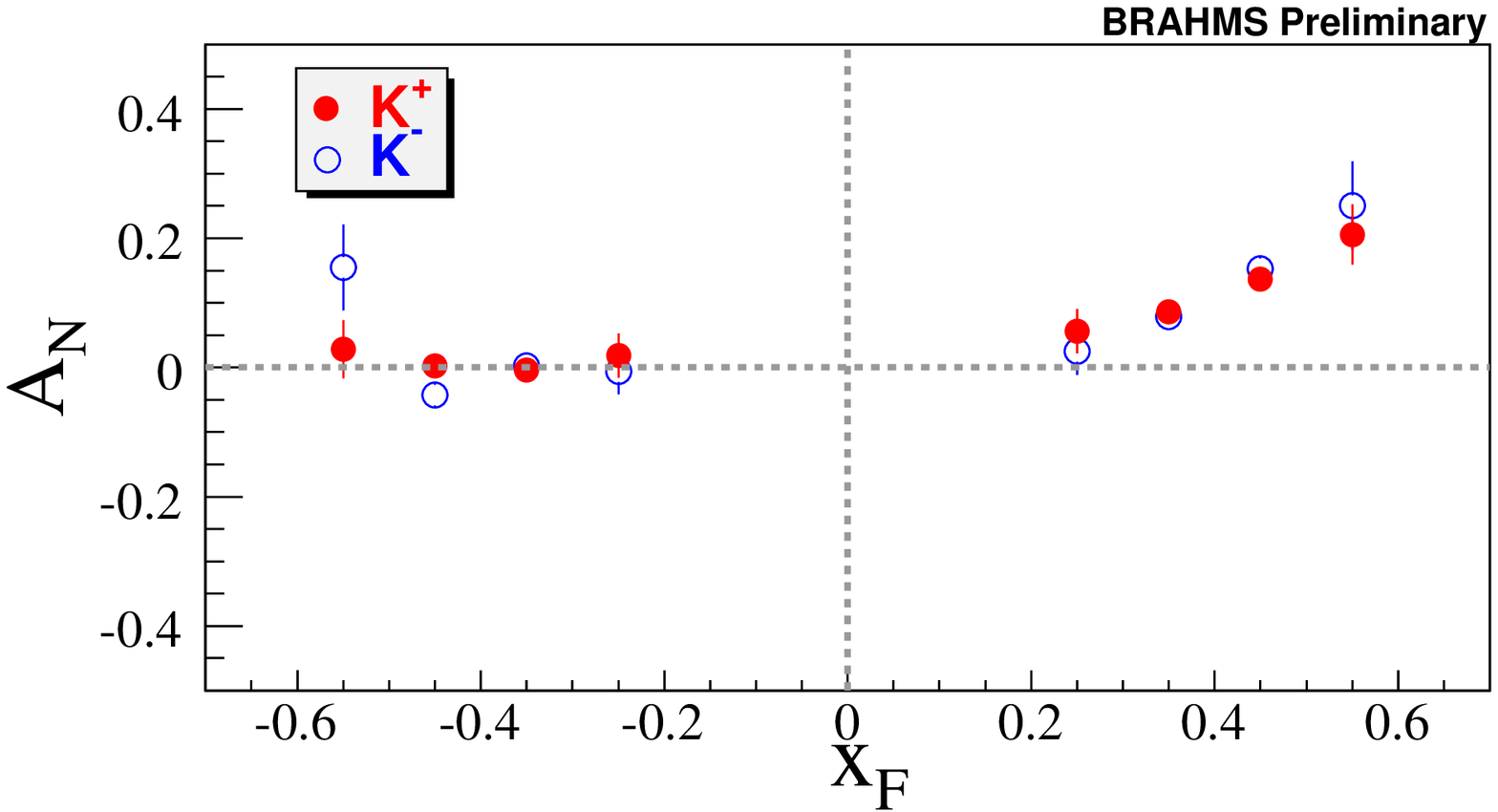,scale=0.4, bb=  100 0 10 370}
\end{minipage}
\begin{minipage}[t]{16.5 cm}
\caption{
 $A_N$ for inclusive charged kaon production
in $pp$ collisions, at $\sqrt s = 200$ GeV (left)
 and $\sqrt s = 62$ GeV  (right)
as a function of $x_F$. Dotted line: GPM, Sivers effect; solid line: twist-three approach.
Data are from \cite{Lee:2007zzh}. \label{pheno:ankaon:brahms}}
\end{minipage}
\end{center}
\end{figure}

To begin with, one can figure out what SSA's would result in the valence-like picture for the
quark Sivers function discussed above. Consider for the sake of clarity the forward region
(large positive $x_F$). In this case, $A_N(K^+)$ should be quite similar to $A_N(\pi^+)$: in
both cases, a valence $u$ quark from the incoming proton fragments dominantly into the
observed meson. This qualitative expectation of the TMD generalized parton model works, in
principle, for the Sivers as well as the Collins effects, and holds true for the twist-three
collinear formalism and for the orbiting valence quark model discussed in section~\ref{theo}.
However, the $K^-$ case is more subtle, since the valence $u$, $d$, quarks from the proton
are both subleading in the fragmentation process. Therefore, the expected SSA is more
sensitive to the details of the approach. We already said in section~\ref{theo} that in the
orbiting valence quark model one would get a zero SSA for $K^-$. This would also be the case
for the Collins effect in the model of Artru~\cite{Artru:1995bh}, because the specific Lund
string model utilized gives a vanishing Collins function for subleading fragmentation. On the
other hand the Sivers effect, involving unpolarized TMD fragmentation functions, would give a
positive and, depending on the FF set adopted, sizable SSA~\cite{Anselmino:1998yz}. This
somehow ``unexpected'' result for $A_N(K^-)$ comes from a non negligible unfavoured $u\to
K^-$ fragmentation function and the fact that, at large $x$, the (positive) $u$ quark Sivers
function is bigger in size than that for a $d$ quark. This expectation seems to be confirmed
by recent preliminary BRAHMS results (see Fig.~\ref{pheno:ankaon:brahms}), although the
theoretical predictions still underestimate the data. This discrepancy could be reduced by
including the sea-quark Sivers effect, or the non negligible contributions of the Collins
effect from favoured and unfavoured FF's.

Similar results have been obtained in the global fit based on the
twist-three approach, adopting chiral-even twist-three distributions
\cite{Kouvaris:2006zy}. See also section~\ref{pheno:sidis:ssa} for
further discussion on the kaon case.

Another useful tool to access specific spin and TMD effects is the study of heavy meson
production in $pp$ collisions. $D$ mesons originate predominantly from the fragmentation of
$c$ or $\bar c$ quarks, which at LO can be created either via $q \bar q$ annihilation, $q
\bar q \to c \bar c$, or via the gluon fusion process,  $gg \to c \bar c$. In this case the
heavy quark mass cannot be neglected. This somehow makes the structure of the kernels in the
helicity formalism of the GPM more involved. The main difference is that now helicity
conservation is broken in the hard scattering amplitudes. Once again, a detailed study
including the exact $\bm{k}_\perp$ kinematics shows that upon $\bm{k}_\perp$ integration the
azimuthal phase factors suppress all allowed terms but the Sivers contribution.

In Ref.~\cite{Anselmino:2004nk} it was shown how at RHIC energies,
$\sqrt{s} = 200$ GeV, and $p_T$ roughly in the range $1\div 3$
GeV$/c$, the gluon fusion mechanism gives the dominant contribution
to the $p^\uparrow p \to D+X$ process, up to $x_F \simeq 0.6$,
opening the possibility for a direct access to the gluon Sivers
function. At lower energies, like those reachable at the proposed
PAX experiment at GSI, $\sqrt s=$ 14 GeV, the $q\bar q$ annihilation
process becomes dominant, giving the opportunity to access directly
the quark Sivers function. For polarized $p\bar p$ collisions at PAX
the potential QSF dominance would be even more
dramatic~\cite{D'Alesio:2006fp}.

Let us finally mention here that the inclusion of a possible intrinsic charm component in the
proton, as originally proposed by Brodsky {\it et al.}~\cite{Brodsky:1980pb} (see
e.g.~Ref.~\cite{Pumplin:2005yf} and references therein), could change drastically this
picture. In this case, in fact, $c$-quark initiated processes and other TMD effects could
play a role. This interesting topic would certainly deserve a more detailed study.

\subsubsection{Prompt photon production}
\label{pheno:gamma}
The inclusive production of direct photons in $pp$ collisions is certainly another useful
tool to access $\bm{k}_\perp$-dependent PDF's. At LO, only two partonic subprocesses,
$q(\bar{q})\,g\to\gamma\, q(\bar{q})$ and $q\bar q\to \gamma g$, contribute. The polarized
differential cross section can be obtained from the general case, Eq.~(\ref{form:genunp}), by
replacing the fragmentation function by a product of two Dirac delta functions, $\hat
D_{C/c}(z,\bm{k}_{\perp C})\to\delta(z)\,\delta(\bm{k}_{\perp C})$.

There are in principle three mechanisms that could contribute to the photon SSA: the quark
Sivers effect for both subprocesses, the gluon Sivers effect via the Compton-like subprocess,
and the Boer-Mulders effect (coupled with the transversity PDF) via $q\bar q$ annihilation.
Notice also that electromagnetic couplings will enhance the $u$-flavour contributions. For
$x_F>0$, and fixed $p_T$, the dominant contribution to $A_N$ should come from the quark
Sivers effect. Adopting the QSF parameterizations extracted from the fit to
$A_N(p^{\uparrow}p\to\pi+X)$, in the GPM one gets a positive SSA, rising at larger $x_F$,
both at $\sqrt{s}=20$ GeV and $\sqrt{s}=200$ GeV~\cite{D'Alesio:2004up}. Notice that the
collinear twist-three approach~\cite{Qiu:1991wg,Qiu:1998ia,Kouvaris:2006zy}, while leading to
a similar description of $A_N$ for pion production, predicts a photon SSA opposite in sign.
The reason is just the opposite sign entering the hard scattering part for the dominant
channel in the twist-three formalism. In this respect, this process can be considered as a
good tool to discriminate between the two approaches.

Concerning the backward rapidity region, in Ref.~\cite{Schmidt:2005gv} it was argued that at
RHIC energies and large $p_T$ values (around 20 GeV$/c$) $A_N$ would be directly sensitive to
the GSF. In fact, in a more quantitative approach, with proper treatment of the noncollinear
kinematics and the relative azimuthal phases, it has been found that the best kinematical
region to look for the GSF at RHIC energies is at negative $x_F$ and $p_T$ values around 5-8
GeV$/c$. In this region the contribution to $A_N$ from the GSF could be as large as 10\%,
whereas the other mechanisms would give at most a 1\% contribution~\cite{D'Alesio:2006fp}.

An even more interesting case is the photon SSA at lower energies, like for instance at the
PAX-GSI experiment. In this case, the QSF gives the main contribution in the forward as well
as in the central rapidity region, whereas again the negative $x_F$ region is potentially
dominated by the GSF. Moreover, by colliding polarized protons off unpolarized antiprotons,
another effect could become accessible: namely the Boer-Mulders function,
$\Delta^{N}f_{q^{\uparrow}/p}$ (or $h_{1}^{\perp}$), coupled to the transversity
distribution. In this configuration the Compton-like subprocess is suppressed because: $i)$
the minimum value of $x$ reached is still large, being of the order of $2p_T/\sqrt s$; $ii)$
the $\bar q$ in $\bar p$ has a valence component. As a result, we have a clear dominance of
the $q\bar q$ subprocess. Moreover, the integration over the $\bm{k}_\perp$-dependent phases
does not wash out the partonic double spin asymmetry. By using the parameterizations so far
extracted for the QSF one gets a photon SSA of the order 5-10\% for $p_T$ around 2-4 GeV/$c$,
while the (maximized) Boer-Mulders effect might give $A_N$ values of the order up to 20\%.
This means that in this kinematical regime a large photon SSA could be a clear signal in
favour of the Boer-Mulders effect, then giving another way to access the transversity
distribution~\cite{D'Alesio:2006fp}.

\subsubsection{Transverse $\Lambda$ polarization in unpolarized $pp$ collisions}
\label{pheno:lam}

%%%%%%ppnp style
\begin{figure}[b!]
\begin{center}
\begin{minipage}[h!t]{16 cm}
\epsfig{figure=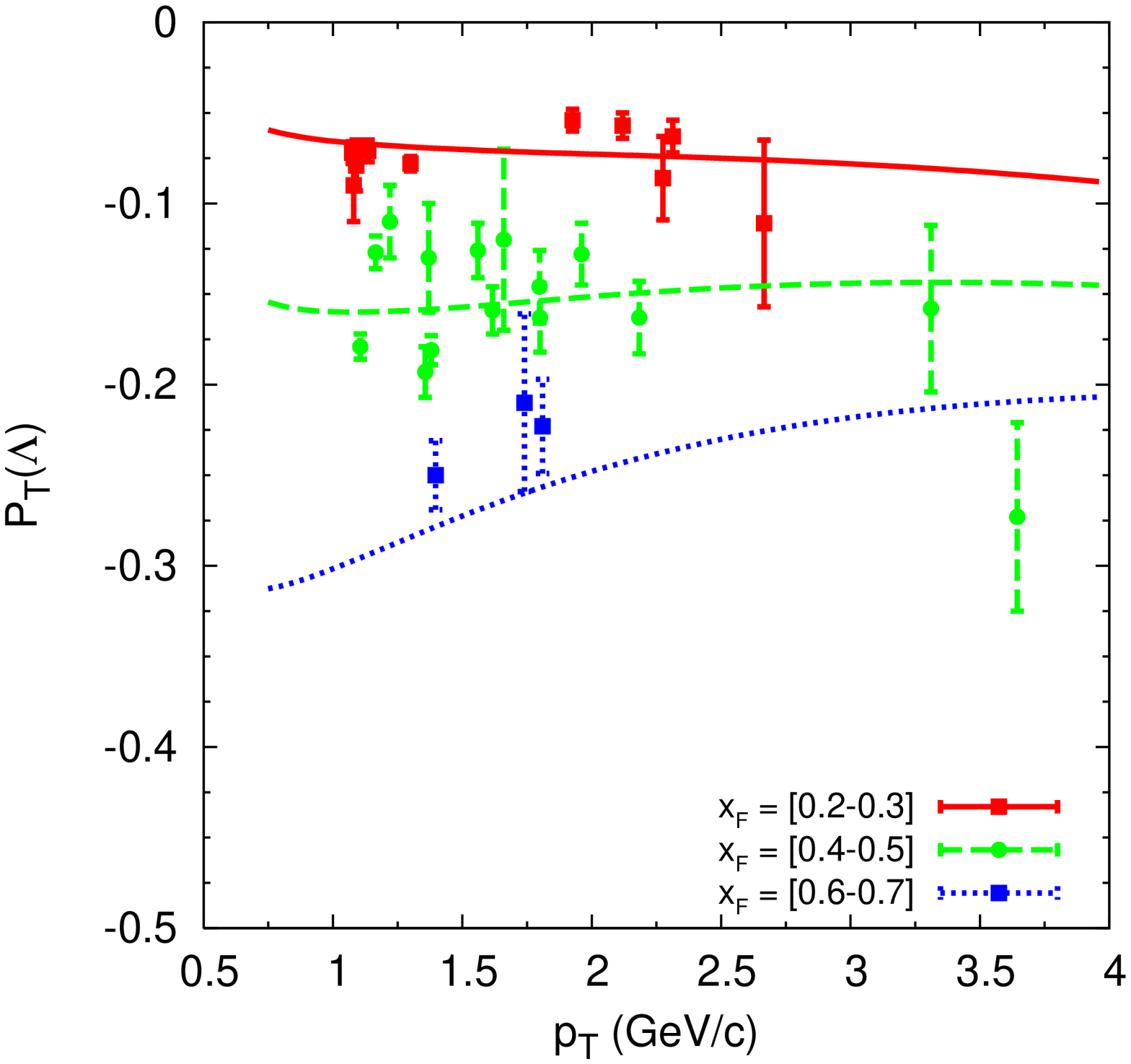,angle=-90,scale=0.35}
%\hspace*{0.5cm}
\epsfig{figure=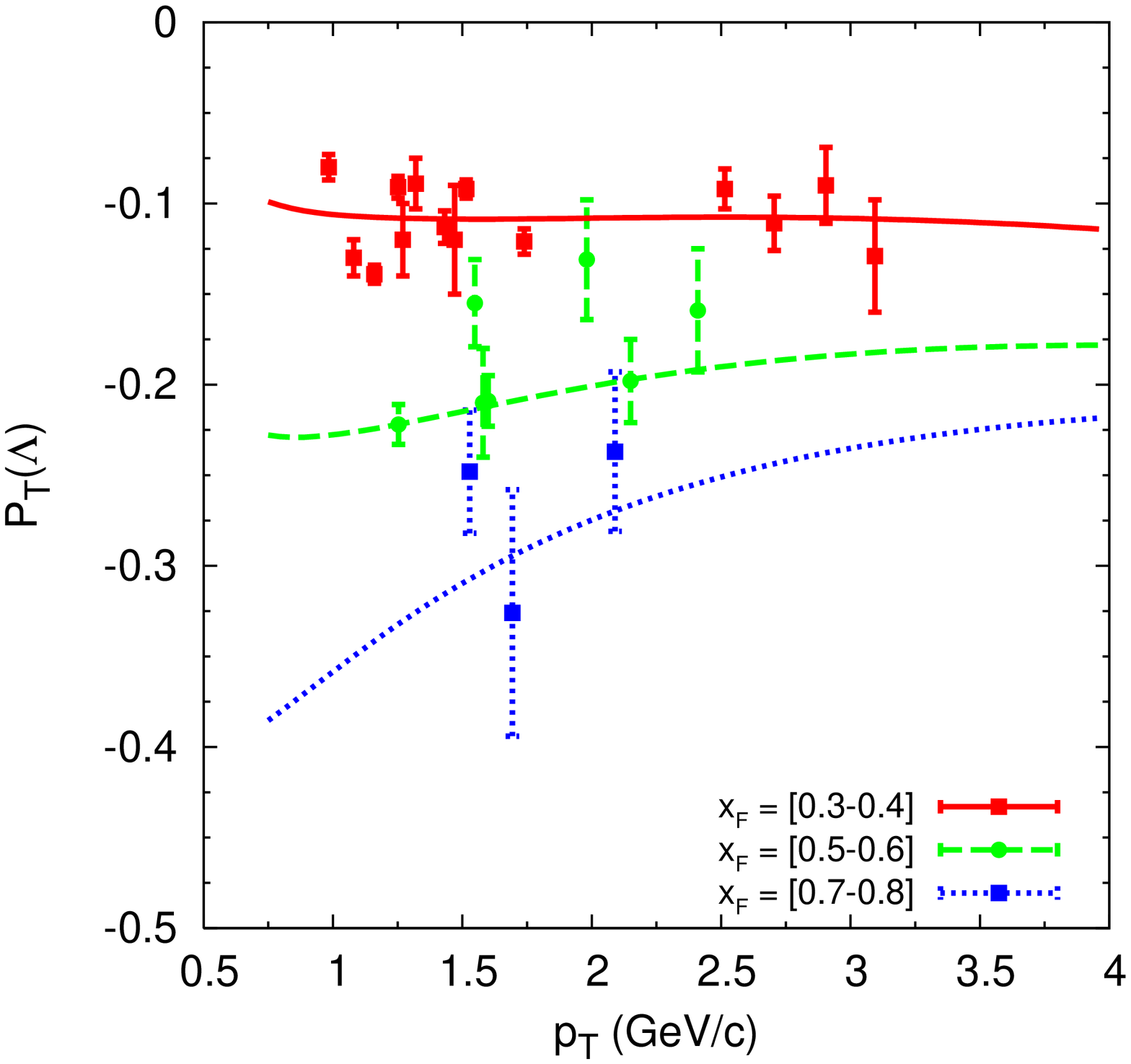,angle=-90,scale=0.35}
\end{minipage}
\begin{minipage}[t]{16.5 cm}
\caption{
   Best fit~\cite{Anselmino:2000vs} to $P_T(\Lambda)$ data from $p$--$Be$ reactions vs.
    $p_T$ and for different $x_F$ bins.
   Data~\cite{Heller:1983ia,Lundberg:1989hw,Ramberg:1994tk}
   are collected at two different c.m.~energies,
   $\sqrt{s_{NN}}\simeq 27$ GeV and $\sqrt{s_{NN}}\simeq 39$ GeV.
   Theoretical curves are evaluated
   at the mean $x_F$ value in the bins, and at $\sqrt{s_{NN}}=27$ GeV;
   the results change very little with the energy.
Unpolarized $\Lambda$ FF set:~\cite{Indumathi:1998am}. \label{pheno:ptlam}}
\end{minipage}
\end{center}
\end{figure}

 Another longstanding problem of spin phenomena in particle physics is the transverse polarization
of hyperons produced in unpolarized hadronic collisions, $P_T(Y)$, that can also be
considered as a SSA. A huge amount of experimental results has been collected, in particular
for the $\Lambda$ particle. These data show that transverse hyperon polarizations can be
quite sizable. Their behaviour as a function of $x_F$ and $p_T$ is very peculiar and
difficult to explain, as well as the differences in size and sign among different hyperons.
Several nonperturbative models have been proposed, but none of them is able to give a
comprehensive description of all data. We will not address in detail this subject. A quick
survey of its experimental and theoretical aspects can be found in
Refs.~\cite{Heller:1996pg}, \cite{Felix:1999tf} respectively, and a more extensive treatment
in the references quoted therein. Most of the experimental data are at fixed target and in
the very forward regime. For this reason, a pQCD approach to $P_T(Y)$ was not seriously
considered until recently. In fact, the description of the unpolarized cross section
data~\cite{Pondrom:1985aw} is difficult in pQCD, even when, as in the GPM approach, full
$\bm{k}_\perp$ effects are included (notice that also in this case the leading term in the
unpolarized cross section is given by the usual convolution of TMD unpolarized PDF's and
FF's). More precisely, data on unpolarized cross sections can be described only partially and
in a limited kinematical range, corresponding to not too low $p_T$ and not too large $x_F$.
Limiting to the $\Lambda$ case, for fixed target experiments the kinematical regions where
data for the unpolarized cross section and for the $\Lambda$ polarization are collected
overlap only partially. A better agreement is found with the recent STAR data on unpolarized
cross sections  at larger energies~\cite{Abelev:2006cs}.

In the GPM approach, focusing on the $qq\to qq$ processes, there are essentially two possible
sources for the $\Lambda$ polarization (see also the end of section~\ref{form}): the
Boer-Mulders function from one of the incoming protons convoluted with a spin transfer
partonic factor and the transverse fragmentation function (the analogue, in the fragmentation
sector, of the transversity function); the so-called (chiral-even) \emph{polarizing}
fragmentation function ($\Delta^N D_{\Lambda^\uparrow/q}$), coupled with unpolarized PDF's
and partonic cross sections. This last mechanism is somewhat the analogue of the Sivers
effect in the fragmentation process. An ongoing preliminary analysis~\cite{Dalesio:2007u}
seems to show that indeed only the polarizing effect can lead to sizable values of
$P_T(\Lambda)$. This study basically confirms the main findings of
Ref.~\cite{Anselmino:2000vs} where, adopting a more simplified planar kinematics and a simple
parameterization of the polarizing FF, the main features of $P_T(\Lambda)$ (including its
size) were reproduced (see Fig.~\ref{pheno:ptlam}).

\subsection{SSA's for double inclusive production in $pp$ collisions}

\label{pheno:pp:phojet}

Another important source of information on TMD distributions could come from the study of
$pp\to h_1\, h_2 + X$ and $pp \to {\rm jet\ jet\/} + X$ processes. Indeed, by considering
specific kinematical configurations  they have been advocated as a tool to extract the gluon
Sivers function. In Ref.~\cite{Boer:2003tx} they considered the almost back-to-back
asymmetric jet correlation in the mid-rapidity region in the GPM approach and  showed that a
sizable asymmetry could be observed at RHIC, mainly driven by the gluon Sivers effect. In
Refs.~\cite{Bacchetta:2005rm,Bomhof:2007su} a more complete treatment was presented,
including gauge-links effects and all possible spin and $\bfk_\perp$ dependent distributions
at leading twist for double inclusive hadron and jet production in $pp$ collisions. Recent
data on jet-jet correlations have been collected at RHIC showing an almost vanishing
asymmetry~\cite{Abelev:2007ii}. A possible explanation of these results could come from the
gauge-links effects properly incorporated in the TMD approach~\cite{Bomhof:2007su}.

A recent phenomenological study for the SSA in  $pp\to\gamma\, {\rm jet\/}+ X$
\cite{Bacchetta:2007sz} (see also~\cite{Vogelsang:2005cs}) has addressed the comparison
between the approach based on the generalized parton model and the colour-gauge-invariant QCD
formalism, where the gauge links are believed to be responsible for initial/final-state
interactions (in other words, to be the source of the phase shifts required to generate
naively T-odd effects). For hadronic processes different from SIDIS and DY, for instance
$p^{\uparrow} p \to$ hadrons, the Wilson line structure becomes more
intricate~\cite{Bomhof:2004aw}. It is therefore more challenging to derive clear-cut
predictions for the sign of the SSA in these processes~\cite{Bomhof:2007su}. Again, in direct
photon production the reduced number of channels and the colourless structure of the photon
greatly simplify the analysis.

Due to the intrinsic parton motions in the initial hadrons, in the process $pp\to\gamma\,
{\rm jet\/}+ X$ the photon and the jet are only approximately back-to-back in the plane
transverse to the beam collision direction. In other words, the azimuthal angles of the
photon and jet transverse momenta, $\bm{K}_{\gamma\,\perp}$ and $\bm{K}_{{\rm j}\,\perp}$,
are related by $\phi_{{\rm j}} - \phi_{\gamma}=\pi+\delta\phi$, where $\delta\phi$ just
originates from $\bm{k}_\perp$ effects.
 By defining suitable weighted azimuthal moments of $\bm{K}_{\gamma\,\perp}/M$,
sensitive to these effects, in Ref.~\cite{Bacchetta:2007sz} it has been shown how one could
check the role of the gauge-links, that in this simple process should only appear as
numerical prefactors for the standard LO partonic cross sections involved.

In a simplified partonic kinematics, keeping only ${\cal O}(k_\perp/{K}_{\gamma\,\perp})$
effects, and suitably choosing the kinematical configuration (that is, large positive values
of the photon pseudorapidity, $\eta_\gamma$, small or negative values for the jet one,
$\eta_{\rm j}$) it has been shown that for RHIC energies the quark-Sivers mechanism
dominates. Moreover, the weighted moment of interest could reach values of 4-5\% in size.
Notice that fixing the sign of the valence quark Sivers functions from SIDIS, the GPM
approach would predict an opposite sign w.r.t.~the colour-gauge-invariant QCD formalism.

The experimental investigation of the sign of the weighted moment, discriminating among these
predictions, seems to be accessible at RHIC and could have the same significance as the
measurement of the relative sign  of the Sivers effect in SIDIS and DY processes. However, a
word of caution is needed here and for the double inclusive processes discussed above, being
the validity of factorization for these processes still under debate, see
Refs.~\cite{Collins:2007jp, Vogelsang:2007jk,Bomhof:2007xt} and section~\ref{theo:develop}.

\subsection{Azimuthal asymmetries and SSA's in Drell-Yan processes}
\label{pheno:dy}

\begin{figure}[h!tb]
\begin{center}
\begin{minipage}[bt]{8 cm}
\epsfig{figure=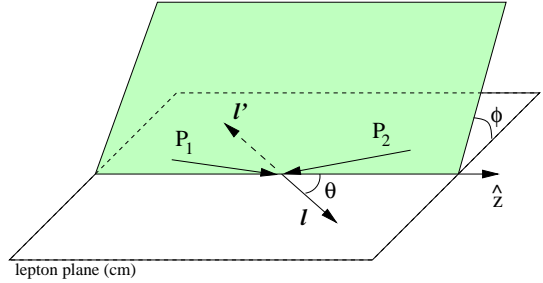,scale=0.4}
\end{minipage}
\begin{minipage}[t]{16.5 cm}
\caption{Kinematics of the Drell-Yan process in
the lepton center of mass frame.
\label{pheno:dy:cs}}
\end{minipage}
\end{center}
\end{figure}

The Drell-Yan process is definitely another valuable tool to access naively T-odd TMD effects
in the PDF sector and discriminate among them. Starting already from the unpolarized cross
section, we can learn on $\bm{k}_\perp$ effects in PDF's by studying the $q_T$
spectrum of the final lepton pair~\cite{D'Alesio:2004up}. But there is much more to learn
from the DY process. Indeed the unpolarized differential cross section, as measured in the
so-called Collins-Soper frame~\cite{Collins:1977iv} for the process $\pi^- \, N \rightarrow
\mu^+ \, \mu^- + X$, where $N$ is either deuterium or tungsten, using a $\pi^-$ beam with
various energies in the range of 150-300
GeV~\cite{Falciano:1986wk,Guanziroli:1987rp,Conway:1989fs},
 \be
 \frac{1}{\sigma}\frac{d\sigma}{d\Omega} = \frac{3}{4\pi}\, \frac{1}{\lambda+3} \, \left( 1+
 \lambda \cos^2\theta + \mu \sin2\theta \cos\phi + \frac{\nu}{2} \sin^2 \theta \cos 2\phi
 \right),
 \label{pheno:unpolxs}
 \ee
shows remarkably large values of $\nu$ (see Fig.~\ref{pheno:dy:cs} for the proper definition
of the angles).

Leading and next-to-leading order perturbative QCD corrections cannot explain these results.
Several explanations have been proposed, e.g. higher twist
effects~\cite{Brandenburg:1994wf,Eskola:1994py} and the role of QCD-vacuum structure in
hadron-hadron scattering~\cite{Nachtmann:1983uz,Brandenburg:1993cj,Boer:2004mv}.

\begin{figure}[h!tb]
\begin{center}
\begin{minipage}[h!t]{14 cm}
\hspace*{-1cm}
\epsfig{file=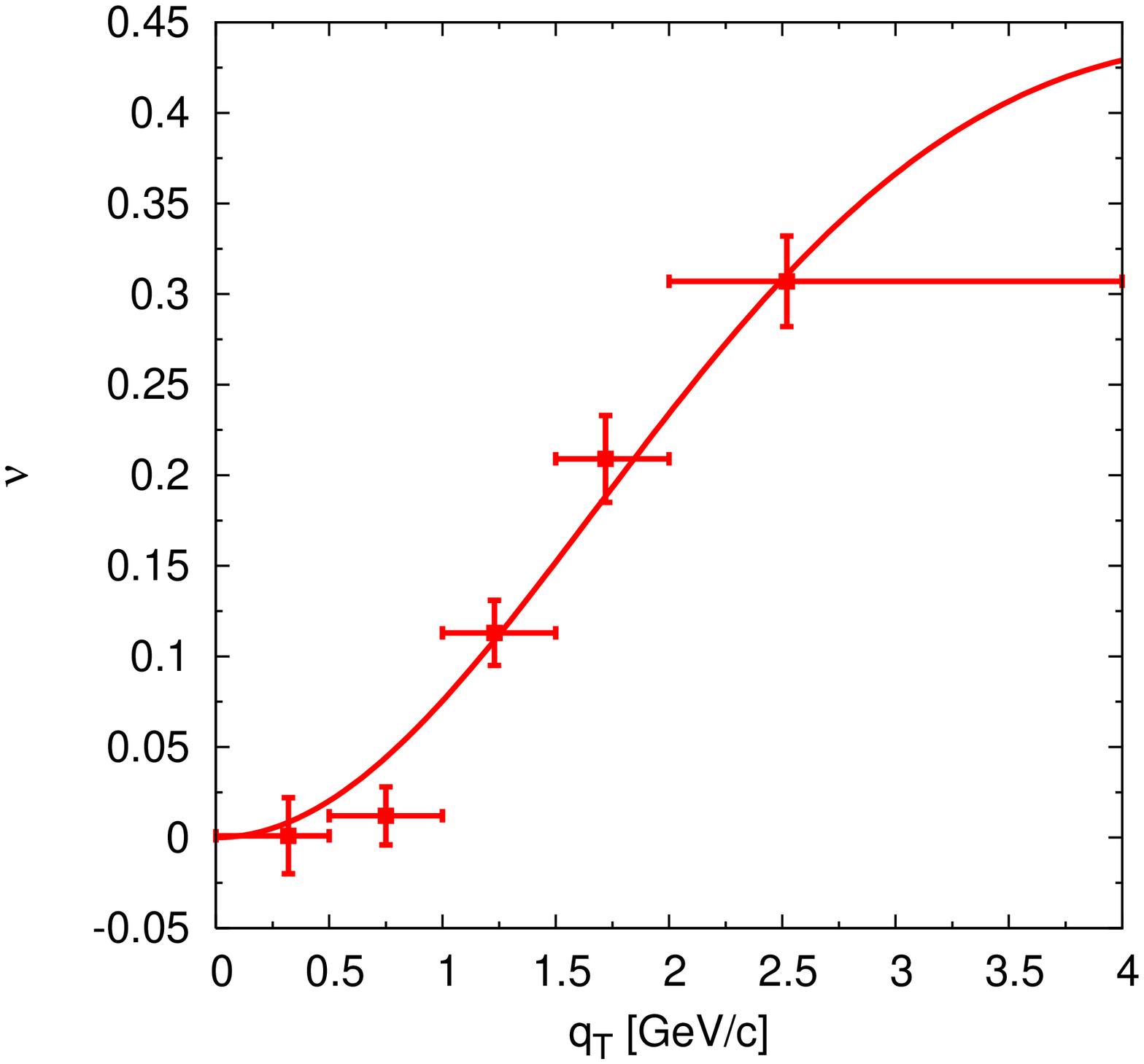,scale=0.35,angle=-90}
\epsfig{file=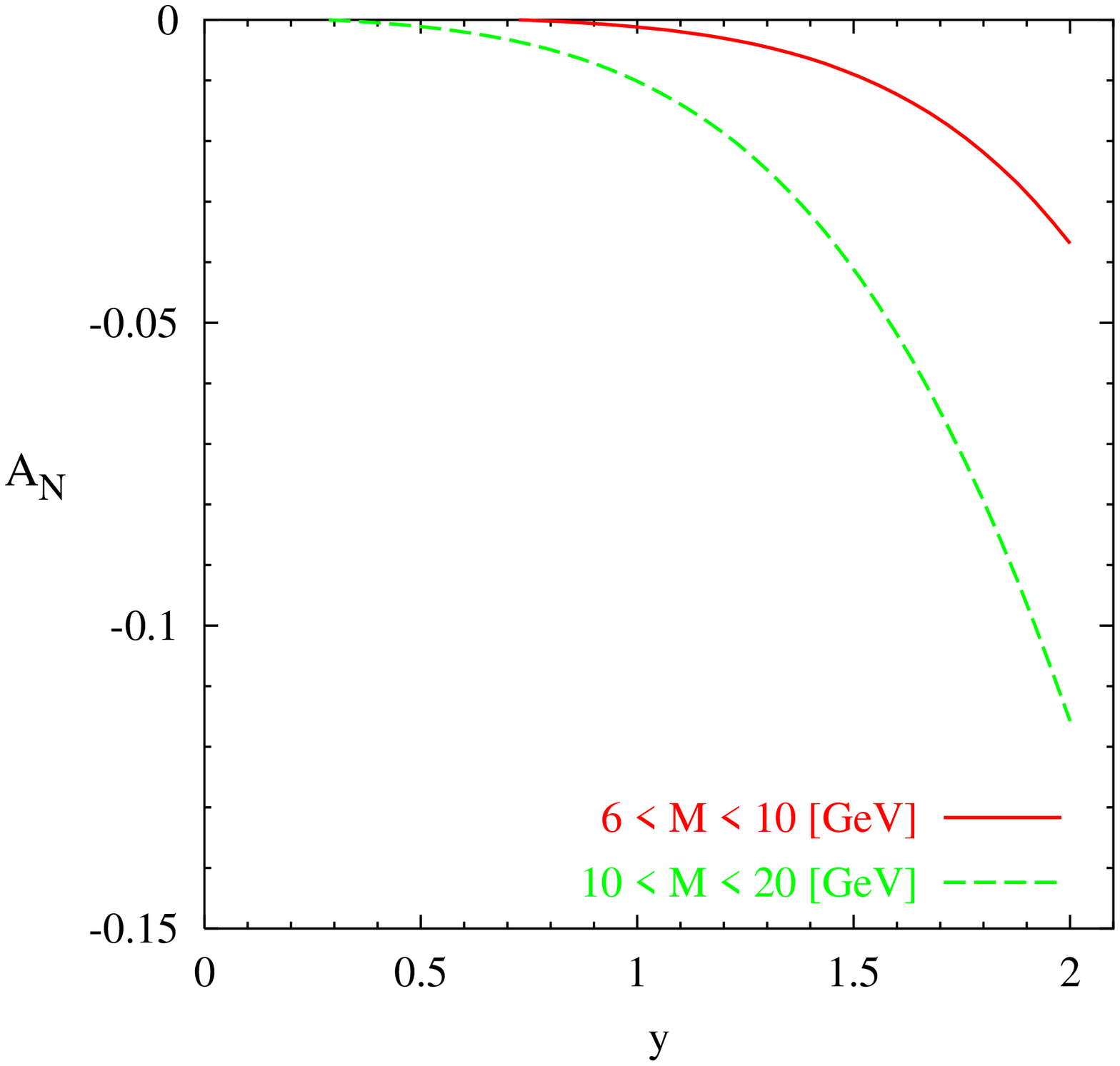,scale=0.35,angle=-90}
\end{minipage}
\begin{minipage}[t]{16.5 cm}
\caption{
Left panel: Data on the $\nu$ coefficient in
Eq.~(\ref{pheno:unpolxs})~\cite{Falciano:1986wk,Guanziroli:1987rp} at 194
GeV, vs. $q_T$,
compared with the fit of Ref.~\cite{Boer:1999mm}.
Right panel: Maximized $A_N$ for the DY process,
at RHIC energies, $\sqrt{s}=200$ GeV, as a function of
the rapidity $y$ and for two ranges of the invariant mass.
The asymmetry is practically negligible in the range $y<0$. PDF set: MRST01.
\label{pheno:unp-an:dy}}
\end{minipage}
\end{center}
\end{figure}

In the context of a pQCD approach with TMD distribution functions, such a large $\cos 2\phi$
azimuthal dependence can arise at {\em leading order\/} only from a convolution of two
Boer-Mulders functions, $\Delta^N f_{s_y/A}$ or $h_1^\perp$. In Ref.~\cite{Boer:1999mm} it
was shown how by a suitable parametrization of this naively T-odd function, under some simple
but reasonable assumptions a description of the $\cos 2\phi$ asymmetry could be obtained, see
Fig.~\ref{pheno:unp-an:dy} (left panel). More recently, the $\cos2\phi$ asymmetry in the
unpolarized DY process $p\bar{p}\to\mu^+\mu^- X$ at $s=50$ GeV$^2$, has been studied in
detail in the same framework but using a more realistic model for the Boer-Mulders function
in the quark-diquark spectator approach~\cite{Gamberg:2005ip}. Notice that the negligible
$\cos 2\phi$ dependence observed recently for DY dimuons in $pD$ collisions~\cite{Zhu:2006gx}
could imply an almost vanishing sea-quark Boer-Mulders function.

Considering the transverse SSA in the Drell-Yan process $p^\uparrow p\to\ell^+\ell^- + X$,
again two TMD mechanisms could play a role: the Sivers effect and the Boer-Mulders effect,
which involves also the transversity distribution. The main difference with respect to the
inclusive processes considered above, like $pp\to h + X$, is that in this case the
measurement of the lepton-pair angular distribution automatically allows to select one
specific effect. In Ref.~\cite{Boer:1999mm} the full structure of the polarized cross
section, at ${\cal O}(k_\perp/M_h)$, has been presented. From there, one can easily see that
integrating over the lepton-plane angular variables, only the Sivers effect survives in the
kinematical regime where $k_\perp \simeq q_T \ll M$ ($M$ here is the invariant mass of the
lepton pair). Notice that other mechanisms that could also generate SSA's in the Drell-Yan
process \cite{Hammon:1996pw,Boer:1997bw,Boer:1999si, Boer:2001tx}, based on higher twist
quark-gluon correlation functions in a generalized pQCD factorization theorem
\cite{Qiu:1998ia}, lead to expressions of $A_N$ which, being dependent on the angle between
the proton polarization direction and the plane containing the final lepton
pair~\cite{Collins:1977iv}, vanish upon corresponding angular integrations. In
Ref.~\cite{Anselmino:2002pd}, considering in the hadronic c.m.~frame the polarized
differential cross section $d\sigma/dM^2dy\,d^2\bm{q}_T$, where $y$ is the rapidity of the
lepton pair, a phenomenological analysis of the Sivers asymmetry has been performed in great
detail. By adopting the Sivers functions as extracted from pion production data (with the
reversed sign, see below)  preliminary predictions for $A_N$ at RHIC were given, see
Fig.~\ref{pheno:unp-an:dy} (right panel). A factorized Gaussian ansatz for the
$\bm{k}_\perp$-dependent PDF's, with some reasonable approximations, allowed to perform
analytically the integration over the transverse parton momenta, resulting in a simple
factorized expression for $A_N$.
%
%%%%%%ppnp style

Let us comment on the sign of the SSA originating from the Sivers effect in the Drell-Yan
process. As already discussed in section~\ref{theo}, if the Sivers effect is due to
initial/final state interactions and colour-gauge invariance (Wilson lines) the corresponding
asymmetry should have opposite sign in Drell-Yan and SIDIS processes, respectively because of
past and future-pointing gauge-links, related, somehow, to $s$-channel and $t$-channel
elementary reactions. On the other hand, the SSA in the $p^\uparrow p\to\pi+ X$ process at
large positive $x_F$ and moderately large $p_T$ is dominated by $t$-channel, quark initiated
processes. Therefore, one can tentatively assume that the Sivers function as extracted from
pion SSA data should contribute with opposite sign to DY processes. Notice that this
assumption is \emph{not} a prediction of the GPM approach, which would consider the Sivers
function as universal, but rather an attempt to incorporate in it additional information
coming from different approaches not yet directly applicable to the $p^\uparrow p\to\pi+ X$
process. Given these considerations, even a simple comparison of the sign of these estimates
with data might be highly significant.

Notice also that, as we will see below, the valence-quark Sivers functions as extracted by
SIDIS data have the same sign as those extracted from $pp\to \pi +X$: this somehow supports
the sign change adopted for DY in this preliminary study.

Predictions for SSA's in DY processes based on the extraction of the Sivers effect from
SIDIS~\cite{Anselmino:2005an} (see also section~\ref{pheno:sidis:ssa}) have been discussed in
Refs.~\cite{Efremov:2004tp,Vogelsang:2005cs,Anselmino:2005ea,Collins:2005rq}. As pointed out
in Ref.~\cite{Collins:2005rq}, the Sivers asymmetry at RHIC should be strongly sensitive to
the sea-quark Sivers distributions.

We finally stress that the study of Ref.~\cite{Anselmino:2002pd} included a combined
description, in the GPM approach, of the unpolarized Drell-Yan cross sections.

Another promising idea in the study of SSA's has been proposed in Ref.~\cite{Brodsky:2002pr},
where they considered electroweak Drell-Yan processes. In this case by looking at a
$\nu_\ell-\ell$ lepton pair one selects a $W$-exchange channel and therefore only one parton
helicity state contributes. The transversity function that
could couple to the Boer-Mulders effect is therefore absent. This process would then give
another way to directly access the Sivers effect.

\subsection{$e^+e^-$ annihilation in two nearly back-to-back hadrons}
\label{pheno:epem}

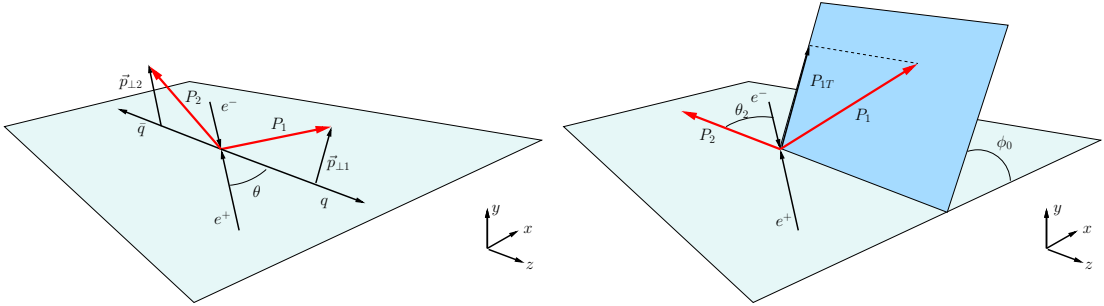
\begin{figure}[h!tb]
\begin{center}
\begin{minipage}[t]{16 cm}
\scalebox{0.3}{\input{figures/epluseminus_paper.pstex_t}}
\scalebox{0.3}{\input{figures/epluseminus_paper1.pstex_t}}
\end{minipage}
\begin{minipage}[t]{16.5 cm}
\caption{Kinematics for the $e^+e^-\to h_1 h_2+ X$ process, according to two different
  configurations considered by Belle and discussed in the text: setup 1 (left) and setup 2 (right).
\label{pheno:kin:epem} }
\end{minipage}
\end{center}
\end{figure}

\begin{figure}[h!tb]
\begin{center}
\begin{minipage}[t]{14 cm}
\hspace*{.3cm}
\epsfig{file=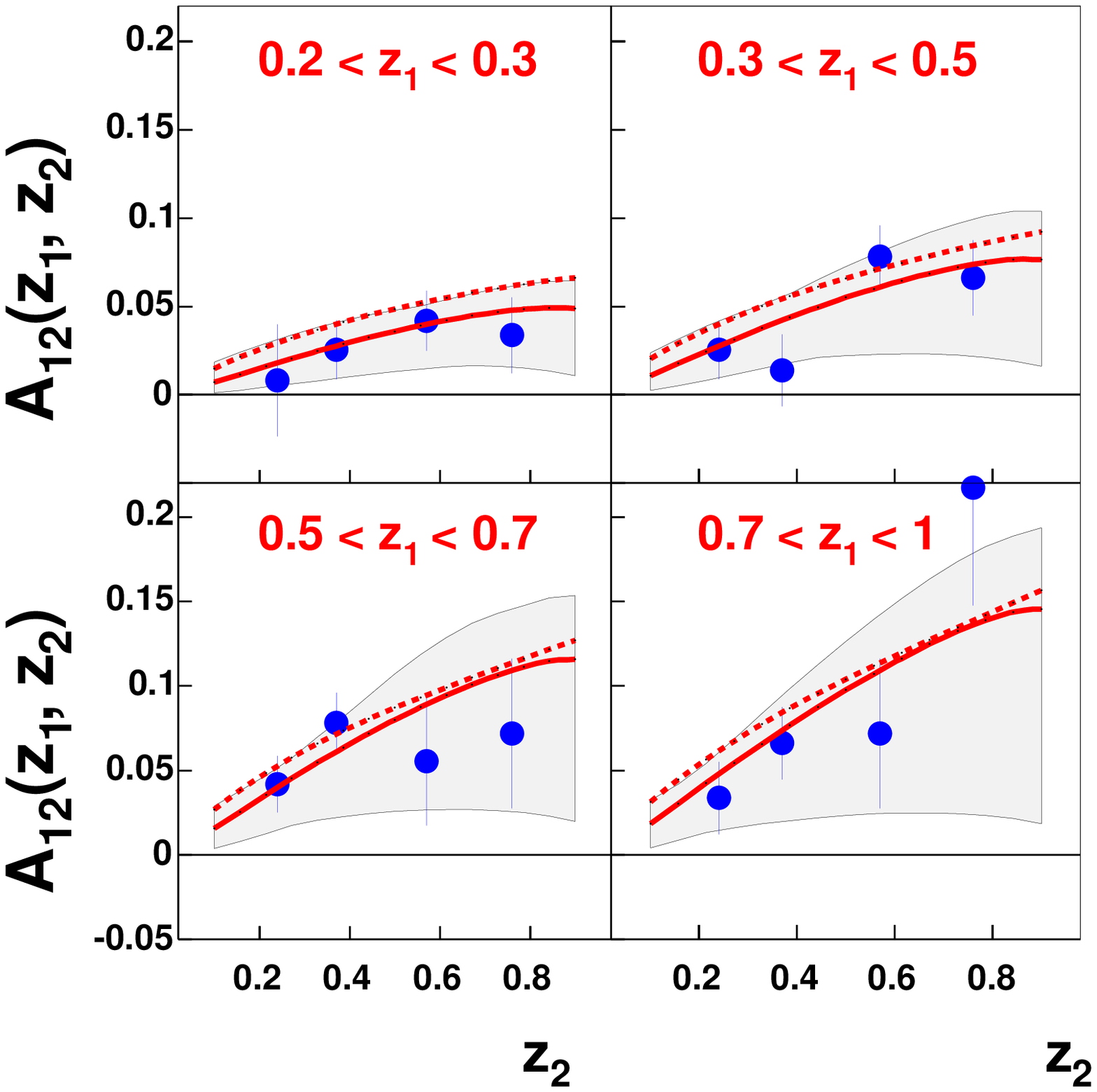,scale=0.4,bb= 10 140 540 660,angle=-90}
\epsfig{file=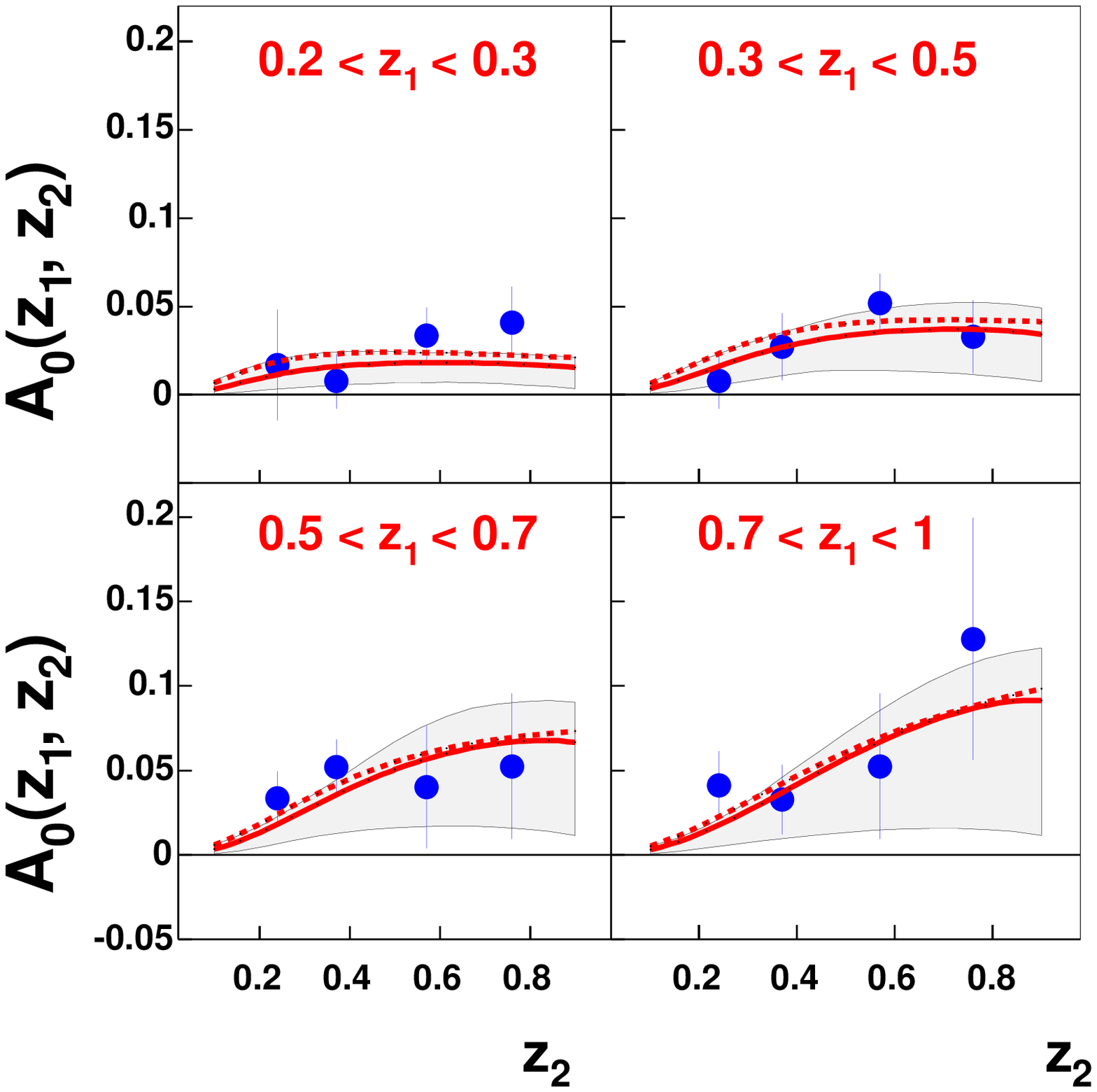,scale=0.4,bb= 10 140 540 660,angle=-90}
\end{minipage}
\begin{minipage}[t]{16.5 cm}
\caption{ Belle data on two azimuthal correlations in $e^+e^- \to \pi \pi +X$
processes~\cite{Abe:2005zx}. Solid lines: results obtained by fitting the $A_{12}$
asymmetry~\cite{Anselmino:2007fs}; dashed lines: results obtained by fitting the $A_0$
asymmetry, according to the setup of Fig.~\ref{pheno:kin:epem} (right
panel)~\cite{Anselmino:2007fs}. The shaded area corresponds to the theoretical uncertainty on
the parameters. FF set: K. \label{pheno:belle} }
\end{minipage}
\end{center}
\end{figure}

Hadron production in $e^+e^-$ collisions would be in principle the most valuable and clean
process for the study of TMD polarized fragmentation functions, like the Collins function,
thanks to the lack of corresponding TMD effects in the initial state. Unfortunately, there is
not in the process $e^+e^-\to q\bar{q}$ any transverse polarization transfer to a single,
on-shell massless quark. Therefore, it is not possible to observe the Collins effect
individually by measuring, e.g., the asymmetry in the distribution around the jet trust axis
(i.e. the fragmenting quark direction) of a hadron species (typically pions) produced in the
quark fragmentation. Indeed, in hadron production in $e^+e^-\to q\bar{q}\to 2\,{\rm jets}$
events, the Collins effect can only be observed when the quark and the antiquark are
considered \emph{simultaneously}. As reported in section~\ref{exp:unp:belle}, the Belle
Collaboration at the KEK-B asymmetric-energy $e^+e^-$ storage rings has in fact performed a
measurement of azimuthal asymmetries in hadron-hadron correlations for inclusive charged
dihadron production, $e^+e^-\to\pi^+\pi^- + X$~\cite{Abe:2005zx}. This asymmetry has been
interpreted as a direct measure of the Collins effect, involving however (and unavoidably)
the convolution of two Collins functions. Two experimental methods have been used to extract
the azimuthal asymmetry from raw data: 1) The first method, requiring the knowledge of the
two-jet thrust axis, corresponds to the kinematical setup shown in the left plot of
Fig.~\ref{pheno:kin:epem} and gives rise to a $\cos(\phi_1+\phi_2)$ modulation in the
dihadron yields, the $A_{12}$ asymmetry discussed below and shown in Fig.~\ref{pheno:belle}
($\phi_1$, $\phi_2$ are the azimuthal angles of the two hadron momenta); 2) The second
method, leading to the asymmetry $A_0$ (see Fig.~\ref{pheno:belle}), does not rely on the
knowledge of the thrust axis and corresponds to the setup shown in the right plot of
Fig.~\ref{pheno:kin:epem}.

%%%%%%%%%%%%%%%%%%%%%%%%%%%%%%%%%%%%%%%%%%%%%%%%%%%%%%%%%%%%%%%%%%%%%%%%%%%

On the theoretical side this azimuthal asymmetries have been studied in
Ref.~\cite{Boer:1997mf}, where their phenomenological interest was pointed out, and more
recently in Refs.~\cite{Efremov:2006qm,Anselmino:2007fs}, where Belle results were used to
extract additional information on the Collins function. Both
Refs.~\cite{Boer:1997mf,Efremov:2006qm} have used in their analysis the setup 2. In
Ref.~\cite{Efremov:2006qm}, utilizing the formalism of Ref.~\cite{Boer:1997mf} and a quark
soliton model for the transversity distribution, the Collins function was extracted by
fitting the HERMES and COMPASS data on azimuthal spin asymmetries in the SIDIS process $\ell
p^\uparrow\to\ell^\prime h + X$ (see next section). The parameterization extracted for the
Collins function was shown to be compatible with Belle results on $e^+e^-\to\pi^+\pi^-+X$. In
Ref.~\cite{Anselmino:2007fs} both kinematical setups were considered, and SIDIS and $e^+e^-$
data were combined to get a simultaneous fit of the Collins and the transversity distribution
functions.

%%%%%%%%%%%%%%%%%%%%%%%%%%%%%%%%%%%%%%%%%%%%%%%%%%%%%%%%%%%%%%%%%%%%%%%%%%%
%%%%%%ppnp style
%%%%%%%%%%%%%%%%%%%%%%%%%%%%%%%%%%%%%%%%%%%%%%%%%%%%%%%%%%%%%%%%%%%%%%%%%%%

As an illustration, let us briefly summarize the case of setup 1, which is more simple to
figure out. Within the GPM approach and the helicity formalism, the differential cross
section for dihadron production can be written as (see Fig.~\ref{pheno:kin:epem} for
notation):
 \bea
 \!\!\!\!\frac{d\sigma ^{e^+e^-\to h_1 h_2+X}}
 {dz_1\,dz_2\,d^2\bm{p}_{\perp 1}\,d^2\bm{p}_{\perp 2}\,d\cos\theta}&=&
 \frac{3\pi\alpha^2}{2s} \, \sum _q e_q^2 \, \left\{
 (1+\cos^2\theta)\,D_{h_1/q}(z_1,{p}_{\perp 1})\,D_{h_2/\bar q}(z_2,{p}_{\perp 2})
 \frac{}{} \right. \nonumber \\ & & + \left. \frac{1}{4}\,\sin^2\theta\,\Delta ^N
 D_{h_1/q^\uparrow}(z_1,{p}_{\perp 1})\,
 \Delta ^N D _{h_2/\bar q^\uparrow}(z_2,{p}_{\perp 2})\,\cos\phi_1 \cos\phi_2\right\}.
 \label{pheno:dsig}
 \eea
The azimuthal asymmetry, as defined by the Belle collaboration (setup 1), can be obtained
from here by the following steps: $i)$ Performing a change of angular variables from
$(\phi_1,\phi_2)$ to $(\phi_1,\phi_1+\phi_2)$ and then integrating over the moduli of the
intrinsic transverse momenta, $p_{\perp 1}$, $p_{\perp 2}$, and over the azimuthal angle
$\phi_1$; $ii)$ Normalizing the result to the azimuthally averaged cross section; $iii)$
Taking the ratio $R$ of unlike-sign (U) to like-sign (L) pion-pair production, in order to
eliminate false asymmetries:
 \be
 R \simeq 1+\cos(\phi_1+\phi_2)\,A_{12}(z_1,z_2)\>, \quad {\rm where}\quad
 A_{12}(z_1,z_2)=\frac{1}{4}\,\frac{\langle \sin^2\theta \rangle}
 {\langle 1+\cos^2\theta \rangle}\,(P_U-P_L)\>,
 \label{pheno:Ra12}
 \ee
the angle $\theta$ is averaged over a range of values given by the detector acceptance,
 \be
 P_{U(L)}=\frac{\sum_q e^2_q \, \Delta ^N D_{\pi^+/q^\uparrow}(z_1)\,
 \Delta ^N D_{\pi^{-(+)}/\bar q^\uparrow}(z_2)}{\sum_q e^2_q D _{\pi^+/q}(z_1)\,
 D _{\pi^{-(+)}/\bar q}(z_2)}\>,
 \label{pheno:pul}
 \ee
and
 \be
 \Delta^ND_{h/q^\uparrow}(z) = \int d^2\bm{p}_\perp \Delta^N D_{h/q^\uparrow}(z,p_\perp)
             = \int d^2\bm{p}_\perp \frac{2p_\perp}{z m_h} \; H_1^{\perp q}(z,p_\perp)
             = 4 \; H_1^{\perp(1/2)q}(z)\>.
 \label{pheno:deltaDz}
 \ee
In this last equation, the relationship among the involved moments of the Collins function,
as defined in Refs.~\cite{Efremov:2006qm,Anselmino:2007fs}, has been also given. As an
example, in Fig.~\ref{pheno:belle} we show the quality of the fit to the asymmetry, as
defined by the Belle Collaboration for the two experimental setups, according to
Ref.~\cite{Anselmino:2007fs}. This allowed to extract the first independent parameterization
of the Collins function, that was then used to extract, in combination with HERMES and
COMPASS results on azimuthal asymmetries, the $u$ and $d$ quark transversity distribution
function (see also section \ref{pheno:sidis}).

Let us finally remind that preliminary Belle data using the unlike-sign over the all charged (C)
pion-pair ratio~\cite{Ogawa:2006bm} have been found in agreement with
the predictions of Ref.~\cite{Efremov:2006qm}.

\subsection{Semi-inclusive deeply inelastic scattering}
\label{pheno:sidis}

Semi-inclusive particle production in deeply inelastic lepton proton scattering, $\ell p\to
\ell' h +X$, is a rich case of study over many points of view. By focusing once again on the
role of intrinsic parton motion, both in the unpolarized and in the single polarized case, we
will show how a consistent picture of various asymmetries observed experimentally can be
given.

At variance with inclusive particle production in proton-proton collisions, the standard
reference frame adopted to deal with SIDIS scattering is not the lepton-proton c.m.~frame,
but rather the virtual boson-proton c.m.~one, for the process $\gamma^* p\to h +X$.
Therefore, one cannot adapt in a straightforward way the kernel structure discussed in
section~\ref{form} to this case. Indeed, a complete treatment of full $k_\perp$ effects in
SIDIS within the generalized parton model and the helicity formalism is still in
progress~\cite{Anselmino:2007aa}. What is currently adopted by many groups, using also
different approaches, is a first order approximation in a $(k_\perp/Q)$ power expansion,
where $Q$ is the virtuality of the exchanged boson. In this context, a complete
classification for SIDIS cross sections in the TMD approach has recently
appeared~\cite{Bacchetta:2006tn}.

As in Drell-Yan processes, in SIDIS the intrinsic parton motions result in azimuthally
dependent terms in the unpolarized and single polarized cross sections for hadron
production~\cite{Cahn:1978se,Cahn:1989yf,Konig:1982uk,Aubert:1983cz,Arneodo:1986cf}.

Let us first of all set the kinematics that we will adopt in the rest of this section. As
already said above we work in the $\gamma^* p$ c.m.~frame, shown in
Fig.~\ref{pheno:planessidis}. We take the photon and the proton colliding along the $z$ axis
with momenta $\bm{q}$ and $\bm{P}$ respectively; following the so-called ``Trento
conventions''~\cite{Bacchetta:2004jz}, the leptonic plane coincides with the $x$-$z$ plane.
We consider the kinematic regime in which $P_T \simeq \Lambda_{\rm QCD} \simeq k_\perp$,
where $P_T = |\bm{P}_T|$ is the final hadron transverse momentum. In this region QCD
factorization holds \cite{Ji:2004wu}
 and leading order elementary processes,
$\ell \, q \to \ell \, q$, are dominating: the soft $P_T$ of the detected hadron is mainly
originating from quark intrinsic motion rather than from higher order pQCD interactions,
which, instead, would dominantly produce large $P_T$
hadrons~\cite{Chay:1991nh,Georgi:1977tv,Daleo:2004pn}.

\subsubsection{Azimuthal dependence in unpolarized SIDIS}
\label{pheno:sidis:cahn}

In the QCD factorization scheme the SIDIS cross section for hadron production reads
 \bea
 && \hspace*{-0.8cm} \frac{d^5\sigma^{\ell p
 \to \ell h+X }}{dx_{\rm B} \, dQ^2 \, dz_h \, d^2 \bm{P} _T} = \sum_q  e_q^2 \int  \! d^2
 \bm{k} _\perp \, f_{q/p}(x,{k} _\perp) \, \frac{2\pi\alpha ^2}{x_{\rm B} ^2
 s^2}\,\frac{\hat s^2+\hat u^2}{Q^4}\, D_{h/q}(z,{p} _\perp) \, \frac{z}{z_h} \, \frac{x_{\rm
 B}}{x}\left( 1 + \frac{x_{\rm B}^2}{x^2}\frac{k_\perp^2}{Q^2} \right)^{\!\!-1},
 \label{pheno:d5unp}
 \eea
where the elementary Mandelstam variables are given by ($\bfl$ being the incoming lepton
three-momentum)
 \be
 \hat s = xs - 2\, \bm{\ell} \cdot \bm{k}_\perp - k_\perp^2 \frac{x_{\rm B}}{x}
 \left( 1 - \frac{x_{\rm B}\, s}{Q^2} \right) \hspace*{1cm}
 \hat u = - x \left( s - \frac{Q^2}{x_{\rm B}} \right) + 2\, \bm{\ell} \cdot \bm{k}_\perp
 - k_\perp^2 \frac{x_{\rm B}^2 s}{x Q^2} \> \cdot
 \label{pheno:su}
 \ee
The on-shell condition for the final quark implies $ x = x_{\rm B}\left [ 1+ \sqrt{1+4
k^2_{\perp}/Q^2}\right ]/2$.

Notice that $\bm{p} _\perp$ is the transverse momentum of the hadron $h$ w.r.t.~the direction
of motion  of the fragmenting quark, and $z$
is the light-cone momentum fraction carried by the produced hadron.
After some algebra one can find an exact expression, at all orders in $(k_\perp / Q)$, for
the light-cone fragmentation variables $z$ and $\bm{p}_\perp$, in terms of the usual SIDIS
and hadronic  variables $x_{\rm B}$, $Q^2$, $\bm{P}_T$, $z_h=(P\cdot P_h)/(P\cdot q)$.
%

%%%%%%%%%%%%%%%%%%%%%%%%%%%%%%%%%%%%%%%%%%%%%%%%%%%%%%%%%%%%%%%%%%%%%%%%%%%
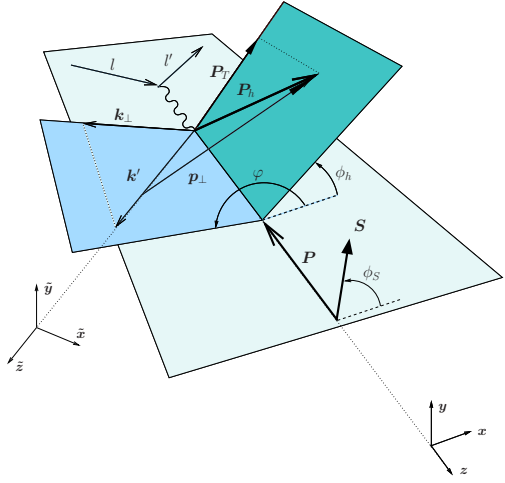
\begin{figure}[h!t]
\begin{center}
\scalebox{0.3}{\input{figures/planessidis.pstex_t}}
\caption{\small Kinematics of the SIDIS
process in the $\gamma^*-p$ c.m.~frame.}
\label{pheno:planessidis}
\end{center}
\end{figure}
%%%%%%%%%%%%%%%%%%%%%%%%%%%%%%%%%%%%%%%%%%%%%%%%%%%%%%%%%%%%%%%%%%%%%%%%%%%

Due to these relations among the kinematical variables, momentum conservation and the
integration over intrinsic $\bm{k}_\perp$ at fixed $\bm{P}_T = P_T(\cos\phi_h, \sin\phi_h,0)$
result in a dependence on the azimuthal angle of the produced hadron, $\phi_h$, that is the
angle between the leptonic and the hadronic planes, see Fig.~\ref{pheno:planessidis}. This
azimuthal dependence also appears in the SIDIS cross section. To see
this, we will adopt for the
$k_\perp$ dependent PDF's and FF's the usual factorized,
flavour independent Gaussian forms (this helps in
finding analytical expressions), that is:
\be
 f_{q/p}(x,k_\perp) = f_{q/p}(x) \, \frac{1}{\pi \langle k_\perp^2\rangle} \,
 e^{-{\kt^2}/{\langle\kt^2\rangle}}\>, \hspace*{1cm}
 D_{h/q}(z,p _\perp) = D_{h/q}(z) \, \frac{1}{\pi \langle p_\perp^2\rangle} \,
 e^{-p_\perp^2/\langle p_\perp^2\rangle}\>.
 \label{pheno:fdgauss}
 \ee
To have a closer look at the azimuthal dependence of the cross section it is instructive, and
often quite accurate, to consider Eq.~(\ref{pheno:d5unp}) at the order  ${\cal
O}(k_\perp/Q)$. In such a case $x \simeq x_{\rm B}, z \simeq z_h$ and
 \bea
 \nonumber
 \frac{d^5\sigma^{\ell p \to \ell h + X }}{dx_B \, dQ^2 \, dz_h \, d^2\bm{P} _T} & \simeq &
 \sum_q \frac{2\pi\alpha^2e_q^2}{Q^4} \> f_q(x_B) \> D_{h/q}(z_h) \\
 &\times & \biggl[
 1+(1-y)^2 %\\
%&& \hskip 36pt
 - 4 \>\frac{(2-y)\sqrt{1-y}\> \langle\kt^2\rangle \, z_h \, P_T}
 {\langle\pt^2\rangle \, Q}\> \cos \phi_h \biggr]
 \frac{1}{\pi\langle\pt^2\rangle} \, e^{-P_T^2/\langle\pt^2\rangle} \, ,
 \label{pheno:cahn:anal:app}
 \eea
where
 \be
 \langle P_T^2 \rangle = \langle \pt^2 \rangle + z_h^2 \langle \kt^2 \rangle\,.
 \label{pheno:eq:meanpt}
 \ee
This approximate result illustrates very clearly the origin of the dependence of the
unpolarized SIDIS cross section on the azimuthal angle $\phi_h$. As first observed by
Cahn~\cite{Cahn:1978se}, such a dependence is directly related to the parton
intrinsic motion and vanishes when $k_\perp =0$.
%%%%%%%%%%%%%%%%%%%%%%%%%%%%%%%%%%%%%%%%%%%%%%%%%%%%%%%%%%%%%%%%%%%%%%%%%%%
%%%%%%ppnp style
\begin{figure}[h!tb]
\begin{center}
\begin{minipage}[t]{10 cm}
\epsfig{file=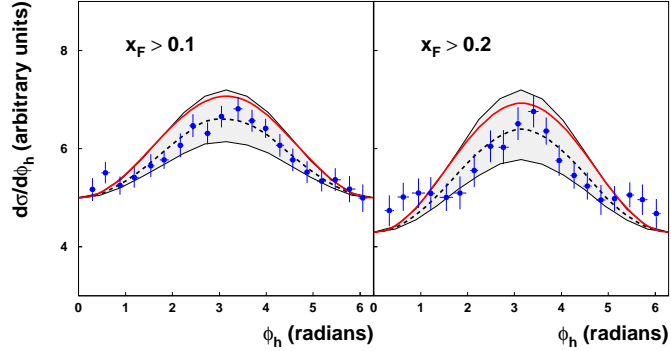,scale=0.5,bb = 10 370 540 670}
\end{minipage}
\begin{minipage}[t]{16.5 cm}
\caption{Fits to data~\cite{Arneodo:1986cf} on
the $\cos\phi_h$ dependence of the unpolarized SIDIS
cross section for charged hadron production. Dashed line:
exact kinematics; solid bold line: ${\cal
O}({\kt}/{Q})$ calculation. The shadowed region corresponds to varying the parameters
$\langle\kt^2\rangle$ and $\langle\pt^2\rangle$, given in Eq.~(\ref{pheno:parameters}) of the
text, by 20\%. Unpolarized PDF and FF sets: MRST01-K.
\label{pheno:cahn}}
\end{minipage}
\end{center}
\end{figure}
%%%%%%%%%%%%%%%%%%%%%%%%%%%%%%%%%%%%%%%%%%%%%%%%%%%%%%%%%%%%%%%%%%%%%%%%%%%

Indeed, this effect (also known as {\em Cahn effect}) has
been measured in different SIDIS
experiments~\cite{Arneodo:1986cf,Aubert:1983cz,Adams:1993hs}. In the TMD approach the
available data can be used to extract information on the Gaussian widths entering the PDF's
and FF's. A systematic analysis, using several sets of experimental data (see
Fig.~\ref{pheno:cahn} as an example) has been carried out in Ref.~\cite{Anselmino:2005nn},
leading to the following best values of the parameters:
 \be
 \langle\kt^2\rangle   = 0.25
 \;({\rm GeV}/c)^2 \quad\quad\quad \langle\pt^2\rangle  = 0.20 \;({\rm GeV}/c)^2 \>.
 \label{pheno:parameters}
 \ee
This study leads to an excellent agreement with data for small values of the final hadron
transverse momentum, $P_T$. However, as somehow expected, it fails in reproducing data at
higher $P_T$, the turning point being around $P_T \sim 1$ GeV/$c$. A similar conclusion was
drawn in Ref.~\cite{Chay:1991nh}. As discussed in Ref.~\cite{Anselmino:2006rv}, by including
also NLO corrections (like hard gluonic radiation and gluon-initiated elementary scatterings)
the agreement with data can be extended at much larger $P_T$ values. This analysis also
confirms that the transition between the regime dominated by nonperturbative, intrinsic
transverse momentum effects and the one mainly governed by NLO radiative effects sits around
$P_T\simeq 1$ GeV$/c$.

We also mention that  preliminary results~\cite{Musch:2007ya} in lattice QCD for the first
$x$-moment of the unpolarized TMD distribution are compatible with a Gaussian $k_\perp$
dependence with $\langle\kt^2\rangle\simeq 0.3$ (GeV$/c)^2$.

In the context of azimuthal asymmetries in unpolarized SIDIS there could be also a $\cos
2\phi_h$ dependence. Indeed a contribution to this term is already present in
Eq.~(\ref{pheno:d5unp}), due to kinematical $(k_\perp/Q)^2$ effects. Moreover, in the TMD
approach including spin effects, an additional contribution appears, coming from the
convolution of the Boer-Mulders and the Collins
functions~\cite{Boer:1997nt,Gamberg:2003ey,Barone:2005kt,Bacchetta:2006tn}.

\subsubsection{Transverse SSA's in SIDIS}
\label{pheno:sidis:ssa}
Let us now consider the single-transverse polarized cross section $d\sigma(\ell
p^\uparrow\to\ell^\prime h + X)$, or $d\sigma_{UT}$. Retaining only leading terms in ${\cal
O} (\kt/Q)$, in the TMD approach at leading twist the differential cross section can be
formally expressed as
 \be
 \frac{d^6\sigma^{\ell p^\uparrow \to \ell^\prime h + X}} {dx_B \, dQ^2
 \, dz_h \, d^2\bm{P} _T \,
 d\phi_S} = A_0 + \frac{1}{2}\left [A_{\rm Sivers}  + A_{\rm Collins} +
 A^\prime_{\rm Collins}\right ]\> .
 \label{pheno:d6pol}
 \ee
The $\phi_S$ dependence originates from the cross section dependence on the angle between the
proton (transverse) polarization vector and the leptonic plane; in the configuration of
Fig.~\ref{pheno:planessidis} this angle is simply $\phi_S - \phi_{\ell'} \equiv \phi_S$,
having chosen $\phi_{\ell'}$ to be zero, with $\phi_S$ varying event by event.

 The $A_0$ term in Eq.~(\ref{pheno:d6pol}) is, apart from a factor of $2\pi$ coming from the $\phi_S$
integration, the unpolarized cross section given in
Eq.~(\ref{pheno:d5unp}) and, in the consistent ${\cal O}(k_\perp/Q)$
approximation, in Eq.~(\ref{pheno:cahn:anal:app}). The terms $A_{\rm
Sivers}$ and $A_{\rm Collins}$ in Eq.~(\ref{pheno:d6pol}) correspond
respectively to the Sivers effect and to the Collins effect
(involving the $k_\perp$ dependent transversity distribution,
$\Delta_T q(x,k_\perp)$) and are related to the
$A_{UT}^{\sin(\phi_h\pm\phi_S)}$ asymmetries discussed in detail
below. The fourth term, $A^\prime_{\rm Collins}$, involving again
the Collins function coupled with the $h_{1T}^\perp$ distribution
(in the notation of the Amsterdam group) provides eventually a
$\sin(3\phi_h-\phi_S)$ azimuthal modulation. Only recent preliminary
results from COMPASS are available on this
observable~\cite{Kotzinian:2007uv}. The measured asymmetry is
compatible with zero within the statistical errors. We will not
consider this contribution in the following fenomenological
analysis.

Exploiting the helicity formalism, the two interesting additional terms, $A_{\rm Sivers}$ and
$A_{\rm Collins}$, read
 \bea
 A_{\rm Sivers} &\equiv & \sum_q e_q^2\int d^2 \bfk _\perp\, \Delta
 ^N f_{q/p^\uparrow} (x,\kt) \sin (\varphi -\phi_S) \; \frac{d
 \hat\sigma}{dQ^2}  \; D_{h/q}(z,p_\perp)\>,
 \label{pheno:sivA}\\
 A_{\rm Collins} &\equiv&
 \sum_q e_q^2   \int \! d^2 \bfk _\perp\,
 \Delta _T q (x,\kt) \,
 \frac{d (\Delta {\hat \sigma})}{dQ^2}
 \Delta^N D_{h/q^\uparrow}(z,\pt) \sin(\phi_S + \varphi +\phi_q^h)\>.
 \label{pheno:colA}
 \eea

Let us explain in more detail the above equations. $\phi_S$ and $\varphi$ identify the
directions of the proton spin, $\bm{S}$, and of the quark intrinsic transverse momentum,
$\bfk_\perp$, see Fig.~\ref{pheno:planessidis}; $\phi_q^h$ is the azimuthal angle of the
final hadron $h$, as defined in the helicity frame of the fragmenting quark. $d{\hat
\sigma}/dQ^2$ is the usual unpolarized elementary cross section for the $\ell q\to \ell q$
process, see also Eq.~(\ref{pheno:d5unp}), while ($y=Q^2/x_B s$)
 \be
 \frac{d(\Delta {\hat \sigma})}{dQ^2} =
 \frac{d{\hat \sigma}^{\ell q^\uparrow \to \ell q^\uparrow}}{dQ^2} - \frac{d{\hat
 \sigma}^{\ell q^\uparrow \to \ell q^\downarrow}}{dQ^2} = \frac{4\pi\alpha^2}{Q^4}\,(1-y)\,
 \cdot
 \label{pheno:dDeltasigma}
 \ee
The Sivers effect comes directly from the possible azimuthal dependence of the number density
for unpolarized quarks inside a transversely polarized proton (generically denoted by
$p^\uparrow$,  with $p^\downarrow$ denoting the opposite polarization state), that can be
written as (see also section~\ref{theo:kt}):
 \be
 \hat{f}_ {q/p^\uparrow} (x,\bm{k}_\perp) = f_ {q/p} (x,k_\perp) +
 \frac{1}{2} \, \Delta^N f_ {q/p^\uparrow}(x,k_\perp)  \;
 {\bm{S}} \cdot (\hat {\bm{P}}  \times
 \hat{\bm{k}}_\perp)\>,
 \label{pheno:Sivers-gen}
 \ee
where the triple product gives in our kinematical configuration the $\sin(\varphi-\phi_S)$
factor entering Eq.~(\ref{pheno:sivA}). Upon integration over $\bfk_\perp$, $A_{\rm Sivers}$
gets the well known $\sin(\phi_h-\phi_S)$ azimuthal dependence.

Because of the kinematics, the Collins case is a bit more involved. Here is the transverse
spin polarization of the fragmenting quark (as inherited by the initial polarized quark) that
originates the possible azimuthal asymmetry in the $\bm{k}_\perp$-dependent fragmentation
process. Namely, for a transversely polarized quark with spin polarization $\hat{\bm{s}}$ and
three-momentum $\bfp_q$,
 \be
 \hat{D}_{h/q,s}(z,\bfp_\perp) = D_{h/q}(z,\pt) + \frac{1}{2} \,
 \Delta^N D_{h/q^\uparrow}(z,\pt)\, \hat{\bm{s}} \cdot(\hat{\bfp}_q \times \hat{\bfp}_\perp)
 \>.
 \label{pheno:Collins-gen}
 \ee
The $\sin(\phi_S + \varphi +\phi_q^h)$ azimuthal dependence in
Eq.~(\ref{pheno:colA}) arises from the combination of the phase
factors appearing in the transversity distribution function, in the
non-planar $\ell\,q \to \ell\,q$ elementary scattering amplitudes,
and in the Collins fragmentation function. Neglecting ${\cal O}(\kt
^2/Q^2)$ terms, one finds
 \be
 \cos\phi_q^h = \frac{P_T}{\pt}\,\cos(\phi_h-\varphi)
 -z\,\frac{\kt}{\pt}\,, \hspace*{1cm} \sin\phi_q^h =
 \frac{P_T}{\pt}\,\sin(\phi_h-\varphi)\>.
 \label{pheno:phases}
 \ee

The integration upon $\bfk_\perp$ results, among other factors,
in the well known $\sin(\phi_h+\phi_S)$ dependence of the Collins effect in SIDIS.

The different azimuthal dependence of the Sivers and the Collins contributions in SIDIS
offers indeed a clear separation technique and the possibility to gain valuable information
on both of them. This is particularly true for the Sivers mechanism, since the other terms in
Eq.~(\ref{pheno:sivA}) are well known. On the other hand, the Collins contribution involves
also the still unknown transversity distribution function. We have seen in
section~\ref{pheno:epem} how a combined analysis of SIDIS results with data on azimuthal
asymmetries in $e^+e^-\to\pi\pi + X$ processes is of great help in this respect.

To separate the contributions in the polarized cross section coming from the two naively
T-odd effects, one defines the moments of the azimuthal asymmetry $A^{\sin (\phi_h \pm
\phi_S)}_{_{UT}}$, given in Eq.~(\ref{exp:sidis:aut}), where in this case $d\sigma(\phi_S)$
corresponds to $(d^6\sigma^{\ell p^{\uparrow} \to \ell^\prime h + X
 })/(dx_B \, dy \, dz_h \, d^2  \bm{P}_T \, d\phi_S)$, see Eq.~(\ref{pheno:d6pol}).

In order to perform analytically all $k_\perp$ integrations one can
choose suitable Gaussian parameterizations for the TMD distributions
appearing in the above equations, see Ref.~\cite{Anselmino:2007fs}:
\bea
\Delta_T q(x, \kt) &=& \frac{1}{2} \, {\cal N}^{\,T}_q(x)\, \left[f_{q/p}(x)+\Delta_L
q(x) \right] \; \frac{e^{-{\kt^2}/{\langle \kt^2\rangle}}}{\pi \langle \kt^2 \rangle}
\label{pheno:tr:funct} \\
\Delta^N f_{q/p^\uparrow}(z,\kt) &=& 2\,{\cal N}^{S}_q(x)\;
f_{q/p}(x)\,\sqrt{2e}\,\frac{\kt}{M'}\,\frac{e^{-\kt^2/{\langle\kt^2\rangle_S}}}{\pi
  \langle \kt^2\rangle}
\label{pheno:siv:funct}\\
\Delta^N D_{h/q^\uparrow}(z,p_\perp) &= &2\,{\cal N}^{C}_q(z)\;
D_{h/q}(z)\,\sqrt{2e}\,\frac{p_\perp}{M}\,\frac{e^{-p_\perp^2/{\langle p_\perp^2\rangle_C}}}{\pi
  \langle p_\perp^2\rangle}\,,
\label{pheno:coll:funct}
 \eea
where $\Delta_L q(x)$ is the usual collinear helicity distribution,
and
 \be
 \langle \kt^2 \rangle_S = \frac{M^{\prime 2} \, \langle \kt^2 \rangle} {M^{\prime 2} +
 \langle \kt ^2 \rangle} \,, \quad
 \langle \pt^2 \rangle_C =
 \frac{M^{ 2} \, \langle \pt^2 \rangle} {M^{2} + \langle \pt ^2 \rangle} \,,
\label{pheno:ktsiv:coll}
 \ee
with $|{\cal N}^{T,S,C}_q|\le 1$ to automatically fulfill proper
bounds. With these choices one gets
 \be
 A_{UT}^{\sin (\phi_h-\phi_S)}(x,y,z, P_T) \simeq \frac{\Delta\sigma_{\rm
 Sivers}}{\sigma_{0}} \; , \label{pheno:siversutapp} \hspace*{1cm}
 A_{UT}^{\sin (\phi_h+\phi_S)}(x, y,z, P_T) \simeq
 \frac{\Delta\sigma_{\rm Collins}}{\sigma_{0}} \; ,
 \label{pheno:collinsutapp}
 \ee
where, omitting a common overall factor, $2 \pi \alpha^2/x y^2 s$, [recall that $x\simeq x_B$
and $z\simeq z_h$ at ${\cal O}(k_\perp/Q)$],
 \bea
 \Delta\sigma_{\rm Sivers} \!\! &=&\!\! \frac{z\,P_T}{M'}\,
 \frac{\sqrt{2e}\, \langle \kt^2 \rangle^2_S}{\langle \kt ^2 \rangle}\,
 \frac{e^{- P_T^2/\langle P_T ^2\rangle _S }}{\langle P_T ^2 \rangle_S^2}
 \left[\, 1+(1-y)^2\, \right]\,
 \sum_q e_q^2 \, 2 {\cal N}^S_q(x) \, f_{q/p}(x) \, D_{h/q}(z)
 \label{pheno:sivers1app}\\
 \Delta\sigma_{\rm Collins} \!\!&=&\!\! \frac{P_T}{M}\,
  \frac{\sqrt{2e}\,\langle p_\perp ^2 \rangle^2_C}{\langle p_\perp ^2 \rangle}
 \, \frac{e^{-P_T^2/{\langle P_T ^2 \rangle} _C}}{{\langle P_T ^2
 \rangle}^2_C}\, (1-y)\,  \sum_q e_q^2 \,
 {\cal N}^{T}_q(x) \left[f_{q/p}(x)+\Delta_L q(x) \right]
 {\cal N}^{C}_q(z)\,D_{h/q}(z) \quad
 \label{pheno:collins1app} \\
 \sigma_{0} \!\!&=&\!\! 2 \pi \,  \frac{1}{\pi \langle P_T ^2 \rangle}\,
 e^{-P_T^2/\langle P_T ^2 \rangle}\,
\left[\, 1+(1-y)^2\, \right]\sum_q e_q^2 \, f_{q/p}(x) \, D_{h/q}(z)\,  ,
 \label{pheno:sivers0app}
 \eea
with
 \be
 \langle P_T^2 \rangle_S = \langle \kt^2 \rangle_S + z^2 \,\langle \kt
 ^2 \rangle \,, \quad
\label{pheno:mean:ptsivcoll2}
\langle P_T^2 \rangle_C = \langle \pt^2 \rangle_C + z^2 \, \langle \kt ^2 \rangle \,,
\ee
and $\langle P_T^2 \rangle$ given in Eq.~(\ref{pheno:eq:meanpt}).
Eq.~(\ref{pheno:sivers1app}) shows that $A_{UT}^{\sin (\phi_h-\phi_S)} = 0$ when $z = 0$ or
$P_T = 0$. Analogously, Eq.~(\ref{pheno:collins1app}) shows that $A_{UT}^{\sin
(\phi_h+\phi_S)} = 0$ for $P_T = 0$.

%%%%%%%%%%%%%%%%%%%%%%%%%%%%%%%%%%%%%%%%%%%%%%%%%%%%%%%%%%%%%%%%%%%%%%%%%%%
%%%%%%ppnp style
\begin{figure}[h!tb]
\begin{center}
\begin{minipage}[t]{17 cm}
\hspace*{-.8cm}
\epsfig{file=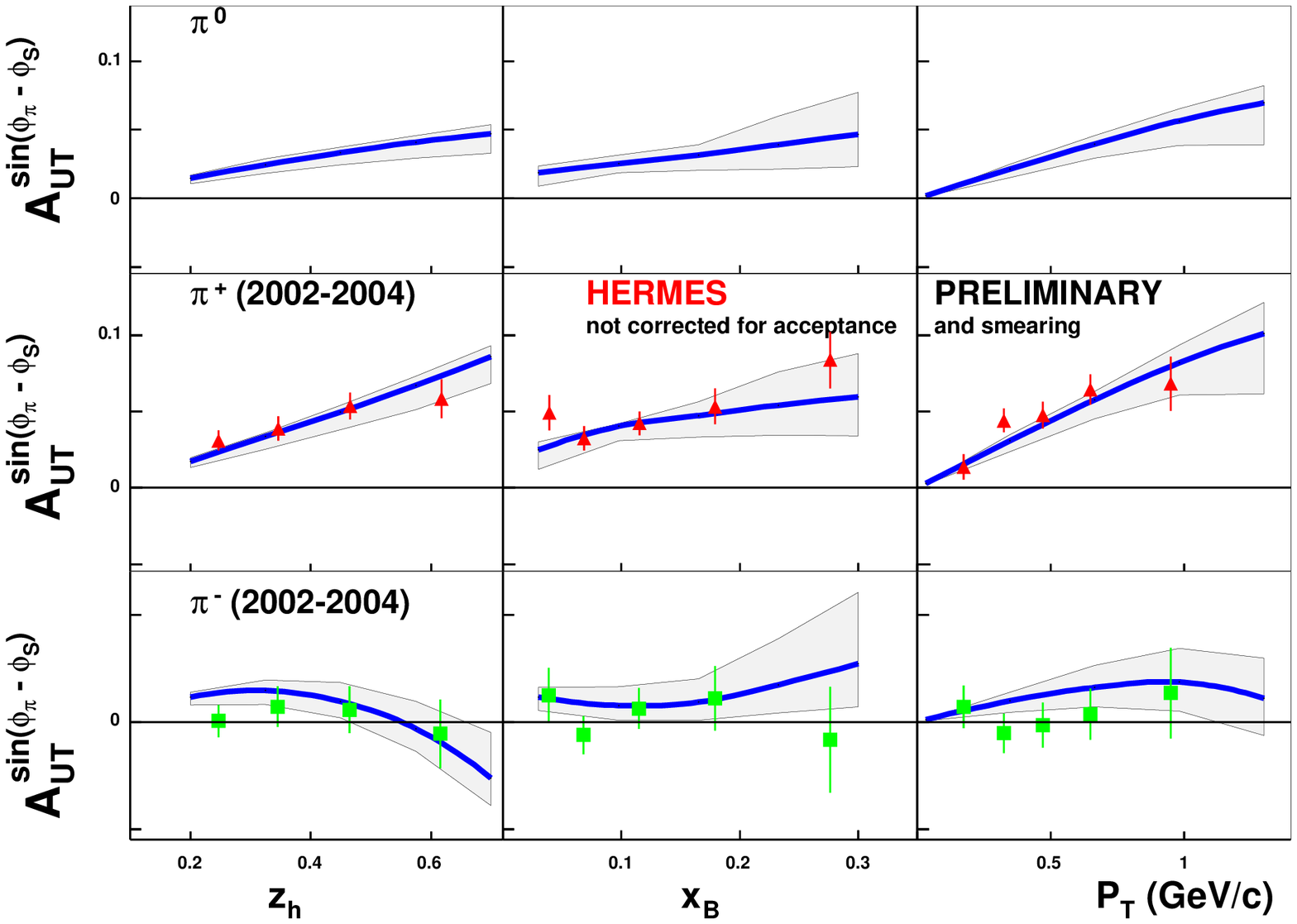,scale=0.45,bb= 20 30 570 450}
\hspace*{.3cm}
\epsfig{file=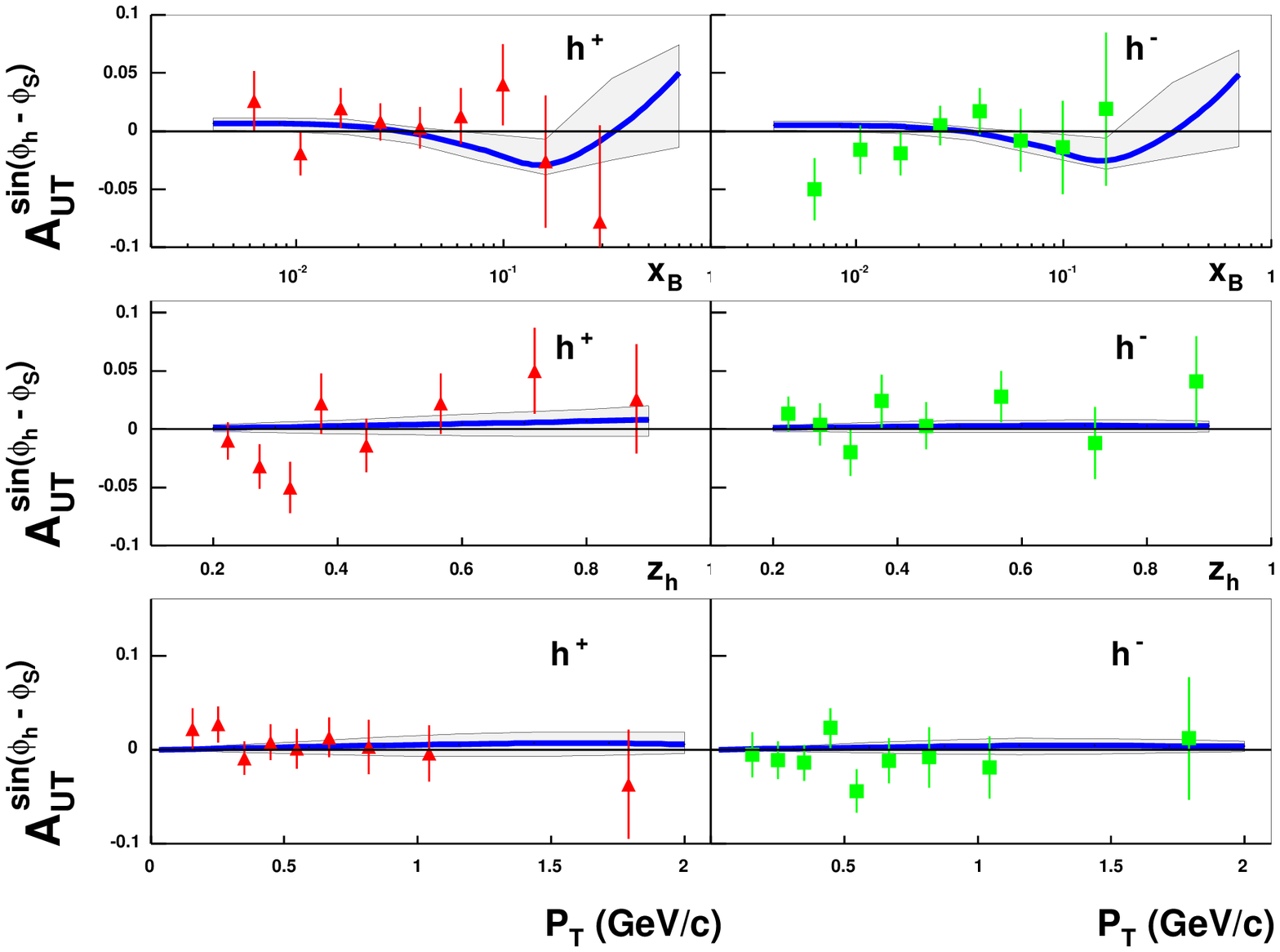,scale=0.45,bb= 20 30 570 450}
\end{minipage}
\begin{minipage}[t]{16.5 cm}
\caption{Left panel: HERMES data on $A_{UT}^{\sin(\phi_\pi-\phi_S)}$
\cite{Diefenthaler:2005gx} for $e^+ p^\uparrow \to e^+ \pi^\pm + X $. Right panel: COMPASS
data~\cite{Alexakhin:2005iw} on $A_{UT}^{\sin(\phi_h-\phi_S)}$ for $\mu D^\uparrow \to \mu\,
h^\pm + X $. Curves are the results of the fit in Ref.~\cite{Anselmino:2005ea}. Predictions
for $\pi^0$ asymmetries (left panel, upper plot) are also shown. The shaded areas span a
region corresponding to one-sigma deviaton at 90\% CL. Unpolarized PDF and FF sets: MRST01-K.
 \label{pheno:siv:hermes-compass}
}
\end{minipage}
\end{center}
\end{figure}
%%%%%%%%%%%%%%%%%%%%%%%%%%%%%%%%%%%%%%%%%%%%%%%%%%%%%%%%%%%%%%%%%%%%%%%%%%%

Concerning the unknown functions, ${\cal N}_q$, in
Refs.~\cite{Anselmino:2005ea,Anselmino:2007fs} forms like \be {\cal
N}_q(x) = N_q x^{a_q} (1-x)^{b_q}(a_q+b_q)^{a_q+b_q}
/a_q^{a_q}b_q^{b_q} \ee have been adopted. In particular, for the
Sivers function only $u$ and $d$ quarks were considered, with a
total of 7 free parameters (including $M'$); for the $u$ and $d$
quark transversity distributions, $a_u=a_d$ and $b_u=b_d$ have been
adopted, keeping the flavour distinction only in $N_q$, besides the
one coming from the flavour structure in Eq.~(\ref{pheno:tr:funct})
(4 free parameters). Finally, for the Collins function, in the pion
case both favoured and unfavoured fragmentation functions have been
considered, with the same $a,b$ but different $N_{\rm fav}$ and
$N_{\rm unf}$ for a total of 5 parameters (including $M$).

%%%%%%%%%%%%%%%%%%%%%%%%%%%%%%%%%%%%%%%%%%%%%%%%%%%%%%%%%%%%%%%%%%%%%%%%%%%
\begin{figure}[h!tb]
\begin{center}
\begin{minipage}[h!t]{17 cm}
\hspace*{0.7cm}
\epsfig{file=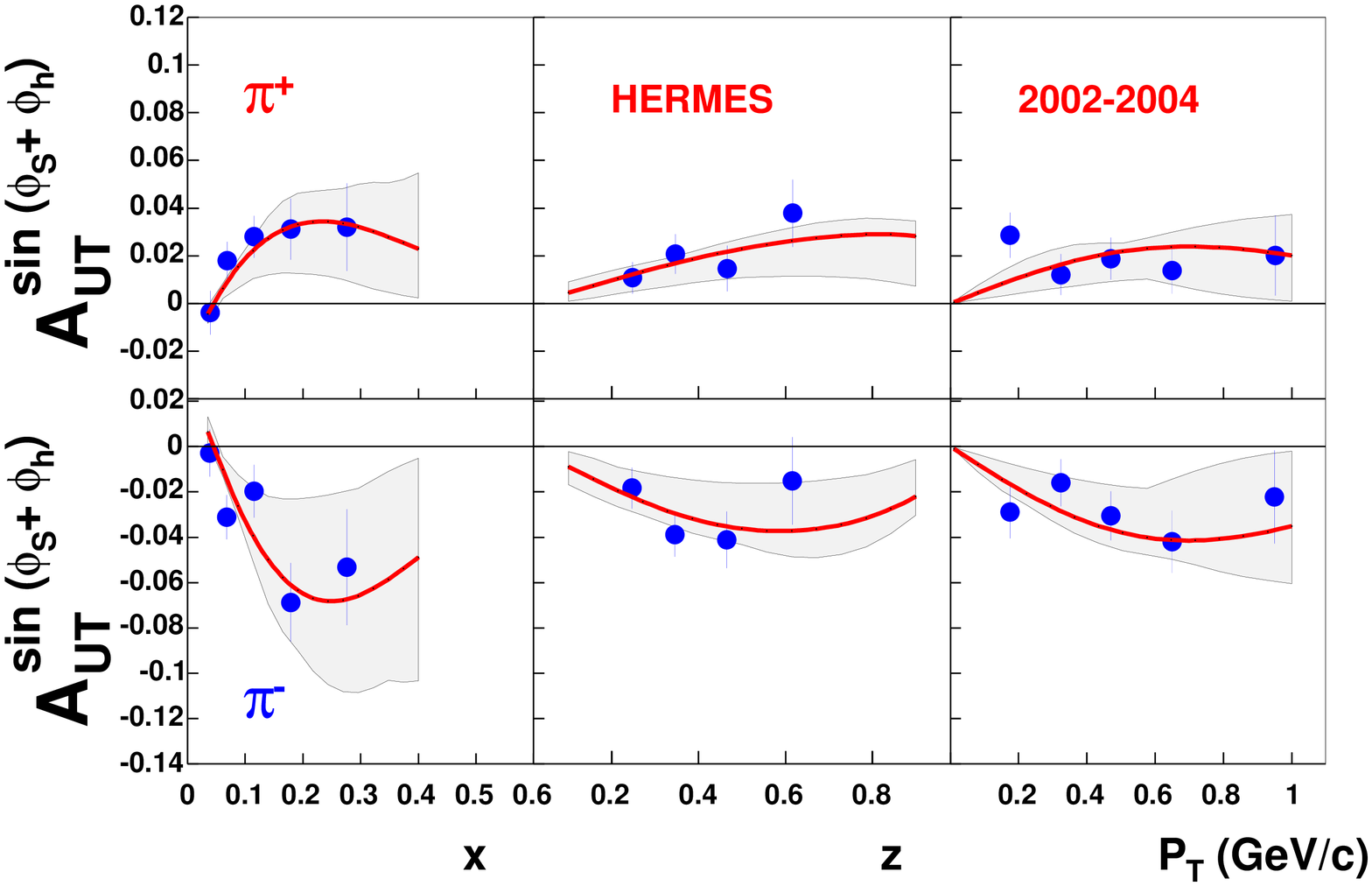,scale=0.35,bb= 60 140 540 660,angle=-90}
\hspace*{2.2cm}
\epsfig{file=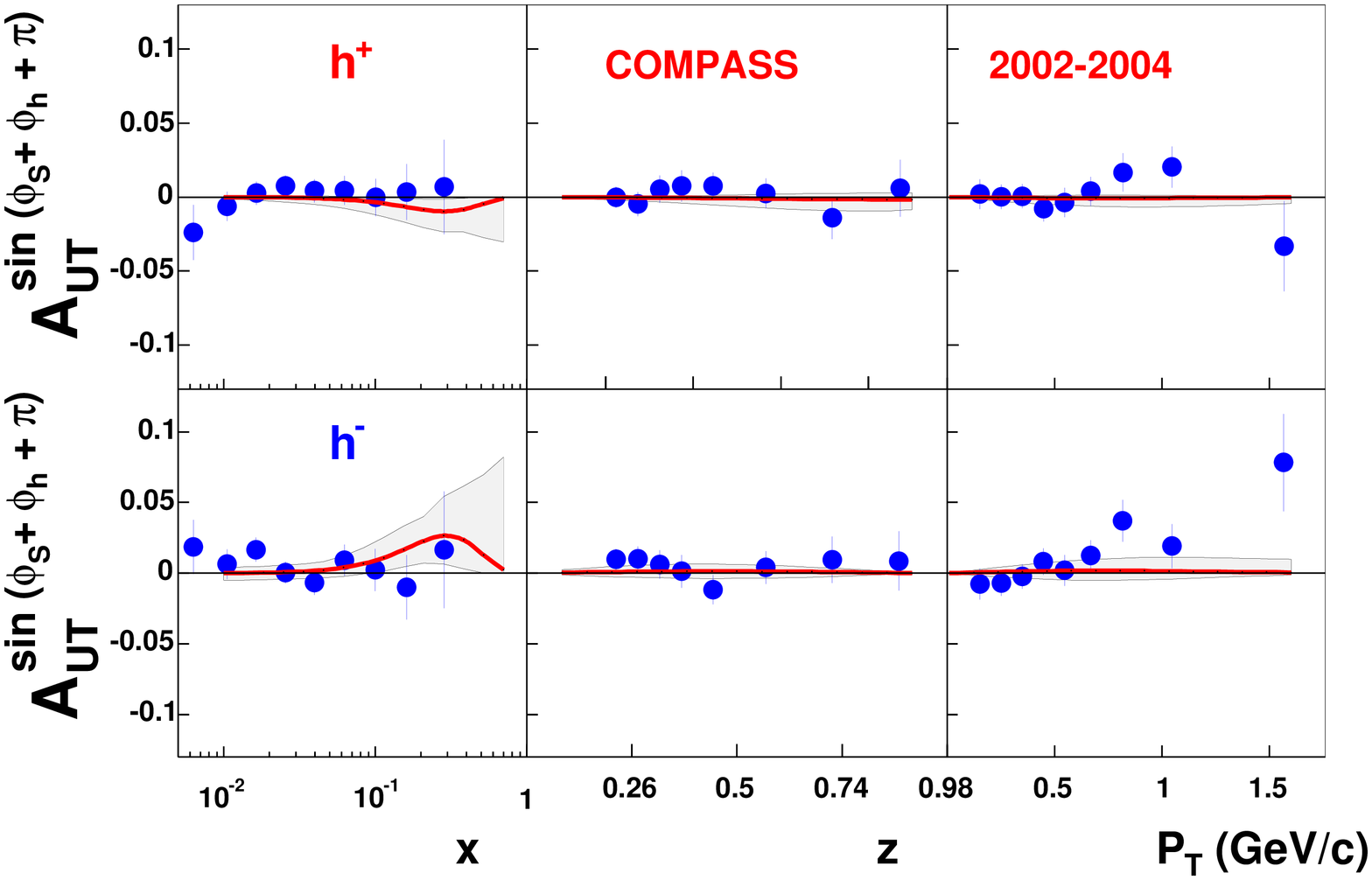,scale=0.35,bb= 60 140 540 660,angle=-90}
\end{minipage}
\begin{minipage}[t]{16.5 cm}
\caption{Left panel: HERMES data~\cite{Diefenthaler:2005gx} on the azimuthal asymmetry
$A_{UT}^{\sin(\phi_S+\phi_h)}$ for $\pi^\pm$ production. Right panel: COMPASS
data~\cite{Ageev:2006da} on $A_{UT}^{\sin(\phi_S+\phi_h)}$, for  charged hadron production in
$\mu D^\uparrow\to \mu\, h^\pm + X$. The extra $\pi$ phase in the figure label keeps into
account the different choice w.r.t.~Trento conventions~\cite{Bacchetta:2004jz}. The curves
are obtained from Eq.~(\ref{pheno:collins1app}) with the parameterizations of
Ref.~\cite{Anselmino:2007fs}. The shaded areas correspond to the theoretical uncertainty on
the parameters. (Un)polarized PDF and FF sets: (GRV98)GRSV2000-K.
\label{pheno:col:hermes-compass} }
\end{minipage}
\end{center}
\end{figure}
%%%%%%%%%%%%%%%%%%%%%%%%%%%%%%%%%%%%%%%%%%%%%%%%%%%%%%%%%%%%%%%%%%%%%%%%%%%

The curves obtained by fitting HERMES (COMPASS) data on $A_{UT}^{\sin (\phi_h-\phi_S)}$ for
pion (charged hadron) production are shown in Fig.~\ref{pheno:siv:hermes-compass}.
Preliminary data from HERMES~\cite{Diefenthaler:2006vn} and
COMPASS~\cite{Martin:2007au,Bradamante:2007ex} on kaon asymmetries are also available. The
corresponding theoretical estimates, adopting the Sivers functions so extracted, are in fair
agreement with HERMES data, even if the $K^+$ results are underestimated in the low $x$
region. Further work is needed, in particular concerning the study of the sea-quark Sivers
functions and flavour decomposition in the fragmentation sector.

For the Collins asymmetry a global fit on HERMES, COMPASS and Belle data in SIDIS and
$e^+e^-$ annihilation has been performed (9 free parameters) allowing the first extraction of
the transversity distribution. The results for the SIDIS case and a comparison with the
HERMES and COMPASS data are presented in Fig.~\ref{pheno:col:hermes-compass}.

%%%%%%%%%%%%%%%%%%%%%%%%%%%%%%%%%%%%%%%%%%%%%%%%%%%%%%%%%%%%%%%%%%%%%%%%%%%
%%%%%%ppnp style
\begin{figure}[h!tb]
\begin{center}
\begin{minipage}[h!t]{8 cm}
\includegraphics[width=.8\textwidth,bb= 10 140 540 660,angle=-90]
{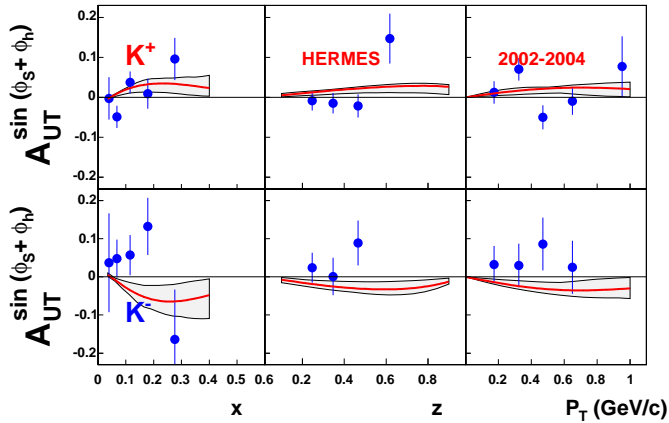}
\end{minipage}
\begin{minipage}[t]{16.5 cm}
\caption{ HERMES data~\cite{Diefenthaler:2006vn} on $A_{_{UT}}^{\sin(\phi_S+\phi_h)}$ for
$K^\pm$, compared with the results of Ref.~\cite{{Anselmino:2007fs}}. (Un)polarized PDF and
FF sets: (GRV98)GRSV2000-K. \label{pheno:col:kaons} }
\end{minipage}
\end{center}
\end{figure}
%%%%%%%%%%%%%%%%%%%%%%%%%%%%%%%%%%%%%%%%%%%%%%%%%%%%%%%%%%%%%%%%%%%%%%%%%%%

Finally, we compare the azimuthal asymmetry $A_{_{UT}}^{\sin(\phi_S+\phi_h)}$ for the
production of $K$ mesons with existing HERMES results~\cite{Airapetian:2004tw}. Preliminary
results are also available from COMPASS~\cite{Martin:2007au,Bradamante:2007ex}. These data
have not been included in the best fit of Ref.~\cite{Anselmino:2005ea}, as they might involve
the transversity distribution of strange quarks in the nucleon, which have been neglected for
SIDIS data on $\pi$ production. We show in Fig.~\ref{pheno:col:kaons} the results obtained
using the extracted $u$ and $d$ transversity distributions of Fig.~\ref{pheno:transv-col} (a,
b panels). For the Collins function the same ${\cal N}^C_q$ and $M$ as extracted for the pion
case were tentatively used, with the appropriate unpolarized FF's, $D_{K^\pm/q}$.

We notice that the above computations are in fair agreement with data concerning the $K^+$
production, which is presumably dominated by $u$ quarks; instead, there seem to be
discrepancies for the $K^-$ asymmetry, for which the role of $s$ quarks might be relevant.
New data on the azimuthal asymmetry for $K$ production, possible from COMPASS and JLab
experiments, might be very helpful in sorting out the eventual importance of the sea quark
transversity distributions in the nucleon.

\subsubsection{Phenomenological extractions of Sivers and Collins functions}
 \label{pheno:sidis:fit}

\begin{figure}[h!tb]
\begin{center}
\begin{minipage}[t]{17 cm}
\hspace*{-1.2cm} \epsfig{file=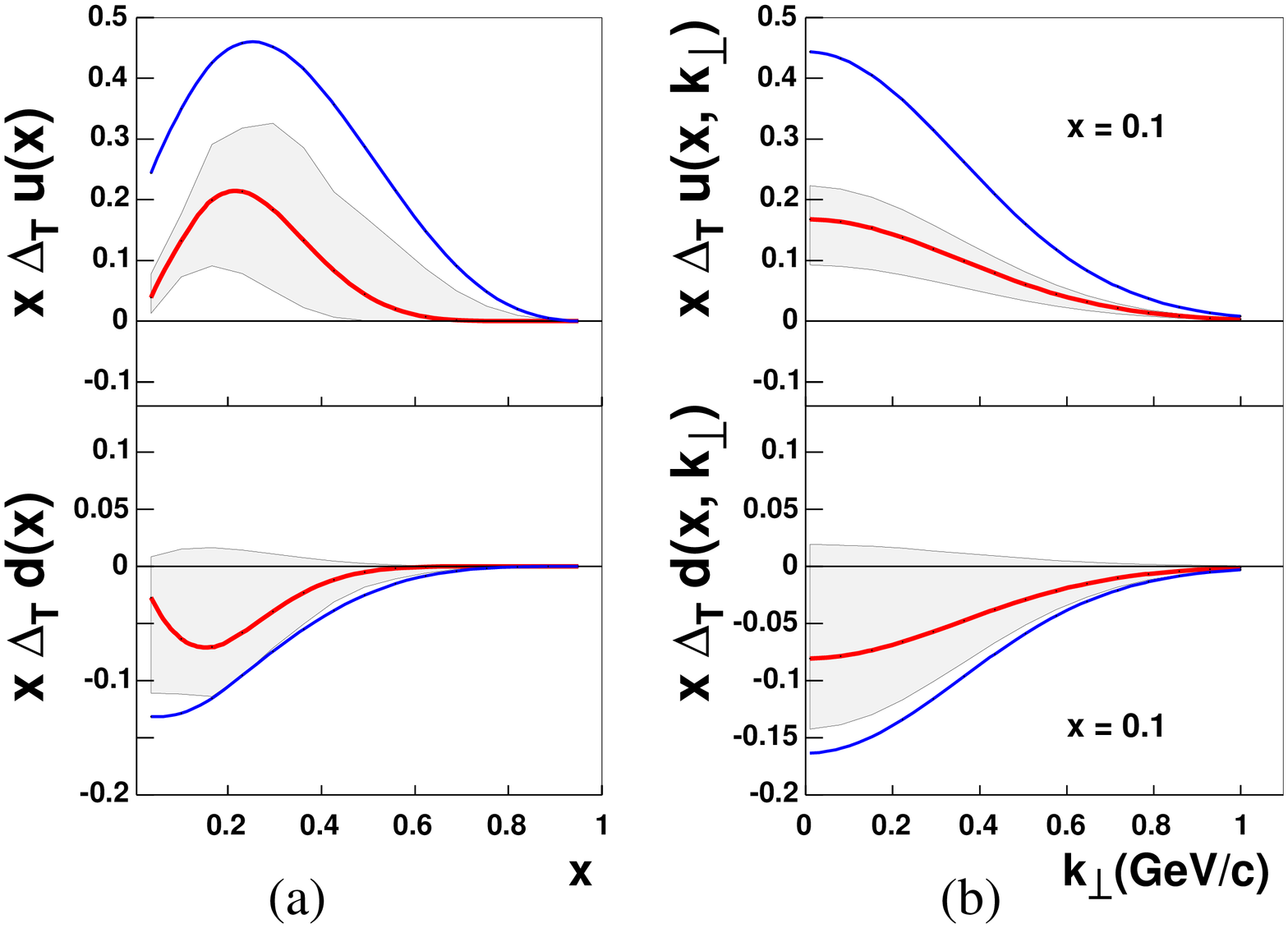,scale=0.35,bb= 100 10 590
660,angle=-90} \hspace*{2.cm} \epsfig{file=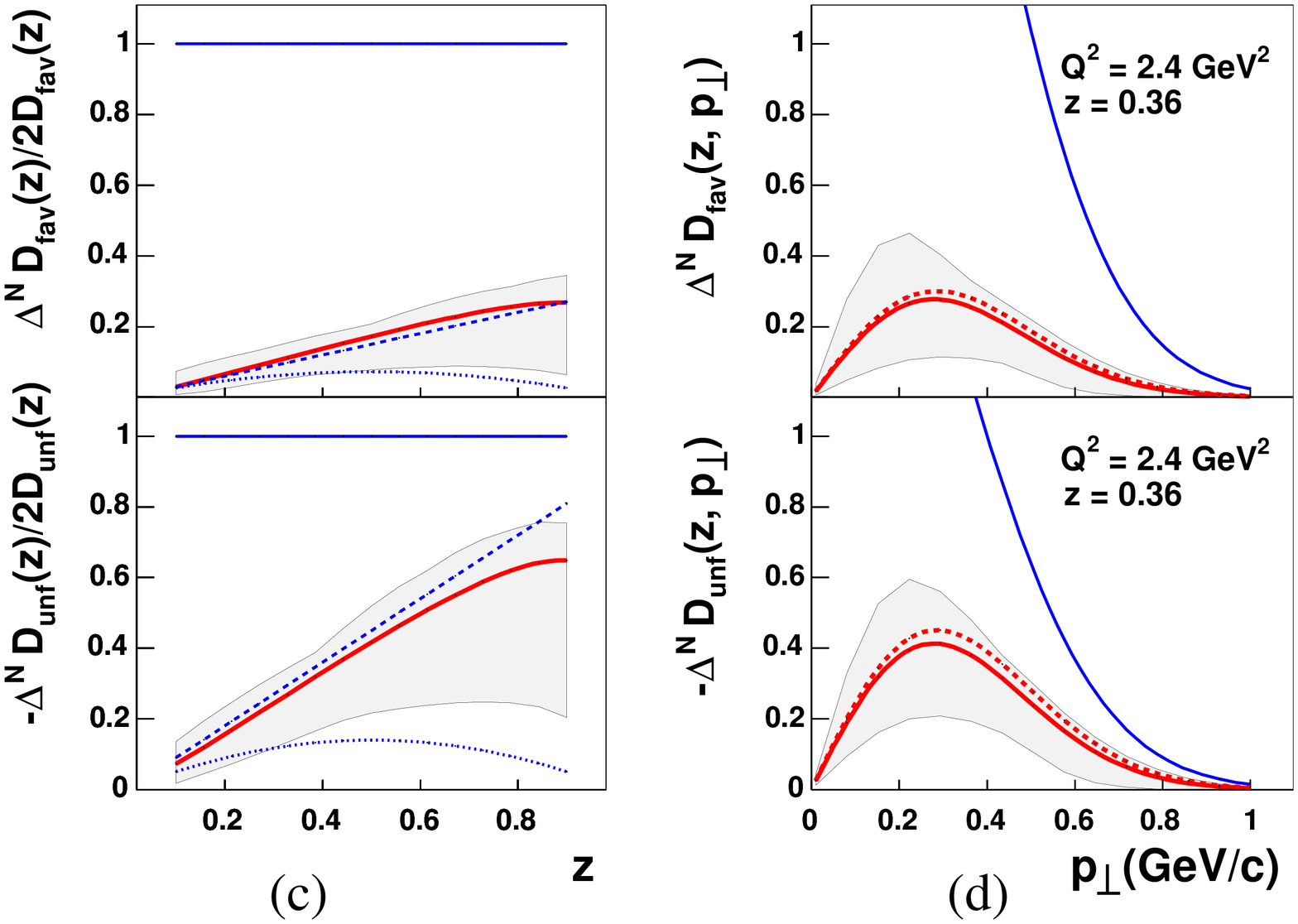,scale=0.35,bb= 100 140 590
660,angle=-90}
\end{minipage}
\begin{minipage}[t]{16.5 cm}
\caption{ The transversity distribution for $u$ and $d$ quarks and
the favoured and unfavoured Collins FF's from
Ref.~\cite{Anselmino:2007fs}. (a): $x\,\Delta _T u(x)$ (upper plot)
and $x\,\Delta _T d(x)$ (lower plot), as functions of $x$ at $Q^2 =
2.4$ (GeV$/c)^2$. The Soffer bound \cite{Soffer:1994ww} is also
shown (solid blue lines). (b): unintegrated transversity
distributions, $x\,\Delta _T u(x,\kt)$ (upper plot) and $x\,\Delta
_T d(x,\kt)$ (lower plot), as defined in Eq.~(\ref{pheno:tr:funct}),
as functions of $\kt$ at $x=0.1$. PDF sets: GRV98-GRSV2000. (c): the
$p_\perp$ integrated Collins functions, Eq.~(\ref{pheno:deltaDz}),
normalized to twice the corresponding unpolarized FF's, vs.~$z$;
results of Refs.~\cite{Efremov:2006qm} (dashed line) and
\cite{Vogelsang:2005cs} (dotted line) are also shown. (d): the
Collins function from the fits of Ref.~\cite{Anselmino:2007fs} vs.
$p_\perp$ at $z=0.36$ and $Q^2= 2.4$ (GeV$/c)^2$ (lower lines). The
positivity bound is also shown (upper blue lines). FF set: K. The
shaded areas correspond to the theoretical uncertainty on the
parameters. \label{pheno:transv-col} }
\end{minipage}
\end{center}
\end{figure}

The phenomenological analysis of HERMES and COMPASS data on moments of the azimuthal
asymmetries in the SIDIS process, with the aim of extracting information on the Sivers and
Collins functions, has been performed independently by three different theoretical
groups~\cite{Efremov:2004tp,Collins:2005ie,Anselmino:2005nn,Anselmino:2005ea,Vogelsang:2005cs,Efremov:2006qm,Anselmino:2007fs}.
Although the three groups have used different theoretical approaches, at the level of
accuracy considered (calculations are performed within LO pQCD, using leading twist TMD
parton distribution and fragmentation functions and keeping only the leading term in a
$(k_\perp/Q)$ power expansion) the results are quite similar and easily comparable. Several
other simplifying approximations were common between these groups, namely the neglect of the
so-called soft-factors \cite{Ji:2004wu,Collins:2004nx} and of Sudakov suppression factors
\cite{Boer:2001he} (see also section~\ref{theo}). However, slightly different approaches were
followed concerning the dependence of the distributions on the transverse parton momenta. As
an example, different functional forms were used. Also, possible constraints coming from
general bounds (e.g.~positivity bounds, the Soffer bound) were imposed and used in different
ways. Regarding the Collins function, in Ref.~\cite{Efremov:2006qm} a quark soliton model was
used for the transversity distribution, while in Ref.~\cite{Vogelsang:2005cs} the Soffer
bound was tentatively adopted. As we have already discussed, in Ref.~\cite{Anselmino:2007fs}
Belle $e^+e^-$ data on azimuthal asymmetries in dihadron production were used to
independently extract the Collins function. {}From there and HERMES and COMPASS data, the
first extraction of the transversity distribution for $u$ and $d$ quarks was performed, see
Fig.~\ref{pheno:transv-col} ((a), (b) panels). Also, different sets for the usual
$\bm{k}_\perp$-integrated PDF's and FF's, in different combinations, were adopted. A useful
summary and more detailed comparisons may be found, in the case of the Sivers function, in
Ref.~\cite{Anselmino:2005an}.

%%%%%%%%%%%%%%%%%%%%%%%%%%%%%%%%%%%%%%%%%%%%%%%%%%%%%%%%%%%%%%%%%%%%%%%%
\begin{figure}[h!tb]
\begin{center}
\begin{minipage}[t]{14 cm}
\epsfig{file=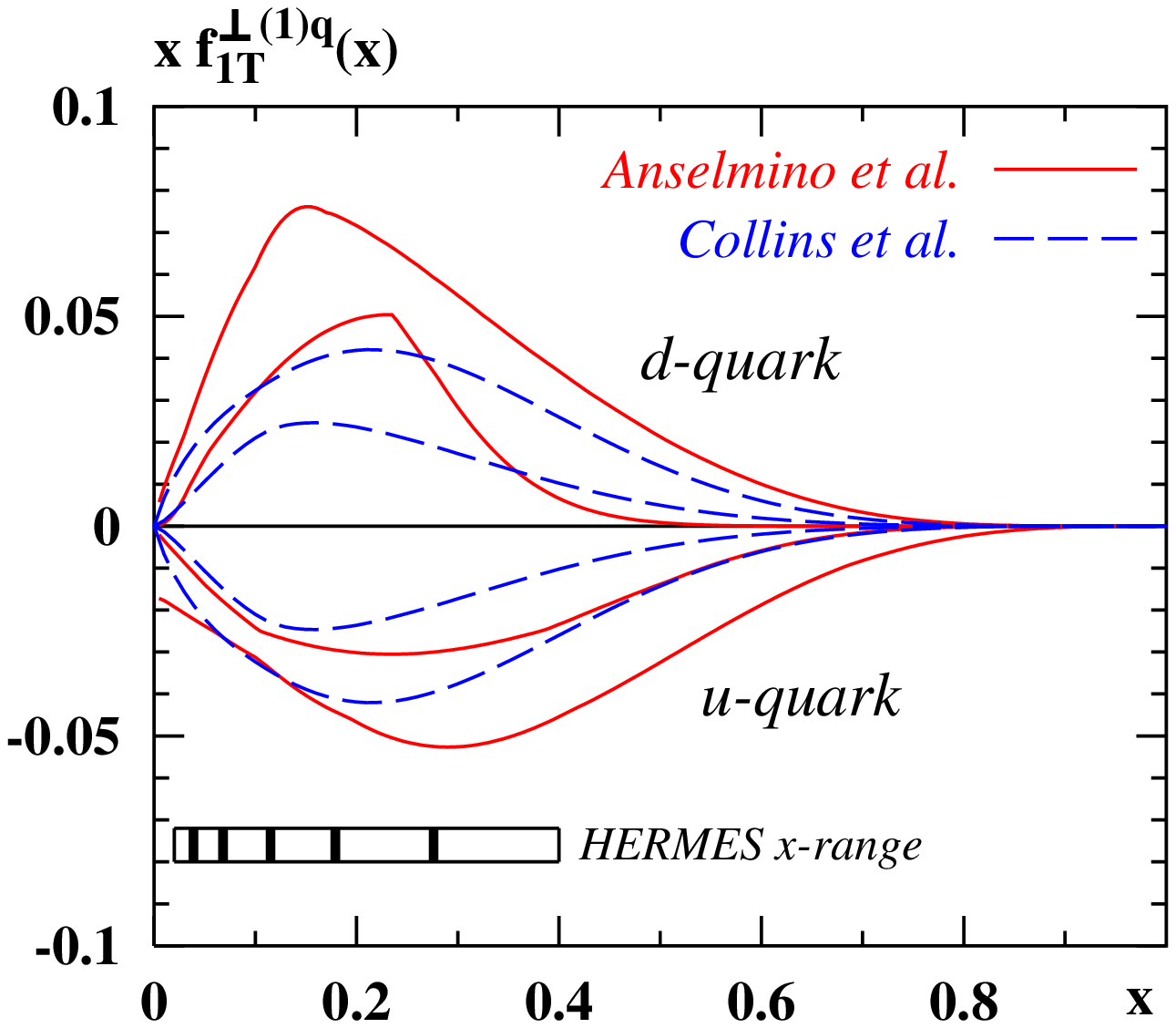,scale=0.4}
\epsfig{file=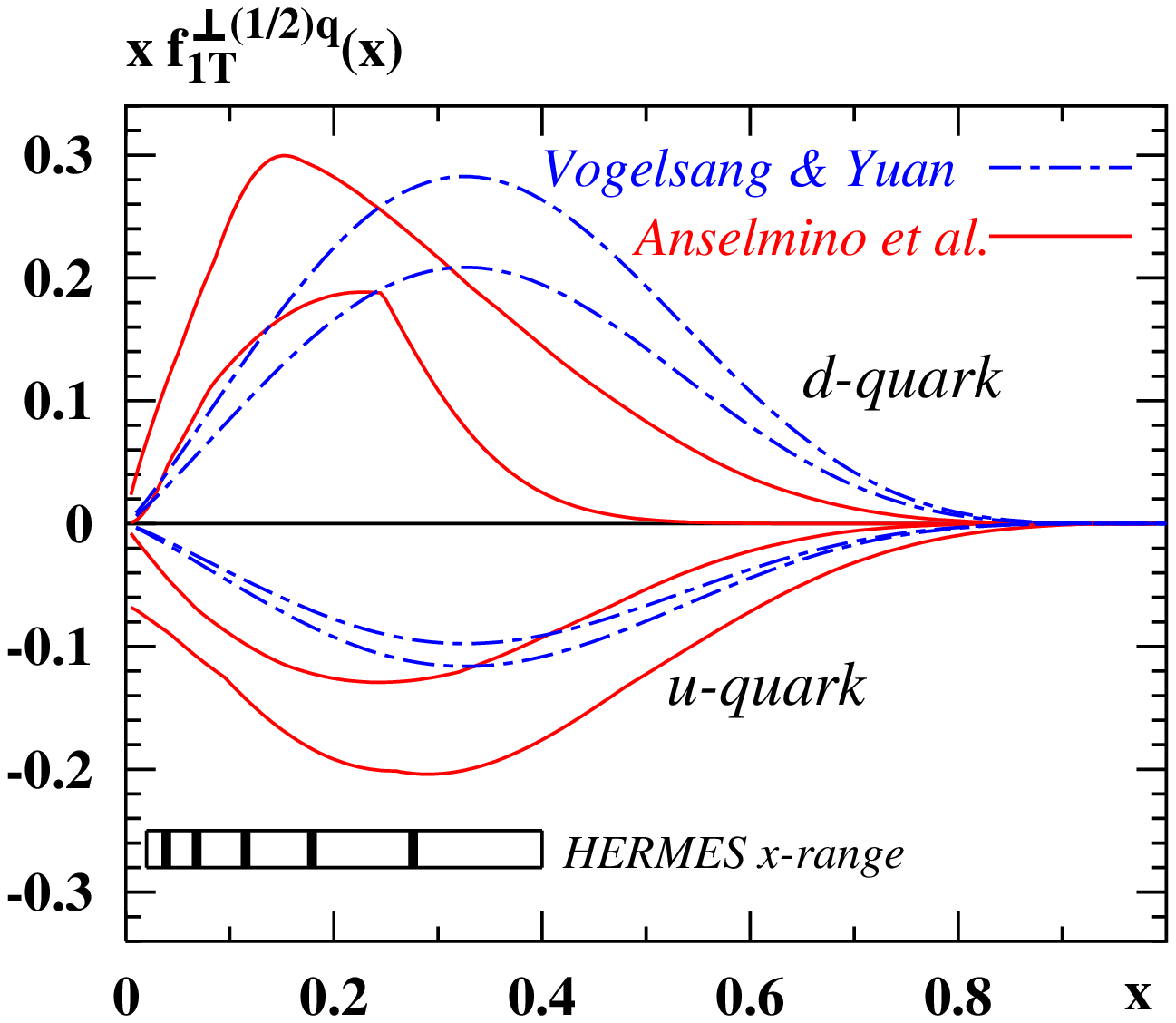,scale=0.4}
\end{minipage}
\begin{minipage}[t]{16.5 cm}
\caption{
    The first and 1/2-transverse moments of the quark Sivers functions,
    see Eq.~(\ref{theo:siv:transv:mom}), as extracted in
    Refs.~\cite{Collins:2005ie,Anselmino:2005ea,Vogelsang:2005cs}.
    The curves indicate the 1-$\sigma$ regions of the various
    parameterizations.
\label{pheno:siv:cfr} }
\end{minipage}
\end{center}
\end{figure}

%%%%%%%%%%%%%%%%%%%%%%%%%%%%%%%%%%%%%%%%%%%%%%%%%%%%%%%%%%%%%%%%%%%%%%%%%%%

In Fig.~\ref{pheno:siv:cfr} the first two moments (in $\bm{k}_\perp$) of the Sivers function
(multiplied by $x$), as extracted by the three theoretical groups, are shown for comparison
(see also Eqs.~(\ref{theo:notation}),~(\ref{theo:siv:transv:mom})). Notice that for each
parameterization the two curves delimiting the one-sigma deviation region are shown. The
range of $x$ values spanned by HERMES experiment is also depicted. Clearly, the
parameterizations extracted suffer of larger uncertainties in the large $x$ sector, which is
not well covered by data. Roughly speaking, all parameterizations are in favour of a
negative(positive) Sivers function ($f_{1T}^\perp$) for $u$($d$) quarks, with shapes and
magnitudes comparable. This behaviour reasonably complies with the
prediction~\cite{Anselmino:2002yx,Drago:2005gz} $f_{1T}^{\perp d}=-f_{1T}^{\perp u}$, also
derived in the large $N_C$ limit of QCD in Ref.~\cite{Pobylitsa:2003ty}. See also the
discussion on the theoretical models in section~\ref{theo:models}.

Fig.~\ref{pheno:transv-col} ((c), (d) panels) shows the first moment of the favoured and
unfavoured Collins fragmentation functions for pions, normalized to twice the corresponding
unpolarized FF, as a function of $z$ ((c) panel) and $p_\perp$ ((d) panel) at fixed $Q^2$.
The three theoretical extractions are compared: Ref.~\cite{Anselmino:2005ea} (solid lines and
the shaded, one-sigma deviation, area); Ref.~\cite{Efremov:2006qm} (dashed line);
Ref.~\cite{Vogelsang:2005cs} (dotted line). For comparison, we also show the positivity bound
(upper solid line). The $p_\perp$ dependence in panel (d) is not a prediction but has been
imposed to the parameterization. Notice also that there is no curve from
Ref.~\cite{Vogelsang:2005cs} here: this is because of some additional simplifying assumptions
performed there.

%% file: figures/epluseminus_paper.pstex_t
\begin{picture}(0,0)%
\includegraphics{figures/epluseminus_paper.pstex}%
\end{picture}%
\setlength{\unitlength}{4144sp}%
\begingroup\makeatletter\ifx\SetFigFont\undefined%
\gdef\SetFigFont#1#2#3#4#5{%
  \reset@font\fontsize{#1}{#2pt}%
  \fontfamily{#3}\fontseries{#4}\fontshape{#5}%
  \selectfont}%
\fi\endgroup%
\begin{picture}(10945,4781)(709,-5833)
\put(11071,-4426){\makebox(0,0)[lb]{\smash{{\SetFigFont{20}{24.0}{\familydefault}{\mddefault}{\updefault}{\color[rgb]{0,0,0}$x$}%
}}}}
\put(11116,-5146){\makebox(0,0)[lb]{\smash{{\SetFigFont{20}{24.0}{\familydefault}{\mddefault}{\updefault}{\color[rgb]{0,0,0}$z$}%
}}}}
\put(4906,-4246){\makebox(0,0)[lb]{\smash{{\SetFigFont{20}{24.0}{\familydefault}{\mddefault}{\updefault}{\color[rgb]{0,0,0}$e^+$}%
}}}}
\put(5041,-1996){\makebox(0,0)[lb]{\smash{{\SetFigFont{20}{24.0}{\familydefault}{\mddefault}{\updefault}{\color[rgb]{0,0,0}$e^-$}%
}}}}
\put(7156,-3121){\makebox(0,0)[lb]{\smash{{\SetFigFont{20}{24.0}{\familydefault}{\mddefault}{\updefault}{\color[rgb]{0,0,0}$\vec p_{\perp 1}$}%
}}}}
\put(6031,-2356){\makebox(0,0)[lb]{\smash{{\SetFigFont{20}{24.0}{\familydefault}{\mddefault}{\updefault}{\color[rgb]{0,0,0}$P_1$}%
}}}}
\put(4321,-1816){\makebox(0,0)[lb]{\smash{{\SetFigFont{20}{24.0}{\familydefault}{\mddefault}{\updefault}{\color[rgb]{0,0,0}$P_2$}%
}}}}
\put(3016,-1501){\makebox(0,0)[lb]{\smash{{\SetFigFont{20}{24.0}{\familydefault}{\mddefault}{\updefault}{\color[rgb]{0,0,0}$\vec p_{\perp 2}$}%
}}}}
\put(3376,-2446){\makebox(0,0)[lb]{\smash{{\SetFigFont{20}{24.0}{\familydefault}{\mddefault}{\updefault}{\color[rgb]{0,0,0}$\bar q$}%
}}}}
\put(7021,-3886){\makebox(0,0)[lb]{\smash{{\SetFigFont{20}{24.0}{\familydefault}{\mddefault}{\updefault}{\color[rgb]{0,0,0}$q$}%
}}}}
\put(5671,-3706){\makebox(0,0)[lb]{\smash{{\SetFigFont{20}{24.0}{\familydefault}{\mddefault}{\updefault}{\color[rgb]{0,0,0}$\theta$}%
}}}}
\put(10441,-4021){\makebox(0,0)[lb]{\smash{{\SetFigFont{20}{24.0}{\familydefault}{\mddefault}{\updefault}{\color[rgb]{0,0,0}$y$}%
}}}}
\end{picture}%

%% file: figures/epluseminus_paper1.pstex_t
\begin{picture}(0,0)%
\includegraphics{figures/epluseminus_paper1.pstex}%
\end{picture}%
\setlength{\unitlength}{4144sp}%
\begingroup\makeatletter\ifx\SetFigFont\undefined%
\gdef\SetFigFont#1#2#3#4#5{%
  \reset@font\fontsize{#1}{#2pt}%
  \fontfamily{#3}\fontseries{#4}\fontshape{#5}%
  \selectfont}%
\fi\endgroup%
\begin{picture}(10945,6009)(709,-5833)
\put(11071,-4426){\makebox(0,0)[lb]{\smash{{\SetFigFont{20}{24.0}{\familydefault}{\mddefault}{\updefault}{\color[rgb]{0,0,0}$x$}%
}}}}
\put(11116,-5146){\makebox(0,0)[lb]{\smash{{\SetFigFont{20}{24.0}{\familydefault}{\mddefault}{\updefault}{\color[rgb]{0,0,0}$z$}%
}}}}
\put(6526,-2176){\makebox(0,0)[lb]{\smash{{\SetFigFont{20}{24.0}{\familydefault}{\mddefault}{\updefault}{\color[rgb]{0,0,0}$P_1$}%
}}}}
\put(4456,-1816){\makebox(0,0)[lb]{\smash{{\SetFigFont{20}{24.0}{\familydefault}{\mddefault}{\updefault}{\color[rgb]{0,0,0}$e^-$}%
}}}}
\put(5626,-1501){\makebox(0,0)[lb]{\smash{{\SetFigFont{20}{24.0}{\familydefault}{\mddefault}{\updefault}{\color[rgb]{0,0,0}$P_{1T}$}%
}}}}
\put(3421,-2581){\makebox(0,0)[lb]{\smash{{\SetFigFont{20}{24.0}{\familydefault}{\mddefault}{\updefault}{\color[rgb]{0,0,0}$P_2$}%
}}}}
\put(4951,-4291){\makebox(0,0)[lb]{\smash{{\SetFigFont{20}{24.0}{\familydefault}{\mddefault}{\updefault}{\color[rgb]{0,0,0}$e^+$}%
}}}}
\put(10486,-4021){\makebox(0,0)[lb]{\smash{{\SetFigFont{20}{24.0}{\familydefault}{\mddefault}{\updefault}{\color[rgb]{0,0,0}$y$}%
}}}}
\put(9361,-2716){\makebox(0,0)[lb]{\smash{{\SetFigFont{20}{24.0}{\familydefault}{\mddefault}{\updefault}{\color[rgb]{0,0,0}$\phi_0$}%
}}}}
\put(4141,-2086){\makebox(0,0)[lb]{\smash{{\SetFigFont{20}{24.0}{\familydefault}{\mddefault}{\updefault}{\color[rgb]{0,0,0}$\theta_2$}%
}}}}
\end{picture}%

%% file: figures/planessidis.pstex_t
\begin{picture}(0,0)%
\includegraphics{figures/planessidis.pstex}%
\end{picture}%
\setlength{\unitlength}{3947sp}%
\begingroup\makeatletter\ifx\SetFigFont\undefined%
\gdef\SetFigFont#1#2#3#4#5{%
  \reset@font\fontsize{#1}{#2pt}%
  \fontfamily{#3}\fontseries{#4}\fontshape{#5}%
  \selectfont}%
\fi\endgroup%
\begin{picture}(10374,9993)(3139,-10492)
\put(12901,-9586){\makebox(0,0)[lb]{\smash{\SetFigFont{17}{20.4}{\familydefault}{\mddefault}{\updefault}{ $\mbox{\boldmath $x$}$}%
}}}
\put(12076,-9061){\makebox(0,0)[lb]{\smash{\SetFigFont{17}{20.4}{\familydefault}{\mddefault}{\updefault}{ $\mbox{\boldmath $y$}$}%
}}}
\put(12526,-10411){\makebox(0,0)[lb]{\smash{\SetFigFont{17}{20.4}{\familydefault}{\mddefault}{\updefault}{ $\mbox{\boldmath $z$}$}%
}}}
\put(8176,-4186){\makebox(0,0)[lb]{\smash{\SetFigFont{20}{24.0}{\familydefault}{\mddefault}{\updefault}{ $\varphi$}%
}}}
\put(5401,-1936){\makebox(0,0)[rb]{\smash{\SetFigFont{20}{24.0}{\familydefault}{\mddefault}{\updefault}{ $l$}%
}}}
\put(6601,-1861){\makebox(0,0)[rb]{\smash{\SetFigFont{20}{24.0}{\familydefault}{\mddefault}{\updefault}{ $l'$}%
}}}
\put(10351,-4186){\makebox(0,0)[rb]{\smash{\SetFigFont{20}{24.0}{\familydefault}{\mddefault}{\updefault}{ $\phi_h$}%
}}}
\put(5776,-2986){\makebox(0,0)[rb]{\smash{\SetFigFont{20}{24.0}{\familydefault}{\mddefault}{\updefault}{ $\bm{k}_\perp$}%
}}}
\put(8326,-2461){\makebox(0,0)[rb]{\smash{\SetFigFont{20}{24.0}{\familydefault}{\mddefault}{\updefault}{ $\bm{P}_h$}%
}}}
\put(5926,-4261){\makebox(0,0)[rb]{\smash{\SetFigFont{20}{24.0}{\familydefault}{\mddefault}{\updefault}{ $\bm{k}'$}%
}}}
\put(9601,-5986){\makebox(0,0)[rb]{\smash{\SetFigFont{20}{24.0}{\familydefault}{\mddefault}{\updefault}{ $\bm{P}$}%
}}}
\put(6826,-4261){\makebox(0,0)[lb]{\smash{\SetFigFont{20}{24.0}{\familydefault}{\mddefault}{\updefault}{ $\bm{p}_\perp$}%
}}}
\put(10951,-6286){\makebox(0,0)[rb]{\smash{\SetFigFont{20}{24.0}{\familydefault}{\mddefault}{\updefault}{ $\phi_S$}%
}}}
\put(10651,-5311){\makebox(0,0)[rb]{\smash{\SetFigFont{20}{24.0}{\familydefault}{\mddefault}{\updefault}{ \mbox{\boldmath $S$}}%
}}}
\put(7801,-2086){\makebox(0,0)[rb]{\smash{\SetFigFont{20}{24.0}{\familydefault}{\mddefault}{\updefault}{ $\bm{P}_T$}%
}}}
\put(3826,-6586){\makebox(0,0)[lb]{\smash{\SetFigFont{17}{20.4}{\familydefault}{\mddefault}{\updefault}{ $\mbox{\boldmath $\tilde y$}$}%
}}}
\put(4501,-7561){\makebox(0,0)[lb]{\smash{\SetFigFont{17}{20.4}{\familydefault}{\mddefault}{\updefault}{ $\mbox{\boldmath $\tilde x$}$}%
}}}
\put(3226,-8236){\makebox(0,0)[lb]{\smash{\SetFigFont{17}{20.4}{\familydefault}{\mddefault}{\updefault}{ $\mbox{\boldmath $\tilde z$}$}%
}}}
\end{picture}

%% file: conclusions-rev.tex
\section{Conclusions and outlook}
\label{conclusion}
In this paper we have presented a thorough review of our present theoretical understanding of
azimuthal and transverse single spin asymmetries in high-energy hadronic processes in the
framework of perturbative QCD. The basic features of the main theoretical approaches to SSA's
and an overview of the available experimental information have been given. An in-depth
treatment of the generalized parton model, including spin and transverse momentum effects
within the helicity formalism, and a detailed discussion of the phenomenology concerning
azimuthal and SSA's complete this analysis. Hopefully, this work will provide the reader with
a useful overview of the present status of this field of research and with a fresh feeling of
its ever-growing development. In fact, in the past few years several important results have
been reached, both from the experimental and theoretical points of view. These results have
confirmed earlier indications that spin and intrinsic transverse momentum effects can be
relevant even at high energies and large $p_T$ and led to a profusion of interesting physical
consequences. In the last years,  a wide dedicated experimental activity has definitely
confirmed previous pioneering results, refining their quality and statistical significance
and enlarging the kinematical regime explored and the number of experimental observables
investigated. From the theoretical side, significant progress has been reached in the
understanding of new spin and TMD functions, like the Sivers distribution and the Collins
function. The crucial role of initial and final state interactions, of a proper account of
gauge links and QCD colour gauge invariance and of their consequences has been clarified.
Related to this, the universality of the new TMD functions both in the distribution and in
the fragmentation sector has been investigated. QCD factorization theorems and collinear
twist-three approaches have been extended to cover several processes and new kinematical
regimes. The theoretical formulation of the generalized parton model has been refined and its
phenomenological applications extended to include most of the new experimental data. The
interplay between twist-three approaches and QCD-improved parton models with inclusion of
intrinsic parton motion has been extensively studied. The large amount of activity devoted to
SSA's and its significant but partial successes have also naturally raised new theoretical
problems and experimental challenges. They will hopefully set up the stage for new
developments in the forthcoming years.

{}From the theoretical side, some of the most relevant open points are: 1) The formal
properties of spin and $\bm{k}_\perp$ dependent functions, their universality and evolution
with scale; 2) The systematic comparison among different theoretical approaches and a deeper
understanding of their interplay and complementarity; 3) The role of Sudakov and soft
factors, and of parton off-shellness and doubly unintegrated parton distributions; 4) A
better understanding of the physical mechanisms involved, according to their twist
classification and their behaviour with the relevant scales;
5) A comprehensive phenomenological analysis
of all experimental data available as a test of universality and of the effectiveness of the
proposed parameterizations for all the new unknown soft functions involved; 6) An improvement
of nonperturbative models and approaches, including lattice calculations, for the same
functions; 7) An improved understanding of the connections among TMD functions, generalized
parton distributions and the orbital angular momentum of partons and its role for hadronic
structure.

{}From the experimental side, interesting developments are awaited in the near future: 1) The
BRAHMS, PHENIX and STAR collaborations at RHIC-BNL will provide us with new results at 62,
200, and 500 GeV c.m.~energies, with larger statistics, an improved kinematical coverage and
including new interesting physical observables; 2) Hopefully, the COMPASS Collaboration will
soon measure azimuthal single spin asymmetries for several charged hadrons with a
transversely polarized proton target; on a more extended time scale they will also  study the
polarized Drell-Yan process with a pion beam colliding on transversely polarized proton and
deuteron targets; 3) The proposed JLAB 12 GeV upgrade, with polarized proton, neutron and
deuteron targets and a large-acceptance detector will provide a detailed kinematical coverage
in both longitudinal and tranverse momentum, with a large $x_B$ experimental reach; 4) The
foreseen fixed-target experiments at $\sqrt{s}\sim 10$ GeV at the JPARC facility will help in
testing spin asymmetries and TMD distributions in the valence region ($x_B>0.1$); 5)~The
proposed 70 GeV upgrade of the transversely polarized proton beam at the IHEP-Protvino
accelerator will provide further information on the moderate energy regime; 6) On a longer
time scale, the PAX experiment at GSI will improve our knowledge of the transversity
distribution and the other spin and TMD functions, provided the capability of efficiently
polarizing antiprotons will be confirmed.

Hopefully, the new forthcoming experimental results and the prosecution of active theoretical
investigation will lead in the near future to significant improvements in our understanding
of SSA's, the spin structure of hadrons and related issues.

\newpage